\documentclass{aastex701}
\newcommand{\nscdr}{NSC~DR2}
\providecommand{\okina}{\textquoteleft}
\newcommand{\dtf}{d25}%
\newcommand{\Pch}{P_{\rm chance}}
\newcommand{\chinu}{\chi^2/\nu}
\newcommand{\vOneObj}{81}%
\newcommand{\vOneDmin}{30}%
\newcommand{\vOneDmax}{150}%
\newcommand{\vOneNights}{four}%
\newcommand{\vOneNearLo}{8}%
\newcommand{\vOneNewNights}{two}%
\newcommand{\dnDmin}{25}%
\newcommand{\dnDmax}{80}%
\newcommand{\dnK}{eight}%
\newcommand{\dnDtmax}{40}%
\newcommand{\dnNmin}{four}%
\newcommand{\dnNminu}{six}%
\newcommand{\dnStat}{0\farcs5}%
\newcommand{\dnExpCap}{100{,}000}%
\newcommand{\dnClumpFloor}{6}%
\newcommand{\dnClumpFac}{5}%
\newcommand{\dnCellA}{30}%
\newcommand{\dnCellB}{120}%
\newcommand{\dnSigFloor}{0\farcs07}%
\newcommand{\dnSigCap}{0\farcs3}%
\newcommand{\dnShiftLo}{20}%
\newcommand{\dnShiftHi}{60}%
\newcommand{\prNscTiles}{23{,}561}%
\newcommand{\prDtTiles}{12{,}598}%
\newcommand{\prDtSixReal}{911}%
\newcommand{\prDtSixCh}{1.8}%
\newcommand{\prDtFiveCh}{2.2}%
\newcommand{\sgBoxes}{29{,}891}%
\newcommand{\sgDevDen}{485}%
\newcommand{\sgDevBase}{111}%
\newcommand{\sgDevBoth}{126}%
\newcommand{\sgTestDen}{446}%
\newcommand{\sgTestBase}{95}%
\newcommand{\sgTestBoth}{107}%
\newcommand{\sgTestBothPct}{24.0}%
\newcommand{\sgTestBothCI}{20.3--28.2}%
\newcommand{\sgRuleGain}{10}%
\newcommand{\sgRuleFalse}{twice}%
\newcommand{\sgRuleCpu}{three}%
\newcommand{\sgFalseDev}{2.0--2.2}%
\newcommand{\myTruthRows}{814}%
\newcommand{\myStripTiles}{540}%
\newcommand{\mySeedN}{three}%
\newcommand{\mySeedWin}{120}%
\newcommand{\mySeedK}{64}%
\newcommand{\myTileBox}{1.3}%
\newcommand{\myBox}{2}%
\newcommand{\myBoxOff}{1}%
\newcommand{\myDeepThr}{100{,}000}%
\newcommand{\myDeepTiles}{24}%
\newcommand{\myKgate}{4}%
\newcommand{\myRcap}{20{,}000}%
\newcommand{\myDmag}{1.0}%
\newcommand{\myMinfar}{three}%
\newcommand{\myFarDt}{150}%
\newcommand{\myGroupSig}{4}%
\newcommand{\myMaxmem}{two}%
\newcommand{\myKatt}{3.5}%
\newcommand{\myAttFloor}{0\farcs5}%
\newcommand{\myAttPasses}{three}%
\newcommand{\myMaxc}{3000}%
\newcommand{\myRcapMin}{1800}%
\newcommand{\myNmin}{seven}%
\newcommand{\myArc}{183}%
\newcommand{\myChiGrow}{4}%
\newcommand{\myChi}{2.0}%
\newcommand{\myPch}{0.02}%
\newcommand{\myDmin}{25}%
\newcommand{\myDmax}{80}%
\newcommand{\myDupRows}{two}%
\newcommand{\myPreregTime}{00:42}%
\newcommand{\myAOneTime}{01:27}%
\newcommand{\myFreezeTime}{03:04}%
\newcommand{\myHeldRead}{04:25}%
\newcommand{\myModN}{230}%
\newcommand{\myModMax}{0\farcs054}%
\newcommand{\myTuneDen}{100}%
\newcommand{\myTuneEnd}{60}%
\newcommand{\myTuneCtlPmin}{0.079}%
\newcommand{\myTuneDevP}{0.0146}%
\newcommand{\pdTiles}{28{,}854}%
\newcommand{\pdBlocks}{165}%
\newcommand{\pdDup}{16}%
\newcommand{\pdCands}{554}%
\newcommand{\pdKnownScreen}{475}%
\newcommand{\pdTol}{300}%
\newcommand{\pdNotKnown}{79}%
\newcommand{\pdKc}{14}%
\newcommand{\pdReg}{31}%
\newcommand{\pdRegDes}{three}%
\newcommand{\pdProp}{34}%
\newcommand{\pdReal}{33}%
\newcommand{\pdVetKnown}{one}%
\newcommand{\pdKnownDistinct}{484}%
\newcommand{\pdKnownAll}{490}%
\newcommand{\pdCtlBlocks}{162}%
\newcommand{\pdCtlCollected}{162}%
\newcommand{\pdCtlMissing}{three}%
\newcommand{\pdSeedUnits}{271}%
\newcommand{\hoN}{226}%
\newcommand{\hoANum}{76}%
\newcommand{\hoADen}{112}%
\newcommand{\hoAPct}{67.9}%
\newcommand{\hoACI}{58.7--75.8}%
\newcommand{\hoAListPct}{31.3}%
\newcommand{\hoGain}{36.6}%
\newcommand{\hoGainCI}{23.7--47.8}%
\newcommand{\hoThrA}{60}%
\newcommand{\hoThrGain}{10}%
\newcommand{\hoBNum}{42}%
\newcommand{\hoBDen}{46}%
\newcommand{\hoBPct}{91.3}%
\newcommand{\hoBCI}{79.7--96.6}%
\newcommand{\hoThrB}{90}%
\newcommand{\hoThrCtl}{10}%
\newcommand{\hoBinAN}{16}\newcommand{\hoBinAMy}{15}\newcommand{\hoBinAPct}{93.8}\newcommand{\hoBinACI}{71.7--98.9}\newcommand{\hoBinAList}{13}\newcommand{\hoBinAListPct}{81.3}%
\newcommand{\hoBinBN}{30}\newcommand{\hoBinBMy}{27}\newcommand{\hoBinBPct}{90.0}\newcommand{\hoBinBCI}{74.4--96.5}\newcommand{\hoBinBList}{13}\newcommand{\hoBinBListPct}{43.3}%
\newcommand{\hoBinCN}{66}\newcommand{\hoBinCMy}{34}\newcommand{\hoBinCPct}{51.5}\newcommand{\hoBinCCI}{39.7--63.2}\newcommand{\hoBinCList}{9}\newcommand{\hoBinCListPct}{13.6}%
\newcommand{\hoBinDN}{97}\newcommand{\hoBinDMy}{7}\newcommand{\hoBinDPct}{7.2}\newcommand{\hoBinDCI}{3.5--14.2}\newcommand{\hoBinDList}{5}\newcommand{\hoBinDListPct}{5.2}%
\newcommand{\hoBinEN}{17}\newcommand{\hoBinEMy}{0}\newcommand{\hoBinEList}{0}%
\newcommand{\hoAllMy}{83}\newcommand{\hoAllPct}{36.7}\newcommand{\hoAllCI}{30.7--43.2}\newcommand{\hoAllList}{40}\newcommand{\hoAllListPct}{17.7}%
\newcommand{\hoEasyHeld}{31}%
\newcommand{\hoEasyDev}{17}%
\newcommand{\hoOldTest}{20}%
\newcommand{\hoMiss}{36}%
\newcommand{\hoMissNights}{25}%
\newcommand{\hoMissRule}{7}%
\newcommand{\hoMissGrow}{4}%
\newcommand{\plCtlUp}{1.23}%
\newcommand{\plPreA}{228}\newcommand{\plPreB}{245}\newcommand{\plPreC}{197}%
\newcommand{\plReal}{167}%
\newcommand{\plCpu}{20.6}%
\newcommand{\plCpuLim}{200}%
\newcommand{\exArea}{486}%
\newcommand{\exPreregTime}{04:38}%
\newcommand{\exN}{47}%
\newcommand{\exMy}{14}%
\newcommand{\exMyPct}{29.8}%
\newcommand{\exMyCI}{18.7--44.0}%
\newcommand{\exList}{5}%
\newcommand{\exListPct}{10.6}%
\newcommand{\puBTen}{12}\newcommand{\puBTenLo}{0.89}%
\newcommand{\puBNine}{5}\newcommand{\puBNineLo}{0.75}%
\newcommand{\puBEight}{11}\newcommand{\puBEightLo}{0.88}%
\newcommand{\puBSeven}{8}\newcommand{\puBSevenLo}{0.84}%
\newcommand{\puBSix}{113}%
\newcommand{\puBSixCa}{109}\newcommand{\puBSixCb}{90}\newcommand{\puBSixCc}{122}%
\newcommand{\puBSixPur}{0.05}%
\newcommand{\puBFive}{14{,}107}%
\newcommand{\puBFiveCa}{13{,}696}\newcommand{\puBFiveCb}{13{,}859}\newcommand{\puBFiveCc}{13{,}923}%
\newcommand{\puBFivePur}{0.02}%
\newcommand{\rxChi}{3}%
\newcommand{\rxPa}{0.05}\newcommand{\rxPb}{0.10}%
\newcommand{\edN}{468}%
\newcommand{\edRule}{371}%
\newcommand{\edRec}{247}%
\newcommand{\edNone}{67}\newcommand{\edNonePct}{18.1}%
\newcommand{\edNoneRec}{1}%
\newcommand{\edOnePct}{79.9}\newcommand{\edOneCI}{73.7--84.9}%
\newcommand{\edLostKnown}{57}%
\newcommand{\edLostNew}{13}\newcommand{\edLostNewCI}{8--19}%
\newcommand{\edBinAPct}{50}%
\newcommand{\edBinBPct}{78}%
\newcommand{\edBinCPct}{94}%
\newcommand{\edBinDPct}{100}%
\newcommand{\tsAll}{9{,}430}%
\newcommand{\tsN}{3{,}500}%
\newcommand{\tsArc}{1095}%
\newcommand{\tsObs}{12}%
\newcommand{\tsCode}{5}%
\newcommand{\tsDev}{1{,}823}\newcommand{\tsTest}{1{,}677}%
\newcommand{\tsEphMed}{0\farcs002}%
\newcommand{\tsSecNsc}{1{,}320}\newcommand{\tsSecDt}{998}%
\newcommand{\tsSecPctNsc}{38}\newcommand{\tsSecPctDt}{28.5}%
\newcommand{\tsRowTol}{2\farcs0}%
\newcommand{\vOneRecall}{7.4}%
\newcommand{\vOneShortPct}{46}%
\newcommand{\vOneSecRec}{25.6}\newcommand{\vOneSecRecCI}{21.7--29.9}%
\newcommand{\lbNscN}{485}\newcommand{\lbDtN}{378}%
\newcommand{\lbNscCad}{257}\newcommand{\lbNscCadPct}{53.0}\newcommand{\lbNscCadCI}{48.5--57.4}%
\newcommand{\lbDtCad}{265}\newcommand{\lbDtCadPct}{70.1}\newcommand{\lbDtCadCI}{65.3--74.5}%
\newcommand{\lbNscTile}{26}\newcommand{\lbNscTilePct}{5.4}\newcommand{\lbDtTile}{13}\newcommand{\lbDtTilePct}{3.4}%
\newcommand{\lbNscDet}{43}\newcommand{\lbNscDetPct}{8.9}\newcommand{\lbNscDetCI}{6.6--11.7}\newcommand{\lbDtDet}{8}\newcommand{\lbDtDetPct}{2.1}\newcommand{\lbDtDetCI}{1.1--4.1}%
\newcommand{\lbNscNoTile}{6}\newcommand{\lbNscNoTilePct}{1.2}\newcommand{\lbDtNoTile}{0}%
\newcommand{\lbNscOrph}{2}\newcommand{\lbNscOrphPct}{0.4}\newcommand{\lbDtOrph}{3}\newcommand{\lbDtOrphPct}{0.8}%
\newcommand{\lbNscLoad}{32}\newcommand{\lbNscLoadPct}{6.6}\newcommand{\lbNscLoadCI}{4.7--9.2}\newcommand{\lbDtLoad}{7}\newcommand{\lbDtLoadPct}{1.9}%
\newcommand{\lbDtLink}{3}\newcommand{\lbDtLinkPct}{0.8}%
\newcommand{\lbDtRule}{29}\newcommand{\lbDtRulePct}{7.7}\newcommand{\lbDtRuleCI}{5.4--10.8}%
\newcommand{\lbNscRec}{111}\newcommand{\lbNscRecPct}{22.9}\newcommand{\lbNscRecCI}{19.4--26.8}%
\newcommand{\lbDtRec}{50}\newcommand{\lbDtRecPct}{13.2}\newcommand{\lbDtRecCI}{10.2--17.0}%
\newcommand{\lbNscAfterCad}{202}\newcommand{\lbDtAfterCad}{100}\newcommand{\lbNscReachRest}{11}\newcommand{\lbDtReachRest}{five}\newcommand{\lbNscReachRec}{six}\newcommand{\lbDtReachRec}{one}%
\newcommand{\lbNscCeilNum}{105}\newcommand{\lbNscCeilDen}{191}\newcommand{\lbNscCeilPct}{55.0}\newcommand{\lbNscCeilCI}{47.9--61.9}%
\newcommand{\lbDtCeilNum}{49}\newcommand{\lbDtCeilDen}{95}\newcommand{\lbDtCeilPct}{51.6}\newcommand{\lbDtCeilCI}{41.7--61.4}%
\newcommand{\lbNscTestNum}{95}\newcommand{\lbNscTestDen}{446}\newcommand{\lbNscTestPct}{21.3}%
\newcommand{\lbDtTestNum}{40}\newcommand{\lbDtTestDen}{309}\newcommand{\lbDtTestPct}{12.9}%
\newcommand{\lbDetPct}{90}%
\newcommand{\lbKnownRuleObj}{21.8}%
\newcommand{\ijPLong}{0.88--0.92}%
\newcommand{\ijPShort}{0.78--0.87}%
\newcommand{\ijPCrowd}{0.04}%
\newcommand{\lrDmin}{20}\newcommand{\lrDmax}{100}%
\newcommand{\lrMinfar}{two}%
\newcommand{\lrN}{four}%
\newcommand{\lrDchi}{3.0}%
\newcommand{\lrP}{0.05}%
\newcommand{\lrSigOld}{0\farcs45}\newcommand{\lrSigNew}{0\farcs20}%
\newcommand{\lrTruth}{367}%
\newcommand{\lrOpp}{four}\newcommand{\lrU}{3}%
\newcommand{\lrWinLo}{20}\newcommand{\lrWinHi}{60}%
\newcommand{\lrHProp}{127}%
\newcommand{\lrHWrongUp}{2.9}%
\newcommand{\lrHWrongThr}{2}%
\newcommand{\lrHCtl}{777}%
\newcommand{\lrHRecPct}{85}%
\newcommand{\lrHArc}{30}%
\newcommand{\lrBCtl}{1{,}545}%
\newcommand{\lrThreeAll}{14}\newcommand{\lrThreeWrong}{two}%
\newcommand{\lrArcShort}{19}%
\newcommand{\lrPSearched}{208}\newcommand{\lrPRec}{five}%
\newcommand{\lrASeeded}{2{,}093}%
\newcommand{\lrAProp}{26}%
\newcommand{\lrASingle}{12}%
\newcommand{\imSnr}{5}%
\newcommand{\imShapeLo}{0.6}\newcommand{\imShapeHi}{1.6}\newcommand{\imRad}{1\farcs5}%
\newcommand{\imOther}{three}%
\newcommand{\imOpp}{two}%
\newcommand{\dtLinks}{674}\newcommand{\dtReal}{126}\newcommand{\dtPoss}{277}\newcommand{\dtNot}{268}%
\newcommand{\dtVerA}{two}\newcommand{\dtVerB}{three}%
\newcommand{\kcClones}{40}%
\newcommand{\kcChi}{5}\newcommand{\kcNear}{0\fdg3}\newcommand{\kcPole}{1}\newcommand{\kcPoleSep}{5}%
\newcommand{\kcSlow}{10}%
\newcommand{\kcLsqRms}{2}%
\newcommand{\kcUsed}{90}\newcommand{\kcRms}{0\farcs4}\newcommand{\kcRowMax}{1\farcs5}%
\newcommand{\kcMatches}{18}%
\newcommand{\kcMissed}{14}%
\newcommand{\slTestable}{41}%
\newcommand{\slExp}{370}%
\newcommand{\slDet}{99.2}\newcommand{\slCtl}{6.1}%
\newcommand{\slPass}{22}\newcommand{\slInc}{19}%
\newcommand{\slTruthFail}{19}\newcommand{\slTruthN}{30}%
\newcommand{\fcImgN}{11{,}467}%
\newcommand{\fcCatNum}{7}\newcommand{\fcCatDen}{320}\newcommand{\fcCatPct}{2.2}\newcommand{\fcCatCI}{1.1--4.4}%
\newcommand{\pfReal}{153}\newcommand{\pfCa}{152}\newcommand{\pfCb}{168}\newcommand{\pfCc}{282}%
\newcommand{\pfMasked}{20}\newcommand{\pfPub}{43}%
\newcommand{\ciLevel}{95}%
\newcommand{\nscYears}{2012--2019}%
\newcommand{\freezeDate}{2026 November 16}%
\newcommand{\kcAuditDate}{2026 November 15}%
\newcommand{\fxDay}{15}%
\newcommand{\myMinfarDigit}{3}%
\newcommand{\ttObjs}{269}%
\newcommand{\lrPstarOff}{2}%
\newcommand{\lrSigOldYr}{2005}\newcommand{\lrSigNewYr}{2010}%
\newcommand{\lrArcA}{10}%
\newcommand{\vOneMatchTol}{2}%
\newcommand{\kcDesYears}{2025--2026}%
\newcommand{\tsDmin}{25}\newcommand{\tsDmax}{80}%
\newcommand{\tsSecTol}{1}%
\newcommand{\hoMagA}{23.5}\newcommand{\hoMagB}{23.0}%
\newcommand{\hoBinALab}{$<22.5$}\newcommand{\hoBinBLab}{22.5--23}\newcommand{\hoBinCLab}{23--23.5}\newcommand{\hoBinDLab}{23.5--24}\newcommand{\hoBinELab}{$>24$}%
\newcommand{\exRegions}{two}\newcommand{\exBoxRa}{25}\newcommand{\exBoxDec}{10}%
\newcommand{\exMag}{23.5}%
\newcommand{\puZero}{0}%
\newcommand{\nlTen}{$\geq10$}\newcommand{\nlNine}{9}\newcommand{\nlEight}{8}\newcommand{\nlSeven}{7}\newcommand{\nlSix}{6}\newcommand{\nlFive}{5}%
\newcommand{\ebA}{7--8}\newcommand{\ebB}{9--11}\newcommand{\ebC}{12--15}%
\newcommand{\ijDmin}{25}\newcommand{\ijDmax}{35}%
\newcommand{\ijCrowdB}{20}%
\newcommand{\tsSecSig}{3}%
\newcommand{\myChanNew}{37}%
\newcommand{\myEarlyNew}{four}%
\newcommand{\mySixN}{two}%
\newcommand{\mySixOppA}{four}\newcommand{\mySixOppB}{three}%
\newcommand{\myYield}{35}%
\newcommand{\dtNew}{117}%
\newcommand{\dtTierB}{32}%
\newcommand{\unObj}{64}\newcommand{\unExp}{179}%
\newcommand{\kcRunDate}{2026 November 17}%
\newcommand{\scTen}{10}\newcommand{\scSixty}{60}%
\newcommand{\scUnderTen}{457}%
\newcommand{\scMed}{0\farcs9}%
\newcommand{\scOverSixty}{seven}%
\newcommand{\scMax}{221}%
\newcommand{\pdCpu}{330}%
\newcommand{\pdCpuLinux}{52}%
\newcommand{\raN}{931}\newcommand{\raVA}{60}\newcommand{\raVB}{183}\newcommand{\raVC}{528}\newcommand{\raVD}{154}\newcommand{\raVE}{six}%
\newcommand{\raVALab}{$<21$}\newcommand{\raVBLab}{21--22}\newcommand{\raVCLab}{22--23}\newcommand{\raVDLab}{23--24}%
\newcommand{\raOne}{208}%
\newcommand{\raVeto}{166}%
\newcommand{\raDense}{206}\newcommand{\raDensePct}{22.1}\newcommand{\raDenseVA}{9}\newcommand{\raDenseVB}{51}\newcommand{\raDenseVC}{122}\newcommand{\raDenseVD}{24}%
\newcommand{\raBoth}{233}\newcommand{\raBothPct}{25.0}\newcommand{\raBothVA}{11}\newcommand{\raBothVB}{56}\newcommand{\raBothVC}{139}\newcommand{\raBothVD}{27}%
\newcommand{\raMy}{247}\newcommand{\raMyPct}{26.5}\newcommand{\raMyVA}{11}\newcommand{\raMyVB}{63}\newcommand{\raMyVC}{150}\newcommand{\raMyVD}{23}%
\newcommand{\raAny}{337}\newcommand{\raAnyPct}{36.2}\newcommand{\raAnyVA}{14}\newcommand{\raAnyVB}{83}\newcommand{\raAnyVC}{199}\newcommand{\raAnyVD}{41}%
\newcommand{\raElig}{468}\newcommand{\raMyEligPct}{52.8}%
\newcommand{\raLatA}{5}\newcommand{\raLatAN}{413}\newcommand{\raLatAAny}{146}%
\newcommand{\raLatC}{10\mbox{--}20}\newcommand{\raLatCN}{234}\newcommand{\raLatCAny}{80}%
\newcommand{\raDtN}{687}\newcommand{\raDtPct}{13.1}%
\newcommand{\vOneNearHi}{35}%
\newcommand{\vOneNearObj}{11}\newcommand{\vOneCrossObj}{12}%
\newcommand{\dnKtri}{3.5}%
\newcommand{\dnSigDiv}{2.557}%
\newcommand{\desCandEight}{760}\newcommand{\desCandSeven}{8{,}321}%
\newcommand{\desSixReal}{52}\newcommand{\desSixAll}{132}%
\newcommand{\desSixDist}{50}\newcommand{\desSixOrbits}{100{,}000}%
\newcommand{\plSeeds}{2.08}%
\newcommand{\myNrms}{2.0}%
\newcommand{\myRcapAtt}{120}\newcommand{\myPad}{1}%
\newcommand{\pdCpuReal}{93}\newcommand{\pdCpuCtl}{234}%
\newcommand{\pdLinuxBlocks}{52}\newcommand{\pdLinuxCtlBlocks}{39}%
\newcommand{\lrDmag}{1.5}\newcommand{\lrMaxRows}{50{,}000}%
\newcommand{\lrSigMid}{0\farcs35}%
\newcommand{\vetMisN}{four}%
\newcommand{\scOne}{1}\newcommand{\scUnderOne}{257}\newcommand{\scMid}{11}\newcommand{\scBeyondTen}{18}%
\newcommand{\lbNscCeilTestNum}{88}\newcommand{\lbNscCeilTestDen}{180}\newcommand{\lbNscCeilTestPct}{48.9}%
\newcommand{\lbDtCeilTestNum}{40}\newcommand{\lbDtCeilTestDen}{86}\newcommand{\lbDtCeilTestPct}{46.5}%
\newcommand{\lbNscLinkR}{8}\newcommand{\lbNscLinkRPct}{1.6}\newcommand{\lbNscLabOnly}{two}%
\newcommand{\hoForcedB}{nine}%
\newcommand{\hoNatBNum}{34}\newcommand{\hoNatBDen}{37}\newcommand{\hoNatBPct}{91.9}\newcommand{\hoNatBCI}{78.7--97.2}%
\newcommand{\hoNatANum}{63}\newcommand{\hoNatADen}{97}\newcommand{\hoNatAPct}{64.9}%
\newcommand{\exNightsBox}{22}\newcommand{\exNightsTyp}{15}%
\newcommand{\plSeedsN}{2{,}081{,}164}\newcommand{\plQuad}{81.8}\newcommand{\plJoint}{14.6}\newcommand{\plPreReal}{1{,}399}%
\newcommand{\plSeedsCa}{2{,}077{,}142}\newcommand{\plSeedsCb}{2{,}066{,}018}\newcommand{\plSeedsCc}{2{,}071{,}112}%
\newcommand{\plQuadCa}{66.2}\newcommand{\plQuadCb}{66.5}\newcommand{\plQuadCc}{65.7}%
\newcommand{\plJointCa}{14.1}\newcommand{\plJointCb}{14.4}\newcommand{\plJointCc}{13.7}%
\newcommand{\pdCtlPartial}{four}%
\newcommand{\pdCtlSetsA}{165}\newcommand{\pdCtlSetsB}{150}\newcommand{\pdCtlSetsC}{159}\newcommand{\pdCtlSets}{474}%
\newcommand{\pdCtlSetsSeven}{463}\newcommand{\pdCtlSetsEight}{11}\newcommand{\pdCtlHeavyPct}{98}%
\newcommand{\pdCtlChiOk}{75}\newcommand{\pdCtlPmin}{0.037}%
\newcommand{\pdRealPmax}{0.0178}%
\newcommand{\pdCtlTailA}{seven}\newcommand{\pdTailA}{0.10}\newcommand{\pdCtlTailB}{three}\newcommand{\pdTailB}{0.05}\newcommand{\pdCtlTailC}{one}\newcommand{\pdTailC}{0.04}%
\newcommand{\pdBlockCands}{376}\newcommand{\pdEarlyCands}{194}\newcommand{\pdVetKnownN}{1}%
\newcommand{\pdKnownTwice}{five}\newcommand{\pdRegDesAll}{Five}\newcommand{\pdRegDesScreen}{two}%
\newcommand{\puPAll}{360}\newcommand{\puPAllLo}{0.9965}\newcommand{\puPTen}{205}\newcommand{\puPTenLo}{0.994}\newcommand{\puPNine}{58}\newcommand{\puPNineLo}{0.978}%
\newcommand{\puPEight}{54}\newcommand{\puPEightLo}{0.977}\newcommand{\puPSeven}{43}\newcommand{\puPSevenLo}{0.971}%
\newcommand{\puPropLo}{0.96}\newcommand{\puNewSeven}{12}\newcommand{\puNewSevenLo}{0.89}\newcommand{\puNewSevenPct}{36}%
\newcommand{\puBFiveExcess}{281}\newcommand{\puBFiveLowPct}{2}%
\newcommand{\puBSixBoth}{$-0.15$, 0.26}\newcommand{\puBFiveBoth}{0.001, 0.039}%
\newcommand{\raAnyCI}{33.2--39.3}\newcommand{\raAnyVDPct}{27}%
\newcommand{\raAnyLab}{376}\newcommand{\raAnyLabPct}{40.4}\newcommand{\raAnyLabVeto}{368}\newcommand{\raAnyLabVetoPct}{39.5}\newcommand{\raLabOnly}{39}%
\newcommand{\raMyDevPct}{28.0}\newcommand{\raMyTestPct}{24.9}\newcommand{\raAnyDevPct}{36.9}\newcommand{\raAnyTestPct}{35.4}%
\newcommand{\ijFarLo}{60}%
\newcommand{\ijPLongPart}{0.62}%
\newcommand{\lrHObj}{84}\newcommand{\lrHObjUp}{4.4}\newcommand{\lrHShort}{three}%
\newcommand{\lrAllSeeded}{2{,}301}%
\newcommand{\pdCtlDoneA}{165}\newcommand{\pdCtlDoneB}{163}%
\newcommand{\pdCtlChiA}{26}\newcommand{\pdCtlChiB}{21}\newcommand{\pdCtlChiC}{28}%
\newcommand{\pdCtlPminAok}{0.044}\newcommand{\pdCtlPminCok}{0.067}%
\newcommand{\pdCtlPminA}{0.042}\newcommand{\pdCtlPminC}{0.047}%
\newcommand{\ttOffset}{+0.35\pm0.15}%
\newcommand{\ttErrShort}{0\farcs04}\newcommand{\ttExpShort}{90}\newcommand{\ttRate}{3}\newcommand{\ttErrLong}{0\farcs2}\newcommand{\ttExpLong}{500}%
\newcommand{\ttRows}{4{,}642}\newcommand{\ttStartRows}{28}\newcommand{\ttStartObj}{six}\newcommand{\ttDa}{0.011}\newcommand{\ttUnmatched}{149}%
\newcommand{\dtRowTol}{1}\newcommand{\dtStrongPct}{10}\newcommand{\dtStrongCtl}{24}%
\newcommand{\dtChanceLo}{9}\newcommand{\dtChanceHi}{32}%
\newcommand{\dtNcProd}{62}\newcommand{\dtNcMeasFail}{27}%
\newcommand{\dtRegReal}{52}\newcommand{\dtRegPoss}{54}\newcommand{\dtRegNot}{five}\newcommand{\dtRegPilot}{five}%
\newcommand{\dtProgrammes}{169}%
\newcommand{\dtFSentA}{370}\newcommand{\dtFSentB}{159}\newcommand{\dtFSentC}{145}%
\newcommand{\gsN}{63}\newcommand{\gsProd}{15}\newcommand{\gsProdDrawn}{16}\newcommand{\gsPoss}{one}\newcommand{\gsProdRealUp}{21.9}\newcommand{\gsProdOkUp}{32.0}\newcommand{\gsOwnN}{24}\newcommand{\gsOwnUp}{14.3}%
\newcommand{\gsRowPct}{9.9}\newcommand{\gsRealRowPct}{86}%
\newcommand{\gsFalseLo}{0.6}\newcommand{\gsFalseHi}{2.2}\newcommand{\gsFalseUp}{10.3}\newcommand{\gsFalseRealUp}{7.1}%
\newcommand{\dtVerRep}{669}\newcommand{\dtVerRepN}{671}\newcommand{\dtProdSingle}{29}%
\newcommand{\dtRealTraced}{111}\newcommand{\dtRealInCat}{84}\newcommand{\dtRealObj}{60}\newcommand{\dtRealObjNew}{54}\newcommand{\dtRealObjDes}{six}\newcommand{\dtRealNotReg}{27}%
\newcommand{\dtExtNights}{two}\newcommand{\dtExtRms}{$0\farcs2$}\newcommand{\dtExtFirst}{43}%
\newcommand{\dtRealDt}{115}\newcommand{\dtFRealA}{113}\newcommand{\dtFRealB}{2}\newcommand{\dtFRealC}{11}\newcommand{\dtFPossA}{195}\newcommand{\dtFPossB}{19}\newcommand{\dtFPossC}{63}%
\newcommand{\dtFNotB}{137}\newcommand{\dtFNotC}{69}\newcommand{\dtFNoneA}{0}\newcommand{\dtFNoneB}{1}\newcommand{\dtFNoneC}{2}\newcommand{\dtNoneN}{3}%
\newcommand{\dtFUnmA}{21}\newcommand{\dtFUnmB}{13}\newcommand{\dtFUnmC}{15}\newcommand{\dtFUnmAll}{49}\newcommand{\dtFTwoA}{14}\newcommand{\dtFTwoB}{4}\newcommand{\dtFTwoC}{27}\newcommand{\dtFTwoAll}{45}%
\newcommand{\dtFFailB}{120}\newcommand{\dtFFailC}{27}\newcommand{\dtFFailAll}{174}%
\newcommand{\dtFCand}{1{,}028}\newcommand{\dtFKnown}{595}\newcommand{\dtFListed}{63}\newcommand{\dtFDesig}{19}\newcommand{\dtFDesigReal}{nine}%
\newcommand{\prFiles}{34}\newcommand{\prPass}{six}\newcommand{\prFail}{eight}\newcommand{\prNoGo}{five}\newcommand{\prNotRun}{six}\newcommand{\prOpen}{three}\newcommand{\prOther}{six}%
\newcommand{\prListDay}{2026 November 17}\newcommand{\prListTime}{04:03}%
\newcommand{\prUsed}{13}%
\newcommand{\prLateFiles}{six}%
\newcommand{\jnJoint}{6}\newcommand{\jnDen}{17}\newcommand{\jnNsc}{7}\newcommand{\jnThr}{ten}%
\newcommand{\cfRec}{two}\newcommand{\cfDen}{11}\newcommand{\cfThr}{35}%
\newcommand{\pfCode}{7}\newcommand{\pfObject}{2026~ED$_{138}$}%
\newcommand{\sgRuleTime}{16:16}%
\newcommand{\exEditTime}{06:03}\newcommand{\exHeadOld}{04:40}\newcommand{\exFileTime}{04:37:43}\newcommand{\exFetchTime}{04:39:01}%
\newcommand{\fxOneVariants}{five}%
\newcommand{\pdWinBlocks}{113}\newcommand{\pdHeaderOnly}{17}%
\newcommand{\ctlReadDay}{2026 November 17}\newcommand{\rxHeldBlocks}{84}%
\newcommand{\fjJoint}{9}\newcommand{\fjCtl}{two}\newcommand{\fjCtlLim}{1.8}\newcommand{\fjCpu}{103}\newcommand{\fjCpuLim}{60}%
\newcommand{\pdFiles}{22{,}563}%
\newcommand{\nhAllTiles}{25{,}607}%
\newcommand{\nhTiles}{25{,}487}\newcommand{\nhArea}{20{,}100}%
\newcommand{\nhSearched}{17{,}962}\newcommand{\nhSearchedArea}{14{,}835}\newcommand{\nhSearchedPct}{73.8}\newcommand{\nhSearchedTilePct}{70}%
\newcommand{\nhAcc}{3{,}170}\newcommand{\nhAccArea}{2{,}696}\newcommand{\nhAccEcl}{485}%
\newcommand{\nhEmpty}{eight}%
\newcommand{\nhDense}{3{,}351}\newcommand{\nhDenseArea}{1{,}792}\newcommand{\nhDenseGalPct}{74}%
\newcommand{\nhDensCut}{60{,}000}%
\newcommand{\nhLat}{10}%
\newcommand{\nhRun}{996}\newcommand{\nhRunArea}{773}%
\newcommand{\nhOut}{120}%
\newcommand{\nhKnownLost}{97}\newcommand{\nhKnownLostAcc}{72}\newcommand{\nhKnownLostDense}{eight}%
\newcommand{\nhKnownSearched}{620}\newcommand{\nhKnownRows}{428}%
\newcommand{\nhExFourTiles}{33{,}203}\newcommand{\nhExFour}{24{,}317}\newcommand{\nhFileFourTiles}{23{,}577}\newcommand{\nhFileFour}{17{,}995}%
\newcommand{\nhAuditDay}{2026 November 7}%
\newcommand{\nhRad}{0.95}%
\newcommand{\exHoleTiles}{eleven}%
\newcommand{\unxLoo}{1}%
\newcommand{\unxComb}{9.44}\newcommand{\unxSa}{5.75}\newcommand{\unxSb}{4.85}\newcommand{\unxSc}{5.96}%
\newcommand{\unxOffLo}{$0\farcs8$}\newcommand{\unxOffHi}{$1\farcs8$}\newcommand{\unxHalf}{1.2}\newcommand{\unxSig}{2.4}%
\newcommand{\unxCtlN}{100}\newcommand{\unxCtlMax}{7.42}%
\newcommand{\unxRms}{$0\farcs07$}%
\newcommand{\nhBoxNoFile}{4{,}194}%
\newcommand{\nhDenseAll}{4{,}193}\newcommand{\nhDenseNoFile}{3{,}711}\newcommand{\nhDenseSearched}{436}%
\newcommand{\nhFilePct}{74}%
\newcommand{\nhPairRad}{1.25}%
\newcommand{\nhRaCol}{359}\newcommand{\nhEmptyRa}{five}\newcommand{\nhEmptyRows}{11}%
\newcommand{\unxModLo}{6}\newcommand{\unxModHi}{9}%
\newcommand{\nhRnCounted}{269}\newcommand{\nhRnAll}{495}\newcommand{\nhRnAff}{171}\newcommand{\nhRnMissPct}{30}\newcommand{\nhRnMiss}{1.62}\newcommand{\nhRnShould}{5.33}\newcommand{\nhRnBlind}{11.9}\newcommand{\nhRnBlindUp}{42}\newcommand{\nhRnSearched}{146}%
\newcommand{\haSubmit}{2026 November 13}\newcommand{\haA}{30.034}\newcommand{\haE}{0.065}\newcommand{\haI}{20.3}%
\newcommand{\haPos}{28}\newcommand{\haNights}{14}\newcommand{\haFirst}{2010}\newcommand{\haLast}{2025}\newcommand{\haZtfSnr}{32}%
\newcommand{\haMyr}{1}\newcommand{\haClones}{200}\newcommand{\haAmp}{8.1^{+0.1}_{-0.2}}\newcommand{\haPer}{9.3}%
\newcommand{\tjEFull}{0.0646}\newcommand{\tjIFull}{20.346}%
\newcommand{\ttStartNew}{25}\newcommand{\ttStartNewObj}{five}
\newcommand{\nhRnFew}{20}\newcommand{\relCatDes}{31}\newcommand{\relCatTroj}{ten}
\newcommand{\stSize}{40}\newcommand{\stStretch}{20}\newcommand{\stSnrTroLo}{14.2}\newcommand{\stSnrTroHi}{42.7}\newcommand{\stSnrDesLo}{7.9}\newcommand{\stSnrDesHi}{14.9}
\newcommand{\stCtlTro}{0.6}\newcommand{\stCtlDes}{0.9}\newcommand{\stPreLo}{2.1}\newcommand{\stPreHi}{13.2}\newcommand{\stDisc}{2026 April 5}
\newcommand{\ctReg}{395}%
\newcommand{\ctNew}{365}%
\newcommand{\ctDes}{30}%
\newcommand{\ctRows}{6{,}708}%
\newcommand{\ctRowsImg}{5{,}201}%
\newcommand{\ctRowsCat}{1026}%
\newcommand{\ctRowsOther}{42}%
\newcommand{\ctRowsOtherObj}{12}%
\newcommand{\ctNewOneOpp}{43}%
\newcommand{\ctOneOppRest}{14}%
\newcommand{\ctTierA}{273}%
\newcommand{\ctTierB}{41}%
\newcommand{\ctTierC}{51}%
\newcommand{\ctTierCOne}{2}%
\newcommand{\ctTierCMulti}{49}%
\newcommand{\ctTierCvOneA}{11}%
\newcommand{\ctTierAifVOne}{284}%
\newcommand{\ctFitKeep}{90}%
\newcommand{\ctFitRms}{0\farcs3}%
\newcommand{\ctSigA}{10}%
\newcommand{\ctTierBImg}{38}%
\newcommand{\ctFitRej}{24}%
\newcommand{\ctAOppTwo}{54}%
\newcommand{\ctAOppThree}{43}%
\newcommand{\ctAOppFour}{62}%
\newcommand{\ctAOppFive}{114}%
\newcommand{\ctANightsLo}{eight}%
\newcommand{\ctANightsMid}{60}%
\newcommand{\ctANightsHi}{205}%
\newcommand{\ctANightsMin}{six}%
\newcommand{\ctARowsImg}{4{,}679}%
\newcommand{\ctARowsCat}{409}%
\newcommand{\ctACatObj}{33}%
\newcommand{\ctAAllImg}{240}%
\newcommand{\ctAOppImgAll}{251}%
\newcommand{\ctTierCMultiImg}{47}%
\newcommand{\ctTierCMultiFit}{two}%
\newcommand{\ctTierCNoImg}{39}%
\newcommand{\ctTierCMultiNightsMed}{8}%
\newcommand{\ctTierCMultiArcMed}{3.9}%
\newcommand{\ctTierCSent}{13}%
\newcommand{\ctTierBArcMed}{61}%
\newcommand{\ctTierBArcMin}{10}%
\newcommand{\ctTierBArcMax}{141}%
\newcommand{\ctTierBLo}{11}%
\newcommand{\ctTierBImgSeven}{29}%
\newcommand{\ctDtDes}{11}%
\newcommand{\ctDtA}{82}%
\newcommand{\ctRowsOtherVOne}{Eight}%
\newcommand{\ctChMy}{58}%
\newcommand{\ctChVOne}{77}%
\newcommand{\ctChVOneDes}{4}%
\newcommand{\ctChDenseNsc}{18}%
\newcommand{\ctChDenseDt}{120}%
\newcommand{\ctChCfht}{6}%
\newcommand{\ctMySix}{three}%
\newcommand{\ctArcAMed}{5.1}%
\newcommand{\ctObjCfht}{37}%
\newcommand{\ctObjHsc}{33}%
\newcommand{\ctObjOth}{5}%
\newcommand{\ctAObjMultiInstr}{48}%
\newcommand{\ctAObjDecA}{172}%
\newcommand{\ctAObjDecB}{216}%
\newcommand{\ctAObjCfht}{31}%
\newcommand{\ctAObjHsc}{31}%
\newcommand{\ctAObjOth}{5}%
\newcommand{\ctAObjBothDec}{120}%
\newcommand{\ctSent}{38}%
\newcommand{\ctSentNew}{36}%
\newcommand{\ctSentDes}{two}%
\newcommand{\ctNotSentNew}{329}%
\newcommand{\ctRowsMine}{6{,}227}%
\newcommand{\ctRowsNewAll}{6{,}269}%
\newcommand{\ctRowsRej}{30}%
\newcommand{\ctArcAFive}{153}%
\newcommand{\ctArcATen}{38}%
\newcommand{\ctArcAMax}{19.0}%
\newcommand{\ctAHMin}{4.4}%
\newcommand{\ctAHMax}{11.9}%
\newcommand{\ctAHMed}{7.3}%
\newcommand{\ctAHLo}{6.6}%
\newcommand{\ctAHHi}{8.1}%
\newcommand{\ctGrMed}{0.08}%
\newcommand{\ctGrNights}{161}%
\newcommand{\ctGrObj}{84}%
\newcommand{\ctRiMed}{$-0.14$}%
\newcommand{\ctRiNights}{98}%
\newcommand{\ctRiObj}{51}%
\newcommand{\ctAVrObj}{69}%
\newcommand{\ctClsSec}{252}%
\newcommand{\ctClsOld}{66}%
\newcommand{\ctClsBoth}{14}%
\newcommand{\ctClsFreeze}{2026 November 25}%
\newcommand{\ctClsShortRun}{all 56}%
\newcommand{\ctClsNew}{207}%
\newcommand{\ctClsTro}{6}%
\newcommand{\ctClsRes}{62}%
\newcommand{\ctClsCla}{108}%
\newcommand{\ctClsDet}{47}%
\newcommand{\ctClsSca}{43}%
\newcommand{\ctClsCen}{7}%
\newcommand{\ctClsPlut}{33}%
\newcommand{\ctClsClones}{50}%
\newcommand{\ctClsMyr}{10}%
\newcommand{\ctClsAgree}{80}%
\newcommand{\ctFar}{ten}%
\newcommand{\ctFarLim}{150}%
\newcommand{\ctFarAMin}{151}%
\newcommand{\ctFarAMax}{310}%
\newcommand{\ctFarQMin}{20.6}%
\newcommand{\ctFarQMax}{41.4}%
\newcommand{\ctFarQThirty}{eight}%
\newcommand{\ctFarATwoFifty}{five}%
\newcommand{\ctFarOppMin}{2}%
\newcommand{\ctFarOppMax}{8}%
\newcommand{\ctFarArcMin}{1.3}%
\newcommand{\ctFarArcMax}{11.0}%
\newcommand{\ctFarSigMax}{0.33}%
\newcommand{\ctFarNonDec}{four}%
\newcommand{\ctFarEMin}{0.73}%
\newcommand{\ctFarEMax}{0.89}%
\newcommand{\ctRetNew}{two}%
\newcommand{\ctHiInc}{nine}%
\newcommand{\ctHiIncLim}{40}%
\newcommand{\ctHiIncMid}{seven}%
\newcommand{\ctIncThirty}{29}%
\newcommand{\ctIncMed}{17}%
\newcommand{\ctQForty}{39}%
\newcommand{\ctQFortySec}{31}%
\newcommand{\ctQMax}{48.8}%
\newcommand{\ctTrA}{30.04}%
\newcommand{\ctTrE}{0.023}%
\newcommand{\ctTrI}{13.2}%
\newcommand{\ctTrH}{8.1}%
\newcommand{\ctTrRms}{0\farcs087}%
\newcommand{\ctTrRows}{22}%
\newcommand{\ctTrNights}{14}%
\newcommand{\ctTrOpp}{six}%
\newcommand{\ctTrArc}{8.4}%
\newcommand{\ctRaA}{207.64}%
\newcommand{\ctRaSigA}{0.06}%
\newcommand{\ctRaE}{0.890}%
\newcommand{\ctRaI}{110.3}%
\newcommand{\ctRaQ}{22.8}%
\newcommand{\ctRaRms}{0\farcs1}%
\newcommand{\ctRaRows}{38}%
\newcommand{\ctRaNights}{25}%
\newcommand{\ctRaOpp}{four}%
\newcommand{\ctRaArc}{8.1}%
\newcommand{\ctRbA}{53.9}%
\newcommand{\ctRbE}{0.572}%
\newcommand{\ctRbI}{138.5}%
\newcommand{\ctRbQ}{23.0}%
\newcommand{\ctRbRms}{0\farcs074}%
\newcommand{\ctRbRows}{38}%
\newcommand{\ctRbNights}{27}%
\newcommand{\ctRbOpp}{seven}%
\newcommand{\ctRbArc}{7.9}%
\newcommand{\ctTrDec}{13}%
\newcommand{\ctTrOth}{nine}%
\newcommand{\ctTroArcMin}{13}%
\newcommand{\ctTroArcMax}{19}%
\newcommand{\ctTroIMin}{20}%
\newcommand{\ctTroIMax}{25}%
\newcommand{\ctRbOther}{14}%
\newcommand{\ctDsI}{142.4}%
\newcommand{\ctDsA}{67.2}%
\newcommand{\ctDsNights}{33}%
\newcommand{\ctTrLibN}{101}%
\newcommand{\ctTrLibClones}{100}%
\newcommand{\ctTrLibFourN}{101}%
\newcommand{\ctRbAgree}{198}%
\newcommand{\ctRbRest}{53}%
\newcommand{\ctRbSecNote}{, so the class is not secure by the \ctClsAgree\% rule}%
\newcommand{\ctRbNorb}{251}%
\newcommand{\ctRbCls}{scattering}%
\newcommand{\ctHbI}{71}%
\newcommand{\ctHbQ}{14}%
\newcommand{\ctHbOpp}{five}%
\newcommand{\ctDsFirst}{2018}%
\newcommand{\ctFarLimB}{250}%
\newcommand{\ctQLimA}{30}%
\newcommand{\ctQLimC}{37}%
\newcommand{\ctQLimD}{42}%
\newcommand{\ctFarQIn}{two}%
\newcommand{\ctFarQLo}{30.1}%
\newcommand{\ctFarXA}{153}%
\newcommand{\ctFarXI}{34}%
\newcommand{\ctQLimB}{40}%
\newcommand{\ctIncLim}{30}%
\newcommand{\ctRaClsSentence}{In the \ctClsMyr-Myr integration it is a scattering object in all 51 orbits.}%
\newcommand{\ctFarLongSentence}{The 100-Myr integrations of the uniform fit give the same class for all ten, in every orbit for nine of them.}%
\newcommand{\ctFarClsSentence}{Nine of the ten are scattering objects. The definition asks for a change of 1.5~au within \ctClsMyr~Myr, which is under one per cent of semimajor axes of this size, so the class says little here.}%
\newcommand{\ctDesRubin}{17}%
\newcommand{\ctDesSubaru}{three}%
\newcommand{\ctDesDecam}{five}%
\newcommand{\ctDesOld}{four}%
\newcommand{\ctDesPs}{one}%
\newcommand{\ctDesPre}{21}%
\newcommand{\ctDesPreNights}{218}%
\newcommand{\ctDesPreMinTxt}{7~days}%
\newcommand{\ctDesPreMax}{13.2}%
\newcommand{\ctDesPreMed}{8.9}%
\newcommand{\ctDesPost}{9}%
\newcommand{\ctDesPostAfter}{seven}%
\newcommand{\ctDesRubinMulti}{two}%
\newcommand{\ctDesRubinOne}{15}%
\newcommand{\ctDesRubMin}{2.9}%
\newcommand{\ctDesRubMax}{13.2}%
\newcommand{\ctDesMultiOpp}{seven}%
\newcommand{\ctDesArcLo}{2}%
\newcommand{\ctDesArcHi}{119}%
\newcommand{\ctMkDisc}{430}%
\newcommand{\ctMkPre}{133}%
\newcommand{\ctMkPreNights}{580}%
\newcommand{\ctMkPub}{56}%
\newcommand{\ctMkPubNights}{275}%
\newcommand{\ctMkNew}{16}%
\newcommand{\ctMkRubin}{15}%
\newcommand{\ctMkPreNotPub}{81}%
\newcommand{\ctMkPreYear}{65}%
\newcommand{\ctMkPreDecam}{65}%
\newcommand{\ctMkNewMin}{6.9}%
\newcommand{\ctMkNewMax}{13.1}%
\newcommand{\ctLostArcLo}{2.0}%
\newcommand{\ctLostArcHi}{21.9}%
\newcommand{\ctAreaOne}{25{,}971}%
\newcommand{\ctAreaFour}{25{,}123}%
\newcommand{\ctAreaNsc}{22{,}163}%
\newcommand{\ctAreaDt}{15{,}197}%
\newcommand{\ctAreaRubin}{5{,}400}%
\newcommand{\ctAreaBoth}{5{,}041}%
\newcommand{\ctAreaNotRubin}{20{,}082}%
\newcommand{\ctNotRubinPct}{80}%
\newcommand{\ctAInRubin}{54}%
\newcommand{\ctAOutRubin}{219}%
\newcommand{\ctDepthMed}{23.65}%
\newcommand{\ctDepthLo}{23.11}%
\newcommand{\ctDepthHi}{23.96}%
\newcommand{\ctRmVA}{18}%
\newcommand{\ctRmVB}{34}%
\newcommand{\ctRmVD}{15}%
\newcommand{\ctDtExp}{149{,}421}%
\newcommand{\ctDtFiles}{182{,}399}%
\newcommand{\ctDtExpT}{30}%
\newcommand{\ctDtLat}{30}%
\newcommand{\ctDtGal}{10}%
\newcommand{\ctDtSig}{5}%
\newcommand{\ctDtAp}{4}%
\newcommand{\ctDtDepG}{23.31}%
\newcommand{\ctDtDepR}{23.14}%
\newcommand{\ctDtDepI}{22.87}%
\newcommand{\ctDtDepZ}{22.27}%
\newcommand{\ctDtDepVR}{24.32}%
\newcommand{\ctDtComp}{90}%
\newcommand{\ctDtCompOff}{0.5}%
\newcommand{\ctDtCompFaint}{44}%
\newcommand{\ctDtAstRa}{0\farcs046}%
\newcommand{\ctDtAstDec}{0\farcs057}%
\newcommand{\ctDtAstMax}{0\farcs22}%
\newcommand{\ctDtLostPct}{29.1}%
\newcommand{\ctDtNoZpPct}{8.1}%
\newcommand{\ctDtCrowded}{40}%
\newcommand{\ctDtCcdLo}{9}%
\newcommand{\ctDtCcdHi}{22}%
\newcommand{\ctDtGaia}{1}%
\newcommand{\ctDtTile}{1}%
\newcommand{\ctDtFirst}{2020 January}%
\newcommand{\ctDtLast}{2026 October}%
\newcommand{\ctDtBoxRa}{60^\circ\mbox{--}140^\circ}%
\newcommand{\ctDtBoxDec}{-10^\circ\mbox{ to }+40^\circ}%
\newcommand{\ctDtMesh}{128}%
\newcommand{\ctDiamLo}{120}%
\newcommand{\ctDiamHi}{210}%
\newcommand{\ctAlbHi}{0.15}%
\newcommand{\ctAlbLo}{0.05}%
\newcommand{\ctKvQ}{21}%
\newcommand{\ctKvI}{104}%
\newcommand{\ctCtlSets}{474}%
\newcommand{\ctCtlMaxN}{eight}%
\newcommand{\ctCtlPmin}{0.037}%
\newcommand{\ctCtlIncomplete}{three}%
\newcommand{\ctNscExp}{412{,}000}%
\newcommand{\ctNscSkip}{30{,}000}%
\newcommand{\ctNscNight}{0.6}%
\newcommand{\ctNscMatch}{1\farcs5}%
\newcommand{\ctVOneUndes}{77}%
\newcommand{\ctHscObj}{ten}%
\newcommand{\ctHscPos}{39}%
\newcommand{\ctHscNights}{18}%
\newcommand{\ctHscGood}{eight}%
\newcommand{\ctHscSig}{1.2}%
\newcommand{\ctHscTieLo}{0\farcs01}%
\newcommand{\ctHscTieHi}{0\farcs07}%
\newcommand{\ctCfhtExp}{218{,}685}%
\newcommand{\ctRubinTno}{380}%
\newcommand{\ctDesTno}{812}%
\newcommand{\ctDesArea}{5000}%
\newcommand{\ctOssosTno}{838}%
\newcommand{\ctHoPct}{68}%
\newcommand{\ctHoListPct}{31}%
\newcommand{\ctEdgePct}{18}%
\newcommand{\ctJoPos}{88}%
\newcommand{\ctJoNights}{54}%
\newcommand{\ctJoArc}{17.22}%
\newcommand{\ctJoTel}{seven}%
\newcommand{\ctJoAmp}{8.7}%
\newcommand{\ctJoClones}{500}%
\newcommand{\ctJoStable}{421}%
\newcommand{\ctJoStablePct}{84}%
\newcommand{\ctJoGyr}{4.5}%
\newcommand{\ctJoHv}{6.96\pm0.10}%
\newcommand{\ctJoSubmit}{2026 November 11}%
\newcommand{\ctVOneDate}{2026 November 10}%
\newcommand{\ctTjA}{30.034}%
\newcommand{\ctTjH}{6.8}%
\newcommand{\ctTjRows}{40}%
\newcommand{\ctTvAmpA}{19\fdg4}%
\newcommand{\ctTvAmpB}{10\fdg6}%
\newcommand{\ctTvAmpC}{8\fdg8}%
\newcommand{\ctTvAmpD}{27\fdg3}%
\newcommand{\ctTvOrb}{51}%
\newcommand{\ctTvOrbD}{101}%
\newcommand{\ctTvGrLo}{0.49}%
\newcommand{\ctTvGrHi}{0.54}%
\newcommand{\ctNtAll}{31}%
\newcommand{\ctNtLFour}{27}%
\newcommand{\ctNtLFive}{four}%
\newcommand{\ctNtRubin}{two}%
\newcommand{\ctThorN}{27{,}500}%
\newcommand{\ctEtDaLo}{3}%
\newcommand{\ctEtDaHi}{145}%
\newcommand{\ctEtQLo}{35.3}%
\newcommand{\ctEtQHi}{36.9}%
\newcommand{\ctEtN}{201}%
\newcommand{\ctRcN}{13}%
\newcommand{\ctRcQ}{15}%
\newcommand{\ctRcShort}{seven}%
\newcommand{\ctRcShortLo}{7}%
\newcommand{\ctRcShortHi}{84}%
\newcommand{\ctRcLong}{six}%
\newcommand{\ctRcLongFar}{two}%
\newcommand{\ctRcLongFarQ}{16.5}%
\newcommand{\ctDcN}{136}%
\newcommand{\ctDcA}{132}%
\newcommand{\ctDcOn}{17}%
\newcommand{\ctDcLo}{13}%
\newcommand{\ctDcLoMax}{0.16}%
\newcommand{\ctDcHi}{four}%
\newcommand{\ctDcHiMin}{0.76}%
\newcommand{\ctDcNear}{two}%
\newcommand{\ctDcLim}{29}%
\newcommand{\ctDcFar}{two}%
\newcommand{\ctDcFarMa}{0.35}%
\newcommand{\ctDcFarMb}{0.07}%
\newcommand{\ctDcFarNa}{13}%
\newcommand{\ctDcFarNb}{14}%
\newcommand{\ctOcIn}{two}%
\newcommand{\ctOcTrip}{one}%
\newcommand{\ctOcR}{23.2}%
\newcommand{\ctOcLimLo}{24.1}%
\newcommand{\ctOcLimHi}{25.2}%
\newcommand{\puPropChancePct}{4}%
\newcommand{\puNewSevenChancePct}{11}%
\newcommand{\ctUnuDone}{37}%
\newcommand{\ctUnuNoExp}{two}%
\newcommand{\ctUnuNoInd}{two}%
\newcommand{\ctUnuTested}{33}%
\newcommand{\ctUnuTwo}{24}%
\newcommand{\ctUnuExp}{207}%
\newcommand{\ctUnuExpDet}{181}%
\newcommand{\ctUnuElig}{451}%
\newcommand{\ctUnuLtNine}{19}%
\newcommand{\ctUnuLtNineTwo}{13}%
\newcommand{\ctUnuGeOne}{32}%
\newcommand{\ctUnuSingle}{eight}%
\newcommand{\ctUnuNone}{one}%
\newcommand{\ctUnuNoneNights}{three}%
\newcommand{\ctUnuNoneExpLo}{4}%
\newcommand{\ctUnuNoneExpHi}{6}%
\newcommand{\ctUnuNoneSnrUp}{2}%
\newcommand{\ctUnuSingleExc}{two}%
\newcommand{\ctUnuRuleFac}{three}%
\newcommand{\ctUnuRuleMag}{1.2}%
\newcommand{\ctUnuRuleDay}{2026 November 17}%
\newcommand{\ctUnuExcExp}{six}%
\newcommand{\ctUnuExcObj}{five}%
\newcommand{\ctUnuExcLeft}{two}%
\newcommand{\ctUnuExcLink}{three}%
\newcommand{\ctUnuLnDet}{211}%
\newcommand{\ctUnuSingleExcNights}{one and two}%
\newcommand{\ctUnuCtlExpPctNoRule}{3.5}%
\newcommand{\ctUnuCtlExc}{40}%
\newcommand{\ctUnuTwoNoRule}{26}%
\newcommand{\ctUnuSingleNoRule}{six}%
\newcommand{\ctUnuExpDetNoRule}{187}%
\newcommand{\ctUnuCtlOneNoRule}{143}%
\newcommand{\ctUnuCtlTwoNoRule}{9}%
\newcommand{\ctTrUnuDetSnr}{14.8}%
\newcommand{\ctTrUnuDetExp}{13.2}%
\newcommand{\ctTrUnuExcSnr}{230.6}%
\newcommand{\ctTrUnuExcExp}{17.3}%
\newcommand{\ctUnuSingleOnly}{four}%
\newcommand{\ctUnuSingleMiss}{two}%
\newcommand{\ctUnuCtlN}{1{,}056}%
\newcommand{\ctUnuCtlOne}{109}%
\newcommand{\ctUnuCtlTwo}{7}%
\newcommand{\ctUnuCtlOnePct}{10}%
\newcommand{\ctUnuCtlTwoPct}{0.7}%
\newcommand{\ctUnuCtlExpPct}{2.8}%
\newcommand{\ctUnuCtlTwoObj}{five}%
\newcommand{\ctUnuCtlTwoObjLo}{3}%
\newcommand{\ctUnuCtlTwoObjHi}{7}%
\newcommand{\ctUnuSixAExp}{five}%
\newcommand{\ctUnuSixANights}{two}%
\newcommand{\ctUnuSixBExp}{seven}%
\newcommand{\ctUnuSixBNights}{three}%
\newcommand{\ctHscNotIn}{two}%
\newcommand{\ctHscOutA}{1.9}%
\newcommand{\ctHscOutB}{4.3}%
\newcommand{\ctHscOutBm}{2.3}%
\newcommand{\ctPropsInList}{27}%
\newcommand{\ctHscTwoNights}{four}%
\newcommand{\ctHscOneNight}{six}%
\newcommand{\ctHscOneExp}{three}%
\newcommand{\ctHscDigest}{seven}%
\newcommand{\ctHscLooseLo}{11}%
\newcommand{\ctHscLooseHi}{18}%
\newcommand{\ctHscLooseOffLo}{7}%
\newcommand{\ctHscLooseOffHi}{10}%
\newcommand{\ctHscCrowd}{eight}%
\newcommand{\ctChkIndNights}{three and five}%
\newcommand{\ctChkNscA}{three}%
\newcommand{\ctChkNscB}{two}%
\newcommand{\ctChkSepLo}{0.05}%
\newcommand{\ctChkSepHi}{0.23}%
\newcommand{\ctChkLooLo}{0.36}%
\newcommand{\ctChkLooHi}{0.55}%
\newcommand{\ctChkLamA}{0.007}%
\newcommand{\ctChkLamB}{0.08}%
\newcommand{\ctChkCtl}{520}%
\newcommand{\ctChkRmsA}{0.08}%
\newcommand{\ctChkRmsB}{0.12}%
\newcommand{\ctChkShift}{0.9}%
\newcommand{\ctChkNscMeas}{Four}%
\newcommand{\ctChkNscAll}{five}%
\newcommand{\ctChkFourAll}{20}%
\newcommand{\ctChkMeasAll}{22}%
\newcommand{\ctChkSnrHi}{11}%
\newcommand{\ctChkFA}{eight of nine}%
\newcommand{\ctChkFB}{eight of 13}%
\newcommand{\ctChkFFour}{eight and twelve}%
\newcommand{\ctChkFNights}{six}%
\newcommand{\ctChkCtlPct}{1.9}%
\newcommand{\ctChkTrkTwo}{none and one}%
\newcommand{\ctChkNotMeas}{two}%
\newcommand{\ctChkPa}{0.64 and 0.21}%
\newcommand{\ctChkPb}{0.93 and 0.57}%
\newcommand{\ctJcRmsHis}{0.14}%
\newcommand{\ctJcRmsMine}{0.13}%
\newcommand{\ctJcSky}{0.06}%
\newcommand{\ctJcSigMax}{1.4}%
\newcommand{\ctJcShared}{34}%
\newcommand{\ctJcDecShared}{ten}%
\newcommand{\ctJcDec}{eleven}%
\newcommand{\ctJcHisElse}{62}%
\newcommand{\ctJcHisElseN}{35}%
\newcommand{\ctJcHisElseTel}{four}%
\newcommand{\ctIndepVExcExp}{13}%
\newcommand{\ctIndepVExcObj}{four}%
\newcommand{\ctTierAvTwo}{189}%
\newcommand{\ctTierAAdded}{84}%
\newcommand{\ctIndepN}{41}%
\newcommand{\ctRareN}{17}%
\newcommand{\ctRareNoTest}{15}%
\newcommand{\ctIndepOne}{25}%
\newcommand{\ctIndepVTwo}{13}%
\newcommand{\ctIndepVRuleOne}{six}%
\newcommand{\ctIndepVRuleNone}{one}%
\newcommand{\ctIndepVIncOne}{nine}%
\newcommand{\ctIndepVIncNone}{six}%
\newcommand{\ctIndepVNone}{seven}%
\newcommand{\ctIndepVOne}{20}%
\newcommand{\ctIndepVOneInc}{15}%
\newcommand{\ctIndepNone}{123}%
\newcommand{\ctEvExt}{30}%
\newcommand{\ctEvUnu}{33}%
\newcommand{\ctARmsMed}{$0\farcs09$}%
\newcommand{\ctMyLtNine}{31}%
\newcommand{\ctASixMy}{two}%
\newcommand{\ctDtTierB}{30}%
\newcommand{\ctFnAReal}{29}%
\newcommand{\ctFnAPoss}{47}%
\newcommand{\ctFnANc}{three}%
\newcommand{\ctFnANo}{three}%
\newcommand{\ctFnBReal}{22}%
\newcommand{\ctFnBPoss}{four}%
\newcommand{\ctFnBNc}{two}%
\newcommand{\ctFnBNo}{two}%
\newcommand{\ctTtStart}{28}%
\newcommand{\ctTtStartObj}{six}%
\newcommand{\ctTtNo}{149}%
\newcommand{\ctTtDa}{0.011}%
\newcommand{\ctTtDe}{0.0025}%
\newcommand{\ctTabChanKey}{TL = single-season tile linker (\texttt{nsc\_tile\_linker}); XT = cross-tile linking (\texttt{nsc\_crosstile}); NM = near mode (\texttt{nsc\_near\_mode}); RV = re-vetted earlier links (\texttt{nsc\_internal\_revetted}); SG = second pass on a shifted grid (\texttt{nsc\_shifted\_grid}); SS = side searches of version~1 (\texttt{side\_search}); DP = first runs of the dense-tile linker and night pools (\texttt{dense\_pools}); DN = dense-tile linker with mask on \nscdr (\texttt{d25\_method\_on\_nsc}); DX = the same on my extraction from the exposures of 2020--2026 (\texttt{d25\_own\_extraction}); MY = multi-year linker (\texttt{multiyear\_linker\_A2}); CF = CFHT MegaCam linker (\texttt{cfht\_linker}); MD = multi-year linker on my 2020--2026 extraction (\texttt{multiyear\_linker\_d25}); M6 = six-night level of the multi-year linker (\texttt{multiyear\_six\_nights}); OD = linkers on other data (Subaru HSC frames, DECam difference images) (\texttt{other\_data\_linkers}); F4 = four-night links on \nscdr (\texttt{nsc\_four\_nights})}%
\newcommand{\ctTabClassKey}{NT = L4 Neptune Trojan; Cen = Centaur; ci = inner classical; cc = main classical, cold; ch = main classical, hot; co = outer classical; det = detached; sca = scattering; a ratio such as 3:2 = resonant, with the resonance as the machine-readable file names it}%
\newcommand{\ctWeak}{92}%
\newcommand{\ctSarcContact}{October 27}%
\newcommand{\ctSarcAsk}{November 5}%
\newcommand{\ctReleaseBy}{2027 June 30}%
\newcommand{\ctHead}{\ctTierA{} undesignated candidates with orbits from two or more oppositions, whose linked detections were re-measured on the images (tier~A); \ctTierCMulti{} further multi-opposition objects that rest on catalogue positions or on fits that reject positions; \ctNewOneOpp{} links seen in a single opposition, which have no orbit; and \ctDes{} objects identified with designated ones}
\newcommand{\ctAbsCatA}{The catalogue has \ctTierA{} undesignated candidates with orbits from two or more oppositions, whose linked detections were re-measured on the images; of the \ctTierAvTwo{} of version~2, \ctIndepN{} were then detected on two or more nights that their fits did not use. Listed apart are \ctWeak{} weaker entries and \ctDes{} designated objects}
\newcommand{\vtCfhtFitUsed}{22}%
\newcommand{\vtCfhtFitRows}{22}%
\newcommand{\vtMyVTwo}{37}%
\newcommand{\vtMyAdded}{21}%
\newcommand{\vtMyAddedA}{four}%
\newcommand{\vtMyAddedC}{17}%
\newcommand{\vtMyDtN}{71}%
\newcommand{\vtSixN}{eight}%
\newcommand{\vtSixA}{one}%
\newcommand{\vtOdN}{seven}%
\newcommand{\vtFarSec}{nine}%
\newcommand{\vtFarQC}{one}%
\newcommand{\vtFarQD}{none}%
\newcommand{\vtDesRel}{19}%
\newcommand{\vtTro}{six}%
\newcommand{\vtTroCap}{Six}%
\newcommand{\vtTnLibN}{101}%
\newcommand{\vtTnLibL}{101}%
\newcommand{\vtTnLibClones}{100}%
\newcommand{\vtTnRows}{14}%
\newcommand{\vtTnNights}{11}%
\newcommand{\vtTnOpp}{six}%
\newcommand{\vtTnArc}{9.1}%
\newcommand{\vtTnRms}{0\farcs065}%
\newcommand{\vtTnA}{30.05}%
\newcommand{\vtTnE}{0.074}%
\newcommand{\vtTnI}{24.2}%
\newcommand{\vtTnH}{7.7}%
\newcommand{\vtLhN}{eight}%
\newcommand{\vtLhDiff}{one}%
\newcommand{\vtLateN}{nine}%
\newcommand{\vtLateFirst}{GCT0400}%
\newcommand{\vtLateLast}{GCT0408}%
\newcommand{\vOneTwoReg}{271}%
\newcommand{\vtSentFirst}{37}%
\newcommand{\vtSentGOneThreeSixRows}{32}%
\newcommand{\vtOtsSha}{\texttt{0850ea63101a0e27\allowbreak fbb3f16e58acf991\allowbreak 3a0ff7dd4a8e0ab8\allowbreak d77c203a71e94f15}}%
\newcommand{\vtHolesN}{five}%
\newcommand{\vtHolesA}{two}%
\newcommand{\vtDcVThree}{114}%
\newcommand{\vtMyYield}{55}%
\newcommand{\vtSentDates}{2026 October 27--30}%
\newcommand{\vtRaDec}{36}%
\newcommand{\vtRaOth}{two}%
\newcommand{\vtDtAllA}{153}
\newcommand{\rvPdAllMiss}{39}%
\newcommand{\rvPdAllOurs}{621}%
\newcommand{\rvPdAllPct}{6.3}%
\newcommand{\rvPdLeMiss}{26}%
\newcommand{\rvPdLeOurs}{515}%
\newcommand{\rvPdLePct}{5.0}%
\newcommand{\rvPdLeCcd}{two}%
\newcommand{\rvPdLeFlag}{four}%
\newcommand{\rvPdLeNone}{20}%
\newcommand{\rvPdLeNonePct}{3.9}%
\newcommand{\rvPdInspMask}{one}%
\newcommand{\rvPdInspCol}{three}%
\newcommand{\rvPdInspSee}{three}
\newcommand{\gzRefound}{nine}%
\newcommand{\gzAll}{thirteen}%
\newcommand{\gzThreeA}{0.434}%
\newcommand{\gzThreeE}{0.0135}%
\newcommand{\gzFourA}{0.003}%
\newcommand{\gzFourE}{0.0002}%
\newcommand{\gzFourI}{0.005}%
\newcommand{\gzTierNineFive}{C}%
\newcommand{\gzThreeTierA}{seven}
\newcommand{\rvKcDay}{2026 November 23}%
\newcommand{\rvKcObj}{387}%
\newcommand{\rvKcObjPairs}{379}%
\newcommand{\rvKcPairs}{3{,}110}%
\newcommand{\rvKcMatch}{22}%
\newcommand{\rvKcNew}{no}%
\newcommand{\rvKcUndec}{11}%
\newcommand{\rvKcRadMulti}{0.5}%
\newcommand{\rvKcRadOne}{15}%
\newcommand{\rvKcNearest}{15}%
\newcommand{\rvKcRobFrac}{95}%
\newcommand{\rvKcRobRes}{1.5}%
\newcommand{\rvKcUsedFrac}{80}%
\newcommand{\rvKcRmsLo}{0\farcs05}%
\newcommand{\rvKcRmsHi}{0\farcs18}%
\newcommand{\rvKcOursMax}{0\farcs53}%
\newcommand{\rvKcTheirsMax}{0\farcs82}%
\newcommand{\rvKcSepMed}{0.7}%
\newcommand{\rvKcSepMax}{4.7}%
\newcommand{\rvKcTruth}{17}%
\newcommand{\rvKcOldMiss}{nine}%
\newcommand{\rvBoxFar}{88.9}%
\newcommand{\rvBoxNear}{22.7}%
\newcommand{\rvInjProd}{0}%
\newcommand{\rvInjAddon}{179}%
\newcommand{\rvInjElig}{242}%
\newcommand{\rvAddLo}{72}%
\newcommand{\rvAddHi}{222}%
\newcommand{\rvFarLo}{180}%
\newcommand{\rvFarHi}{1{,}111}%
\newcommand{\rvNscBlocks}{167}%
\newcommand{\rvDtBlocks}{85}%
\newcommand{\rvUzNights}{19}%
\newcommand{\rvUzR}{92.5}%
\newcommand{\rvUzMpc}{74}%
\newcommand{\rvUzRms}{0\farcs13}%
\newcommand{\rvUzRowMax}{0\farcs20}%
\newcommand{\rvGgR}{88.1}%
\newcommand{\rvChNights}{44}%
\newcommand{\rvChR}{79.6}%
\newcommand{\rvNscAThreeA}{57}%
\newcommand{\rvNscAThreeAn}{78}%
\newcommand{\rvNscAThreeB}{46}%
\newcommand{\rvNscAThreeBn}{85}%
\newcommand{\rvDtAFiveA}{12}%
\newcommand{\rvDtAFiveAn}{17}%
\newcommand{\rvDtAFiveB}{7}%
\newcommand{\rvDtAFiveBn}{12}%
\newcommand{\rvDtWideB}{4}%
\newcommand{\rvGeomNsc}{57--60}%
\newcommand{\rvRubArea}{1{,}906}%
\newcommand{\rvRubTiles}{1{,}992}%
\newcommand{\rvRubLinks}{70}%
\newcommand{\rvRubKnown}{66}%
\newcommand{\rvRubReg}{three}%
\newcommand{\rvRubNstar}{5}%
\newcommand{\rvRubDev}{13/15}%
\newcommand{\rvRubHeld}{20/20}%
\newcommand{\rvRubLimA}{0.029}%
\newcommand{\rvRubLimB}{0.024}%
\newcommand{\rvRubLimC}{0.022}%
\newcommand{\rvRubAreaA}{102.5}%
\newcommand{\rvRubAreaC}{139.3}%
\newcommand{\rvRubElig}{96--99}%
\newcommand{\rvRubPlaced}{3--7}%
\newcommand{\rvHscFarRecA}{0.57}%
\newcommand{\rvHscFarRecB}{0.67}%
\newcommand{\rvHscFar}{250}%
\newcommand{\rvCsMatched}{0 of 134}%
\newcommand{\rvCsMissing}{65 of 121}%
\newcommand{\rvCsForced}{24 of 24}%
\newcommand{\rvCsNscLoss}{12--14}%
\newcommand{\rvCsDtLoss}{7--18}%
\newcommand{\rvCsSlowNsc}{6--10}%
\newcommand{\rvCsSlowDt}{15--24}%
\newcommand{\rvCsGeoNsc}{80}%
\newcommand{\rvCsGeoDt}{47}%
\newcommand{\rvCsEdOld}{19}%
\newcommand{\rvCsEdAll}{42}%
\newcommand{\rvPdVisits}{two}%
\newcommand{\rvPdAstMed}{0\farcs008}%
\newcommand{\rvPdAstSd}{0\farcs015--0\farcs025}%
\newcommand{\rvPdZp}{0.02}%
\newcommand{\rvPdMfiveA}{0.05}%
\newcommand{\rvPdMfiveB}{0.09}%
\newcommand{\rvPdHiSnr}{90}%
\newcommand{\rvPdTnoSnr}{26.5}%
\newcommand{\rvPdInsp}{nine}%
\newcommand{\rvPdUnexpl}{two}%
\newcommand{\rvDfDepth}{22.58}%
\newcommand{\rvDfGate}{22.76}%
\newcommand{\rvAcTargets}{28}%
\newcommand{\rvAcCtl}{20}%
\newcommand{\rvAcComa}{20--60}%
\newcommand{\rvAcAper}{3\farcs2}%
\newcommand{\rvAcAfLo}{15}%
\newcommand{\rvAcAfHi}{50}%
\newcommand{\rvSoTruth}{69/69}%
\newcommand{\rvSoCtl}{0/1{,}093}%
\newcommand{\rvSoUnk}{118}%
\newcommand{\rvSxRec}{55/111}%
\newcommand{\rvSxCost}{18{,}000}%
\newcommand{\rvCsFacNsc}{1.25}%
\newcommand{\rvCsFacDt}{2}%
\newcommand{\rvMdStageReal}{17}%
\newcommand{\rvMdStageRealN}{217}%
\newcommand{\rvMdStageCtl}{none}%
\newcommand{\rvMdStageCtlN}{110}%
\newcommand{\rvMdStack}{twelve}%
\newcommand{\rvSixCtlP}{0}%
\newcommand{\rvSixCtlE}{1}%
\newcommand{\rvSixCtlLim}{6}%
\newcommand{\rvSixKnown}{five}%
\newcommand{\rvHscQRec}{0.62}%
\newcommand{\rvCsSlowBlocks}{85}%
\newcommand{\rvCsSlowCtl}{no}%
\newcommand{\rvSfNscMy}{9.8}%
\newcommand{\rvSfNscD}{8.1}%
\newcommand{\rvSfDtMy}{7.2}%
\newcommand{\rvSfDtD}{5.9}%
\newcommand{\rvSfOverMy}{10}%
\newcommand{\rvSfOverD}{19}%
\newcommand{\rvSfNightsSim}{8.3}%
\newcommand{\rvSfNightsObs}{6.3}%
\newcommand{\rvKcScreen}{10{,}335}%
\newcommand{\rvKcSelVOne}{1{,}153}%
\newcommand{\rvKcSelDrift}{1{,}021}%
\newcommand{\rvKcSelPole}{936}%
\newcommand{\rvItTiles}{5{,}912}%
\newcommand{\rvItAll}{5{,}972}%
\newcommand{\rvItCtl}{2{,}119}
\newcommand{\papertitle}{Linking Outer Solar System Objects in Archival DECam Imaging: A Multi-year Linker Tested on Held-out Known Objects, and \ctTierA{} Multi-opposition Candidates Including Six L4 Neptune Trojans, Two Retrograde Objects and Ten with $a>150$~au}
\newcommand{\ctAbsCat}{\ctAbsCatA}
\newcommand{\absconf}{, in confidence}
\newcommand{\paperabstract}{%
I link single-night detections in public DECam imaging: \nhFilePct\% of the NOIRLab Source Catalog DR2 (\nscYears) sky with four or more nights, and my extraction from \ctDtExp{} exposures (2020--2026). Of \raN{} secure known objects, single-season linking recovers \raDensePct--\raBothPct\%, a multi-year linker \raMyPct\%, together \raAnyPct\%. The multi-year rule, fixed before its held-out tests, recovers \ctHoPct\% of DES objects brighter than $r=\hoMagA$ in a strip of its footprint and \exMy{} of \exN{} known objects outside it. Its shifted-sky controls, complete for \pdCtlCollected{} of \pdBlocks{} blocks, return no candidate. This limits pure chance alignments, not links that mix real and chance detections. An identity check by orbit with \rvKcPairs{} joint fits finds no designated object among the candidates. Searches beyond 79~au recover, blind, only 2014~UZ$_{224}$, Gonggong and Chiminigagua. \ctAbsCat, \ctDesRubin{} of them Rubin discoveries with \ctDesRubMin--\ctDesRubMax~yr of arc before discovery. \vtTroCap{} are candidate L4 Neptune Trojans whose orbits and clones librate for \ctClsMyr~Myr (one reported independently by Jongma 2026, mostly from other data), two are retrograde candidates and \ctFar{} have $a>\ctFarLim$~au. There is no follow-up or validated selection function. The Minor Planet Center has astrometry of \ctSent{} objects; unreleased positions are available on request\absconf.}
\hypersetup{pdftitle={\papertitle},pdfauthor={Gioele Cerati}}%

\begin{document}

\title{\papertitle}

\author[0009-0002-5933-3806]{Gioele Cerati}
\affiliation{Independent researcher, Italy}
\email{gioele@cerati.org}

\begin{abstract}
\paperabstract
\end{abstract}


\section{Introduction}\label{sec:intro}

A trans-Neptunian object (TNO) at 30--50~au moves a few arcseconds per hour. In one night it looks like any other faint source, and it is recognised only when detections from different nights are connected. Surveys designed for this choose their cadence for linking and measure their detection efficiency: the Outer Solar System Origins Survey \citep[OSSOS;][]{bannister2018}, which released \ctOssosTno{} outer Solar System discoveries tracked over several oppositions; the searches of the Dark Energy Survey (DES) by \citet{bernardinelli2020} and \citet{bernardinelli2022}, which found \ctDesTno{} TNOs over \ctDesArea~deg$^2$ by linking detections across the six years of the survey; and the DECam Ecliptic Exploration Project \citep{trilling2024,smotherman2024}, which stacks images along trial motions. The linking algorithms differ: orbit fits for sparse astrometry \citep{bernstein2000}, heliocentric transformation and clustering of tracklets \citep[HelioLinC;][]{holman2018}, test orbits without tracklets \citep[THOR;][]{moeyens2021}, and clustering of single detections propagated over trial orbits \citep{ruch2026}.

Much public DECam imaging was taken for other purposes and has no such cadence. The NOIRLab Source Catalog DR2 \citep[\nscdr;][]{nidever2021} lists every detection in the public exposures of \nscYears. It has been searched for moving objects before. \citet{fasbender2021} searched its first release for tracklets without fitting heliocentric orbits, and THOR was run on \nscdr{} itself by the Asteroid Institute, which reported \ctThorN{} candidate asteroids, most of them in the main belt \citep{b6122024}. The distant object 2017~OF$_{201}$ was found in archival DECam data \citep{cheng2026}. Version 1 of this paper \citep{cerati2026v1} linked the single-night detections of \nscdr{} within one season and one sky tile and reported \vOneObj{} objects. Its estimate of its own completeness was crude. Its check of identity with known objects was made by position, and that kind of check misses objects designated on short arcs: none of the objects published in version 1 as undesignated has turned out to be designated, but \kcMissed{} objects added to my working list afterwards were already designated (Section~\ref{sec:cat_known}).

This version replaces it. The search now covers fourteen years of DECam exposures, with my own extraction of the detections of 2020--2026, and it adds a linker that works across years, modelled on the DES searches. The paper reports, in this order:

\begin{enumerate}
\item \emph{Recall on known objects, and where they are lost} (Sections~\ref{sec:valid_budget}--\ref{sec:valid_all}). On one set of \raN{} known objects with secure orbits and by one credit rule, each channel recovers \raDensePct--\raMyPct\% and all together \raAnyPct\%. The first loss is that most objects never have the nights that a single-season rule asks for.
\item \emph{A multi-year rule fixed before its held-out tests, and its controls} (Sections~\ref{sec:link_my}, \ref{sec:valid_held} and \ref{sec:valid_ctl}). In a strip inside the DES footprint, where the cadence is favourable, the rule recovers \ctHoPct\% of the held-out DES objects brighter than $r=\hoMagA$; outside that footprint it recovers \exMy{} of \exN{} known objects. The same code on shifted copies of the sky returns no candidate with \myNmin{} or more nights in the \pdCtlCollected{} of \pdBlocks{} blocks whose controls are complete, which limits pure chance alignments and not links that mix real and chance detections; with six nights the real sky and the controls cannot be told apart.
\item \emph{Identification by orbit} (Section~\ref{sec:cat_known}). Each object is compared with the designated ones at their epochs and not at mine; this found \kcMissed{} identities that my earlier check by position had missed. For this version the check was repeated on the whole catalogue with a repaired screen, and it finds no further identity (Section~\ref{sec:cat_known}).
\item \emph{A catalogue} (Sections~\ref{sec:catalogue} and \ref{sec:rare}) of \ctReg{} objects: \ctHead. Of the \ctTierAvTwo{} tier-A candidates of version~2, \ctIndepN{} have been detected on two or more nights that their fits did not use, and \ctIndepOne{} more on one such night; \ctUnuNone{} candidate that was tested on \ctUnuNoneNights{} such nights is not detected at the predicted position, and a later search outside the rule of that test finds it and is not counted (Section~\ref{sec:valid_indep}); the \ctTierAAdded{} tier-A candidates added since have not been through these tests. Tier~A includes \vtTro{} candidate L4 Neptune Trojans, one of which was reported independently by \citet{jongma2026}, whose orbit rests mostly on data that I did not use and agrees with mine, and was later matched provisionally by \citet{han2026} to an object linked in TESS images (Section~\ref{sec:rare_trojan}), \ctRetNew{} retrograde candidates and \ctFar{} candidates with $a>\ctFarLim$~au, one of them retrograde.
\item \emph{Searches for distant objects} (Section~\ref{sec:far}). Searches for slow, distant objects in the same data find no new object, recover known distant objects that they were not told about, and give upper limits on the number of such objects.
\end{enumerate}

The Vera C.\ Rubin Observatory \citep{ivezic2019} has begun to discover outer Solar System objects at a rate no earlier survey reached: its early data alone gave roughly \ctRubinTno{} TNOs in less than two months \citep{rubin2026news}, and the ten-year survey is predicted to multiply the known population \citep{kurlander2025,murtagh2026}. Most objects reported here that lie in its footprint will be found by it, if they have not been already. What a DECam archive adds beside Rubin is measured here in two ways. First, it holds years of earlier positions: for the \ctDesRubin{} Rubin discoveries that are in my catalogue the archive gives \ctDesRubMin--\ctDesRubMax~yr of arc before discovery (Section~\ref{sec:cat_known}),  which for the \ctDesRubinOne{} of them with a single-opposition arc can turn a discovery arc of weeks into a multi-year orbit; my positions of those identified in version~2 are published with this version and the others are available on request.{} Second, in 2026 November \ctNotRubinPct\% of the cells of Figure~\ref{fig:coverage} with at least \dnNmin{} nights, which cover more than the sky that was read, and \ctAOutRubin{} of the \ctTierA{} tier-A objects lay in sky that Rubin had not yet observed in the public record (Section~\ref{sec:data_cov}). Rubin will cover most of it. A third use is only proposed: recoveries of \lrAProp{} designated objects whose orbits are too poor to find them again, none yet verified on images (Section~\ref{sec:valid_seed}).

\emph{Changes in version 3.} The catalogue grows from \ctTierAvTwo{} to \ctTierA{} tier-A candidates and from \vOneTwoReg{} to \ctReg{} objects, mostly through the multi-year linker run on my 2020--2026 extraction and through a second pass of the same linker on \nscdr{} (Section~\ref{sec:cat_channels}). The tests of Sections~\ref{sec:vet} and \ref{sec:valid} are those of version~2 and concern its sample; the objects added since rest on the same rules of linking and vetting, but not yet on the tests that do not depend on the link. New in this version are the repeated check of identity, the searches beyond 80~au and an estimate of what a mask of the DECam pipeline removes from the detections of slow objects. An independent search of \nscdr{} by \citet{giorgi2026} also found \gzAll{} objects of this catalogue, eleven of version~2 and two added since; Section~\ref{sec:cat_giorgi} compares the two.

No object was followed up, no validated selection function is derived, and the paper makes no inference about populations; Section~\ref{sec:limits} lists these and the other limits. Section~\ref{sec:data} describes the data, Sections~\ref{sec:link} and \ref{sec:vet} the linking and the vetting, Section~\ref{sec:valid} the tests, among them those that do not depend on the link (Section~\ref{sec:valid_indep}), Sections~\ref{sec:catalogue} and \ref{sec:rare} the catalogue and its objects of sparsely populated classes, and Section~\ref{sec:discussion} what the archives still yield. The positions of undesignated objects are not printed, except those of GCT0044, which are released with this version; Section~\ref{sec:data_avail} says why, what has been submitted to the Minor Planet Center, and how to obtain them.
\section{Data and coverage}\label{sec:data}

The search uses two sets of single-exposure DECam detections, which together span fourteen years, and two other archives in which arcs are extended. This imaging was largely not taken for moving objects, and its cadence is whatever the original programmes needed. Some of it was: the public archive also holds the exposures of Solar System programmes, among them the DECam Ecliptic Exploration Project, which observes in the wide $VR$ filter \citep{trilling2024,trujillo2024}, and at least \ctAVrObj{} of the \ctTierA{} tier-A objects of the catalogue have a position on a $VR$ exposure. I have not classified the exposures by programme, so ``largely'' is not a measured fraction.

\subsection{NSC DR2, 2012--2019}\label{sec:data_nsc}
The NOIRLab Source Catalog DR2 \citep[\nscdr;][]{nidever2021} lists the detections in about \ctNscExp{} public exposures of DECam \citep{flaugher2015} and of two other NOIRLab cameras, taken in \nscYears{} and processed by the DECam Community Pipeline \citep{valdes2014}. I accessed it through the Astro Data Lab \citep{fitzpatrick2014,nikutta2020}. \nscdr{} groups repeated detections at one position into objects, so a moving object appears in it as a set of unrelated objects, each seen within one night. The input of every linker is these single-night objects, which I call orphan detections: \nscdr{} objects whose detections all fall within \ctNscNight~d, each with a flag that says whether it has a Gaia DR3 \citep{gaia2016,gaia2023} or Legacy Surveys DR10 \citep{dey2019} counterpart within \ctNscMatch. The single-season linkers ran on \prNscTiles{} tiles of about one square degree and the multi-year linker on the same tile files (Section~\ref{sec:link}). These are the tiles for which I had downloaded the catalogue, not all of \nscdr{} (Section~\ref{sec:data_cov}). Version 1 skipped tiles with more than \ctNscSkip{} orphan detections per square degree, that is the Galactic plane and bulge; the later linkers bound their cost in crowded tiles instead, but injected objects are rarely recovered there (Section~\ref{sec:valid_inj}), and most tiles with a predicted density of more than \nhDensCut{} detections per square degree (\nhDenseNoFile{} of the \nhDenseAll{} with \dnNmin{} or more nights) are not in my copy of the catalogue, so no linker read them. The catalogue is not complete to its stated depth for moving objects: a known object is detected in about \lbDetPct\% of the exposures only when it is at least a magnitude brighter than that depth (Section~\ref{sec:valid_budget}).

\subsection{My extraction of DECam exposures of 2020--2026}\label{sec:data_dt}
No public catalogue of single-exposure detections covers the DECam exposures taken after \nscdr. I extracted one, which I call \dtf.

\emph{Selection.} From the NOIRLab Astro Data Archive I took the calibrated images of the Community Pipeline with their data-quality masks: science exposures in $g$, $r$, $i$, $z$ and $VR$ of at least \ctDtExpT~s, public at the time of the query, taken between \ctDtFirst{} and \ctDtLast, within $\ctDtLat^\circ$ of the ecliptic and more than $\ctDtGal^\circ$ from the Galactic plane, or inside a second region at RA~$\ctDtBoxRa$, Dec~$\ctDtBoxDec$. These are \ctDtExp{} exposures (\ctDtFiles{} archive files, since some exposures were processed more than once).

\emph{Detection.} On each CCD the background is estimated on a mesh of \ctDtMesh~pixels and sources are the local maxima above $\ctDtSig\sigma$ of the smoothed image, away from flagged pixels. Fluxes are measured in an aperture of \ctDtAp~pixels, and positions and zero points use the astrometric solution and the calibration of the Community Pipeline, whose astrometry, like that of \nscdr, is tied to Gaia DR2 \citep{gaia2018}. All sources are kept. Within tiles of $\ctDtTile^\circ$, detections that repeat at one position on other nights are static and are removed, as are those within $\ctDtGaia''$ of a Gaia source; the rest are the orphan detections of \dtf.

\emph{Depth and accuracy.} The median $5\sigma$ depth of an exposure is \ctDtDepG{} in $g$, \ctDtDepR{} in $r$, \ctDtDepI{} in $i$, \ctDtDepZ{} in $z$ and \ctDtDepVR{} in $VR$. Measured on known objects at their computed positions, detection is \ctDtComp\% complete or better for objects at least \ctDtCompOff~mag brighter than that depth and \ctDtCompFaint\% in the half magnitude next to it. Per exposure, the positions are offset from Gaia by a median of \ctDtAstRa{} in right ascension and \ctDtAstDec{} in declination, and by \ctDtAstMax{} at most. Times are those of mid-exposure.

\emph{Known defects.} An audit of the extraction found defects that cost detections, and I have not yet re-extracted. Exposures processed twice enter as pairs of files, and where one of the pair has no zero point the pair is lost to the linker: \ctDtLostPct\% of the exposures, against the \ctDtNoZpPct\% that have no zero point in any version. One CCD produces \ctDtCcdLo--\ctDtCcdHi\% of all orphan detections. The pipeline's flag for transient pixels, which the extraction masks, sits on real moving objects (Section~\ref{sec:null}). No catalogue exists for \ctDtCrowded{} files with more than a million sources. These losses are inside the recall measured in Section~\ref{sec:valid_budget}; they lower it and are not corrected for anywhere.

\subsection{Other archives}\label{sec:data_ext}
\emph{CFHT MegaCam.} For arc extensions I use MegaCam \citep{boulade2003} frames from the Canadian Astronomy Data Centre, located with SSOIS \citep{gwyn2012}, as in version 1; \ctObjCfht{} undesignated objects have positions measured on them. The archive holds \ctCfhtExp{} public MegaCam exposures, typically deeper than a DECam exposure. A linker that I ran on a catalogue of MegaCam detections failed its test on recall (\cfRec{} of \cfDen{} known objects, where \cfThr\% was required; Section~\ref{sec:null}). Of its \ctChCfht{} objects, five (GCT0239, GCT0246, GCT0282, GCT0283 and GCT0284) rest on CFHT positions alone, with image-verified positions in one opposition; they are in tier~C and are not counted. The sixth, GCT0285, was later found on DECam exposures of several seasons and is in tier~A (Section~\ref{sec:cat_channels}).

\emph{Subaru HSC.} Raw Hyper Suprime-Cam frames \citep{miyazaki2018} come from SMOKA \citep{baba2002}. I reduce them myself, with an astrometric solution on Gaia DR3 for each CCD (rms \ctHscTieLo--\ctHscTieHi{} on the frames used). The frozen catalogue has HSC positions for \ctObjHsc{} undesignated objects. A test in which \ctHscObj{} objects were searched on HSC frames at positions fixed beforehand is described in Section~\ref{sec:valid_indep}.

\emph{Others.} For \ctObjOth{} undesignated objects I measured positions on frames of other public archives, among them SDSS \citep{york2000} and Pan-STARRS1 \citep{chambers2016,flewelling2020}.

\emph{A check of my HSC reduction against the survey's own catalogues.} The HSC-SSP third public data release \citep{aihara2022} publishes source catalogues of single visits. For \rvPdVisits{} visits that I had reduced from the raw frames, I compared my catalogues with them. The astrometry agrees: the median offset is \rvPdAstMed{} or less in each coordinate, with a scatter of \rvPdAstSd. The zero points agree within \rvPdZp~mag, and my limiting magnitude at a signal-to-noise ratio of 5 is \rvPdMfiveA{} and \rvPdMfiveB~mag brighter than theirs, so my catalogues are slightly shallower. Above a signal-to-noise ratio of 7, about \rvPdHiSnr\% of my sources are in their catalogue. The public catalogues are not a complete list of the moving objects on the frames, though. Of the \rvPdAllOurs{} detections of known objects with good orbits that I make on the two visits, \rvPdAllMiss{} (\rvPdAllPct\%) have no counterpart in the public catalogue, and \rvPdLeMiss{} of the \rvPdLeOurs{} at $V\leq24$ (\rvPdLePct\%). Of these \rvPdLeMiss, \rvPdLeCcd{} fall on a detector that is not in the release and \rvPdLeFlag{} have a flagged source of the release within 2$''$; for the other \rvPdLeNone{} (\rvPdLeNonePct\% of the detections at $V\leq24$) the release has no source at all. Several of the missing ones are bright, among them a designated trans-Neptunian object at a signal-to-noise ratio of \rvPdTnoSnr. I looked at the pixels of \rvPdInsp{} of them: \rvPdInspCol{} lie on a weakly defective column, \rvPdInspSee{} are in a visit of exceptional seeing, \rvPdInspMask{} is under a mask of my own reduction, and for \rvPdUnexpl{} I found no cause. For a search for moving objects the per-visit catalogues are therefore a useful cross-check of a reduction, but not a replacement for it.

\subsection{Coverage}\label{sec:data_cov}
\begin{figure*}
\centering
\includegraphics[width=0.8\textwidth]{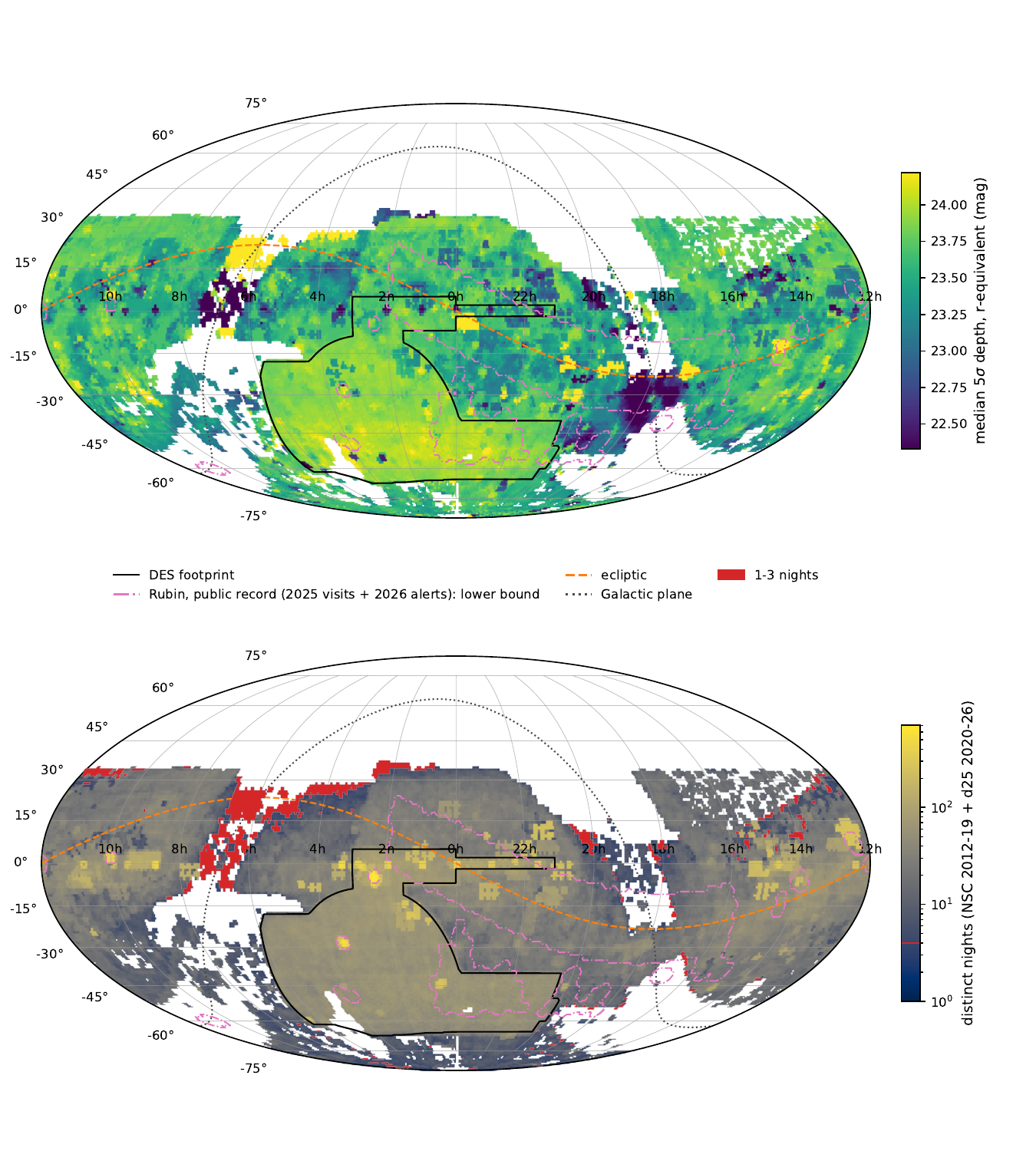}
\caption{Coverage of the two DECam data sets, \nscdr{} (\nscYears) and \dtf{} (2020--2026), in cells of one square degree (equatorial coordinates). Top: median $5\sigma$ depth of the exposures of a cell, $r$-equivalent. Bottom: number of distinct nights; cells with fewer than \dnNmin{} nights are red. Black: DES footprint. Dash-dotted: sky observed by Rubin in the public record (a lower bound). Dashed: ecliptic. Dotted: Galactic plane. The \nscdr{} layer is drawn from the catalogue files that I had downloaded, in cells offset by half a degree, and is neither all the sky that \nscdr{} covers nor only the sky that was read (Section~\ref{sec:data_cov}). No catalogue object is shown.\label{fig:coverage}}
\end{figure*}

Figure~\ref{fig:coverage} maps the two DECam data sets. At least one night exists over \ctAreaOne~deg$^2$, and at least \dnNmin{} nights, the least a single-season link needs, over \ctAreaFour~deg$^2$: \ctAreaNsc~deg$^2$ in \nscdr{} and \ctAreaDt~deg$^2$ in \dtf, which overlap. For \nscdr{} these areas are not counted on the list of exposures. They are counted on the catalogue files that I had downloaded, in the half-degree-shifted boxes of Section~\ref{sec:link_shift}: a cell counts when its box, which takes detections from four tiles, has them on enough nights. They are therefore neither the sky that \nscdr{} covers nor the sky that was searched: a box is drawn on the cell of one of its four tiles whether or not that tile has a file, and \nhBoxNoFile~deg$^2$ of the \ctAreaNsc~deg$^2$ are cells whose own tile has none. By the list of exposures, \nscdr{} has DECam exposures on \dnNmin{} or more nights in \nhExFourTiles{} tiles of one square degree (\nhExFour~deg$^2$; a night counts for a tile when an exposure is centred within $\nhRad^\circ$ of its centre). My copy of the catalogue has a file for \nhFileFourTiles{} of them (\nhFileFour~deg$^2$; the count includes files that hold no row, Section~\ref{sec:link_my}), and no linker of this paper read the others: some downloads failed and were not retried, and most tiles above a density cut were left out of the download. Section~\ref{sec:link_my} gives the numbers for the multi-year linker. The median depth of the exposures of a cell, expressed in $r$ for an object of solar colour, is \ctDepthMed{} (\ctDepthLo--\ctDepthHi{} for 80\% of the cells). The number of nights varies by two orders of magnitude from cell to cell, and the cadence within them is irregular; both matter more than the area (Section~\ref{sec:valid_budget}). Table~\ref{tab:data} summarises the data sets.

The figure also shows the sky that the Rubin Observatory had observed in the public record I could assemble: the science visits of 2025 and the cells with alerts in 2026. That is \ctAreaRubin~deg$^2$ and is a lower bound, because visits of 2026 that produced no alert are missing. Of the cells of Figure~\ref{fig:coverage} with at least \dnNmin{} nights, which cover more than the sky that was read (previous paragraph), \ctAreaBoth~deg$^2$ lie inside it and \ctAreaNotRubin~deg$^2$ (\ctNotRubinPct\%) outside. Of the \ctTierA{} tier-A objects, \ctAInRubin{} are inside and \ctAOutRubin{} outside. This is the state of 2026 November and will not last: the Rubin survey will cover most of this sky in its first years.

\begin{deluxetable}{llrr}
\tabletypesize{\footnotesize}
\tablecaption{Data sets\label{tab:data}}
\tablehead{\colhead{Data set} & \colhead{Years} & \colhead{Exposures} & \colhead{Area (deg$^2$)}}
\startdata
\nscdr{} (DECam and others) & \nscYears & \ctNscExp & \ctAreaNsc \\
\dtf{} (DECam, this work) & 2020--2026 & \ctDtExp & \ctAreaDt \\
Both & 2012--2026 & & \ctAreaFour \\
CFHT MegaCam (extension) & \nodata & \ctCfhtExp & \nodata \\
Subaru HSC (extension) & \nodata & per object & \nodata \\
\enddata
\tablecomments{Area: sky with at least \dnNmin{} nights; for \nscdr{} it is counted on the downloaded catalogue files and is neither the area of the catalogue nor the area searched: it is larger than the area of the tiles that were read (Section~\ref{sec:data_cov}). The exposure count of \nscdr{} includes its other cameras; that of CFHT is the public archive, not the frames used.}
\end{deluxetable}
\section{Linking}\label{sec:link}

Four linkers on DECam detections produced the catalogue, apart from the \ctChCfht{} objects of a linker on CFHT frames (Section~\ref{sec:data_ext}) and the channels added in this version (Section~\ref{sec:link_my_d25}); a further mode starts from the published astrometry of a designated object. All take as input the single-night detections that have no counterpart on other nights (the orphan detections of Section~\ref{sec:data}), in tiles of one square degree read with a margin ($\myTileBox^\circ$ on a side). Three words are used in one sense throughout: a \emph{link} is a set of detections that a linker returns, a \emph{candidate} is a link that passes the rule of its linker, and an \emph{object} is a candidate entered in the catalogue. A \emph{row} is one detection in a detection catalogue; a \emph{position} is what enters an orbit fit, whether measured on an image or taken from a row. Three of the four linkers ask for \vOneNights{} nights inside one season and one tile (Sections~\ref{sec:link_v1}--\ref{sec:link_shift}), the fourth for \myNmin{} nights over several years (Section~\ref{sec:link_my}). No ingredient is new: linking with tracklets and trees \citep{kubica2007,denneau2013}, links of distant objects across years in the Pan-STARRS catalogues \citep{weryk2016,holman2018ps1}, single detections gathered along orbits without tracklets \citep{moeyens2021}, shift-and-stack on pixels \citep{whidden2019} or on detection catalogues \citep{stetzler2025}, and the DES searches that the multi-year linker follows. Appendix~\ref{app:prereg} gives the frozen parameters and Section~\ref{sec:data_avail} the status of the code.

\begin{figure*}[t]
\centering\includegraphics[width=\textwidth]{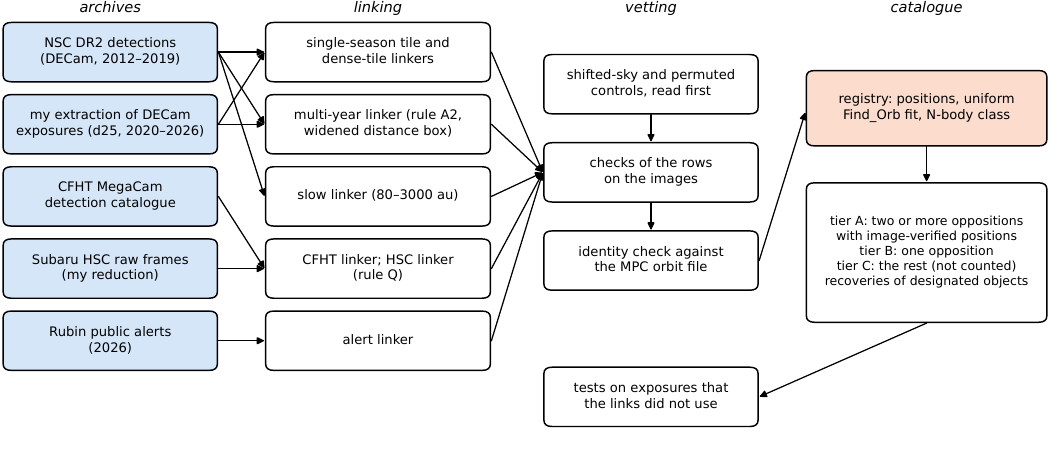}
\caption{Stages from the archives to the catalogue. Every linker hands its candidates to the same vetting: control runs on a shifted or permuted sky, read before the real run, checks of the rows on the images, and the identity check against the MPC orbit file. Objects that pass enter the registry, which holds a uniform orbit fit for each and an N-body class once the orbit has been integrated; the tiers of Section~\ref{sec:cat_tiers} follow from the number of oppositions with image-verified positions. Tier-A objects then go through the tests on exposures that their links did not use (Section~\ref{sec:valid_indep}).}\label{fig:pipeline_v3}
\end{figure*}

Figure~\ref{fig:pipeline_v3} shows the stages from the archives to the tiers of the catalogue.

\subsection{The single-season tile linker of version 1}\label{sec:link_v1}
The first \vOneObj{} objects came from the linker described in version 1: the tangent-plane model of \citet{bernstein2000}, a bound orbit at \vOneDmin--\vOneDmax~au, \vOneNights{} nights in one tile, and a step through trial distances, the device that HelioLinC applies to tracklets \citep{holman2018}. A mode for \vOneNearLo--\vOneNearHi~au (the near mode) and a pass over merged tiles (cross-tile linking) gave \vOneNearObj{} and \vOneCrossObj{} of the objects. Its candidates were tested one by one (Section~\ref{sec:valid_v1}), not by a control run of the whole chain.

\subsection{The dense-tile linker and the over-density mask}\label{sec:link_dense}
The dense-tile linker searches the same pattern at \dnDmin--\dnDmax~au, at a cost that stays bounded in tiles with hundreds of nights, with its own controls on every tile.

\emph{Rows.} Before linking, an over-density mask drops exposures and sky cells that hold too many rows (ghosts, halos, stray light, trails; Table~\ref{tab:param}). Each row carries one magnitude, built with a fixed colour term per band, and a per-axis error between \dnSigFloor{} and \dnSigCap; neither was calibrated on known objects, and the magnitude gates therefore depend on the colour of the object.

\emph{Links.} A row can start a link if no row of another night lies within \dnStat{} of it and if either another exposure of the same night confirms it or no other exposure of that night reaches it. It is paired with rows on the nearest nights within \dnDtmax~days, a third night gives $(\gamma,\dot\gamma)$, and a triplet that fits is extended night by night along a six-parameter fit. A link needs \dnNmin{} nights, and \dnNminu{} if none has a same-night pair.

\emph{Controls and candidate rule.} Every tile is linked on the real rows and on five controls, with the starting conditions recomputed: three copies in which each night is moved rigidly by $\dnShiftLo$--$\dnShiftHi''$ per axis, the tile rotated by half a turn, and time reversed about the opposition of the tile centre. A link is a candidate when no control link of its own tile is at least as good in nights, nights with a same-night pair, extra detections and rms (the own-tile rule). In \dtf{} the same must also hold against the control links of every tile of its density class; the tiles are divided into three such classes by their number of rows. The linker ran on the \prNscTiles{} tiles of \nscdr{} that had an orphan file in my copy of the catalogue and linked without error, and on \prDtTiles{} tiles of \dtf. That copy does not hold all of \nscdr{} (Section~\ref{sec:link_my}).

\subsection{A second pass on a shifted tile grid}\label{sec:link_shift}
Because an object near \vOneDmin~au crosses a tile within a season, I ran the dense-tile linker on \nscdr{} again on \sgBoxes{} boxes shifted by half a degree and merged the two grids, which about doubles the chance links. The condition for adopting it was written before the test: \sgRuleGain{} more known objects on the development half, at most \sgRuleFalse{} the expected false candidates on random tiles, and at most \sgRuleCpu{} times the CPU (Section~\ref{sec:valid_budget}). It was not run on \dtf.

\subsection{The multi-year linker}\label{sec:link_my}
More than half of the known objects in the footprint never have \dnNmin{} nights in one season and one tile, but many have \myNmin{} or more over the eight years of \nscdr. The multi-year linker searches for those. It copies in outline the DES searches \citep{bernardinelli2020,bernardinelli2022}: triplets of nights are grown along their orbit, and the survivors are cut on the number of distinct nights, on the arc and on the fit. The names of those cuts and the values \myNmin{} nights and six months are those of \citet{bernardinelli2022}; a limit of \myChiGrow{} on $\chinu$ is in both papers.

The data differ in what stands behind the transient list. DES built its transient catalogue by matching single-epoch detections to its own coadded catalogue, on uniformly processed exposures of one planned survey, with the detection efficiency of every exposure measured and synthetic objects injected. My input is the orphan rows of \nscdr: unplanned exposures of many programmes, a static-sky veto that depends on what other catalogues hold at that position, and no efficiency per exposure. The reality test differs more. With these cuts DES was left with \desCandEight{} candidates of eight or more nights and \desCandSeven{} of seven, and confirmed about eight hundred with a stack of the exposures that are not in the link \citep{bernardinelli2022}; at six nights its first search confirmed \desSixReal{} of \desSixAll{} linkages \citep{bernardinelli2020}. My linker had no such stage when it ran; a test of the same kind was made afterwards, on exposures that no link used, for all its candidates (Section~\ref{sec:valid_indep}). In its place the linker has a narrower distance range, a tighter fit, three far nights before any joint fit, a cut on a chance score, a control run of the whole chain on a shifted sky (Section~\ref{sec:valid_ctl}), which is my own device and untested by others, and measurements on the images (Section~\ref{sec:vet_img}).

\emph{Seeds.} A seed is a link of \mySeedN{} nights inside \mySeedWin~days made by the core of Section~\ref{sec:link_dense}, with no same-night pair required: three rows are a seed when they fit an orbit of the distance range within $\dnKtri\sigma$. Seeds are made on every tile and on two lattices of $\myBox^\circ$ boxes (each tile or box is a seed unit); the pilot strip of Section~\ref{sec:valid_held} gave \plSeeds~million. To bound the cost, a tile with more than \myDeepThr{} rows (\myDeepTiles{} of \myStripTiles{} in the pilot strip) is seeded with the single-season parameters and a box above that threshold is not seeded; I did not count these over the whole sky.

\emph{Growth.} For every seed the code fits six parameters and propagates an exact two-body barycentric orbit with light time, as seen from Cerro Tololo; fitted to $N$-body ephemerides of \myModN{} known objects over \nscYears{} this model leaves an rms of at most \myModMax. The fit carries Gaussian priors on $\gamma$ (centred on the distance range, with half its width) and on $\dot\gamma$; $\chi^2$ is that of the rows alone, $\nu$ is twice the number of rows minus six, and ``bound'' is a test on the fitted orbit. Rows of other nights inside the error ellipse of the seed orbit and within \myDmag~mag of the seed are tried one at a time in a four-row fit. Rows of different nights that give the same $(\gamma,\dot\gamma)$ are grouped, and a group goes to a joint fit only if it holds \myMinfar{} nights more than \myFarDt~days from the seed. With the joint orbit every night of the block is searched and rows are attached inside a gate of at most $\myRcapAtt''$. This is the idea of THOR with a seed orbit in place of a test orbit; Table~\ref{tab:param} gives every parameter.

\emph{Final rule.} A set of rows is a candidate if (i) it has detections on at least \myNmin{} distinct nights; (ii) the arc that remains when any one night is removed is longer than \myArc~days; (iii) its orbit is bound and lies inside \myDmin--\myDmax~au; (iv) the joint fit has $\chinu\leq\myChi$; (v) its chance score is $\Pch\leq\myPch$; (vi) it shares fewer than \myDupRows{} rows with any candidate of more nights. For the score, let
\begin{equation}\label{eq:pchance}
\lambda=\sum_{k}n_k A_k,\qquad \Pch=P\left[{\rm Poisson}(\lambda)\geq N-(n_{\rm seed}+\myMinfarDigit)\right],
\end{equation}
where the sum runs over the nights of the block on which the set has no member, $n_k$ is the surface density of rows within \myDmag~mag of the set near the prediction, $A_k$ the area of the attachment gate, $N$ the number of distinct nights and $n_{\rm seed}$ the number of seed nights. $\Pch$ is small when few rows of similar magnitude lie near the track on the nights where nothing was attached. It is a score, not the probability that the link is false: it has no factor for the millions of seeds tried, the \myMinfar{} far nights are exempt although they are picked in the widest gate, and its threshold was placed between the values of real objects and of control sets seen in tuning. It is not the false-positive rate of DES, which is the expected number of spurious transients in the search area of the last linked detection. The two thresholds have the same value, \myPch, and different meanings. The score rises with the number of covered nights, so the effective night floor is higher where coverage is best; I have not measured recall against coverage.

\emph{Time tags and frames.}\label{sec:link_time} \nscdr{} gives the start of the exposure as the time of a row, while the position belongs to mid-exposure: numbered asteroids agree with Horizons at the start plus half the exposure time to $\ttOffset$~s. The dense-tile and multi-year linkers use the row time uncorrected: an along-track error of \ttErrShort{} for a \ttExpShort~s exposure at $\ttRate''$ per hour and \ttErrLong{} for \ttExpLong~s, whose effect on $\chinu$ I have not measured. The \dtf{} rows and all image measurements are at mid-exposure. An audit of the \ttRows{} DECam positions of the \ttObjs{} objects of version~2 that have them found \ttStartRows{} at exposure start, in \ttStartObj{} objects, and \ttUnmatched{} that could not be matched to an exposure. Correcting the \ttStartRows{} moves $a$ by \ttDa~au for one object, a fifth of its uncertainty, and by 0.001~au or less for the others. The correction was applied on 2026 November 18, together with other corrections of time tags, and the fits of this version use the corrected times. Catalogue rows and image measurements are tied to different reference solutions, whose offsets I have not measured.

\emph{Freeze and production.} The rule was tuned on one half of a set of known objects and frozen before the other half was read (Section~\ref{sec:valid_held}); I call it A2. The same code then ran on the tiles of \nscdr{} for which I had downloaded the catalogue, and not on all of \nscdr: \pdFiles{} tile files in \pdBlocks{} blocks, cells of $30^\circ$ in right ascension by $10^\circ$ in declination that are subdivided where they hold too many rows. The blocks do not overlap, so a track that straddles an edge can be lost or found in part (Section~\ref{sec:valid_edge}). The real pass ran first, then three shifted-sky controls block by block, for about \pdCpu{} core-hours, most of them estimated (notes to Table~\ref{tab:ctl}).

\emph{Sky that was not searched.} The runner reads a tile only if its file is on my tile server. An audit made after the run, on \ctlReadDay, showed what that leaves out. \nscdr{} has DECam exposures on \myNmin{} or more nights in \nhAllTiles{} tiles of one square degree (nights counted as in Section~\ref{sec:data_cov}). Of these, \nhTiles{} (\nhArea~deg$^2$) lie inside the block grids of this linker, which include the pilot strip and the regions of the earlier runs. Of them, \nhSearched{} (\nhSearchedArea~deg$^2$; \nhSearchedPct\% of that area, \nhSearchedTilePct\% of those tiles) had a file with rows and were searched. A further \nhAcc{} tiles (\nhAccArea~deg$^2$) were never put on the tile server, by accident: for most of them download errors were stored as empty results and never retried; \nhAccEcl{} of these tiles have an ecliptic latitude under $\nhLat^\circ$. An earlier audit, on \nhAuditDay, had found failed downloads of this kind; the tiles were not downloaded again, and the production ran on the copy as it was. A further \nhEmpty{} tiles have a file with no row (one of them has \nhEmptyRows{} rows). Of them, \nhEmptyRa{} are in the column at right ascension $\nhRaCol^\circ$, where the catalogue stores part of the detections on the other side of $0^\circ$ from the one my query asked for; searched tiles near right ascension $0^\circ$ lack rows for the same reason. A count on \nhRnCounted{} of the \nhRnAll{} tile files nearest to it, not complete when this paper was written, found \nhRnAff{} files that together lack \nhRnMissPct\% of their rows (\nhRnMiss{} of \nhRnShould{} million; from under 1\% to all the rows of a file). Of these files, \nhRnSearched{} are searched tiles, \nhEmptyRa{} are the files with no row just counted and \nhRnFew{} are tiles with four to six nights. The numbers of files and of missing rows can only grow when the count is completed; the percentage can move either way. A row missing from one file can be in the file of the next column, where the two overlap. For the files counted, at least \nhRnBlind~deg$^2$ have no row in any file, and the rows stored with a negative right ascension are in no file over at most \nhRnBlindUp~deg$^2$. Another \nhDense{} tiles (\nhDenseArea~deg$^2$), \nhDenseGalPct\% of them at a Galactic latitude under $\nhLat^\circ$, were left out by a density cut: the download list, made for the linker of version 1, omits the tiles with a predicted density of more than \nhDensCut{} detections per square degree; \nhDenseSearched{} tiles above that value had a file all the same and are among the searched ones. The \nhOut{} tiles outside the block grids are all above that cut and have no file. The cut is not in the file written before the production, and that file does not say which tiles had a file (Appendix~\ref{app:prereg}). The last \nhRun{} tiles (\nhRunArea~deg$^2$) had no file either. The search described here is the one that stood on \ctlReadDay, and the tiles without a file on that date are not part of it; a later search of those tiles, with the same multi-year linker and its six-night level, is the source of \vtHolesN{} objects of the catalogue (\vtHolesA{} of them in tier~A), which are counted with their channels (Section~\ref{sec:cat_channels}). The other linkers on \nscdr{} read the same files (Sections~\ref{sec:link_v1}--\ref{sec:link_shift}), so I take the tiles without a file as unsearched by them too; I did not check whether the linker of version 1, which ran inside its download job, read a tile whose file is no longer there. Section~\ref{sec:valid_edge} counts the known objects in those tiles.

\subsection{Linkers added after version 2}\label{sec:link_new}
\emph{The multi-year linker on my 2020--2026 extraction.}\label{sec:link_my_d25} Rule A2 (Section~\ref{sec:link_my}) was run unchanged on the orphan detections of \dtf, which version~2 had searched only one season at a time. The seeds are those of the whole-sky seed run, and the growth ran in \rvDtBlocks{} overlapping blocks, each tile grown in exactly one of them, with a shifted-sky control of every block. Vetting has two steps. First, unused nights are searched for in both DECam catalogues along the orbit of each candidate, within its uncertainty, and a candidate goes on only if at least three such nights fit with it at an rms of $0\farcs12$ or less; then every position is measured on its own exposure. In the first round of this vetting the shifted-sky candidates were mixed blind with the real ones under opaque names: \rvMdStageReal{} of \rvMdStageRealN{} real candidates passed, and \rvMdStageCtl{} of \rvMdStageCtlN{} control candidates. A candidate is registered when its link positions pass on their own exposures in at least two oppositions, at least one unused night is confirmed on its own exposures, and a fresh fit keeps every position at an rms of $0\farcs12$ or less. For \rvMdStack{} of the objects of this channel the unused night is confirmed only by the forced photometry of several exposures together, not on a single one; their records say so, and I accepted this weaker confirmation as a deliberate choice.

\emph{A six-night level of the multi-year linker.}\label{sec:link_six} On the second pass of the multi-year linker on \nscdr{} I added a level for sets of six nights, one fewer than rule A2 asks, with its own rule frozen before the real sets were read. A six-night set becomes a candidate only if it then passes the image vetting of Section~\ref{sec:vet_img}. Its shifted-sky controls gave \rvSixCtlP{} and \rvSixCtlE{} sets in the two groups of blocks against limits of \rvSixCtlLim, and it recovered \rvSixKnown{} designated objects of the held-out half. The same level was run on the tiles of \nscdr{} that version~2 had never read (Section~\ref{sec:link_my}), once they had been downloaded. Most objects of this level rest on catalogue positions, which the tier rule of Section~\ref{sec:cat_tiers} does not count as image-verified.

\emph{A linker on Subaru HSC frames.}\label{sec:link_hsc} I reduce raw Hyper Suprime-Cam frames myself (Section~\ref{sec:data_ext}). HSC fields are often observed on three or more nights within a few weeks, and a linker of three-night groups on my catalogues of those frames, with an rms cut fixed on the development half, gave no group in its held-out shifted-sky controls and recovered \rvHscQRec{} of the held-out synthetic objects at 30--80~au and $r=22.5$--24. A group of one season is not an object of the catalogue: two groups in different seasons must fit one orbit, and at least two nights that neither group used must be confirmed on the images with a clean static check. During this work three defects of the linker were found and corrected under frozen amendments: the fit of three nights stalled on spans of about four weeks, the grid of trial distances for pairs of nights was too coarse beyond 300~au, and the merging of tracklets discarded true ones. The corrections raised the synthetic recall in both seasons and left the controls at zero.

\emph{Difference images of a crowded field.}\label{sec:link_crowded} In crowded fields the catalogues lose faint moving objects among the stars (Section~\ref{sec:valid_inj}). In one pilot I subtracted reference images from DECam exposures of a crowded field and linked the residual detections over the nights. On the direct images the row test of Section~\ref{sec:vet} cannot work there, because its control places fall on stars; for this pilot I accepted a test on differences of pairs of exposures instead, with the same thresholds on signal-to-noise ratio, shape, distance from the prediction and control places, and a static check on other epochs. This is an exception to the image rule of the catalogue, and I state it as such. A second version of the pilot did not reach a fainter depth and was stopped (Section~\ref{sec:null}).

\subsection{Seeded recovery of designated objects}\label{sec:link_seed}
The growth stage alone can follow a designated object with a poor orbit: the seed is then its published astrometry, downloaded from the Minor Planet Center by designation. Six settings differ from the blind linker, all fixed before the held-out test of Section~\ref{sec:valid_seed} (Table~\ref{tab:param}). A recovery is proposed when a set holds at least \lrN{} of my nights, the increase of $\chi^2$ over the fit of the seed alone, divided by two less than twice the number of my rows, is at most \lrDchi, and $P^*\leq\lrP$, where
\begin{equation}\label{eq:pstar}
P^*=1-\exp\left\{-T\,P\left[{\rm Poisson}(\lambda)\geq n-\lrPstarOff\right]\right\},
\end{equation}
$n$ is the number of my nights in the set, $\lambda$ is as in Equation~(\ref{eq:pchance}) with the magnitude window of this mode (Table~\ref{tab:param}), and $T$ is the number of sets the same seed returned, which counts outputs and not trials. A seed gives one proposal, or none if two passing sets share no row. In the controls my rows are shifted night by night and the seed is left in place. Such a control contains no other real object near the corridor, so it cannot produce the wrong match that a short seed is most exposed to (Section~\ref{sec:valid_seed}).
\section{Vetting}\label{sec:vet}

``Verified on images'' does not mean the same thing in every channel; the machine-readable catalogue gives for each object the positions measured on images, the positions taken from a catalogue, and the oppositions that hold image-verified positions. The check against designated objects is described with its results in Section~\ref{sec:cat_known}, and the tests that do not use the link of an object in Section~\ref{sec:valid_indep}.

\subsection{Rules of version 1}\label{sec:vet_v1}
The first \vOneObj{} objects were vetted as described in version 1: five vetoes on the link, then a forced PSF measurement on at least \vOneNewNights{} nights not used in the link, judged against off-source control positions. Not all of them went through one uniform chain; the evidence table lists what each went through.

\subsection{Image rule for the multi-year candidates}\label{sec:vet_img}
Every candidate of the multi-year linker that is not already a known object is checked on the images before it is registered. On each night the exposure must show a point source at the place of the detection: a PSF fit within \imRad{} with a signal-to-noise ratio of at least \imSnr{} and a width between \imShapeLo{} and \imShapeHi{} times that of the stars. The same sky position must be empty on \imOther{} exposures of other years, for which the rule sets no depth. An object is accepted when the rule holds on every night that has an image, a Find\_Orb fit \citep{findorb} uses every row kept, and the rows span at least \imOpp{} oppositions. For the candidates of version~2 the registered positions are the image measurements; \vtMyAddedC{} of the \vtMyAdded{} candidates registered since on \nscdr{} keep catalogue positions in my files and are in tier~C (Section~\ref{sec:cat_channels}).

\emph{Who applied it.} The measurements are made by a script. The verdict on each object, including the decision to drop a night, was given by an AI coding agent working from a written brief that states the rule and asks that every judgement call be listed. It was not blind: every item was a real-sky proposal and no false link was mixed in.

\emph{What it tests.} The rule tests the detections and not the link. It removes artefacts and static sources; a chance link of real moving sources, such as main-belt asteroids, would pass it night by night. Its outcome is the one expected whether the links are real or not: of the \pdProp{} proposals of the production blocks, \pdReal{} passed, \pdVetKnown{} was identified as a comet and none was rejected. I have not put control sets through this rule. Their rows are real detections at their true image positions, so I expect the rule to accept most of them, and I do not use it as evidence that a link is real. The evidence that does not depend on the link is the test on exposures that no link used (Section~\ref{sec:valid_indep}).

\emph{Judgement calls.} An object should keep \myNmin{} nights after vetting. GCT0243 and GCT0271 were accepted with six: in each, one night was dropped because an exposure of another year shows a faint static source at that place (under the predicted position for GCT0243; an extended source $0\farcs57$ from one position for GCT0271). They span \mySixOppA{} and \mySixOppB{} oppositions with image-verified positions and a clean fit, stay in the catalogue and are not counted in the yield of the rule. Both have since been measured on exposures that their links did not use (Section~\ref{sec:valid_indep}): GCT0243 is detected at S/N$\,\geq5$ on all \ctUnuSixAExp{} of them, on \ctUnuSixANights{} nights that hold no position of the link, and GCT0271 on all \ctUnuSixBExp, on \ctUnuSixBNights{} such nights; none of the 32 control tracks of either object reaches two such nights. That test has no static veto. In at least \vetMisN{} proposals some rows of my copy of \nscdr{} name an exposure pointed one to three degrees away from the detection; the object was then measured at the orbit prediction on the correct, neighbouring exposures. Whether this comes from the catalogue or from my join, and how many rows of the linker input carry it, I have not measured.

\subsection{Image checks of the single-season candidates}\label{sec:vet_dt}
Links of the single-season channel were checked by a scripted chain that re-measures each row on its image and looks for the object on exposures the link did not use. The verdict (real, possible or not confirmed) is computed by the script, and recomputed from the stored measurements it reproduces \dtVerRep{} of \dtVerRepN{} recorded verdicts; the definitions are in the notes to Table~\ref{tab:dtfunnel}.

The \dtLinks{} checked links mix three selections, which the table separates: the production rule on \dtf{} (Section~\ref{sec:link_dense}); a second pass on dense zones of \dtf, in which a link had to beat the controls of its own tile only; and the same method on \nscdr, also with the own-tile rule alone. Over the whole survey the production rule expects about four chance links (\prDtSixCh{} of six or more nights, Section~\ref{sec:valid_inj}, and \prDtFiveCh{} of five). That figure does not describe what was sent to the images: the run was made in several passes, each with its own pool of controls, and counted pass by pass the chance links expected are \dtChanceLo--\dtChanceHi. Of its \dtNcProd{} links not confirmed, at most \dtNcMeasFail{} have measured rows that fail, which is of the same order. The other two selections are at the level of chance, the second by its own controls and the third because most of its links have four nights, and they hold most of the links not confirmed. Many frames could not be measured because of defects of the chain that were fixed later; the verdicts were not redone.

\emph{Control links through the chain.} To measure how often the chain accepts a chance link, \gsN{} control links of the same linker on \dtf{} were sent through it with the same thresholds. They are links of five to seven nights on shifted copies of the same rows, in three groups that I do not pool. The first group is the control links that pass the production rule: the shifted controls hold \gsProdDrawn, and \gsProd{} were tested (two share a file name, and one was overwritten). None was called real and \gsPoss{} was called possible, so in this group the fraction called real is below \gsProdRealUp\% and the fraction called real or possible below \gsProdOkUp\% (two-sided \ciLevel\% Clopper--Pearson limits, here and below). The other two groups, \gsOwnN{} links of five nights and \gsOwnN{} of six or more, pass the own-tile rule alone, which is weaker than the production rule; none was called real or possible (below \gsOwnUp\% in each group). The links were regenerated with the frozen linker, which reproduced the control tables of the production run, and every predicted position was moved back to its place on the real image. Of the rows of the \gsN{} links that could be measured, \gsRowPct\% are verified, against \gsRealRowPct\% for the links of the production rule called real. With \dtChanceLo{} to \dtChanceHi{} chance links among the \dtFSentA{} of the production rule, the first group puts the number expected to be called real or possible at \gsFalseLo--\gsFalseHi, and below \gsFalseUp{} at \ciLevel\%; the number expected to be called real is below \gsFalseRealUp. The limits are wide because the group has \gsProd{} links. The sample was drawn with a fixed seed and without a rule written beforehand. It does not cover the other two selections, whose control links were not tested, nor the extension into other seasons described below.

\emph{From links to objects.} The \dtFRealA{} links of the production rule that were called real are not \dtFRealA{} objects. Of the \dtRealTraced{} whose verdict I could recompute, \dtRealInCat{} are links of \dtRealObj{} catalogue objects, \dtRealObjNew{} undesignated and \dtRealObjDes{} designated: an object linked in more than one season or tile has more than one link. The other \dtRealNotReg{} were not entered in the catalogue before it was frozen. I have no record that any of them was rejected by a later check, so they are not a measure of false acceptance. Counted by the first link of each object and not by any of its links, the verdict was real for \dtRegReal{} of the \dtNew{} undesignated catalogue objects of this channel, possible for \dtRegPoss{} and not confirmed for \dtRegNot.

\emph{Extension into other seasons.} Links that were possible or not confirmed entered the catalogue only after a search for the same object in other seasons. The orbit of the link and its clones are propagated to the exposures of \dtf{} and of \nscdr{} on which the cloud of clones is small enough to search, the detections inside the cloud are collected, and each new night is measured on its images. For \nscdr{} this step fetched from the Astro Data Lab the tiles of my tile list that my copy lacked, so it is not confined to the downloaded tiles; a tile whose fetch failed was treated as missing. For \dtf{} it read the tiles that had been extracted at that time. An object was accepted when at least \dtExtNights{} new nights were measured on images and kept by a Find\_Orb fit with an rms of at most \dtExtRms, and when the same search with the cloud rotated by $\pm3^\circ$ and $\pm6^\circ$ returned fewer new nights than the real one. In the first batch of \dtExtFirst{} links, the rotated clouds returned at most two new nights for the objects that were accepted, which had 2 to 21; one or two nights of an extension can therefore be chance. The rate at which the whole step accepts a false extension was not measured: the only measured rate for a step of this kind is that of version 1 (Section~\ref{sec:valid_indep}). A few links were instead checked a second time with the corrected chain. Of the \dtNew{} objects, \dtTierB{} have all their positions in one opposition and should be read as single-opposition candidates.

\begin{deluxetable*}{lcccc}
\tabletypesize{\footnotesize}
\tablecaption{Image checks of the single-season channel, by selection\label{tab:dtfunnel}}
\tablehead{\colhead{} & \colhead{Production rule,} & \colhead{Own-tile rule,} & \colhead{Same method,} & \colhead{All} \\
\colhead{} & \colhead{\dtf} & \colhead{dense zones} & \colhead{\nscdr} & \colhead{}}
\startdata
Links sent to the images & \dtFSentA & \dtFSentB & \dtFSentC & \dtLinks \\
Real & \dtFRealA & \dtFRealB & \dtFRealC & \dtReal \\
Possible & \dtFPossA & \dtFPossB & \dtFPossC & \dtPoss \\
Not confirmed & \dtNcProd & \dtFNotB & \dtFNotC & \dtNot \\
\quad fewer than two measurable nights & \dtFUnmA & \dtFUnmB & \dtFUnmC & \dtFUnmAll \\
\quad two verified nights only & \dtFTwoA & \dtFTwoB & \dtFTwoC & \dtFTwoAll \\
\quad rows measured, test failed & \dtNcMeasFail & \dtFFailB & \dtFFailC & \dtFFailAll \\
No output & \dtFNoneA & \dtFNoneB & \dtFNoneC & \dtNoneN \\
\enddata
\tablecomments{Fewer than two measurable nights: fewer than two nights with a frame that could be measured. Two verified nights only: no third verified night and no strong extra. A catalogue row is verified when a PSF fit on its own exposure, at the position predicted by the other rows, has a signal-to-noise ratio of at least \imSnr, a width between \imShapeLo{} and \imShapeHi{} times that of the stars, no static catalogue source under it, and lies within $2''$ to $6''$ of the prediction (three times its uncertainty) and within $\dtRowTol''$ of the row. A night is verified when it holds a verified row. An extra is an exposure away from every night of the link on which the orbit predicts a detection; it is detected when it passes the same test within 1.5~mag of the predicted magnitude, and strong when at most \dtStrongPct\% of \dtStrongCtl{} control positions on that frame pass the test. A link is possible when at least \dtVerA{} nights are verified and either \dtVerB{} nights are verified from more than \dtVerA{} rows or one extra is strong; it is real when at least \dtVerA{} nights are verified and at least two strong extras fall on \dtVerA{} or more nights. The production rule gave \dtFCand{} candidate entries; \dtFKnown{} were known objects and \dtFListed{} already listed, and the rest went to the images. Links later matched to objects designated since: \dtFDesig, all called real (\dtFDesigReal) or possible. Of the \dtNew{} undesignated catalogue objects, one more had no first verdict and \dtRegPilot{} come from an earlier pilot. The catalogue also holds \ctDtDes{} designated objects of this channel.}
\end{deluxetable*}
\section{Validation}\label{sec:valid}

Proportions carry \ciLevel\% Wilson intervals \citep{wilson1927}, counts exact Poisson intervals \citep{garwood1936}, and differences of proportions the interval of \citet{newcombe1998} for independent samples, wider than the paired one that would apply. A ``development'' set was looked at while parameters were chosen; a ``held-out'' or ``test'' set was read after the rule and its pass criteria were written down, once for each variant fixed beforehand, with the exceptions listed in Appendix~\ref{app:prereg}. The runs were executed by an AI coding agent from the frozen files. Freezes and reads lie hours apart within single days. No third party holds a timestamp of the rule files (the position files of 2026 November 25 are timestamped: Section~\ref{sec:data_avail}). Recall on known objects is not completeness for unknown ones.

\subsection{Known objects: truth set and loss budget}\label{sec:valid_budget}
\emph{Truth set.} Of the \tsAll{} trans-Neptunian objects and Centaurs of the JPL Small-Body Database I keep the \tsN{} with $e<1$, a data arc of at least \tsArc~days, at least \tsObs{} observations and a condition code of at most \tsCode, split by the parity of a hash of the designation (development \tsDev, test \tsTest). Their positions come from my own $N$-body integration \citep{park2021}, which agrees with JPL Horizons \citep{giorgini1996} to \tsEphMed{} (median over objects of the largest difference). An orbit is secure when its $\tsSecSig\sigma$ uncertainty in Horizons stays below $\tsSecTol''$ in the years a catalogue covers: \tsSecNsc{} objects for \nscdr{} and \tsSecDt{} for \dtf, \tsSecPctNsc\% and \tsSecPctDt\% of the multi-opposition objects, so every recall below is measured on the best-observed third. An object is recovered when its own rows (orphan rows within \tsRowTol{} of the ephemeris), by row identifier, make at least half of a candidate. This one rule is used for every channel; the pipeline's own label for known objects, which is wrong at some epoch for \lbKnownRuleObj\% of these objects, is quoted only where it is named.

\emph{The recall of version 1 is withdrawn.} Version 1 gave an end-to-end recall of \vOneRecall\% on a faulty denominator: \vOneShortPct\% of the objects counted as detectable have arcs shorter than one year and cannot be predicted to the $\vOneMatchTol''$ used for the match. For secure orbits the same pipeline recovers \vOneSecRec\% (\vOneSecRecCI; development half, its own label).

\emph{Loss budget.} Table~\ref{tab:budget} assigns every secure object at \tsDmin--\tsDmax~au that falls on a live CCD to the first stage at which the dense-tile linker loses it. A rule that asks for \dnNmin{} nights in one season and one tile cannot reach \lbNscCadPct\% of the objects in \nscdr{} and \lbDtCadPct\% in \dtf, whatever the detection: a property of the rule, not of the sky. After that the losses are the catalogue in \nscdr{} and the control rule in \dtf. The catalogue loses many objects because, on nights with a single exposure, \nscdr{} holds a detection of a known object in about \lbDetPct\% of the exposures only when the object is at least a magnitude brighter than the stated depth of the exposure. Of the objects the rule can reach, the chain recovers \lbNscCeilPct\% and \lbDtCeilPct\% on the development half and \lbNscCeilTestPct\% and \lbDtCeilTestPct\% on the test half; of all objects, \lbNscTestPct\% and \lbDtTestPct\% on the test half.

\begin{deluxetable}{lcc}
\tablecaption{First loss of secure known objects in single-season linking (development half)\label{tab:budget}}
\tablehead{\colhead{First stage at which the object is lost} & \colhead{\nscdr{} (\lbNscN{} objects)} & \colhead{\dtf{} (\lbDtN{} objects)}}
\startdata
Rule cannot apply: no pattern of nights even with perfect detection & \lbNscCad{} (\lbNscCadPct\%) & \lbDtCad{} (\lbDtCadPct\%) \\
Rule cannot apply: pattern only across a tile boundary & \lbNscTile{} (\lbNscTilePct\%) & \lbDtTile{} (\lbDtTilePct\%) \\
Not detected in the catalogue on enough nights & \lbNscDet{} (\lbNscDetPct\%) & \lbDtDet{} (\lbDtDetPct\%) \\
No orphan tile in the local copy & \lbNscNoTile{} (\lbNscNoTilePct\%) & \lbDtNoTile \\
Orphan selection (static veto, merged with another night) & \lbNscOrph{} (\lbNscOrphPct\%) & \lbDtOrph{} (\lbDtOrphPct\%) \\
Loader cuts (counterpart flags, magnitude, shape) & \lbNscLoad{} (\lbNscLoadPct\%) & \lbDtLoad{} (\lbDtLoadPct\%) \\
Rows linkable, no candidate credited by row identifiers & \lbNscLinkR{} (\lbNscLinkRPct\%) & \lbDtLink{} (\lbDtLinkPct\%) \\
Linked, not a candidate (control rule) & \nodata & \lbDtRule{} (\lbDtRulePct\%) \\
Recovered & \lbNscRec{} (\lbNscRecPct\%) & \lbDtRec{} (\lbDtRecPct\%) \\
\hline
Recovered, of those the rule can reach & \lbNscCeilNum/\lbNscCeilDen{} (\lbNscCeilPct\%) & \lbDtCeilNum/\lbDtCeilDen{} (\lbDtCeilPct\%) \\
\enddata
\tablecomments{Each object is counted once. For \nscdr{} only the candidates were stored, so links that are not candidates cannot be separated; its seventh row includes \lbNscLabOnly{} objects that the pipeline's own label credits and row identifiers do not. Test half: recovered \lbNscTestNum{} of \lbNscTestDen{} and \lbDtTestNum{} of \lbDtTestDen; of those the rule can reach, \lbNscCeilTestNum{} of \lbNscCeilTestDen{} and \lbDtCeilTestNum{} of \lbDtCeilTestDen. \ciLevel\% intervals: first row \lbNscCadCI\% and \lbDtCadCI\%; not detected \lbNscDetCI\% and \lbDtDetCI\%; loader cuts (\nscdr) \lbNscLoadCI\%; control rule (\dtf) \lbDtRuleCI\%; recovered \lbNscRecCI\% and \lbDtRecCI\%; last row \lbNscCeilCI\% and \lbDtCeilCI\%. The denominator of the last row is the objects that have the pattern of nights in one tile on exposures predicted to be deep enough for them. The first two rows leave more, \lbNscAfterCad{} and \lbDtAfterCad: the \lbNscReachRest{} and \lbDtReachRest{} objects of the difference have no such pattern and were detected all the same on shallower exposures (\lbNscReachRec{} and \lbDtReachRec{} of them are recovered), and they are in neither the numerator nor the denominator of the last row.}
\end{deluxetable}

\emph{Second grid.} With the shifted grid the development half goes from \sgDevBase{} to \sgDevBoth{} of \sgDevDen{} objects and the test half from \sgTestBase{} to \sgTestBoth{} of \sgTestDen{} (\sgTestBothPct\%, \sgTestBothCI), none lost. The condition on false candidates was written for random tiles, where the counts are too small to tell one from two; on the development tiles the ratio is \sgFalseDev, at or above the limit of \sgRuleFalse. I adopted the second grid on that basis.

\subsection{Held-out test of the multi-year linker}\label{sec:valid_held}
\emph{Design.} The truth is the table of the six-year DES search \citep{bernardinelli2022} as distributed by VizieR \citep{ochsenbein2000}, which has \myTruthRows{} rows. This truth is selected by the rule under test: DES linked these objects with \myNmin{} nights and six months of arc, largely in exposures that \nscdr{} also holds. The test measures what is lost in going from the DES transient catalogue to my orphan rows on the DES cadence, nothing more. The data are the orphan rows in a strip of \myStripTiles{} tiles inside the DES footprint. Objects were split by the hash of Section~\ref{sec:valid_budget}, except that objects of the test half of that section went to the held-out half whatever their parity. Four criteria were fixed (notes to Table~\ref{tab:heldout}), with every held-out object of the magnitude range in the denominator.

\emph{Result.} The held-out half holds \hoN{} objects and was read once. All four criteria pass by the letter (Table~\ref{tab:heldout}): (a) \hoANum{} of \hoADen{} objects, \hoAPct\% (\hoACI), against \hoAListPct\% for the lists, a gain of \hoGain{} points (\hoGainCI); (b) \hoBNum{} of \hoBDen, \hoBPct\% (\hoBCI); (c) no candidate in any control; (d) \plCpu{} core-hours. 

\begin{deluxetable}{lcccc}
\tablecaption{Held-out DES objects recovered in the pilot strip, by magnitude\label{tab:heldout}}
\tablehead{\colhead{DES $r$} & \colhead{Objects} & \colhead{Multi-year linker} & \colhead{\ciLevel\% interval} & \colhead{Single-season lists}}
\startdata
\hoBinALab & \hoBinAN & \hoBinAMy{} (\hoBinAPct\%) & \hoBinACI\% & \hoBinAList{} (\hoBinAListPct\%) \\
\hoBinBLab & \hoBinBN & \hoBinBMy{} (\hoBinBPct\%) & \hoBinBCI\% & \hoBinBList{} (\hoBinBListPct\%) \\
\hoBinCLab & \hoBinCN & \hoBinCMy{} (\hoBinCPct\%) & \hoBinCCI\% & \hoBinCList{} (\hoBinCListPct\%) \\
\hoBinDLab & \hoBinDN & \hoBinDMy{} (\hoBinDPct\%) & \hoBinDCI\% & \hoBinDList{} (\hoBinDListPct\%) \\
\hoBinELab & \hoBinEN & \hoBinEMy & \nodata & \hoBinEList \\
All & \hoN & \hoAllMy{} (\hoAllPct\%) & \hoAllCI\% & \hoAllList{} (\hoAllListPct\%) \\
\enddata
\tablecomments{Rule A2, frozen before the read; one read. Criteria fixed beforehand: (a) recall at DES $r<\hoMagA$ of at least \hoThrA\%, and \hoThrGain{} points above that of the single-season candidate lists; (b) recall at $r<\hoMagB$ of at least \hoThrB\%; (c) a mean number of candidates in three shifted-sky controls of at most \hoThrCtl\% of the real candidates that are not known objects; (d) at most \plCpuLim{} core-hours. The single-season lists are the candidates of the first grid of the dense-tile linker on the same tiles, credited by the same rule. Of the \hoMiss{} objects missed at $r<\hoMagA$, \hoMissNights{} have fewer than \myNmin{} nights of usable rows, \hoMissRule{} lack the pattern the rule asks for, and \hoMissGrow{} had a seed that did not grow or was cut.}
\end{deluxetable}

\emph{How far the result carries.} Criterion (b) passes by one object. The override put \hoOldTest{} objects into the held-out half, \hoForcedB{} of them brighter than $r=\hoMagB$; without them the recall is \hoNatBNum{} of \hoNatBDen{} at $r<\hoMagB$ (\hoNatBPct\%, \hoNatBCI), again a pass by one object, and \hoNatANum{} of \hoNatADen{} at $r<\hoMagA$ (\hoNatAPct\%). Criterion (c) has no power: the real candidates that are not known objects are so few that the threshold is a fraction of one candidate per control, below the upper limit of \plCtlUp{} that three empty controls allow. The held-out half is the easier half (the lists find \hoEasyHeld\% of it and \hoEasyDev\% of the development half, on which the frozen rule finds \myTuneEnd{} of \myTuneDen), so the gain over the single-season lists is the number to carry to other fields, not the recall.

\emph{Outside the DES footprint.} I tested the frozen rule again on \exRegions{} regions of $\exBoxRa^\circ\times\exBoxDec^\circ$ outside the DES footprint, chosen by their number of secure objects; they have more nights than the typical sky outside DES (a median of \exNightsBox{} per box against \exNightsTyp). In one of them \exHoleTiles{} tiles with \myNmin{} or more nights had no file and were not searched (Section~\ref{sec:link_my}); the objects of the test are chosen by their ephemerides, so an object there stays in the denominator. The criteria were written before any tile was fetched, with one correction recorded in Appendix~\ref{app:prereg}. Of \exN{} secure objects at $V<\exMag$ the linker recovers \exMyPct\% (\exMyCI) against \exListPct\% for the lists of both grids; the controls are empty and criterion (c) again has no power. So \hoAPct\% is the recall of bright DES objects in one DES strip, against \exMyPct\% in these regions and \raMyPct\% of all secure objects in \nscdr{} (Table~\ref{tab:recall}); most misses lack the nights, and given the nights recall is \edBinAPct\% at seven or eight (Section~\ref{sec:valid_edge}). Section~\ref{sec:disc_des} examines why DES, which is deeper in its footprint, does not list the undesignated candidates that fall there.

\subsection{Shifted-sky controls and chance alignments}\label{sec:valid_ctl}
A control is the whole pipeline, seeds included, run on a copy of the catalogue in which every night is moved rigidly by $\dnShiftLo$--$\dnShiftHi''$ per axis. The shift destroys every real track and keeps the density of the rows of each night and all structure within a night. It does not keep correlations between nights below the scale of the shift: rows that recur at one sky position (static sources, artefacts tied to bright stars) are moved apart, so chance links through such rows are under-counted, most of all slow single-season links. Links that mix real and chance rows cannot occur at all, because a control holds no real track. The control bounds pure coincidences only: in the seeded mode two of fourteen three-night sets were wrong where the controls predicted none (Section~\ref{sec:valid_seed}). I have not measured the fraction of foreign rows in the candidates of recovered known objects, which the credit rule allows up to one half, nor run a control that keeps real seeds and moves whole seasons.

\emph{Pilot strip.} The three controls returned no candidate (Table~\ref{tab:ctl}). Before the last two cuts of the rule (on $\chinu$ and on $\Pch$) they hold about two hundred sets each, so the zero rests entirely on those cuts. They were tuned on the development half and on a quarter of one control, where the smallest $\Pch$ of a control set was \myTuneCtlPmin{} (a set that also fails the first cut) and the largest of a development object \myTuneDevP; the threshold was placed between the two after seeing both.

\begin{figure}
\centering\includegraphics[width=0.47\linewidth]{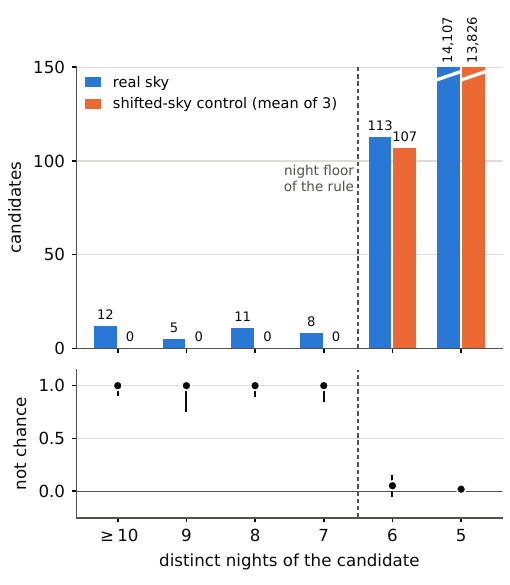}\hfill\includegraphics[width=0.5\linewidth]{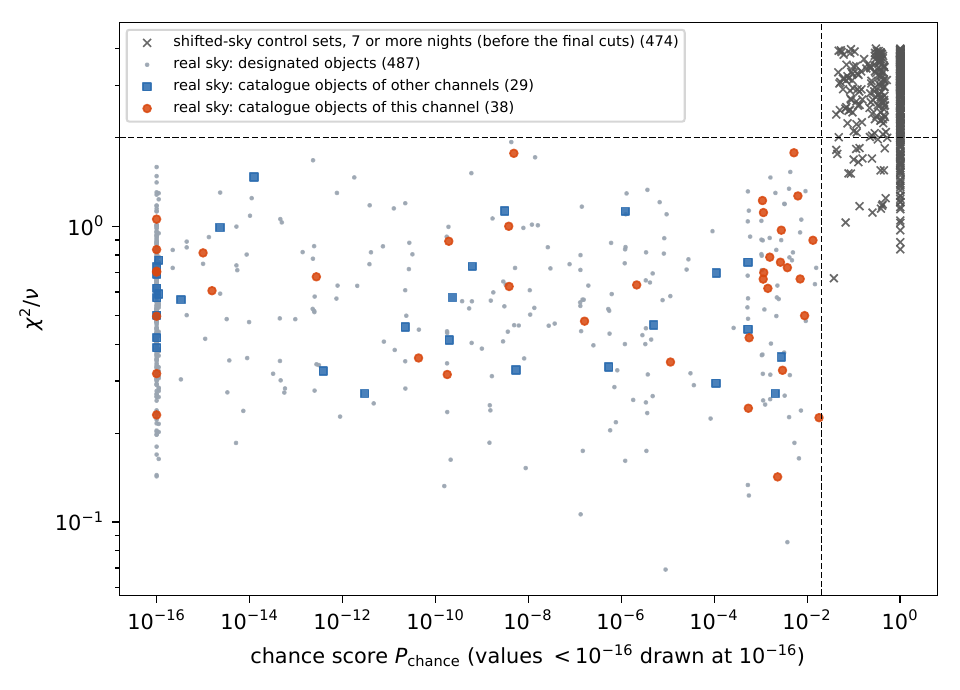}
\caption{Left: real and control candidates of the multi-year linker by number of distinct nights with the night floor and the final cuts removed (Part~B of Table~\ref{tab:purity}); bottom panel, the fraction not explained by chance alignments (``not chance''). From \myNmin{} nights up the controls are empty in these regions; below, the real counts exceed the controls by a few per cent at most. Right: $\chinu$ against the chance score $\Pch$ for the \pdCands{} real candidates and for the \pdCtlSets{} control sets of \myNmin{} or more nights before the two final cuts (dashed lines). The legend separates the real candidates whose detections belong to catalogue objects of this channel (38 candidates: 36 of its \myChanNew{} objects and two objects first found on the shifted grid), those that belong to catalogue objects of other channels (29), and the others, which are designated objects (487); this grouping is by detections and is not that of Table~\ref{tab:ctl}. Real sets are stored only after the cuts, so those that the cuts removed cannot be shown.\label{fig:purity}\label{fig:chipch}}
\end{figure}

\emph{Production.} Of the \pdCands{} candidates, \pdBlockCands{} come from the \pdBlocks{} production blocks themselves and the rest from the earlier runs (Table~\ref{tab:ctl}). The three controls returned no candidate. They are complete on \pdCtlBlocks{} of \pdBlocks{} blocks; the other \pdCtlMissing{} hold no real candidate. Before the last two cuts the controls hold \pdCtlSets{} sets, all with seven or eight nights and \pdCtlHeavyPct\% of them in two blocks, which hold no undesignated candidate. None has $\Pch\leq\myPch$ (Figure~\ref{fig:chipch}), but the margin is thin: the smallest control value is \pdCtlPmin, for a seven-night set that passes the first cut easily, and the largest real value is \pdRealPmax, for a seven-night undesignated candidate. Extrapolated to \myPch, the tail of the control sets that pass the first cut gives about one set in the three controls together, that is a few tenths of a false candidate per pass. Against \pdReal{} undesignated candidates that is not negligible.

\begin{deluxetable}{lcccc}
\tablecaption{The multi-year linker on the real sky and on three shifted-sky controls\label{tab:ctl}}
\tablehead{\colhead{} & \colhead{Real sky} & \colhead{Control 1} & \colhead{Control 2} & \colhead{Control 3}}
\startdata
\multicolumn{5}{l}{Pilot strip} \\
Seeds & \plSeedsN & \plSeedsCa & \plSeedsCb & \plSeedsCc \\
Four-row fits (millions) & \plQuad & \plQuadCa & \plQuadCb & \plQuadCc \\
Joint fits (millions) & \plJoint & \plJointCa & \plJointCb & \plJointCc \\
Sets before the two final cuts & \plPreReal & \plPreA & \plPreB & \plPreC \\
Candidates & \plReal & \puZero & \puZero & \puZero \\
\hline
\multicolumn{5}{l}{The \pdBlocks{} production blocks} \\
Blocks with a complete pass & \pdBlocks & \pdCtlDoneA & \pdCtlDoneB & \pdCtlBlocks \\
Sets of \myNmin{} or more nights before the two final cuts & \nodata & \pdCtlSetsA & \pdCtlSetsB & \pdCtlSetsC \\
\quad of which with $\chinu\leq\myChi$ & \nodata & \pdCtlChiA & \pdCtlChiB & \pdCtlChiC \\
\quad of which with $\Pch\leq\myPch$ & \nodata & \puZero & \puZero & \puZero \\
Smallest $\Pch$ & \nodata & \pdCtlPminA & \pdCtlPmin & \pdCtlPminC \\
Candidates & \pdBlockCands & \puZero & \puZero & \puZero \\
\hline
\multicolumn{5}{l}{Candidates after the merge with the \pdEarlyCands{} of the earlier runs, less \pdDup{} that shared rows: \pdCands} \\
Known by the position screen (Section~\ref{sec:vet_known}) & \pdKnownScreen & & & \\
Known by the joint-fit check & \pdKc & & & \\
Comet, identified at image vetting & \pdVetKnownN & & & \\
Already in my catalogue & \pdReg & & & \\
Undesignated candidates, re-measured on the images & \pdReal & & & \\
\enddata
\tablecomments{On \pdCtlMissing{} blocks a restart of one machine interrupted the controls; the \pdCtlPartial{} shifts that had finished there are empty as well. The production took about \pdCpu{} core-hours, about \pdCpuReal{} for the real pass and \pdCpuCtl{} for the controls: \pdCpuLinux{} were measured on the Linux machine, which ran the real pass of \pdLinuxBlocks{} blocks and the controls of \pdLinuxCtlBlocks, and the rest is estimated from the wall time of the tasks on two Windows machines. Of the \pdCtlSets{} control sets, \pdCtlSetsSeven{} have seven nights and \pdCtlSetsEight{} eight; of the \pdCtlChiOk{} that pass the first cut, \pdCtlTailA{} lie below $\Pch=\pdTailA$, \pdCtlTailB{} below \pdTailB{} and \pdCtlTailC{} below \pdTailC. The smallest $\Pch$ of each control is taken over all its sets; among the sets that pass the first cut the smallest values are \pdCtlPminAok, \pdCtlPmin{} and \pdCtlPminCok. One of the \pdSeedUnits{} seed units of one block failed in the real pass and in its first control. \pdRegDesAll{} designated objects of my catalogue were found again, \pdRegDes{} through the joint fit and \pdRegDesScreen{} through the screen. With the \myEarlyNew{} undesignated candidates that the same linker had found in the earlier runs the channel has \myChanNew{} candidates in the catalogue; \mySixN{} keep six nights after vetting (Section~\ref{sec:vet_img}), so the yield of the rule is \myYield.}
\end{deluxetable}

\emph{Chance alignments by number of nights.} The last column of Table~\ref{tab:purity} is the fraction of candidates not explained by chance alignments of shifted rows, not a purity against every kind of false link. For the production as run (Part~A) it is above \puPAllLo. But most candidates are known objects, and a false one can only be among the others: against the \pdProp{} proposals the limit is \puPropLo, and against the \puNewSeven{} undesignated candidates that had seven nights \puNewSevenLo. Seven nights is where the control sets before the final cuts lie and where \puNewSevenPct\% of these undesignated candidates sit.

For Part~B, in the two regions outside the DES footprint, the growth was run again keeping every set it returns, with \lrMinfar{} far nights instead of \myMinfar{} and no final cut. From \myNmin{} nights up the controls are empty; at six nights real and controls agree; at five nights the real sky has \puBFiveExcess{} sets more than the controls, a two-sigma excess: partial sets of real objects, or controls that run \puBFiveLowPct\% low. This is consistent with what DES reported: its second search set the floor at seven nights after about \desSixOrbits{} six-night orbits beyond \desSixDist~au gave one real source \citep{bernardinelli2022}, and its first kept six nights only with the sub-threshold test \citep{bernardinelli2020}, which this linker did not have. Where the counts of the real sky and of the controls separate depends on the field: the pilot strip and the two heavy blocks, which contain deep fields, have seven- and eight-night control sets that only the final cuts remove. Part~B covers two regions, \exArea~deg$^2$ in all, and its rule was not written beforehand. No candidate with fewer than \myNmin{} catalogue nights was ever measured on images, and looser values of the two final cuts gain nothing (Section~\ref{sec:null}).

\begin{deluxetable}{lccccc}
\tablecaption{Real and control candidates of the multi-year linker by number of nights\label{tab:purity}}
\tablehead{\colhead{Nights} & \colhead{Real} & \colhead{Control 1} & \colhead{Control 2} & \colhead{Control 3} & \colhead{Not chance}}
\startdata
\multicolumn{6}{l}{A. Production as run (rule A2), all \pdBlocks{} production blocks} \\
\nlTen & \puPTen & \puZero & \puZero & \puZero & $>\puPTenLo$ \\
\nlNine & \puPNine & \puZero & \puZero & \puZero & $>\puPNineLo$ \\
\nlEight & \puPEight & \puZero & \puZero & \puZero & $>\puPEightLo$ \\
\nlSeven & \puPSeven & \puZero & \puZero & \puZero & $>\puPSevenLo$ \\
All, after the merge & \puPAll & \puZero & \puZero & \puZero & $>\puPAllLo$ \\
Not known and not in my catalogue & \pdProp & \puZero & \puZero & \puZero & $>\puPropLo$ \\
Undesignated, seven nights & \puNewSeven & \puZero & \puZero & \puZero & $>\puNewSevenLo$ \\
\hline
\multicolumn{6}{l}{B. A different growth (two far nights), no night floor, no final cut, two regions (\exArea~deg$^2$)} \\
\nlTen & \puBTen & \puZero & \puZero & \puZero & $>\puBTenLo$ \\
\nlNine & \puBNine & \puZero & \puZero & \puZero & $>\puBNineLo$ \\
\nlEight & \puBEight & \puZero & \puZero & \puZero & $>\puBEightLo$ \\
\nlSeven & \puBSeven & \puZero & \puZero & \puZero & $>\puBSevenLo$ \\
\nlSix & \puBSix & \puBSixCa & \puBSixCb & \puBSixCc & \puBSixPur{} (\puBSixBoth) \\
\nlFive & \puBFive & \puBFiveCa & \puBFiveCb & \puBFiveCc & \puBFivePur{} (\puBFiveBoth) \\
\enddata
\tablecomments{Last column: one minus the ratio of the mean control count to the real count, the fraction not explained by chance alignments of shifted rows. Lower limits use the two-sided \ciLevel\% Poisson upper limit of an empty control, \plCtlUp{} per pass (the mean of three passes), with the real count fixed, and are rounded down; the intervals at six and five nights include the Poisson error of both counts. Part~A: candidates after all cuts; the controls are complete on \pdCtlBlocks{} of the blocks. Part~B: every set returned by the growth ($\chinu\leq\myChiGrow$, arc condition, bound orbit), each object counted once.}
\end{deluxetable}

\subsection{Loss at block edges}\label{sec:valid_edge}
To measure what the block edges cost I applied the rule as geometry, without any fit, to the \edN{} secure objects at \myDmin--\myDmax~au with at least \myNmin{} nights of their own rows in the tiles of \nscdr{} that I had downloaded. Rows exist only in those tiles: the nights that an object spent in a tile without a file are not counted, and an object that has its nights only there is not in this set (Section~\ref{sec:link_my}). With the whole sky as one block the rule holds for \edRule{} objects, and the production recovered \edRec. For \edNone{} of them (\edNonePct\%) the rule holds in no single block, and \edNoneRec{} of these was found; of the objects whose rows all lie in one block, \edOnePct\% (\edOneCI) were. That fraction is \edBinAPct\% for objects with \ebA{} nights, \edBinBPct\% with \ebB, \edBinCPct\% with \ebC{} and \edBinDPct\% with 16 or more. The edges thus cost about \edLostKnown{} known objects and, if undesignated objects cross edges as known objects with the same number of nights do, about \edLostNew{} undesignated ones (\edLostNewCI{} from the known objects alone). Other losses of the production (tiles and boxes above the row threshold, one failed seed unit) are not counted, and no first-loss table like Table~\ref{tab:budget} exists for this linker. The tiles without a file are a loss of another kind, since the objects in them have no rows to count. I counted them by prediction instead: a night counts for an object when its predicted position lies within $\nhPairRad^\circ$ of the centre of an exposure deep enough for it; the detector footprint is not applied. Of the \raN{} objects of Table~\ref{tab:recall}, \nhKnownLost{} have \myNmin{} or more such nights but fewer than \myNmin{} in the tiles with a file; \nhKnownLostAcc{} of them reach \myNmin{} when the tiles missing by accident are added (Section~\ref{sec:link_my}), and \nhKnownLostDense{} when the tiles above the density cut are added. The same count gives \nhKnownSearched{} objects with \myNmin{} such nights in the searched tiles, of which \nhKnownRows{} have \myNmin{} nights of rows there, so a predicted night is not always a row.

\subsection{All channels on one set of known objects}\label{sec:valid_all}
Table~\ref{tab:recall} puts the channels that work on \nscdr{} on one denominator, the \raN{} secure objects at \tsDmin--\tsDmax~au that fall on a live CCD, and one credit rule, row identifiers. This denominator is set by the exposures and not by the tiles I had downloaded: it contains the objects of the tiles that no linker read (Section~\ref{sec:link_my}), and they count as not recovered unless they were found through their rows elsewhere. The channels recover \raDensePct\% to \raMyPct\% each and \raAnyPct\% (\raAnyCI) together; of the \raElig{} objects with \myNmin{} nights of rows, which exist only in the downloaded tiles, the multi-year linker recovers \raMyEligPct\%. Recall falls to \raAnyVDPct\% at $V$ of \raVDLab{} and to none of the \raVE{} objects beyond. The tile linker of version 1 is not in the table, because its links were stored without row identifiers. Its own label credits it with \raOne{} objects, and with \raVeto{} after its vetoes. If those were added, the union of all channels would hold \raAnyLab{} objects (\raAnyLabPct\%), or \raAnyLabVeto{} (\raAnyLabVetoPct\%) with the vetoes; \raLabOnly{} of them would be present through that label alone. On \dtf, with its own denominator of \raDtN{} secure objects, the single-season channel recovers \raDtPct\%.

The table is a description, not a held-out result or a completeness: it contains the development objects of every channel (on the development and test halves the multi-year linker has \raMyDevPct\% and \raMyTestPct\%, the three channels together \raAnyDevPct\% and \raAnyTestPct\%). I make no population inference.

\begin{deluxetable}{lccccc}
\tablecaption{Secure known objects recovered by each channel on \nscdr{}, one denominator and one credit rule\label{tab:recall}}
\tablehead{\colhead{Channel} & \colhead{All} & \colhead{$V$ \raVALab} & \colhead{\raVBLab} & \colhead{\raVCLab} & \colhead{\raVDLab}}
\startdata
Objects & \raN & \raVA & \raVB & \raVC & \raVD \\
\hline
Dense-tile linker & \raDense{} (\raDensePct\%) & \raDenseVA & \raDenseVB & \raDenseVC & \raDenseVD \\
Dense-tile linker, both grids & \raBoth{} (\raBothPct\%) & \raBothVA & \raBothVB & \raBothVC & \raBothVD \\
Multi-year linker (rule A2) & \raMy{} (\raMyPct\%) & \raMyVA & \raMyVB & \raMyVC & \raMyVD \\
Any of the three & \raAny{} (\raAnyPct\%) & \raAnyVA & \raAnyVB & \raAnyVC & \raAnyVD \\
\enddata
\tablecomments{$V$ is the median predicted magnitude in \nscYears. An object is recovered when its own rows, by row identifier, make at least half of a candidate. Within $\raLatA^\circ$ of the ecliptic the three channels recover \raLatAAny{} of \raLatAN{} objects, at $\raLatC^\circ$ \raLatCAny{} of \raLatCN.}
\end{deluxetable}

\begin{figure*}[t]
\centering\includegraphics[width=\textwidth]{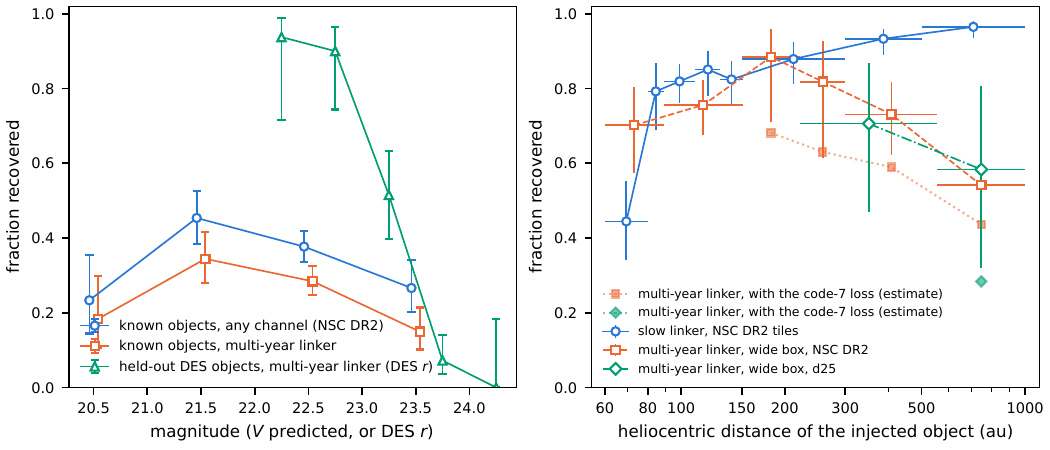}
\caption{Left: fraction of secure known objects recovered from \nscdr{} detections by any channel and by the multi-year linker, against median predicted $V$ (Table~\ref{tab:recall}), and fraction of held-out DES objects recovered by the multi-year linker in the pilot strip, against DES $r$ (Table~\ref{tab:heldout}). Right: recall of synthetic distant objects (semimajor axis 200--1000~au, perihelion 40--90~au) against heliocentric distance, for the slow linker on four \nscdr{} tiles (objects detectable on four or more nights) and for the multi-year linker with the distance box widened to 80--1000~au (eligible objects of one \nscdr{} block and one \dtf{} block); horizontal bars give the bin widths and vertical bars the 95\% Wilson intervals. The injections add catalogue rows and so cannot lose a detection to the code-7 mask of the community pipeline; the dotted grey points scale the multi-year recall by the fraction of eligible objects that keep the seven-night geometry once the mask is applied to each night with the rate measured on known objects. This correction is an estimate.}\label{fig:recall_v3}
\end{figure*}

Figure~\ref{fig:recall_v3} gathers the recalls of this section against magnitude and against distance.

\subsection{Injections}\label{sec:valid_inj}
Synthetic bound objects were added to real tiles as catalogue rows, at \ijDmin--\ijDmax~au only. They test the dense-tile linker and its candidate rule, not detection on the images. Of the injected objects with \dnNmin{} nights in one season, the fractions that reach a proposal are \ijPLong{} in tiles with many nights (\ijPLongPart{} in one that the catalogue covers in part), \ijPShort{} in tiles with few nights, and \ijPCrowd{} in a crowded tile at Galactic latitude below $\ijCrowdB^\circ$: in crowded fields the single-season channel is not effective. The survey-wide count for \dtf, \prDtSixReal{} links of six or more nights against \prDtSixCh{} expected by chance, does not describe what was sent to the images (Section~\ref{sec:vet_dt}). The multi-year linker has no injection test: at \ijFarLo--\myDmax~au, at high inclination and at high eccentricity its recall is not measured. Injections on Sedna-like orbits at larger distances are made in Section~\ref{sec:far}; other orbits are not tested there.

\subsection{Seeded recovery}\label{sec:valid_seed}
\emph{Truth test.} In the pilot strip and the two regions there are \lrTruth{} designated objects with at least \lrOpp{} oppositions and an MPC uncertainty code of at most \lrU. Each was treated as lost: only the published astrometry of its discovery opposition, or of the first \lrWinLo--\lrWinHi~days of it, was given as the seed, so an object can enter twice. The rule and four criteria were frozen before any held-out seed was run. On the held-out half, read once, the code made \lrHProp{} proposals from the seeds of \lrHObj{} distinct objects, all right in every linked night. With objects as the unit the \ciLevel\% upper limit on wrong proposals is \lrHObjUp\% (\lrHWrongUp\% if seeds are counted; the criterion was \lrHWrongThr\% on the point estimate). The same seeds on the shifted sky gave no proposal in \lrHCtl{} seed-passes. Of the seeds with an arc of \lrHArc~days or longer whose object has \lrN{} far nights of my rows, \lrHRecPct\% are recovered.

\emph{Limits.} Only \lrHShort{} of the \lrHProp{} proposals come from seeds shorter than \lrArcA~days, the kind the real targets have, and completeness there is \lrArcShort\%. Three nights are not safe: of \lrThreeAll{} three-night sets that pass the same cuts \lrThreeWrong{} are wrong, one of them a true row with two chance rows, where the controls gave none in \lrBCtl{} seed-passes. With \dtf{} rows added in one region a control produced a false four-night recovery. Over the downloaded tiles of \nscdr, \lrAllSeeded{} designated objects with poor orbits were seeded (\lrPSearched{} in a pilot on the regions of the truth test and \lrASeeded{} over the rest of the sky) and the code returns \lrAProp{} proposals, \lrPRec{} of them from the pilot; they are unvetted, no image has been looked at, and they are not results of this paper.

\subsection{Detections lost to a data-quality mask}\label{sec:valid_code7}
Section~\ref{sec:null} notes that the Community Pipeline marks the transients of its own image differencing with data-quality value \pfCode, and that \nscdr{} and my extraction both drop detections on those pixels. After version~2 I measured how often this removes a moving object, on designated objects only. On the masks of the exposures, value \pfCode{} lies on the core of \rvCsMatched{} detections of known objects that the catalogue holds, and on the core of \rvCsMissing{} bright detections that it lacks, at every magnitude. On the images the objects are there: the forced signal-to-noise ratio at the predicted place is 5 or more for \rvCsForced{} of the flagged points I measured. The mask fires when the same field was observed in the same band on another night or earlier in the night, which is common in repeated fields.

The loss is an estimate from that sample. It removes about \rvCsNscLoss\% of the detections of known objects that should be detectable in \nscdr{} and \rvCsDtLoss\% in \dtf. For objects farther than 80~au, which always move several widths of the image between nights so that every revisit counts, I estimate a loss of about \rvCsSlowNsc\% of all detections in \nscdr{} and \rvCsSlowDt\% in \dtf. What matters for the multi-year rule is the number of nights: in a simulation in which each night is dropped with the probability of its exposures, about \rvCsGeoNsc\% of the synthetic distant objects of \nscdr{} that have the geometry of rule A2 keep it, and about \rvCsGeoDt\% of those of \dtf.

This changes how the recall numbers of this paper should be read. The recalls measured on known objects (Sections~\ref{sec:valid_budget}--\ref{sec:valid_all}) use the real catalogues and already include the loss. The recalls measured on synthetic objects (Sections~\ref{sec:valid_inj} and \ref{sec:far}) do not, because the synthetic detections are added as catalogue rows and are never masked: they are optimistic, by my estimate by a factor of about \rvCsFacNsc{} for the distant searches on \nscdr{} and about \rvCsFacDt{} on \dtf. A re-extraction that keeps the masked pixels recovers the lost detections: on a pilot of whole exposures it restored the designated object \pfObject{} in \rvCsEdAll{} of \rvCsEdAll{} predicted points, \rvCsEdOld{} of which had been absent.

The distant searches of Section~\ref{sec:far} on \nscdr{} were made without the masked detections. I repeated them on \rvCsSlowBlocks{} blocks with the masked detections of my re-extraction added to the catalogue rows, with the same rule and the same shifted-sky controls. The controls give \rvCsSlowCtl{} candidate, and the real sky gives the same single candidate as before, a link at about 80~au already among the production links, and no new one. Adding those detections therefore does not change the null result beyond 80~au in these blocks; how much it raises the recall was not measured.

\subsection{Searches beyond 80~au}\label{sec:far}
\emph{Why the linkers of version~2 stop near 89~au.} Every multi-year and dense-tile search of version~2 seeds and grows its links in a box of inverse geocentric distance that, with its tolerances, spans \rvBoxNear--\rvBoxFar~au, so an object farther than about \rvBoxFar~au cannot even form a seed. I found this when I looked for the reason why no candidate of those runs lies beyond 80~au. On synthetic objects on Sedna-like orbits added to one block of \nscdr{} as catalogue rows, the frozen multi-year chain recovers \rvInjProd{} of the \rvInjElig{} objects at 89--151~au that have the night and arc geometry of rule A2; the same chain with the box moved outwards recovers \rvInjAddon{}, with no candidate in the shifted-sky controls. The static veto, which I suspected first, plays no part.

\emph{The searches.} I then ran rule A2 unchanged, with boxes moved outwards, on both DECam data sets, and read the shifted-sky controls of each run before its real candidates.
On \nscdr, a box covering about \rvAddLo--\rvAddHi~au returned no control candidate in \rvNscBlocks{} blocks and three real ones. Two were already in the production lists, one of them Gonggong at \rvGgR~au. The third, a link of \rvUzNights{} nights at \rvUzR~au, is 2014~UZ$_{224}$, a designated object of the development half that the production run had lost at the edge of its box: a joint fit with its \rvUzMpc{} published positions has an rms of \rvUzRms, and none of my positions is farther than \rvUzRowMax{} from it. A second box on \nscdr, at about \rvFarLo--\rvFarHi~au, returned no candidate in the controls and none on the real sky.
On \dtf{} the same two kinds of box ran on \rvDtBlocks{} blocks. The box at \rvFarLo--\rvFarHi~au returned no candidate in its control and none on the real sky. A single wide box reaching from the edge of production to about \rvFarHi~au returned one candidate in its control and one on the real sky, a link of \rvChNights{} nights at \rvChR~au that shares all its rows with a production candidate already identified as (532037) Chiminigagua.
The Rubin alerts were searched beyond 80~au as well (Section~\ref{sec:disc_rubin}): no link passes the rule there. A separate slow linker, written for 80--3000~au and run on the \nscdr{} tiles that the linkers of version~2 had not read, had covered \rvItTiles{} of its \rvItAll{} tiles when this version was written: its main rule (three seasons with at least two nights each and an rms of at most 0\farcs2) passed no link in \rvItCtl{} control runs and none on the real sky, and its weaker rule passed only links that the controls show to be dominated by chance. The run is not finished, and no candidate from it is reported.

\emph{What the null results mean.} Three designated distant objects were found blind (2014~UZ$_{224}$, Gonggong and Chiminigagua), and no new object passes rule A2 beyond the old box in either DECam data set. The searches are not complete at these distances. On the synthetic objects of the far box on \nscdr, \rvNscAThreeA{} of \rvNscAThreeAn{} eligible objects are recovered at 300--556~au and \rvNscAThreeB{} of \rvNscAThreeBn{} at 556--1000~au; on \dtf{} the recoveries are \rvDtAFiveA{} of \rvDtAFiveAn{} at 222--556~au and \rvDtAFiveB{} of \rvDtAFiveBn{} at 556--1000~au, and \rvDtWideB{} of \rvDtAFiveBn{} for the wide box. These fractions are of objects that already have the geometry of the rule, and more than half of the distant objects do not have it: \rvGeomNsc\% of the synthetic objects with rows at 300--1000~au never reach \myNmin{} nights over more than six months. The synthetic rows are also added to the catalogue, so they do not lose detections to the data-quality mask of Section~\ref{sec:valid_code7}; by my estimate the real recall is lower by about a fifth on \nscdr{} and by about half on \dtf. In the Subaru HSC search, after the corrections of its linker described in Section~\ref{sec:link_hsc}, no link lies beyond \rvHscFar~au in either season, the controls give none, and the held-out synthetic objects on Sedna-like orbits at 300--1000~au are recovered at fractions of \rvHscFarRecA{} and \rvHscFarRecB{} in the two seasons.

\begin{figure*}[t]
\centering\includegraphics[width=\textwidth]{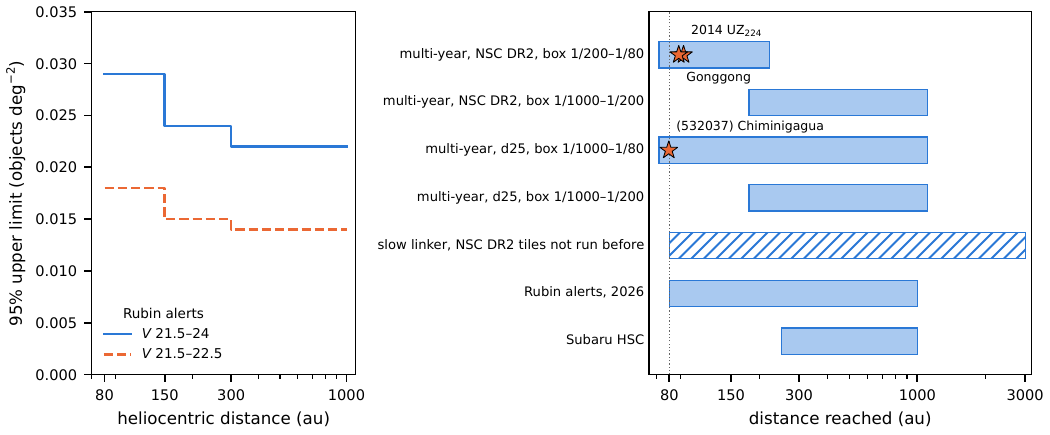}
\caption{Searches for objects beyond 80~au. Left: 95\% upper limits on the surface density of objects at 80--1000~au in the public Rubin alert stream of 2026 (1{,}906~deg$^2$), $3/(A\,\epsilon)$ with the effective area $A\,\epsilon$ of each distance bin measured with injected movers, for all injected magnitudes and for the brightest bin; no link of an unidentified object passes the rule beyond 80~au. Right: distance range of each search, as set by its distance box. No search found an undesignated object beyond 80~au. Stars mark the three designated objects that searches recovered without being pointed at them: 2014~UZ$_{224}$ at 92.5~au and Gonggong at 88.1~au in \nscdr{} and (532037) Chiminigagua at 79.6~au in \dtf. The hatched bar is a search whose run was not complete when this version was written. Only the Rubin search has an effective area in its files; for the others the recall by distance is in Figure~\ref{fig:recall_v3}.}\label{fig:limits_v3}
\end{figure*}

\subsection{Tests that do not depend on the link}\label{sec:valid_indep}\label{sec:valid_v1}
The image rules of Section~\ref{sec:vet} re-measure the detections that a linker chose. A test of the link itself needs data that the link did not use. Four such tests exist, each for a part of the catalogue. One standard is applied to all four: an object counts when it is detected on two or more independent nights, that is nights that hold no position of its fit. Objects detected on one such night are reported apart and not counted. In the three tests on DECam exposures a detection also has an upper limit, which I added after the results had been seen; it is described with the test of the multi-year linker. Table~\ref{tab:indep} gives the counts by channel where the split does not help to locate an object.

\emph{Version 1: unused exposures with control positions.} These numbers apply to the \vOneObj{} objects of version 1 \citep{cerati2026v1}. Forced photometry at fixed predicted positions, on \slExp{} exposures that contributed no position to any fit, detects the \slTestable{} testable objects in \slDet\% of cases against \slCtl\% at control positions on the same exposures. The pass rule written beforehand failed its own validity criterion (\slTruthFail{} of \slTruthN{} designated objects passed it). The rule adopted instead is post hoc: it was fixed on the designated objects and the controls before the candidates were inspected, except six pilot candidates inspected earlier. Under it \slPass{} objects pass, \slInc{} are inconclusive and none fails; \ctIndepVOne{} of those that pass and \ctIndepVOneInc{} of the inconclusive ones are tier-A candidates of this version. That rule counts exposures, and some of them lie on nights that hold positions of the fit. Recounted by nights for this version, with a detection at S/N$\,\geq5$ and the night defined as in the next paragraph, \ctIndepVTwo{} of the \ctIndepVOne{} are detected on two or more independent nights, \ctIndepVRuleOne{} on one and \ctIndepVRuleNone{} on none; of the \ctIndepVOneInc{} inconclusive ones, \ctIndepVIncOne{} are detected on one such night and \ctIndepVIncNone{} on none. The upper limit on the S/N described in the next paragraph is applied to this recount: it excludes \ctIndepVExcExp{} exposures of \ctIndepVExcObj{} objects and changes none of these numbers. The test shows flux at every predicted position, not that the flux is one moving body. Version 1 also measured how often its confirmations accept a false link: run on shifted predictions that cross the same exposures, the archival image search confirmed none of \fcImgN{} control orbits, and the catalogue precovery behind most scripted confirmations confirmed \fcCatNum{} of \fcCatDen{} shifted links (\fcCatPct\%, \fcCatCI\%), all in the fields of four candidates, so the links are not independent and the interval is too narrow.

\emph{Multi-year linker: unused exposures with control tracks.} The same test was made for the \myChanNew{} candidates of the multi-year linker, with the code of version 1. The orbit is fitted to the positions of the link alone. Forced photometry is made at its prediction on archival DECam exposures that the link did not use, that are at least 0.3~mag deeper than the object and on which the prediction is good to $3''$ (\ctUnuElig{} exposures for the \myChanNew{} objects), and on 32 control tracks per object on the same exposures. Only nights that hold no position of the link are counted. A detection is a forced S/N of at least 5 that is not above \ctUnuRuleFac{} times the S/N expected from the depth measured on the exposure and the predicted magnitude of the object. The upper limit is post hoc. I added it on \ctUnuRuleDay, after the results had been seen, because the rule of the test had no upper limit, so that the flux of a neighbouring star or galaxy at the predicted position would count as a detection. A measurement more than about \ctUnuRuleMag~mag brighter than the prediction is taken to be dominated by another source. The limit is applied to the objects and to the control tracks alike, and it restores no detection. The test is complete: all \ctUnuDone{} objects are finished. Of these, \ctUnuNoExp{} have no eligible exposure and \ctUnuNoInd{} have eligible exposures only on nights of their link. Of the other \ctUnuTested, \ctUnuGeOne{} are detected at the predicted position on at least one independent night: \ctUnuTwo{} on two or more and \ctUnuSingle{} on one. Without the upper limit these two numbers were \ctUnuTwoNoRule{} and \ctUnuSingleNoRule: \ctUnuExcExp{} exposures of \ctUnuExcObj{} objects are excluded by it, and \ctUnuExcLeft{} of those objects are left with one independent night. The remaining \ctUnuNone{} of the \ctUnuTested{} tested candidates is not detected at the predicted position on any of its \ctUnuNoneNights{} independent nights: the depth model of the test, which made those exposures eligible, expected an S/N of \unxModLo--\unxModHi{} on them; its own magnitudes measured on the nights of its link, used afterwards, give \ctUnuNoneExpLo--\ctUnuNoneExpHi, and by them none of the three exposures is 0.3~mag deeper than the object; the measured S/N is below \ctUnuNoneSnrUp; on two of those nights the prediction is uncertain by nearly $3''$, the limit of the test. On every unused exposure of the nights of its own link it is detected. Under the rule of this test it fails, and no count of this section is changed by what follows. Afterwards I examined this one object again. No night of its fit decides the prediction: leaving out any one night moves the three predictions by less than $\unxLoo''$. I then searched a set of sampled orbits that reaches beyond the central 95\% region of the fit (the clones of the fit, of the fits with one night left out and of an $N$-body fit), adding the three independent exposures along each orbit. The best orbit has a combined S/N of \unxComb{} (\unxSa, \unxSb{} and \unxSc{} on the single exposures). It lies \unxOffLo--\unxOffHi{} from the nominal prediction, at about \unxHalf{} times the half-width of the central 95\% interval of the fit along its long axis on all three exposures: outside that interval, about \unxSig{} standard deviations for a Gaussian cloud. None of \unxCtlN{} control tracks reaches that value; the largest has \unxCtlMax. A fit with these three positions keeps all the positions of the object with an rms of \unxRms. This second search was not part of the test written beforehand. Its acceptance rule was written down after the peaks on the single exposures had been seen, and it was applied to this object only, not to the other candidates and not to control objects. On single exposures the peaks are not significant, and the result rests on their combination. I do not count it: in every count of this paper the object is detected on no independent night. It stays in the catalogue with its earlier orbit, and the three positions are not in the catalogue; its flag is in the withheld per-object results of this test and not in the released tables. Of the \ctUnuSingle{} candidates detected on one independent night only, \ctUnuSingleOnly{} have only one such night, \ctUnuSingleMiss{} miss a second night on which the prediction is uncertain by about $2''$, and \ctUnuSingleExc{} have further nights (\ctUnuSingleExcNights) on which the measurement is excluded by the upper limit. In all, \ctUnuExpDet{} of the \ctUnuExp{} exposures of the tested objects on independent nights show the object, against \ctUnuCtlExpPct\% of the control positions (\ctUnuExpDetNoRule{} exposures and \ctUnuCtlExpPctNoRule\% without the upper limit, which excludes \ctUnuCtlExc{} control measurements). A control track reaches one such night in \ctUnuCtlOne{} of \ctUnuCtlN{} cases (\ctUnuCtlOnePct\%) and two such nights in \ctUnuCtlTwo{} (\ctUnuCtlTwoPct\%); without the upper limit the numbers were \ctUnuCtlOneNoRule{} and \ctUnuCtlTwoNoRule. One night alone is therefore weak evidence, and I count an object as passing when it has two. The \ctUnuCtlTwo{} control tracks that reach two nights belong to \ctUnuCtlTwoObj{} of the objects that pass, one or two of the 32 tracks of each: for those objects the rate is \ctUnuCtlTwoObjLo--\ctUnuCtlTwoObjHi\%, not the pooled \ctUnuCtlTwoPct\%. Among the tested objects, \ctUnuLtNine{} have fewer than nine nights in their link, and \ctUnuLtNineTwo{} of those pass. The test has no static veto. The definition of a night was corrected after the first numbers had been read, because a night at Cerro Tololo was being split after local midnight; the counts given here use the corrected definition. The upper limit on the S/N is a second change made after the numbers had been read. The expected S/N rests on a predicted magnitude, so the limit can also remove a true detection: on the nights of the links, which hold positions of the object and which no count uses, it excludes \ctUnuExcLink{} of the \ctUnuLnDet{} unused exposures with S/N$\,\geq5$. The expected S/N is also not calibrated by filter. Two of the \ctUnuExcExp{} exposures excluded on independent nights are between 3.4 and 3.9 times the expected S/N; the other four are more than five times the expected S/N. I did not inspect the images of the excluded exposures: the limit decides by the numbers alone.

\emph{Subaru frames at positions written down beforehand.} For \ctHscObj{} tier-A objects of several channels, the predicted position on archival Subaru HSC frames and its uncertainty were written to a file before any frame of the object was reduced; for \ctHscDigest{} of them a digest of that file was written at the same time. The digests are not published, so a reader cannot verify the order; the files are available on request. All \ctHscObj{} were found, with \ctHscPos{} positions on \ctHscNights{} nights: \ctHscTwoNights{} objects on three nights each and \ctHscOneNight{} on a single night, \ctHscOneExp{} of these on a single exposure. Under the standard of this section the \ctHscTwoNights{} count; for the \ctHscOneNight{} the prediction is the only test. The positions found were then added to the fits of the catalogue for these \ctHscTwoNights{} objects: their nights are independent of the fit that made the prediction, not of the fit of the catalogue. The upper limit on the S/N of the previous paragraph is not applied here: the files of this test hold no depth of the exposures, and a position is accepted on a fit of the point-spread function with a cut on its shape, not on a forced S/N alone. For \ctHscGood{} objects every position lies within $\ctHscSig\sigma$ of the prediction; the exceptions are one position of one object at $\ctHscOutA\sigma$ and one position of another at $\ctHscOutB\sigma$ of an orbit uncertainty of $0\farcs05$ ($\ctHscOutBm\sigma$ once the measurement error is included). For two of the single-night objects the prediction was loose, with a $1\sigma$ size of $\ctHscLooseLo''$ and $\ctHscLooseHi''$, and the object was found $\ctHscLooseOffLo''$ and $\ctHscLooseOffHi''$ from it; on one exposure of the second, \ctHscCrowd{} point-like sources lay inside the search ellipse, and only one of them is found again, at the same offset from the prediction, on another exposure of that night. The requirement of version 1 that the forced signal exceed that of the control positions was reported and not required, because on these deep frames control positions often fall on faint sources; the change was made before the first fit, and every accepted position but one meets the requirement. One exposure of one object, taken in poor seeing and $0\farcs5$ from the prediction, was dropped by hand. For \ctHscNotIn{} of the single-night objects the new positions lie inside the existing arc and were not entered in the frozen catalogue. Which objects these are is withheld, because the public pointing list of the telescope would narrow the sky where they can be.

\emph{Two candidates on DES exposures.} The consistency test of Section~\ref{sec:disc_des} singles out two tier-A candidates that DES should have linked according to my orbits and magnitudes. In version~2 my positions of them all came from exposures of 2020--2026. Both are in the DES data. \nscdr{} holds \ctChkNscA{} single-night detections of the first, $\ctChkSepLo''$--$\ctChkSepHi''$ from its predicted positions. It holds \ctChkNscB{} of the second; each lies $\ctChkLooLo''$--$\ctChkLooHi''$ from the position predicted when the other is added to my positions. From the density of single-night detections on the same exposures, \ctChkLamA{} and \ctChkLamB{} such detections are expected by chance, and none lies at the \ctChkCtl{} control positions. Fits that include these detections keep all of them (rms $\ctChkRmsA''$ and $\ctChkRmsB''$) and move $a$ and $e$ by at most $\ctChkShift\sigma$. Forced photometry at the positions of those fits, on the DES $r$ and $i$ exposures, gives S/N$\,\geq5$ on \ctChkFA{} and on \ctChkFB{} exposures (S/N$\,\geq4$ on \ctChkFFour), on \ctChkFNights{} nights for each object. On the same exposures \ctChkCtlPct\% of the control positions reach S/N$\,\geq5$, and of the 32 control tracks of each object \ctChkTrkTwo{} reach two nights; \ctChkNotMeas{} exposures of the second object could not be measured. \ctChkNscMeas{} of the \ctChkNscAll{} exposures of those catalogue detections are among the exposures measured, so the forced measurements are not all independent of the fits; without the nights of the catalogue detections, \ctChkIndNights{} nights with S/N$\,\geq5$ remain. No exposure of these two objects is above the upper limit of the previous paragraphs. Both objects count under the standard of this section. The check was made while version~2 was being written, and its numbers were recomputed from the measurement files by a second script. Those detections are now part of the fits of the catalogue.

\emph{Detections without controls.} For \unObj{} of the objects added after version 1 the catalogue records \unExp{} exposures, never used by the link, on which the object was measured without control positions: detections, not a controlled test.

\begin{deluxetable}{lccccc}
\tablecaption{Tier-A candidates of version~2 and the tests on DECam exposures that their links did not use\label{tab:indep}}
\tablehead{\colhead{Channel} & \colhead{Tier A} & \colhead{Tested} & \colhead{Detected on two} & \colhead{Detected on} & \colhead{Without} \\
\colhead{} & \colhead{} & \colhead{with controls} & \colhead{or more nights} & \colhead{one night} & \colhead{controls}}
\startdata
Version 1 (five channels) & 57 & 35 & 13 & 15 & 0 \\
Dense-tile linker, other runs & 14 & --- & --- & --- & 0 \\
Dense-tile linker with mask, \dtf & 81 & --- & --- & --- & 33 \\
Multi-year linker & 37 & 33 & 24 & 8 & 0 \\
\hline
All & 189 & 68 & 37 & 23 & 33 \\
\enddata
\tablecomments{Second row: the rows of Table~\ref{tab:channels} for the first runs and night pools, for the linker with mask on \nscdr{} and for the second pass on a shifted grid. Tested: objects with a forced measurement on at least one exposure that their fit did not use (version 1) or on at least one independent night (multi-year linker). The next two columns count independent nights with a detection, by the one standard of this section: S/N$\,\geq5$ and, by the limit added post hoc on \ctUnuRuleDay, not above \ctUnuRuleFac{} times the expected S/N (without that limit the row of the multi-year linker would read \ctUnuTwoNoRule{} and \ctUnuSingleNoRule); for version 1 the \ctIndepVTwo{} are among the \ctIndepVOne{} objects that pass its post hoc rule. Tested objects in neither column are detected at the predicted position on no independent night: \ctIndepVNone{} of version 1 and \ctUnuNone{} of the multi-year linker. Last column: objects with detections on unused exposures that were measured without control positions. The two objects detected on DES exposures and the \ctHscObj{} of the Subaru test are not split by channel, because the split would help to locate them. With them, \ctIndepN{} distinct tier-A objects are detected on two or more independent nights, and \ctIndepOne{} more on one night only. Of the tier-A objects, \ctEvExt{} have positions measured on Subaru or CFHT frames, nearly all found while an arc was extended and therefore part of the fit.}
\end{deluxetable}

\emph{What the candidates rest on.} Of the \ctTierAvTwo{} tier-A candidates of version~2, \ctIndepN{} are detected on two or more nights that their fits did not use, and \ctIndepOne{} more on one such night. For the other \ctIndepNone{} the evidence is the orbit, a fit to positions in two or more oppositions with a median rms of \ctARmsMed; \ctUnuNone{} of them is the candidate of the multi-year linker that was tested and is not detected at the predicted position on any of its \ctUnuNoneNights{} independent nights; a later search outside the rule of that test finds it and is not counted. For the \ctTierAAdded{} tier-A candidates added since version~2 the evidence is, for now, the orbit alone. For the multi-year linker the shifted-sky controls limit pure chance links (Section~\ref{sec:valid_ctl}). The rate at which the image checks accept a false link is measured for the production rule on \dtf{} alone (Section~\ref{sec:vet_dt}).
\section{The catalogue}\label{sec:catalogue}

The catalogue is the state of my working list of objects when it was frozen for this paper: \ctReg{} objects with \ctRows{} positions; two identifications with designated objects are withheld. Of these, \ctDes{} are identified with designated objects (Section~\ref{sec:cat_known}). The other \ctNew{} had no designation on \rvKcDay, and I call them undesignated. That word does not mean unseen by others: \ctRowsOtherObj{} of them have isolated tracklets of other observers in the public file of the Minor Planet Center (MPC), one corresponds to an object reported by \citet{jongma2026} and, provisionally, to one linked in TESS images by \citet{han2026} (Section~\ref{sec:rare_trojan}), \ctVOneUndes{} were in version 1, and eleven of version~2 and two added since were also found, independently, by \citet{giorgi2026} (Section~\ref{sec:cat_giorgi}). Objects that lie on exposures of programmes searching for outer Solar System objects may also be known to those teams and not yet reported; at least \ctAVrObj{} tier-A objects have a position on a $VR$ exposure (Section~\ref{sec:data}). Discovery credit for any object is for the MPC to assign. This section says what an entry rests on, how the entries divide by channel and by evidence, and what can be said of their orbits. Most counts below can be recomputed from the machine-readable tables (Section~\ref{sec:cat_mrt}); those that need the withheld positions, such as the split by Rubin's sky, cannot.

\subsection{Tiers}\label{sec:cat_tiers}
Version 1 graded its \vOneObj{} objects by evidence classes that needed a re-measurement of each object. That standard exists for those objects only, so I replace it by three tiers that are computed from the positions of an object and from its uniform orbit fit (Section~\ref{sec:cat_orbits}).

\emph{Tier~A}: image-verified positions in at least two oppositions; a fit that keeps at least \ctFitKeep\% of the positions with an rms below \ctFitRms; and a semimajor axis determined to $\sigma_a/a\leq\ctSigA\%$.

\emph{Tier~B}: all positions lie in one opposition, with the same fit quality. Such an object has a measured motion and inclination, but no orbit: $a$, $e$ and $q$ are not quoted, and it is not classified.

\emph{Tier~C}: neither of the two. Either the object has two or more oppositions but fewer than two of them contain image-verified positions, or its fit is not clean.

A position is image-verified when it is my own measurement on the frame, tied to Gaia DR3 \citep{gaia2023}, or a forced measurement at the predicted place, or when the record of the object states that every position was checked on the frames at registration. A position taken from \nscdr{} or from my first pipeline, with no such record, is a catalogue position; it is tied to Gaia DR2 through the Community Pipeline. The tier is computed from these labels in the files, not from a new inspection.

\emph{Tier~A} has \ctTierA{} objects, and they are the candidates that this paper counts. They have two (\ctAOppTwo), three (\ctAOppThree), four (\ctAOppFour) and five or more (\ctAOppFive) oppositions; in \ctAOppImgAll{} of them every opposition holds image-verified positions.

\emph{Tier~C} has \ctTierC{} objects, and \ctTierCMulti{} of them are multi-opposition objects with a median of \ctTierCMultiNightsMed{} nights over \ctTierCMultiArcMed~yr. Of these, \ctTierCMultiImg{} have fewer than two oppositions with image-verified positions, \ctTierCNoImg{} of them because all their positions are catalogue positions, and for \ctTierCMultiFit{} the fit rejects more than a tenth of the positions. Of the \ctTierCMultiImg, \ctTierCvOneA{} were in the highest evidence class of version 1, where their detections had been examined on the images without a record per position; if those checks were accepted, tier~A would have \ctTierAifVOne{} objects. I keep the stricter count. The tier therefore records which positions have an image measurement on file, and not how well an object is observed. Of the tier-C objects, \ctTierCSent{} are among those submitted to the MPC. The other \ctTierCOne{} are one-opposition links with a fit that is not clean.

\emph{One-opposition links} are \ctNewOneOpp: the \ctTierB{} of tier~B and the \ctTierCOne{} of tier~C. The arcs of tier~B run from \ctTierBArcMin{} to \ctTierBArcMax{} days (median \ctTierBArcMed), and \ctTierBLo{} of its objects have fewer than seven nights. Each link passed the rule of its channel, and for \ctTierBImg{} of the \ctTierB{} every position was measured on the images (\ctTierBImgSeven{} of these have seven or more nights). But a single opposition gives no independent test of a link. Of these links, \dtProdSingle{} come from the production rule on \dtf, whose image chain was tested on control links (Section~\ref{sec:vet_dt}): among all \dtFSentA{} links that it checked, multi-opposition objects included, \gsFalseLo--\gsFalseHi{} chance links are expected to pass, and fewer than \gsFalseUp{} at \ciLevel\%. For the other \ctOneOppRest{} no such rate is measured. I therefore list these links as objects to be followed and do not count them as candidates with orbits.

\emph{Positions.} The \ctTierA{} tier-A objects have \ctARowsImg{} image-verified positions and \ctARowsCat{} catalogue positions, the latter in \ctACatObj{} objects; the other \ctAAllImg{} have none. Of the \ctTierA{} objects, \ctANightsLo{} have \ctANightsMin{} nights, \ctANightsMid{} have seven to nine and \ctANightsHi{} have ten or more. For the \ctASixMy{} six-night objects of the multi-year linker on \nscdr{} of version~2, GCT0243 and GCT0271, the reason they stay in the tier is the test on exposures that their links did not use (Sections~\ref{sec:vet_img} and \ref{sec:valid_indep}); three tier-A objects of the multi-year linker on \dtf, GCT0347, GCT0388 and GCT0391, also have six nights, and that channel has not been through the test. The median arc is \ctArcAMed~yr; \ctArcAFive{} arcs are longer than five years, \ctArcATen{} longer than ten, and the longest is \ctArcAMax~yr. DECam exposures of \nscYears{} contribute positions to \ctAObjDecA{} of them and DECam exposures of 2020--2026 to \ctAObjDecB{} (\ctAObjBothDec{} have both); \ctAObjCfht{} have positions from CFHT MegaCam, \ctAObjHsc{} from Subaru HSC and \ctAObjOth{} from other archives, so that \ctAObjMultiInstr{} tier-A objects are measured with more than one instrument.

The files of all \ctNew{} undesignated objects hold \ctRowsNewAll{} positions: \ctRowsMine{} are mine (\ctRowsImg{} image-verified, \ctRowsCat{} from catalogues) and \ctRowsOther{} are positions of other observers, published by the MPC as isolated tracklets and used in the fits of \ctRowsOtherObj{} objects. These are never counted as my nights or oppositions. \ctRowsOtherVOne{} of the \ctRowsOtherObj{} objects were listed with their tracklets in version 1 \citep{cerati2026v1}. For the objects added after version 1 I have no complete record of a search of the isolated-tracklet file, so other observers may hold tracklets of those as well. The fits reject \ctRowsRej{} of the \ctRowsNewAll{} positions.

\begin{figure*}[t]
\centering\includegraphics[width=\textwidth]{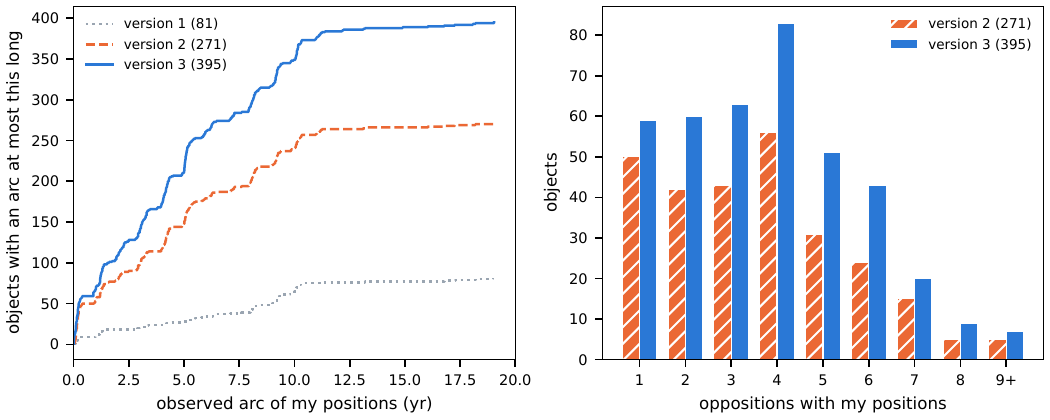}
\caption{Left: cumulative distribution of the observed arc of my own positions for the objects of version~1 (81), version~2 (271) and this version (395); medians 8.0, 4.3 and 4.3~yr. Right: number of oppositions with my positions, versions~2 and~3 (version~1 did not count them). Of the 271 objects of version~2, 12 have a longer arc and 13 more oppositions here, from positions added after that version.}\label{fig:arcs_v3}
\end{figure*}

Figure~\ref{fig:arcs_v3} shows how the arcs and the numbers of oppositions grew from version~1 to this version.

\subsection{Channels}\label{sec:cat_channels}
\begin{figure*}[t]
\centering\includegraphics[width=0.62\textwidth]{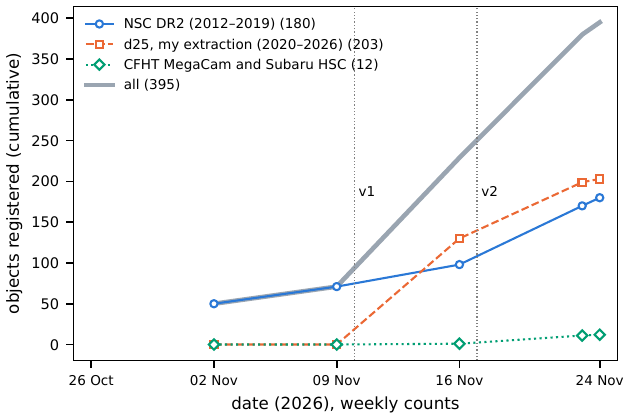}
\caption{Cumulative number of catalogue objects by week of registration, by the archive that the discovery channel searched: \nscdr{} for the linkers on \nscdr{} detections, \dtf{} for the linkers on my 2020--2026 extraction and the difference-image pilot, and the Subaru HSC and CFHT MegaCam linkers. Objects later extended with positions from other archives stay with the archive of their discovery. For 140 objects the registration date is known only as an upper bound and they are counted at that bound. Dotted vertical lines mark versions~1 and~2 of this paper. No catalogue object was first found in the Rubin alert stream.}\label{fig:timeline_v3}
\end{figure*}

Table~\ref{tab:channels} divides the objects by the linker that first produced them. Figure~\ref{fig:timeline_v3} shows when they were registered, by the archive of their first detections. The channel of an object is the route recorded when it was registered; an object found again by a later channel keeps its first one.

\begin{deluxetable*}{llcrrrrr}
\tabletypesize{\footnotesize}
\tablecaption{Catalogue objects by discovery channel and tier\label{tab:channels}}
\tablehead{\colhead{Channel} & \colhead{Detections} & \colhead{Section} & \colhead{Undes.} & \colhead{A} & \colhead{B} & \colhead{C} & \colhead{Designated}}
\startdata
Single-season tile linker (v1) & \nscdr & \ref{sec:link_v1} & 42 & 27 & 3 & 12 & 3 \\
Cross-tile linking (v1) & \nscdr & \ref{sec:link_v1} & 11 & 10 & 0 & 1 & 1 \\
Near mode (v1) & \nscdr & \ref{sec:link_v1} & 11 & 8 & 3 & 0 & --- \\
Earlier links, re-vetted (v1) & \nscdr & \ref{sec:link_v1} & 7 & 7 & 0 & 0 & --- \\
Side searches (v1) & various & \ref{sec:link_v1} & 6 & 5 & 0 & 1 & --- \\
Dense-tile linker, first runs and night pools & \nscdr, \dtf & \ref{sec:link_dense} & 5 & 5 & 0 & 0 & 1 \\
Dense-tile linker with mask & \nscdr & \ref{sec:link_dense} & 12 & 5 & 5 & 2 & 4 \\
Second pass on a shifted grid & \nscdr & \ref{sec:link_shift} & 4 & 4 & 0 & 0 & --- \\
Dense-tile linker with mask & \dtf & \ref{sec:link_dense} & 117 & 82 & 30 & 5 & 11 \\
Multi-year linker, rule A2 & \nscdr & \ref{sec:link_my} & 58 & 41 & 0 & 17 & --- \\
CFHT MegaCam linker & CFHT & \ref{sec:data_ext} & 6 & 1 & 0 & 5 & --- \\
Multi-year linker, rule A2 & \dtf & \ref{sec:link_my_d25} & 71 & 71 & 0 & 0 & --- \\
Multi-year linker, six nights & \nscdr & \ref{sec:link_six} & 8 & 1 & 0 & 7 & --- \\
Linkers on other data & HSC, DECam & \ref{sec:link_hsc} & 7 & 6 & 0 & 1 & --- \\
Four-night links & \nscdr & \ref{sec:cat_known} & 0 & 0 & 0 & 0 & 10 \\
\hline
All & & & 365 & 273 & 41 & 51 & 30 \\
\enddata
\tablecomments{Undes.: objects without a designation on \rvKcDay; A, B, C: their tiers (Section~\ref{sec:cat_tiers}). Designated: objects identified with designated objects (Table~\ref{tab:designated}). (v1): channels of version 1, which together give its \vOneObj{} objects. Detections: \nscdr, DECam \nscYears; \dtf, my extraction of DECam exposures of 2020--2026 (Section~\ref{sec:data}). The multi-year linker works on \nscdr{} and, since this version, on \dtf; a few of its objects were later extended with positions from other data.}
\end{deluxetable*}

The five channels of version 1 give \ctChVOne{} undesignated objects and \ctChVOneDes{} designated ones, the \vOneObj{} objects of that version. The dense-tile linker gives \ctChDenseNsc{} undesignated objects on \nscdr{} and \ctChDenseDt{} on \dtf, \dtNew{} of them from its production run there. That run is the largest channel and holds \ctDtTierB{} of the \ctTierB{} tier-B objects, because a \dtf{} tile is often covered in one season only. The catalogue objects of that run, \dtNew{} undesignated and \ctDtDes{} designated, do not correspond one to one to the \dtRealDt{} links on \dtf{} that the image chain called real at the first check (Section~\ref{sec:vet_dt}). Of the \ctDtA{} tier-A objects of the channel, \ctFnAReal{} come from links called real, \ctFnAPoss{} from links called possible and \ctFnANc{} from links not confirmed, and \ctFnANo{} have no record of that check. Those not called real entered the catalogue after an extension of the arc into other seasons or after a second check with a corrected chain, and all now have image-verified positions in at least two oppositions. Of the \ctDtTierB{} tier-B links of the channel, \ctFnBReal{} were called real, \ctFnBPoss{} possible and \ctFnBNc{} not confirmed, and \ctFnBNo{} have no record there. The same chain called \gsPoss{} of the \gsProd{} control links that pass the production rule possible and none real (Section~\ref{sec:vet_dt}); that test covers the production rule and not the extension into other seasons. The test of the CFHT linker failed on recall (Sections~\ref{sec:data_ext} and \ref{sec:null}); five of its \ctChCfht{} objects are in tier~C, with no DECam position and image-verified positions in one opposition only. The sixth, GCT0285, was later found on DECam exposures of several seasons and is in tier~A; its fit is described in Section~\ref{sec:cat_orbits}.

On \nscdr{} the multi-year linker contributed \ctChMy{} objects: the \vtMyVTwo{} of version~2, all in tier~A, and \vtMyAdded{} registered since, most of them from its second pass, of which \vtMyAddedA{} are in tier~A and \vtMyAddedC{} in tier~C because their positions are catalogue positions only. The image vetting left \ctMySix{} of them with six nights, GCT0243 and GCT0271 of version~2 and one added since (Section~\ref{sec:vet_img}). Rule A2 asks for \myNmin{} nights, so I do not count them in its yield, which is \vtMyYield{} objects. With the cuts that this linker copies, DES kept candidates of nine or more nights without a further test and tested the others on exposures that the link had not used \citep{bernardinelli2022}; \ctMyLtNine{} of these \ctChMy{} objects have fewer than nine nights. The image rule of this channel re-measures the linked detections and would accept a false link (Section~\ref{sec:vet_img}). The test on exposures that no link used was made on the \vtMyVTwo{} objects of version~2 and is complete for them (Section~\ref{sec:valid_indep}): \ctUnuTwo{} of the \ctUnuTested{} objects tested passed it (\ctUnuTwoNoRule{} before an upper limit on the S/N was added post hoc) and \ctUnuNone{} was not detected at the predicted position; a search of a wider set of orbits made afterwards, outside the rule of the test, finds it and is not counted. The objects added since version~2 have not been through that test.

Five channels were added after version~2 (Section~\ref{sec:link_my_d25}), the fifth, four-night links of \nscdr{} within single seasons, with recoveries of designated objects only (Section~\ref{sec:cat_known}). The same multi-year linker run on \dtf{} gave \vtMyDtN{} objects, all in tier~A; its six-night level on \nscdr{} gave \vtSixN{}, of which \vtSixA{} is in tier~A; and the linkers on other data, on Subaru HSC frames and on DECam difference images, gave \vtOdN{} undesignated objects. Table~\ref{tab:channels} gives them one row; Section~\ref{sec:cat_mrt} says why.

\subsection{Orbits, dynamical classes and absolute magnitudes}\label{sec:cat_orbits}
\emph{Orbits.} For this paper every object was fitted again in one pass, from its current positions, with Find\_Orb \citep{findorb} and the JPL DE440 ephemeris \citep{park2021}, with the settings of version 1 \citep{cerati2026v1}. Elements are barycentric osculating elements (J2000 ecliptic); the uncertainties come from clones of the Find\_Orb covariance carried to the barycentric frame. All \ctReg{} fits converged. For two objects of the CFHT linker, GCT0283 and GCT0285, Find\_Orb started from its default initial orbit kept only a few positions and gave an unbound orbit; started from the two-body orbit of their registration, the same program and settings keep all positions (\vtCfhtFitUsed{} of \vtCfhtFitRows{} for GCT0285), and these are the fits used. For \ctFitRej{} undesignated objects the fit rejects at least one position. The elements quoted in this paper come from this fit and from no earlier one. Their epoch lies within the years of observation and is withheld with the positions.

\begin{figure*}[t]
\centering\includegraphics[width=\textwidth]{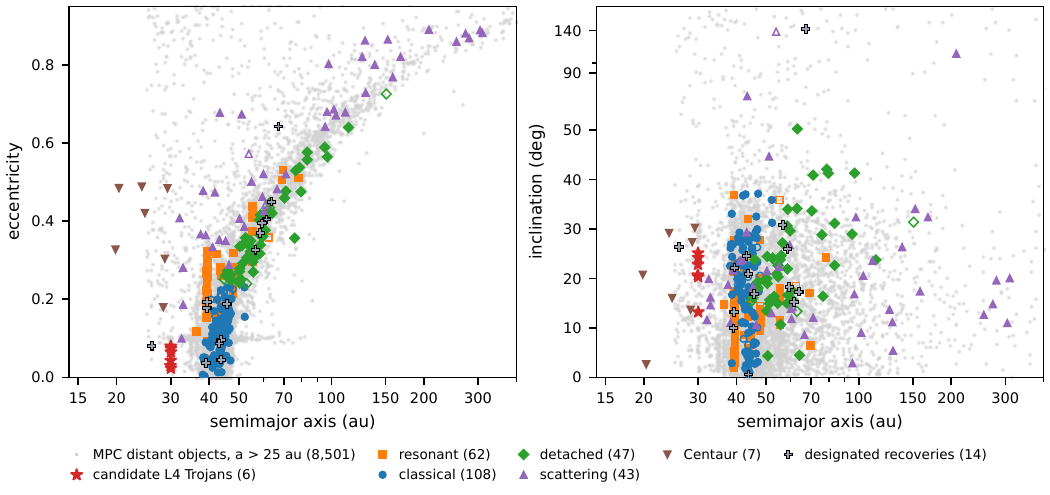}
\caption{Eccentricity (left) and inclination (right) against barycentric semimajor axis for the 273 tier-A candidates of the catalogue, with the rounded elements of Table~\ref{tab:cat271}, and for the 14 designated objects of Table~\ref{tab:designated} with two or more oppositions. Filled symbols mark a secure dynamical class and open symbols a class that is not secure. Grey points are the 8,501 objects with $a>25$~au in the MPC list of distant objects (heliocentric osculating elements), shown for orientation only. The inclination axis is linear below 50\arcdeg{} and logarithmic above.}\label{fig:orbits_v3}
\end{figure*}

\emph{Dynamical classes.} Classes are assigned as in version 1: the nominal orbit and \ctClsClones{} clones are integrated for \ctClsMyr~Myr with REBOUND \citep{rein2012,rein2019} under the four giant planets, each orbit is classified with the scheme of \citet{gladman2008}, and the object takes the class of the majority, called secure when at least \ctClsAgree\% of the orbits agree. The classes were frozen on \ctClsFreeze, after every integration that can change a class had finished: all \ctTierA{} tier-A objects have a class, and \ctClsSec{} of the classes are secure. The \ctTierAAdded{} objects added after version~2 were integrated with the same scripts, clones and rules as those of version~2, including the 100-Myr integrations for the \vtLhN{} of them with $a>100$~au or a class that is not secure. Those longer runs do not change any class, which is taken from the \ctClsMyr-Myr runs; they give another class for \vtLhDiff{} of the \vtLhN, GCT0378, whose \ctClsMyr-Myr class (the 5:2 resonance) is not secure and which the 100-Myr runs call detached. Of these classes, \ctClsOld{} come from integrations made before the catalogue was frozen, on the orbit of their date and not on the uniform fit, some of them with 100 or 200 clones; the \ctClsBoth{} objects that were integrated on both orbits have the same class. For the \ctClsNew{} classes added since version~1 a class is secure when \ctClsAgree\% of the orbits agree and, for arcs of the fitted positions under three years (positions of other observers included), when clones drawn from the covariance inflated to three sigma give the same class in \ctClsAgree\% of the cases, as in version 1; that second test has been run for \ctClsShortRun{} objects that need it. Table~\ref{tab:classes} gives the classes and Figure~\ref{fig:orbits_v3} the elements. The table says which classes are present and how firmly, and nothing about their proportions on the sky, for which a selection function would be needed.

\begin{deluxetable*}{lrrl}
\tabletypesize{\footnotesize}
\tablecaption{Dynamical classes of the tier-A objects\label{tab:classes}}
\tablehead{\colhead{Class} & \colhead{$N$} & \colhead{Secure} & \colhead{Members by resonance or sub-class}}
\startdata
Candidate L4 Neptune Trojan & 6 & 6 & --- \\
Resonant & 62 & 56 & 3:2 (33), 2:1 (8), 7:4 (7), 5:2 (6), 5:3 (2), 7:2 (2), 21:5 (1), 3:1 (1), 4:3 (1), 9:5 (1) \\
Classical & 108 & 99 & inner (8), cold (19), hot (79), outer (2) \\
Detached & 47 & 43 & --- \\
Scattering & 43 & 41 & --- \\
Centaur & 7 & 7 & --- \\
\hline
All classified & 273 & 252 & \\
\enddata
\tablecomments{REBOUND, \ctClsMyr~Myr, nominal orbit and clones; classes after \citet{gladman2008}. Secure: at least \ctClsAgree\% of the orbits agree and, for arcs of the fitted positions under three years, clones of the covariance inflated to three sigma give the same class. The counts are not class fractions of a population.}
\end{deluxetable*}

Among the tier-A objects there are \ctClsTro{} candidate L4 Neptune Trojans, \ctClsRes{} resonant objects (\ctClsPlut{} of them in the 3:2), \ctClsCla{} classical, \ctClsDet{} detached and \ctClsSca{} scattering objects, and \ctClsCen{} Centaurs.

\emph{Absolute magnitudes.} $H$ is the value Find\_Orb derives from the magnitudes reported with the positions. These are in different bands ($g$, $r$, $i$, $z$, $VR$) and were not brought to one system, no phase function beyond the Find\_Orb default was fitted, and $H$ has no error bar. The tier-A objects have $H$ from \ctAHMin{} to \ctAHMax, with a median of \ctAHMed{} and the central 68\% between \ctAHLo{} and \ctAHHi. For albedos of \ctAlbHi--\ctAlbLo{} the median corresponds to diameters of about \ctDiamLo--\ctDiamHi~km. $H$ is not on one photometric system. The band attached to a position is not always the filter of its exposure, and where one object has magnitudes in two bands on the same night the median difference over tier~A is \ctGrMed~mag between $g$ and $r$ (\ctGrNights{} nights of \ctGrObj{} objects) and \ctRiMed~mag between $r$ and $i$ (\ctRiNights{} nights of \ctRiObj{} objects), whereas the photometry of version 1 gave $g-r$ of \ctTvGrLo--\ctTvGrHi{} for four of these objects and $r-i$ is positive for trans-Neptunian objects. The values of $H$ are therefore uncertain by 0.5~mag or more, and the brightest, $H=\ctAHMin$ for GCT0010, is unverified; I give no size for any single object. The distribution of $H$ is the one the search is sensitive to, set by the depth of the exposures (Section~\ref{sec:data}) and by the distances searched; it is not corrected for any selection and is not a luminosity function.

\subsection{Designated objects in the catalogue and in the multi-year run}\label{sec:cat_known}\label{sec:vet_known}
\emph{The check.} Version 1 compared each link with known objects at my epochs \citep{giorgini1996,berthier2006}. That fails for designated objects with short arcs: an audit on \kcAuditDate{} found \kcMatches{} identifications, \kcMissed{} of which that check had missed. They were of objects added to my working list after version 1 was written, and none of them was published as an undesignated object.

The check now used is made at the designated object's own epochs, where its orbit is good. My orbit, a Find\_Orb fit with \kcClones{} clones drawn from its covariance, is screened against the MPC list of distant objects and the MPC orbit file, for objects slower than $\kcSlow''$ per hour. A pair passes the screen if my orbit agrees with the other one at the first, middle and last date of the other arc within \kcChi{} times the spread of the clones, or if the other orbit passes within \kcNear{} of my positions at my epochs, or if the poles lie within $\kcPole^\circ$ and the objects less than $\kcPoleSep^\circ$ apart. Pairs that pass go to a two-body least-squares fit of my positions with five positions of the other orbit inside its own arc, which must have an rms below $\kcLsqRms''$. For pairs that link, the published astrometry is downloaded by designation and fitted with my positions in one Find\_Orb fit: the objects are the same if at least \kcUsed\% of the positions are used, the rms is below \kcRms{} and every residual is below \kcRowMax. Version~2 did not measure the miss rate of this check on objects of known identity; the repeated check below does. No position of an undesignated object leaves my machines: the catalogues are downloaded whole, and the requests name designated objects only. The image measurements of Section~\ref{sec:vet} do send requests: cutouts are fetched from the NOIRLab archive at the positions and dates of the candidates.

\emph{A repaired check, and the catalogue checked again.} The check described above has a weakness that I found after version~2 was posted. It compared my orbit with a catalogue orbit only at the dates of the catalogue arc, and a pair reached the joint fit only if a two-body fit linked the two first. When either arc is short neither test is reliable, so a true pair could be dropped before the one step that decides. It did happen: among the single-season links of a later search, two that I had taken for new objects with a second opposition are the designated objects 2015~KJ$_{176}$ and (523643) 2010~TY$_{53}$, and a search for second oppositions in \nscdr{} found \rvKcOldMiss{} more designated objects that the old screen had not matched (Section~\ref{sec:null}). None of them was in the catalogue of version~2.

The repaired screen propagates every orbit of the MPC files to my own epochs. It keeps a catalogue orbit if it passes within $\rvKcRadMulti^\circ$ of my positions (within $\rvKcRadOne^\circ$ for orbits of a single opposition, whose positions years away from their arc are poorly known), and it always keeps the \rvKcNearest{} nearest orbits of each kind. The pairs chosen from those kept, as listed below, go to a joint Find\_Orb fit with the published astrometry, downloaded by designation. The objects are the same if at least \rvKcRobFrac\% of the published positions and all of mine lie within $\rvKcRobRes''$ of the joint orbit and at least \rvKcUsedFrac\% of all positions are used. A fit that fails is recorded as undecided, never as no match. Before it was used on the catalogue, the repaired check was run on \rvKcTruth{} designated objects of known identity that my linkers had found, and it identified all of them.

I ran it on all \rvKcObj{} objects of the catalogue on \rvKcDay. The screen kept \rvKcScreen{} pairs of an object and a catalogue orbit, for \rvKcObjPairs{} objects, and \rvKcPairs{} of them were fitted jointly: \rvKcSelVOne{} pairs of the orbits within 0.2$^\circ$ and the three nearest of each kind, \rvKcSelDrift{} more for the three one-opposition orbits nearest by an offset scaled to their drift, and \rvKcSelPole{} chosen by the orbital pole. It identifies \rvKcMatch{} objects with designated ones, with joint rms of \rvKcRmsLo--\rvKcRmsHi{} and no residual above \rvKcOursMax{} for my positions or \rvKcTheirsMax{} for the published ones: they are exactly the \rvKcMatch{} that the catalogue already carried. It finds \rvKcNew{} new identity. The joint fit failed for \rvKcUndec{} pairs; I fitted these again with a full n-body orbit, and none of them is the same object. For the matched objects the screen separation at my epochs has a median of $\rvKcSepMed^\circ$ and reaches $\rvKcSepMax^\circ$: GCT0212 (2000~GM$_{147}$) was kept only because its pole is close to the pole of my orbit, which is the case that a cone of fixed radius would have lost. As before, the check cannot see an object whose orbit has not been published, and its result holds for the MPC files of that day. The \vtLateN{} objects added on the next day, \vtLateFirst--\vtLateLast, came from links whose candidates had been run through the same check before registration; each was identified with one designated object by a joint fit (Table~\ref{tab:designated}), and none is counted as undesignated. GCP0032, added the same day, was identified with 2001~FJ$_{228}$ in the same way.

\begin{figure*}[t]
\centering\includegraphics[width=\textwidth]{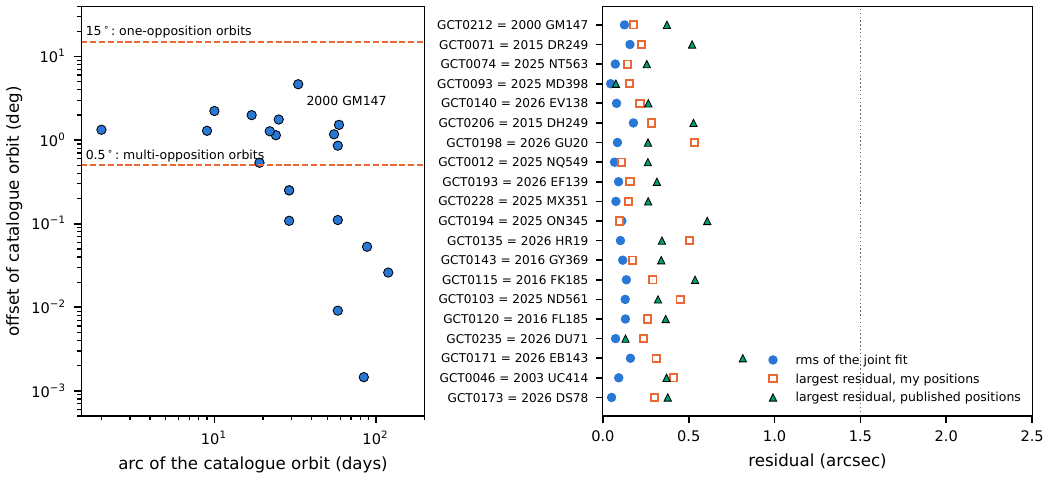}
\caption{Identity check of the whole catalogue against the MPC orbit file, repeated for this version. Every catalogue orbit is propagated to the epochs of my positions and kept when it passes within 0.5\arcdeg{} (multi-opposition orbits) or 15\arcdeg{} (one-opposition orbits); the orbits within 0.2\arcdeg, the three nearest of each kind, the three one-opposition orbits nearest by an offset scaled to their drift, and those chosen by orbital pole go to a joint Find\_Orb fit of my positions with the published ones, 3{,}110 fits in all. Left: offset between the catalogue orbit and my positions for the 20 objects identified that this version lists (two others are not public), all of them with one-opposition catalogue orbits, against the arc of that orbit; 12 lie beyond 0.5\arcdeg, and 2000~GM147 was reached only through the ranking by pole. Right: rms of each joint fit and the largest residuals of my positions and of the published ones; the dotted line is the 1\farcs5 limit of the match rule. The check finds the 22 identities already in the catalogue and no new one; the 11 pairs whose first fit failed were all rejected by a second fit.}\label{fig:identity_v3}
\end{figure*}

\begin{figure*}[t]
\centering\includegraphics[width=\textwidth]{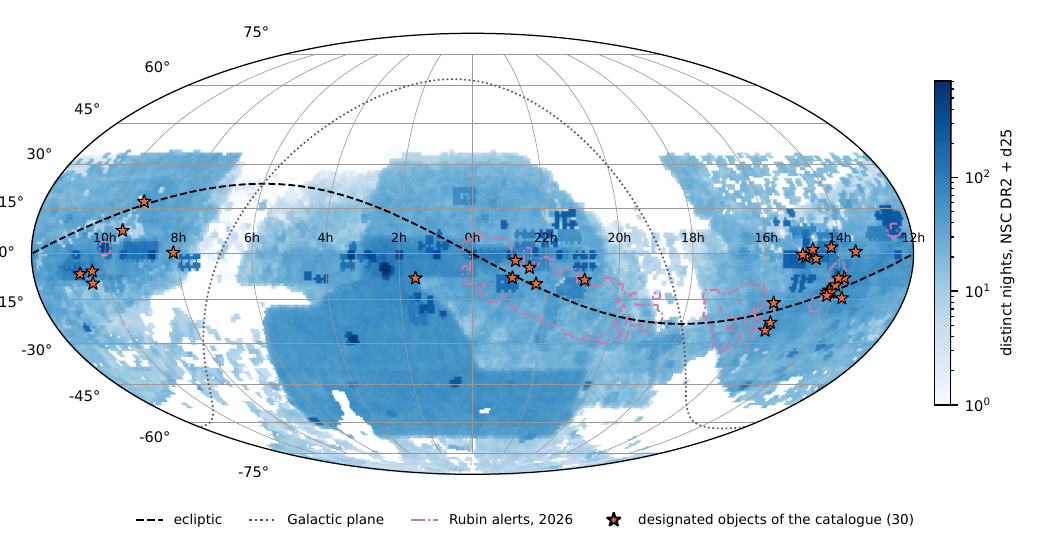}
\caption{Sky coverage of the two DECam data sets, \nscdr{} (2012--2019) and \dtf{} (2020--2026), as the number of distinct nights in cells of one square degree (equatorial coordinates, Mollweide projection), with the cells of the 2026 Rubin alert stream outlined. Stars mark the 30 designated objects of the catalogue at one of my positions. The positions of the undesignated objects are withheld until they are reported to the Minor Planet Center, so they are neither drawn nor counted per cell.}\label{fig:sky_v3}
\end{figure*}

\begin{deluxetable*}{lllcrrcccr}
\tabletypesize{\footnotesize}
\tablecaption{Catalogue objects identified with designated objects\label{tab:designated}}
\tablehead{\colhead{ID} & \colhead{Designation} & \colhead{Discovery} & \colhead{My years} & \colhead{Nights} & \colhead{Opp.} & \colhead{Joint fit} & \colhead{rms} & \colhead{Before} & \colhead{Nights} \\
\colhead{} & \colhead{} & \colhead{} & \colhead{} & \colhead{} & \colhead{} & \colhead{(used/all)} & \colhead{($''$)} & \colhead{(yr)} & \colhead{before}}
\startdata
GCT0012 & 2025~NQ$_{549}$ & 2025, Rubin & 2018 & 6 & 1 & 19/19 & 0.07 & 6.9 & 6 \\
GCT0046 & 2003~UC$_{414}$ & 2003, CTIO & 2015--18 & 14 & 4 & 39/40 & 0.09 & --- & --- \\
GCT0071 & 2015~DR$_{249}$ & 2015, DECam & 2014--23 & 9 & 4 & 25/25 & 0.16 & 1.0 & 5 \\
GCT0074 & 2025~NT$_{563}$ & 2025, Rubin & 2016--24 & 10 & 3 & 20/20 & 0.07 & 8.9 & 10 \\
GCT0093 & 2025~MD$_{398}$ & 2025, Rubin & 2014--19 & 11 & 5 & 26/26 & 0.05 & 11.2 & 11 \\
GCT0103 & 2025~ND$_{561}$ & 2025, Rubin & 2022 & 5 & 1 & 33/34 & 0.13 & 2.9 & 5 \\
GCT0115 & 2016~FK$_{185}$ & 2016, Subaru & 2021--26 & 18 & 2 & 49/49 & 0.14 & --- & --- \\
GCT0120 & 2016~FL$_{185}$ & 2016, Subaru & 2021 & 8 & 1 & 22/22 & 0.13 & --- & --- \\
GCT0135 & 2026~HR$_{19}$ & 2026, Rubin & 2016--26 & 31 & 5 & 79/84 & 0.10 & 10.3 & 31 \\
GCT0140 & 2026~EV$_{138}$ & 2026, Rubin & 2013--24 & 29 & 9 & 43/43 & 0.08 & 12.9 & 29 \\
GCT0143 & 2016~GY$_{369}$ & 2016, Subaru & 2021 & 9 & 1 & 21/21 & 0.12 & --- & --- \\
GCT0171 & 2026~EB$_{143}$ & 2026, Rubin & 2021--25 & 8 & 2 & 20/20 & 0.16 & 4.9 & 8 \\
GCT0173 & 2026~DS$_{78}$ & 2026, Rubin & 2018--26 & 33 & 3 & 777/780 & 0.05 & 7.8 & 22 \\
GCT0193 & 2026~EF$_{139}$ & 2026, Rubin & 2018 & 7 & 1 & 19/19 & 0.09 & 8.1 & 7 \\
GCT0194 & 2025~ON$_{345}$ & 2025, Rubin & 2018 & 6 & 1 & 27/27 & 0.11 & 7.0 & 6 \\
GCT0198 & 2026~GU$_{20}$ & 2026, Rubin & 2013--24 & 28 & 8 & 40/42 & 0.08 & 13.2 & 28 \\
GCT0206 & 2015~DH$_{249}$ & 2015, DECam & 2022--26 & 11 & 3 & 22/22 & 0.18 & --- & --- \\
GCT0212 & 2000~GM$_{147}$ & 2000, Maunakea & 2020--25 & 10 & 4 & 17/17 & 0.13 & --- & --- \\
GCT0228 & 2025~MX$_{351}$ & 2025, Rubin & 2014--17 & 13 & 4 & 34/34 & 0.08 & 11.2 & 13 \\
GCT0235 & 2026~DU$_{71}$ & 2026, Rubin & 2016--23 & 9 & 4 & 22/22 & 0.07 & 9.8 & 9 \\
GCT0400 & 2015~HM$_{196}$ & 2014, DECam & 2014 & 6 & 1 & 38/40 & 0.04 & 0.0 & 4 \\
GCT0401 & 2026~GJ$_{24}$ & 2026, Rubin & 2014 & 4 & 1 & 76/76 & 0.08 & 12.0 & 4 \\
GCT0402 & 2014~HL$_{208}$ & 2014, DECam & 2014 & 6 & 1 & 20/20 & 0.08 & 0.0 & 4 \\
GCT0403 & (612549) 2003~HG$_{57}$ & 2003, Maunakea & 2014 & 4 & 1 & 35/35 & 0.22 & --- & --- \\
GCT0404 & 2026~EF$_{144}$ & 2026, Rubin & 2014 & 4 & 1 & 45/45 & 0.09 & 11.9 & 4 \\
GCT0405 & 2015~HD$_{195}$ & 2015, DECam & 2013 & 4 & 1 & 21/21 & 0.06 & 2.0 & 4 \\
GCT0406 & 2026~EL$_{144}$ & 2026, Rubin & 2014 & 4 & 1 & 22/22 & 0.05 & 11.9 & 4 \\
GCT0407 & 2025~MD$_{399}$ & 2025, Rubin & 2015 & 4 & 1 & 16/16 & 0.08 & 10.1 & 4 \\
GCT0408 & (602714) 2014~ON$_{394}$ & 2014, Pan-STARRS & 2018 & 4 & 1 & 115/115 & 0.16 & --- & --- \\
GCP0032 & 2001~FJ$_{228}$ & 2001, Kitt Peak & 2014 & 6 & 1 & 33/33 & 0.20 & --- & --- \\
\enddata
\tablecomments{Discovery: year and site of the discovery observation in the MPC record. My years, Nights, Opp.: positions of this work. Joint fit: positions used and available in the Find\_Orb fit of my positions with the published astrometry, and its rms. Before: time from my first position to the discovery observation, and the number of my nights before it; a dash where my positions are all later. The published arcs are single-opposition arcs of \ctDesArcLo--\ctDesArcHi~days, except for \ctDesMultiOpp{} of the objects \vtLateFirst--\vtLateLast.}
\end{deluxetable*}

\emph{The \ctDes{} objects of the catalogue} (Table~\ref{tab:designated}; Figure~\ref{fig:sky_v3} shows where most of them lie). Of the \ctDes, \ctDesRubin{} are discoveries of the Vera C.\ Rubin Observatory of 2025--2026, \ctDesSubaru{} were found with Subaru, \ctDesDecam{} with DECam, \ctDesPs{} with Pan-STARRS and \ctDesOld{} earlier. All but \ctDesMultiOpp{} have a published arc of \ctDesArcLo--\ctDesArcHi~days in a single opposition, the kind of orbit on which an identity check made by position at my epochs fails. The other \ctDesMultiOpp{} are among the \vtLateN{} objects \vtLateFirst--\vtLateLast, identified on 2026 November 24 from four-night links of \nscdr{} within single seasons; their published arcs span several oppositions, and my positions extend them backwards or fill an opposition that they lack. For \ctDesPre{} objects, the \ctDesRubin{} Rubin discoveries among them, my positions precede the discovery by \ctDesPreMinTxt{} to \ctDesPreMax~yr (median \ctDesPreMed~yr; \ctDesPreNights{} nights in all): for the Rubin objects the archive holds \ctDesRubMin--\ctDesRubMax~yr of earlier arc. For the other \ctDesPost, my positions come after the published arc and extend it forward by years for \ctDesPostAfter{} of them; for (612549) 2003~HG$_{57}$ and (602714) 2014~ON$_{394}$ they lie inside the published arc and fill an opposition that it lacks. GCT0103, a tier-B object of version~2, is now identified with the Rubin discovery 2025~ND$_{561}$. GCP0032 is 2001~FJ$_{228}$, a Jupiter Trojan whose published astrometry covers seven days of 2001; my positions, from a four-night link of \nscdr{} in 2014, give it a second opposition. Since 2026 November 24 the objects identified with designated ones carry identifiers GCP$nnnn$ in my list; this paper keeps the GCT identifiers under which they were first listed, and GCP0032 had none. Discovery credit belongs to the discoverers.  My positions of the \vtDesRel{} objects identified in version~2, published again with this version (Section~\ref{sec:data_avail}), give each of them a multi-year arc, which its published arc does not have; those of the others are available on request.{} These identifications rest on joint fits with astrometry measured by others, which makes them the part of the catalogue that an outside reader can test most directly, for the \vtDesRel{} whose positions are published{} (Section~\ref{sec:data_avail}).

\emph{The candidates of the multi-year run.} Here ``known'' has a second, weaker meaning. Of the \pdCands{} candidates that the multi-year linker returned on \nscdr, \pdKnownAll{} are known objects (\pdKnownDistinct{} distinct designated minor planets, \pdKnownTwice{} of them returned as two candidates, and the comet C/2014~UN$_{271}$ \citep{bernardinelli2021}, identified at image vetting), \pdReg{} were already in my catalogue and \pdReal{} were registered (Section~\ref{sec:valid_ctl}). For \pdKnownScreen{} of the known candidates the identification is a screen and not a joint fit: the candidate lies within $\pdTol''$ of an MPC orbit, propagated as a two-body orbit, at its first, middle and last detection. The largest of the three separations is below $\scOne''$ for \scUnderOne{} of them and below $\scTen''$ for \scUnderTen{} (median \scMed); it lies between $\scTen''$ and $\scSixty''$ for \scMid{} and above $\scSixty''$ for \scOverSixty{} (at most $\scMax''$). The \scBeyondTen{} beyond $\scTen''$ are unverified identifications. The other \pdNotKnown{} candidates went through the full check, which identified \pdKc{} with objects designated in \kcDesYears. None of the known candidates was examined on images. With that caution, the matches show what the archive holds of objects that others discovered. For \ctMkPub{} of the \pdKnownDistinct{} my nights precede the first published observation of the object (\ctMkPubNights{} nights): precoveries in the strict sense. A discovery record was found for \ctMkDisc, and for \ctMkPre{} my nights precede the discovery observation (\ctMkPreNights{} nights). That larger number needs care: for \ctMkPreNotPub{} of the \ctMkPre{} other observers have published observations older than my first night, \ctMkPreYear{} precede the discovery observation by less than a year, and at least \ctMkPreDecam{} are discoveries made with DECam, for which my earlier nights are exposures of the same camera. Of the \pdKnownDistinct, \ctMkNew{} were discovered in 2025--2026, \ctMkRubin{} of them by Rubin, and for these my nights are \ctMkNewMin--\ctMkNewMax~yr older than the discovery.

\subsection{An independent search of \nscdr}\label{sec:cat_giorgi}
\citet{giorgi2026} searched the northern sky of \nscdr{} (Bok and Mayall exposures north of declination $+30^\circ$ and DECam exposures north of $+3^\circ$, largely outside the Rubin footprint) for distant objects with a linker of his own, with thresholds fixed before the search, mirror-parallax and night-shuffle controls, injected objects and a check of every new object in other archives. His detection selection, linking and vetting are independent of mine. He reports ten new trans-Neptunian objects and lists, among the objects his search recovered, \gzRefound{} candidates of this paper: GCT0010, GCT0011, GCT0016, GCT0018, GCT0022, GCT0025, GCT0027, GCT0068 and GCT0069 (\gzThreeTierA{} of them in tier~A). His elements of these \gzRefound{} agree with mine within \gzThreeA~au in $a$ and \gzThreeE{} in $e$, and his inclinations equal mine to the precision he prints. I take them as an independent recovery of these candidates from the same catalogue by a different method.

Four of his ten new objects are also in my catalogue, with the same orbits: his decam\_north\_0040 and decam\_north\_0051 are GCT0128 and GCT0275, both listed with their elements in version~2 of this paper (2026 November 17); his decam\_north\_0032 and decam\_north\_0030 (which he calls Enyo) are GCT0295 and GCT0390, found independently in my pipeline and entered in my list on 2026 November 18 and 21; they appear in this paper for the first time. For these four, his elements and mine differ by at most \gzFourA~au in $a$, \gzFourE{} in $e$ and \gzFourI$^\circ$ in $i$. GCT0295 is in tier~\gzTierNineFive{} here, because fewer than two of its oppositions have positions re-measured on the images; his detections of it in other archives are an independent confirmation. I have not submitted my positions of these four objects to the Minor Planet Center. In all, \gzAll{} objects of this catalogue, eleven of version~2 and two added since, were also found, independently, by a search that shares only the archive with mine.

\subsection{Machine-readable tables}\label{sec:cat_mrt}
Three tables are released with the paper: the catalogue (one row per object: tier, channel, oppositions, nights, positions measured on images and taken from catalogues, arc, rounded barycentric elements with uncertainties, $H$ and class; Table~\ref{tab:rare} shows the same columns for the objects of Section~\ref{sec:rare}), the evidence per object (image-verified and catalogue positions, oppositions with image-verified positions, positions from DECam and from other telescopes, detections on exposures that the link never used, and the results of version 1 where they exist), and the designated objects of this section. Table~\ref{tab:cat271} in Appendix~\ref{app:catalogue} prints fourteen columns of the first of these tables for all \ctReg{} objects. A fourth file lists one digest per undesignated object (Section~\ref{sec:data_avail}), and my positions of GCT0044 are released in a file of their own. For undesignated objects the tables contain no position, date, year of observation, epoch, node, argument of perihelion or anomaly, and they do not say which telescope other than DECam observed an object, except for the \ctChCfht{} objects of the CFHT linker, whose channel names it. The objects of the linkers on Subaru HSC frames and on DECam difference images share one channel in these tables, because each of those searches covered a small part of the sky and its name would locate its objects. This withholds the positions; it does not prevent their recovery. The detections of most objects are in public catalogues and the method is described here, so the rounded elements shorten a search considerably, and the objects of a class that fixes the longitude, such as the Trojans, are confined to a part of the ecliptic by their class alone. Section~\ref{sec:data_avail} says how the positions can be obtained.
\section{Objects of sparsely populated classes}\label{sec:rare}

Table~\ref{tab:rare} lists the tier-A objects of classes with few known members, by their rounded elements, with what each orbit rests on: the oppositions that hold image-verified positions and the numbers of image-verified and catalogue positions (Section~\ref{sec:cat_tiers}). The orientation angles are withheld with the positions (Section~\ref{sec:data_avail}), which limits what this section can claim.

\emph{Tests independent of the link.} The \vtTro{} candidate Trojans, the two retrograde candidates and the \ctFar{} objects with $a>\ctFarLim$~au of this section are \ctRareN{} objects (GCT0136 is in two of the groups). None of them is detected on two or more independent nights in a test on DECam exposures (Section~\ref{sec:valid_indep}). GCT0276 and GCT0014 are each detected on one such night, GCT0276 in the test of the multi-year linker (Section~\ref{sec:rare_trojan}) and GCT0014 in the test of version~1, and the other \ctRareNoTest{} were in none of the three tests. Whether one of them is among the \ctHscObj{} of the Subaru test is withheld (Section~\ref{sec:valid_indep}).

\begin{deluxetable*}{lccccccccccl}
\tabletypesize{\footnotesize}
\tablecaption{Tier-A objects of sparsely populated classes: rounded barycentric elements and the data behind them\label{tab:rare}}
\tablehead{\colhead{ID} & \colhead{$a$} & \colhead{$e$} & \colhead{$i$} & \colhead{$q$} & \colhead{$H$} & \colhead{Opp.} & \colhead{Nights} & \colhead{Arc} & \colhead{$N_{\rm img}/N_{\rm cat}$} & \colhead{Instr.} & \colhead{Class} \\
\colhead{} & \colhead{(au)} & \colhead{} & \colhead{(deg)} & \colhead{(au)} & \colhead{(mag)} & \colhead{(img)} & \colhead{} & \colhead{(yr)} & \colhead{} & \colhead{} & \colhead{}}
\startdata
\cutinhead{Candidate L4 Neptune Trojans}
GCT0014 & 29.99 & 0.043 & 22.8 & 28.7 & 7.6 & 8 (4) & 22 & 13.2 & 25/21 & D+ & L4 Trojan cand. \\
GCT0015 & 30.01 & 0.080 & 25.1 & 27.6 & 7.8 & 11 (5) & 28 & 19.0 & 14/34 & D+ & L4 Trojan cand. \\
GCT0044 & 30.03 & 0.065 & 20.3 & 28.1 & 6.8 & 10 (9) & 23 & 16.8 & 30/10 & D+ & L4 Trojan cand. \\
GCT0077 & 29.94 & 0.030 & 20.7 & 29.0 & 7.8 & 9 (9) & 22 & 18.2 & 47/0 & D+ & L4 Trojan cand. \\
GCT0276 & 30.04 & 0.023 & 13.2 & 29.3 & 8.1 & 6 (6) & 14 & 8.4 & 22/0 & D+ & L4 Trojan cand. \\
GCT0339 & 30.05 & 0.074 & 24.2 & 27.8 & 7.7 & 6 (6) & 11 & 9.1 & 14/0 & D & L4 Trojan cand. \\
\cutinhead{Retrograde}
GCT0084 & 53.85 & 0.572 & 138.5 & 23.0 & 8.5 & 7 (7) & 27 & 7.9 & 38/0 & D+ & scattering$^\dagger$ \\
GCT0136$^{a}$ & 207.6 & 0.890 & 110.3 & 22.8 & 8.3 & 4 (4) & 25 & 8.1 & 38/0 & D+ & scattering \\
\cutinhead{$a>\ctFarLim$~au}
GCT0098 & 273.2 & 0.880 & 14.8 & 32.8 & 7.3 & 8 (8) & 36 & 10.1 & 44/0 & D & scattering \\
GCT0104 & 280.6 & 0.868 & 19.6 & 36.9 & 7.3 & 6 (6) & 28 & 10.0 & 45/0 & D+ & scattering \\
GCT0124 & 310.1 & 0.881 & 20.1 & 36.7 & 6.8 & 8 (8) & 21 & 11.0 & 25/0 & D+ & scattering \\
GCT0175 & 255.3 & 0.859 & 12.7 & 35.9 & 7.5 & 5 (5) & 15 & 5.3 & 18/0 & D & scattering \\
GCT0182 & 304.5 & 0.888 & 11.0 & 34.0 & 7.2 & 5 (5) & 15 & 8.2 & 17/0 & D & scattering \\
GCT0233 & 152.6 & 0.865 & 34.1 & 20.6 & 8.7 & 6 (6) & 12 & 8.0 & 17/0 & D & scattering \\
GCT0245 & 158.3 & 0.768 & 17.4 & 36.7 & 6.8 & 5 (5) & 10 & 4.9 & 10/0 & D & scattering \\
GCT0344 & 167.9 & 0.821 & 32.4 & 30.1 & 7.9 & 4 (4) & 12 & 3.3 & 14/0 & D & scattering \\
GCT0355 & 150.9 & 0.726 & 31.4 & 41.4 & 7.1 & 2 (2) & 13 & 1.3 & 21/0 & D+ & detached$^\dagger$ \\
\cutinhead{$\ctHiIncLim^\circ<i<90^\circ$}
GCT0006 & 78.73 & 0.537 & 42.1 & 36.4 & 6.6 & 3 (3) & 8 & 6.3 & 4/5 & D+ & detached \\
GCT0085 & 51.05 & 0.673 & 44.7 & 16.7 & 9.6 & 5 (5) & 12 & 4.2 & 15/0 & D & scattering \\
GCT0087 & 43.37 & 0.678 & 71.0 & 14.0 & 10.1 & 5 (5) & 18 & 5.3 & 38/0 & D & scattering \\
GCT0217 & 71.04 & 0.477 & 40.9 & 37.2 & 7.7 & 4 (4) & 11 & 5.1 & 13/0 & D & detached \\
GCT0234 & 96.99 & 0.564 & 41.3 & 42.3 & 6.2 & 2 (2) & 9 & 4.1 & 9/0 & D & detached \\
GCT0336 & 79.61 & 0.475 & 41.2 & 41.8 & 6.6 & 4 (4) & 12 & 5.9 & 13/0 & D & detached \\
GCT0345 & 63.14 & 0.425 & 50.6 & 36.3 & 6.9 & 4 (4) & 9 & 10.1 & 11/0 & D & detached \\
\enddata
\tablecomments{Elements of the uniform fit (Section~\ref{sec:cat_orbits}). Opp., Nights, Arc: positions of this work; in parentheses, the oppositions that hold image-verified positions. $N_{\rm img}/N_{\rm cat}$: image-verified and catalogue positions. Instr.: D, DECam only; D+, DECam and at least one other telescope, which is not named because its public pointing list would narrow the sky where the object can be. Class: from the \ctClsMyr-Myr integrations (Section~\ref{sec:cat_orbits}); $^\dagger$: not secure. $^{a}$: GCT0136 also has $a>\ctFarLim$~au, so that \ctFar{} objects do.}
\end{deluxetable*}

\subsection{\vtTroCap{} candidate L4 Neptune Trojans}\label{sec:rare_trojan}
Four of the \vtTro{} were in version 1 \citep{cerati2026v1}: GCT0014, GCT0015, GCT0044 and GCT0077, with arcs of \ctTroArcMin--\ctTroArcMax~yr and inclinations of $\ctTroIMin^\circ$--$\ctTroIMax^\circ$. Their libration was tested in version 1, on the orbits of that version, and the test was not repeated on the uniform fit: all \ctTvOrb{} integrated orbits of GCT0014, GCT0015 and GCT0044 and all \ctTvOrbD{} of GCT0077 librate about L4 for \ctClsMyr~Myr, with half-ranges of $\lambda-\lambda_{\rm N}$ of \ctTvAmpA, \ctTvAmpB, \ctTvAmpC{} and \ctTvAmpD. Version 1 measured $g-r$ between \ctTvGrLo{} and \ctTvGrHi{} for the four, the colour range of most known Neptune Trojans \citep{lin2019,markwardt2023}. Two of the four rest in part on catalogue positions: GCT0014 and GCT0015 have image-verified positions in four of eight and in five of eleven oppositions (Table~\ref{tab:rare}).

\emph{An independent report.} GCT0044 corresponds to the L4 Trojan reported by \citet{jongma2026}, who found it in a stack of ZTF images and measured it on archival frames of \ctJoTel{} other telescopes: \ctJoPos{} positions on \ctJoNights{} nights over \ctJoArc~yr. His report is independent of this work, and his orbit rests mostly on data that I did not use: in his table of telescopes, \ctJcHisElse{} of his \ctJoPos{} positions, on \ctJcHisElseN{} of his \ctJoNights{} nights, come from \ctJcHisElseTel{} telescopes from which I have no position of this object. Most of my positions, in turn, are on exposures that he also lists: \ctJcShared{} of my \ctTjRows, among them \ctJcDecShared{} of my \ctJcDec{} DECam positions, lie within 90~s of a frame in his table of frames. The two orbits agree. His published elements, propagated with the giant planets, fit my \ctTjRows{} positions with an rms of $\ctJcRmsHis''$, against $\ctJcRmsMine''$ for my own orbit; the two orbits differ on the sky by less than $\ctJcSky''$ between 2008 and 2026, and in each element by at most $\ctJcSigMax\sigma$ of my fit. I have made no joint fit: his paper says that its positions are released with it, but no file of them accompanied its arXiv version when this was written. His libration half-range is $\ctJoAmp^\circ$, against the \ctTvAmpC{} of version 1, and his $H_V=\ctJoHv$ stands against my $H=\ctTjH$. His paper adds what mine does not have: \ctJoStable{} of \ctJoClones{} clones (\ctJoStablePct\%) remain L4 Trojans for \ctJoGyr~Gyr, and he notes that the object may be the brightest known L4 Neptune Trojan.

\begin{figure*}
\centering
\includegraphics[width=\textwidth]{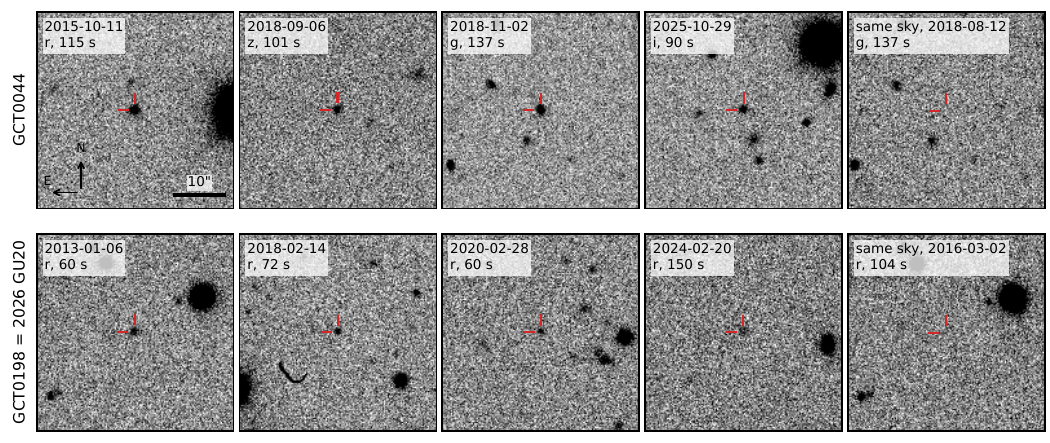}
\caption{DECam images of GCT0044 (top) and of GCT0198 = 2026~GU$_{20}$ (bottom), a designated object of Table~\ref{tab:designated}, from the public NOIRLab archive. Each panel is $\stSize''$ on a side, with north up and east to the left. The first four panels of each row show the object on four nights in time order, with the UT date, the filter and the exposure time. The ticks mark the position given in the released file; for the first three panels of GCT0044 this is an \nscdr{} catalogue position that I did not re-measure. All four images of 2026~GU$_{20}$ predate its discovery observation of \stDisc, by \stPreLo{} to \stPreHi~yr. The last panel of each row shows the sky position of one of the four (the third for GCT0044, the first for 2026~GU$_{20}$) on another night, on an exposure of equal or greater depth (2018 August 12 and 2016 March 2): nothing is at the ticks. The aperture signal-to-noise ratios measured for this figure are \stSnrTroLo{} to \stSnrTroHi{} for GCT0044, \stSnrDesLo{} to \stSnrDesHi{} for 2026~GU$_{20}$, and \stCtlTro{} and \stCtlDes{} at the ticks of the two control panels. The pixels are those of the Community Pipeline images, rotated by 90 degrees and not resampled. The display is a linear stretch between the 1st and 99th percentiles of the central $\stStretch''$ of each panel, in inverted grey scale; no other processing was applied.\label{fig:stamps}}
\end{figure*}

\emph{A second report.} Version~1 of this paper was submitted to arXiv on \ctVOneDate{} and the paper of \citet{jongma2026} on \ctJoSubmit. On \haSubmit{} \citet{han2026} submitted a paper on a blind search of TESS full-frame images, in which they link an object whose orbit, in their words, ``closely resembles'' GCT0044. They give $a=\haA$~au, $e=\haE$ and $i=\haI^\circ$. These are the elements of my orbit at the digits they print (mine are $a=\ctTjA$~au, $e=\tjEFull$, $i=\tjIFull^\circ$); they print no uncertainty and no other element for this object, so the three numbers alone do not show that it is the same one. They also report it from the ground: \haPos{} positions on \haNights{} nights between \haFirst{} and \haLast{} from Pan-STARRS1, DECam and 90Prime, and a ZTF stack with a signal-to-noise ratio of \haZtfSnr. In their integration of \haMyr~Myr with the four giant planets all \haClones{} clones of their posterior librate around L4, with an amplitude of $\haAmp$~degrees and a period of \haPer~kyr; they state that this establishes libration over the tested interval only. They call the identity provisional because the report they cite, which was then version~1 of this paper, withholds positions.
My positions of GCT0044 are published with this version (Section~\ref{sec:data_avail}), so the identity can now be tested by anyone who holds positions of their object. Figure~\ref{fig:stamps} shows the object on four of the DECam exposures of that file, beside one of the designated objects of Section~\ref{sec:cat_known}, whose positions are published as well.
I have made no joint fit with their positions. Their report is external to this work: it is not one of the tests of Section~\ref{sec:valid_indep} and adds nothing to the counts given there. GCT0044 is the only undesignated object of this catalogue that, to my knowledge, other authors have reported.

\emph{GCT0276}, the fifth, is a candidate of the multi-year linker. It has \ctTrRows{} positions on \ctTrNights{} nights in \ctTrOpp{} oppositions over \ctTrArc~yr: \ctTrDec{} from DECam and \ctTrOth{} from other telescopes. The fit keeps all of them with an rms of \ctTrRms{} and gives $a=\ctTrA$~au, $e=\ctTrE$ and $i=\ctTrI^\circ$, with $H=\ctTrH$. The libration test is that of version 1: the nominal orbit and \ctTrLibClones{} clones drawn from the covariance were integrated for \ctClsMyr~Myr with the four giant planets, and in all \ctTrLibN{} orbits the angle $\lambda-\lambda_{\rm N}$ librates about L4 for the whole run. The same test on the orbit fitted from four oppositions, before the last positions were added, gives the same result in all \ctTrLibFourN{} orbits, so it does not depend on the last extension of the arc. I do not print the centre and the amplitude of the libration, which would confine the object in longitude; they are in the withheld files. No colour has been measured for it. In the test on DECam exposures that its link did not use (Section~\ref{sec:valid_indep}) it has two independent nights, with one exposure on each. On one it is detected with an S/N of \ctTrUnuDetSnr, against \ctTrUnuDetExp{} expected. On the other the measured S/N is \ctTrUnuExcSnr, against \ctTrUnuExcExp{} expected, and it is excluded by the upper limit that I added on \ctUnuRuleDay, which takes such a measurement to be the flux of another source. This test, which the object passed before the limit was added, therefore shows it on one independent night, one short of the two that the standard of that section asks for.

\emph{GCT0339}, the sixth, was added after version~2 and is a candidate of the multi-year linker on \dtf. It has \vtTnRows{} positions on \vtTnNights{} nights in \vtTnOpp{} oppositions over \vtTnArc~yr, all from DECam. The uniform fit keeps all of them with an rms of \vtTnRms{} and gives $a=\vtTnA$~au, $e=\vtTnE$ and $i=\vtTnI^\circ$, with $H=\vtTnH$. Its libration test is the one made for GCT0276, on the uniform fit: the nominal orbit and \vtTnLibClones{} clones drawn from the covariance were integrated for \ctClsMyr~Myr with the four giant planets, and in all \vtTnLibL{} of the \vtTnLibN{} orbits the angle $\lambda-\lambda_{\rm N}$ librates about L4 for the whole run. As for GCT0276, I do not print the centre and the amplitude of the libration. It has not been through any of the tests on exposures that its link did not use.

Stability beyond \ctClsMyr~Myr is established for one of the \vtTro, GCT0044, by the integrations of \citet{jongma2026}; I have not tested it for the others.

\emph{Context.} All \vtTro{} candidates are at L4. The MPC listed \ctNtAll{} Neptune Trojans in late 2025, \ctNtLFour{} at L4 and \ctNtLFive{} at L5 \citep{murtagh2026}, and the Rubin Observatory has since reported \ctNtRubin{} long-term stable L5 Trojans, which doubled the number of those known \citep{schwamb2026}. Known Neptune Trojans come from targeted searches and from wide surveys \citep{sheppard2006,gerdes2016,lin2016,bernardinelli2022}, and the Rubin survey is predicted to find many more \citep{murtagh2026}. The absence of L5 objects here says nothing about L5. In the years of \nscdr{} that region lay in front of the Galactic centre, in tiles that the linkers of version 1 skip and where the later linkers recover few injected objects (a fraction of \ijPCrowd{} in a crowded tile; Section~\ref{sec:valid_inj}). In 2020--2026 it has moved away from the Galactic plane, but I have not measured what \dtf{} covers there. \vtTroCap{} candidates from a search without a selection function do not constrain the size or the inclination distribution of the population.

\subsection{Two retrograde candidates}\label{sec:rare_retro}
Tier~A has \ctRetNew{} objects with $i>90^\circ$.

GCT0084 was in version 1 with an arc of one year and a class that was not secure. It now has \ctRbRows{} positions on \ctRbNights{} nights in \ctRbOpp{} oppositions over \ctRbArc~yr, \ctRbOther{} of them from instruments other than DECam, all kept by the fit (rms \ctRbRms): $a=\ctRbA$~au, $e=\ctRbE$, $i=\ctRbI^\circ$, $q=\ctRbQ$~au. On this orbit it is a \ctRbCls{} object in \ctRbAgree{} of \ctRbNorb{} integrated orbits and a detached one in the other \ctRbRest\ctRbSecNote.

GCT0136 has $a=\ctRaA\pm\ctRaSigA$~au, $e=\ctRaE$, $i=\ctRaI^\circ$ and $q=\ctRaQ$~au, from \ctRaRows{} positions on \ctRaNights{} nights in \ctRaOpp{} oppositions over \ctRaArc~yr, \vtRaDec{} measured on DECam images and \vtRaOth{} on frames of another telescope, all kept by the fit (rms \ctRaRms). It was found by the dense-tile linker on \dtf.
\ctRaClsSentence

A third retrograde object of the catalogue, GCT0173, is the Rubin discovery 2026~DS$_{78}$ ($i=\ctDsI^\circ$, $a=\ctDsA$~au; Table~\ref{tab:designated}), for which the archive holds \ctDsNights{} nights back to \ctDsFirst.

For scale, the MPC orbit file of \kcRunDate{} lists \ctRcN{} minor planets with $i>90^\circ$ and $q>\ctRcQ$~au. Of these, \ctRcShort{} have single-opposition arcs of \ctRcShortLo--\ctRcShortHi~days, 2026~DS$_{78}$ among them, and \ctRcLong{} have multi-opposition arcs. The first found, 2008~KV$_{42}$, has $i=\ctKvI^\circ$ and $q\sim\ctKvQ$~au \citep{gladman2009}, and 2011~KT$_{19}$ has a similar orbit \citep{chen2016}. The two candidates here also have perihelia between Uranus and Neptune and arcs of \ctRbArc{} and \ctRaArc~yr, so they would add two orbits to the \ctRcLong{} with multi-opposition arcs. Of those \ctRcLong, \ctRcLongFar{} have $a>\ctFarLim$~au, with $q$ of at most \ctRcLongFarQ~au; GCT0136 has $a=\ctRaA$~au with $q=\ctRaQ$~au. I make no statement on the orientation of the orbital planes: the nodes are withheld, and two objects selected by an uneven footprint could not support one.

\subsection{Candidates with $a>\ctFarLim$~au}\label{sec:rare_far}
Tier~A has \ctFar{} objects with a barycentric semimajor axis above \ctFarLim~au, \vtFarSec{} of them by more than three times its uncertainty. They have \ctFarOppMin{} to \ctFarOppMax{} oppositions, arcs of \ctFarArcMin--\ctFarArcMax~yr and $\sigma_a/a\leq\ctFarSigMax\%$; \ctFarNonDec{} have positions from a second instrument. Their semimajor axes run from \ctFarAMin{} to \ctFarAMax~au and their eccentricities from \ctFarEMin{} to \ctFarEMax.

Their perihelia run from \ctFarQMin{} to \ctFarQMax~au. Of the \ctFar, \ctFarQThirty{} have $q$ between \ctFarQLo{} and \ctFarQMax~au, \ctFarATwoFifty{} of them with $a>\ctFarLimB$~au. The other \ctFarQIn{} have perihelia inside the orbit of Neptune: the retrograde GCT0136, and GCT0233 with $q=\ctFarQMin$~au. \ctFarClsSentence{} For one of them, a 100-Myr integration of \ctEtN{} orbits made on an earlier fit shows the kind of motion: $a$ changes by \ctEtDaLo--\ctEtDaHi~au in \ctClsMyr~Myr while $q$ stays between \ctEtQLo{} and \ctEtQHi~au. \ctFarLongSentence

Whether these objects belong to the samples in which an alignment of distant orbits has been discussed depends on the cut: \ctFarQThirty{} of the \ctFar{} satisfy the $a>\ctFarLim$~au, $q>\ctQLimA$~au of \citet{trujillo2014}, \citet{shankman2017} and \citet{bernardinelli2020psj}, \ctFarATwoFifty{} the stricter $a>\ctFarLimB$~au of \citet{bernardinelli2020psj}, \vtFarQC{} the $q>\ctQLimC$~au of that paper, and \vtFarQD{} the $q>\ctQLimD$~au of \citet{brown2021}. GCT0233 ($a=\ctFarXA$~au, $q=\ctFarQMin$~au, $i=\ctFarXI^\circ$) is a Neptune-crossing orbit of long period, of the kind discussed by \citet{batygin2024}, and I do not assign it to their sample. This paper cannot test an alignment: the sample is selected by where and when the archive holds enough nights, a footprint that is uneven in ecliptic longitude and has no selection function (Section~\ref{sec:limits}), which is enough to produce or to hide an apparent alignment in samples of this size \citep{shankman2017,bernardinelli2020psj,napier2021}, and the angles that such a test needs are withheld here. I give no longitude of perihelion, no pole and no statistic on them.

\subsection{High inclinations and large perihelia}\label{sec:rare_inc}
Tier~A has \ctHiInc{} objects with $i>\ctHiIncLim^\circ$: the two retrograde ones and \ctHiIncMid{} between $\ctHiIncLim^\circ$ and $90^\circ$ (Table~\ref{tab:rare}). GCT0087 ($i=\ctHbI^\circ$, $q=\ctHbQ$~au) was a single-opposition object in version 1 and now has \ctHbOpp{} oppositions; it is a scattering object in all its clones. Among the \ctTierA{} tier-A objects the median inclination is $\ctIncMed^\circ$ and \ctIncThirty{} have $i>\ctIncLim^\circ$. Of the tier-A objects, \ctQForty{} have $q>\ctQLimB$~au, \ctQFortySec{} of them by more than three times the uncertainty of $q$; the largest of these is still that of GCT0069 ($q=\ctQMax$~au), described in version 1. These numbers are not an inclination or a perihelion distribution, and none is inferred from them.
\section{Discussion}\label{sec:discussion}

\subsection{What public archives still yield, and why}\label{sec:disc_yield}
The exposures searched here have been public for years, and parts of them were searched before: by DES in its own footprint, by \citet{fasbender2021} for tracklets, and with THOR by the Asteroid Institute, which reported \ctThorN{} candidate asteroids from the same catalogue, most of them in the main belt \citep{b6122024}. This search adds \ctTierA{} multi-opposition candidates and matches to \pdKnownDistinct{} designated objects. If the candidates are real, the tests of Section~\ref{sec:valid} suggest why earlier searches left them.

First, the nights. More than half of the known objects with secure orbits never have \dnNmin{} nights within one season and one tile (\lbNscCadPct\% in \nscdr, \lbDtCadPct\% in \dtf), so a single-season rule cannot reach them however good the linker is. Linking across years reaches part of them: in the test outside the DES footprint, where no survey planned the cadence, the multi-year linker recovered \exMy{} of \exN{} known objects against \exList{} for the single-season lists. Over the tiles of \nscdr{} that it read it contributed \ctChMy{} candidates, of which \vtMyYield{} keep \myNmin{} nights after vetting. Second, the exposures of 2020--2026 had no public catalogue of detections at all; \vtDtAllA{} tier-A objects come from the one extracted here (\ctDtA{} from the dense-tile linker and \vtMyDtN{} from the multi-year linker). Third, the channels are complementary: on the same \raN{} known objects the best single channel recovers \raMyPct\% and all together \raAnyPct\%. Fourth, an object designated on a short arc is not recognised by a check made by position at the archival epochs, and its archival positions stay unlinked. That was a fault of my earlier check, and the check by orbit found \kcMissed{} identities that it had missed.

Two limits are measured as well. The recall of the multi-year linker on known objects is \ctRmVA\% for $V<21$, \ctRmVB\% at $V=21$--22 and \ctRmVD\% at $V=23$--24. Faint objects are lost because single exposures do not detect them, and no linker changes that. Objects brighter than $V=21$ are also found less often than those of $V=21$--23. One cause is known: the loader shared by the dense-tile and multi-year linkers drops rows brighter than magnitude 20 in their band. I have not measured how much of the loss it explains. Below \myNmin{} nights the real sky and the shifted controls give counts that cannot be told apart, so sparser objects cannot be separated from chance without an independent test on images.

\subsection{Consistency with DES and OSSOS}\label{sec:disc_des}
DES and OSSOS characterise their detection efficiency and release survey simulators \citep{bannister2018,bernardinelli2022}. This search does not, so its yield cannot be compared with theirs. What can be tested is consistency: whether those surveys should have found an object that I report. Version 1 propagated each of its objects to every exposure of both surveys and found no contradiction except one case that it could not explain \citep{cerati2026v1}. I repeated that test, with the same procedure and the uniform orbits, for the \ctDcN{} undesignated multi-opposition objects added since (\ctDcA{} of them in tier~A) by version~2; it was not repeated for the objects of version 1, nor for the \vtDcVThree{} undesignated multi-opposition objects added in this version.

Of the \ctDcN, \ctDcOn{} fall on a DES CCD at least once. For \ctDcLo{} of them the probability of meeting the DES requirements (at least seven detection nights, an arc beyond 180 days and three nights within 90 days) is at most \ctDcLoMax. For the other \ctDcHi{} it is \ctDcHiMin{} or more. Of these, \ctDcNear{} were closer than the inner limit of the DES search in the DES years \citep[$d>\ctDcLim$~au;][]{bernardinelli2022}. The other \ctDcFar{} are predicted to be detectable on DES CCDs on \ctDcFarNa{} and \ctDcFarNb{} nights, at a median of \ctDcFarMa{} and \ctDcFarMb~mag brighter than the 50\% completeness of those exposures, and they are not in the DES catalogue. Both are in the DES data: the single-exposure catalogue holds detections of them at the predicted places, and forced photometry on the DES exposures shows them (Section~\ref{sec:valid_indep}). They are objects that DES did not link, and I cannot say why. They lie close to the single-exposure limit, with a forced S/N between 4 and \ctChkSnrHi{} on \ctChkFourAll{} of the \ctChkMeasAll{} exposures measured. Their magnitudes are not on one photometric system (Section~\ref{sec:cat_orbits}), and with magnitudes fainter by 0.3 and 0.5~mag the probability of meeting the DES requirements falls to \ctChkPa{} for one and to \ctChkPb{} for the other, so the magnitudes are not good enough to call either case a contradiction. My own linkers did not link these detections either: they are single detections on separate nights.

For OSSOS, \ctOcIn{} objects fall inside a field outline, and \ctOcTrip{} of them has a night with three or more exposures there, at a predicted $r=\ctOcR$, brighter than the limits of the OSSOS blocks ($r=\ctOcLimLo$--\ctOcLimHi). That night is not one of the nights on which the objects of that block were discovered \citep{bannister2018}; on those nights the object is inside a field outline on a single exposure. The case is explained by the geometry of the survey, at the level of field outlines and not of CCDs. Which objects these are is withheld with the positions.

\subsection{Beside Rubin}\label{sec:disc_rubin}
Within its footprint Rubin will find most of what a 4-m archive holds. What an archive can add is earlier positions. Of the \ctDesRubin{} Rubin discoveries in the catalogue, \ctDesRubinOne{} have a published single-opposition arc and \ctDesRubinMulti, 2026~GJ$_{24}$ and 2026~EF$_{144}$, have arcs of three oppositions; my positions give each \ctDesRubMin--\ctDesRubMax~yr of arc before its discovery;  these positions (published with this version for the objects of version~2, and on request for the others; Section~\ref{sec:data_avail}) can turn a discovery arc of weeks into a multi-year orbit.{} Among the known candidates of the multi-year run, which are matches in the catalogue detections and have not been examined on images, \ctMkNew{} were discovered in 2025--2026 (\ctMkRubin{} by Rubin), with nights \ctMkNewMin--\ctMkNewMax~yr older than their discovery. The linker does this blind, so the same run that yields candidates yields the earlier arcs of objects that will be discovered later, as long as their orbits are compared at the right epochs (Section~\ref{sec:vet_known}). In 2026 November, \ctAreaNotRubin~deg$^2$ of the cells of Figure~\ref{fig:coverage} with at least \dnNmin{} nights, which cover more than the sky that was read, also lay outside the sky that Rubin had observed in the public record; that advantage is temporary (Section~\ref{sec:data_cov}).

Two by-products are proposals and not results. The first is the precovery of designated objects: \ctMkPub{} objects with archival nights before their first published observation, and \ctMkPre{} with nights before their discovery observation (Section~\ref{sec:cat_known}). The second is the recovery of designated objects with poor orbits, by seeding the linker with their published astrometry (Section~\ref{sec:valid_seed}): of \lrAllSeeded{} such objects, \lrAProp{} have a proposed recovery, \lrASingle{} of them objects seen in one opposition only, with new arcs of \ctLostArcLo--\ctLostArcHi~yr. The positions behind both are catalogue positions. None has been examined on images, and until they are I report counts and not astrometry.

\emph{The public alerts.} Rubin removes from its alerts the detections that it attributes to known objects, and it needs several detections in a night to link a new one. A slow object seen once or twice per night on several nights therefore stays in the alert stream as unassociated detections. I ran a linker on those detections, as distributed by the Fink broker \citep{moller2021}, over \rvRubTiles{} tiles of about one square degree (\rvRubArea~deg$^2$) with all the public alerts of 2026 up to the middle of November. The candidate rule asks for a bound orbit with an rms of at most $0\farcs2$ at a distance of at least 25~au, and its night floor was set from five shifted- and permuted-sky controls before the real links were read: with \rvRubNstar{} or more nights the controls give no link. The rule recovers \rvRubDev{} designated objects of the development half and \rvRubHeld{} of the held-out half. On the real sky \rvRubLinks{} links pass. \rvRubKnown{} of them are designated objects, all of them Rubin discoveries of 2025--2026, and \rvRubReg{} are objects of my catalogue found again from the alerts alone. The one remaining link was examined on DECam exposures of two later seasons that its orbit predicts: it is not detected on them, so it is not a candidate and is not part of the catalogue.

No link beyond 80~au passes the rule. For synthetic objects added to the same tiles, \rvRubElig\% of those that have enough nights in a tile are recovered, but only \rvRubPlaced\% of all those placed, because most of the alert footprint has fewer than \rvRubNstar{} nights per tile in 2026. From the effective areas of the synthetic test (\rvRubAreaA--\rvRubAreaC~deg$^2$ at 80--1000~au) the 95\% upper limits on the surface density of such objects at $V=21.5$--24 are \rvRubLimA, \rvRubLimB{} and \rvRubLimC~deg$^{-2}$ at 80--150, 150--300 and 300--1000~au. They hold for the synthetic population and do not model Rubin's own image differencing. What limits this search is the number of nights in the alerts, not the linker, and the limits will drop as the survey goes on.

\subsection{Limits}\label{sec:limits}
\begin{enumerate}
\item \emph{No follow-up, one pipeline.} No object was observed on purpose. Apart from the joint fits of the \ctDes{} designated objects with the astrometry of their discoverers, the tracklets of other observers in the fits of \ctRowsOtherObj{} objects, the independent report of GCT0044 by \citet{jongma2026}, whose orbit rests mostly on data that I did not use, and the object that \citet{han2026} match to it provisionally, every position was measured by one pipeline, mine, on archival frames.
\item \emph{No validated selection function.} Recall is measured on known objects, by magnitude and latitude. Known objects were found by surveys with their own footprints and depths, so this recall is not a completeness for unknown ones. A first simulation, in which synthetic objects were injected into the frozen multi-year and dense-tile linkers on \nscdr{} and on \dtf, gives mean efficiencies over its sampled population of \rvSfNscMy\%, \rvSfNscD\%, \rvSfDtMy\% and \rvSfDtD\% for the four chains, and its shifted-sky controls agree with the real sky. It failed its own held-out tests, however: its linking model does not pass on held-out blocks, and on the test half of the known objects it predicts \rvSfOverMy\% (multi-year linker) and \rvSfOverD\% (dense-tile linker on \nscdr) more recoveries than are observed, mainly because it gives each object too many nights with detections (\rvSfNightsSim{} against \rvSfNightsObs). Its efficiencies are therefore not used here. Nothing in this paper supports an inference on class fractions, size distributions or orbital orientations.
\item \emph{Most candidates have no test of their link.} The image rules re-measure the detections of a link on their frames. They remove artefacts and static sources and would pass a chance link of real moving sources, and for the multi-year channel the rate at which they accept one is not measured. For the production rule of the single-season channel on \dtf{} it is: the chain called \gsPoss{} of the \gsProd{} control links that pass that rule possible and none real (Section~\ref{sec:vet_dt}); the other selections of that channel and the extension of arcs into other seasons have no such test. Of the \ctTierAvTwo{} tier-A candidates of version~2, \ctIndepN{} are detected on two or more nights that their fits did not use and \ctIndepOne{} more on one such night (Section~\ref{sec:valid_indep}). These counts apply an upper limit that I added to the tests on DECam exposures on \ctUnuRuleDay, after their results had been seen: a measurement with an S/N above \ctUnuRuleFac{} times the expected one is not a detection, because the rule written beforehand had no upper limit, so that the flux of a neighbouring source would count. In the test of the multi-year linker, \ctUnuExcExp{} exposures of \ctUnuExcObj{} objects are excluded, and \ctUnuTwo{} of the \ctUnuTested{} tested objects pass instead of \ctUnuTwoNoRule. The limit is post hoc and its factor is my choice (with a factor of five, 25 of the \ctUnuTested{} would pass); the expected S/N rests on a predicted magnitude, not on a measurement of that night, so the limit can also remove true detections, and it is not applied to the Subaru test. One tier-A candidate of the multi-year linker fails this test: it is not detected at the predicted position on any of its \ctUnuNoneNights{} independent nights. A search of a wider set of orbits made afterwards finds it on the three exposures combined, outside the central 95\% interval of its fit, with no control track as high; that search was not part of the test, its rule was written after the single exposures had been looked at, and it was applied to this object only, so it is not counted. It stays in the catalogue. For \ctEvUnu{} objects there are detections on unused exposures without control positions. The \ctNewOneOpp{} links seen in one opposition are listed and not counted: their orbits and classes are not determined. In version~2, \dtTierB{} such links came from the dense-tile linker on \dtf, \dtProdSingle{} of these from its production rule, for which \gsFalseLo--\gsFalseHi{} chance links are expected to pass the image chain among the \dtFSentA{} links of the whole selection, and fewer than \gsFalseUp{} at \ciLevel\%.
\item \emph{What the controls show.} The three controls of the multi-year production are complete on \pdCtlCollected{} of \pdBlocks{} blocks, all without a candidate; on \ctCtlIncomplete{} blocks a restart of one machine interrupted them, and one of \pdSeedUnits{} seed units of one block failed in the real pass. Before the two final cuts \ctCtlSets{} control sets reach \myNmin{} nights (none more than \ctCtlMaxN), and none has a chance score below \ctCtlPmin, against the limit of \myPch. The zero therefore depends on those cuts, which were tuned on the development half of the pilot and on a quarter of one control. The controls limit pure chance alignments and not links that mix real and chance detections: set against the \pdProp{} proposals that were not known objects, the fraction not explained by chance alignments is above \puPropLo, and above \puNewSevenLo{} for the \puNewSeven{} that had seven nights. At six nights real and control counts cannot be told apart (\puBSix{} against \puBSixCa, \puBSixCb{} and \puBSixCc), and in the production run no candidate of the multi-year linker with fewer than \myNmin{} catalogue nights was examined on images (the six-night level added since has its own rule and controls, Section~\ref{sec:link_six}); the \ctMySix{} six-night objects of that channel lost a night at vetting. The held-out half was the easier one (the single-season lists find \hoEasyHeld\% of it against \hoEasyDev\% of the development half), and one of the criteria fixed in advance was met by a single object (Section~\ref{sec:valid_held}).
\item \emph{Gaps in the recall.} The multi-year linker has no injection test: its recall is known only from known objects (Section~\ref{sec:valid_inj}), which do not sample large distances, high inclinations or high eccentricities well. Of the known objects that satisfy its rule, \ctEdgePct\% do so in no single block and cannot be found, and by the same proportion an estimated \edLostNew{} objects (\edLostNewCI) were lost.
\item \emph{Sky that was not searched.} No linker read all of \nscdr. All of them took their tiles from one local copy of the catalogue, which lacks tiles missing by accident, most of them after failed downloads, and most tiles above a density cut (Section~\ref{sec:link_my}). The multi-year linker searched \nhSearched{} of the \nhTiles{} tiles that have \myNmin{} or more nights inside its block grids, \nhSearchedArea{} of \nhArea~deg$^2$. The tiles missing by accident left out \nhAccArea~deg$^2$, with \nhAccEcl{} tiles at ecliptic latitudes under $\nhLat^\circ$, and the density cut \nhDenseArea~deg$^2$, mostly near the Galactic plane. That cut was not in the file written before the production. The failed downloads had been found by an audit on \nhAuditDay{} and were not repaired before the run; what the multi-year linker missed was counted after it, on \ctlReadDay. The recall on the \raN{} known objects counts the objects of those tiles as lost. The recall given the nights (\raMyEligPct\% of \raElig{} objects) and the loss at block edges refer to the downloaded tiles only, because an object has rows only there: by prediction, \nhKnownLost{} of the \raN{} objects have their \myNmin{} nights only when tiles without a file are counted, \nhKnownLostAcc{} of them through the tiles missing by accident (Section~\ref{sec:valid_edge}). The areas and the map of Section~\ref{sec:data_cov} are drawn from the downloaded files of \nscdr{} in shifted boxes: they are not its footprint, and they cover more than the sky that was read. Some searched tiles also lack rows: near right ascension $0^\circ$ my query missed detections that the catalogue stores on the other side of $0^\circ$ from the one it asked for, and a count of \nhRnCounted{} of the \nhRnAll{} tile files nearest to it, not complete when this paper was written, gives \nhRnAff{} files that together lack \nhRnMissPct\% of their rows, \nhRnSearched{} of them searched tiles (Section~\ref{sec:link_my}). The search described here is the one that stood on \ctlReadDay, and the tiles without a file on that date are not part of it; a later search of those tiles, with the same multi-year linker and its six-night level, is the source of \vtHolesN{} objects of the catalogue (\vtHolesA{} of them in tier~A), which are counted with their channels (Section~\ref{sec:cat_channels}). I have not made the same audit of \dtf; in Table~\ref{tab:budget} no object of that set is lost to a missing tile.
\item \emph{Identification by screen.} Most known candidates of the multi-year run are identified by a match within $\pdTol''$ and not by a joint fit, and none of these matches was examined on images (Section~\ref{sec:cat_known}).
\item \emph{DES and OSSOS.} By my orbits and magnitudes DES should have linked \ctDcFar{} tier-A objects that it does not list. Both are in the DES data, and why they were not linked is not established (Section~\ref{sec:disc_des}). The test was not repeated for the objects of version 1 with their present orbits, and it does not cover the \vtDcVThree{} undesignated multi-opposition objects added in this version.
\item \emph{Quality of the data.} In tier~A, \ctARowsCat{} positions of \ctACatObj{} objects come from catalogues and were not re-measured on the frames, and the label image-verified is taken from the files and not from a new inspection (Section~\ref{sec:cat_tiers}). $H$ is not on one photometric system and its brightest values are unverified; \ctClsOld{} classes come from integrations of earlier orbits (Section~\ref{sec:cat_orbits}). The extraction of 2020--2026 has known defects that lose exposures (Section~\ref{sec:data_dt}), the test of the CFHT linker failed on recall, and the join of \dtf{} rows to the \nscdr{} seeds failed both its tests (Section~\ref{sec:null}), so the two DECam sets are linked separately. An audit of the time tags of the DECam positions of version~2 found \ctTtStart{} at exposure start, in \ctTtStartObj{} objects, and \ctTtNo{} that could not be matched to an exposure; the \ctTtStart{} were moved to mid-exposure before the fits of this version (a change in $a$ of at most \ctTtDa~au and in $e$ of at most \ctTtDe), and not all of the others have been resolved.
\item \emph{Positions.} The astrometry of the undesignated objects is not public (Section~\ref{sec:data_avail}), so an independent check of an object requires a request to me, and the rounded elements are given without their epoch.
\end{enumerate}

\subsection{Null results}\label{sec:null}
Several things did not work; they are limits of the methods I tried, not of the archives. Of the tests of this project that had a rule written beforehand (Appendix~\ref{app:prereg}), \prFail{} failed and \prNoGo{} ended in a decision not to proceed. Two of them qualify the main result.

\emph{Joining \dtf{} rows to the \nscdr{} seeds.} A test of the multi-year linker on \nscdr{} and \dtf{} rows together, in one region, failed its first criterion: it recovered \jnJoint{} of \jnDen{} secure objects against \jnNsc{} for \nscdr{} alone, where a gain of \jnThr{} points was required. A second test, with a corrected row key and seeds of both eras, recovered \fjJoint{} of the \jnDen{} but failed too: under rule A2 its shifted-sky control returned \fjCtl{} candidates where the limit was \fjCtlLim, and it cost \fjCpu{} core-hours against a limit of \fjCpuLim. On the joined catalogue, then, the two final cuts do not keep the control empty.

\emph{Looser cuts in the multi-year linker.} Relaxing the two final cuts of rule A2 gains no real candidate in the regions outside the DES footprint ($\chinu\leq\rxChi$, or $\Pch\leq\rxPa$ or \rxPb), and with \rxPb{} a control candidate already passes. The night floor cannot be lowered with this rule unless a test independent of the link is added (Section~\ref{sec:valid_ctl}).

Two other failures are cited above. A multi-apparition linker on detections of the CFHT MegaCam catalogue failed its recall criterion on held-out stripes (\cfRec{} of \cfDen{} known objects, where \cfThr\% was required, with no wrong link). The DECam Community Pipeline \citep{valdes2014} marks the transients of its own image differencing with data-quality value \pfCode, and both \nscdr{} and my extraction mask those pixels; the designated object \pfObject{} loses \pfMasked{} of its \pfPub{} published positions to this mask. A linker on detections that keep the flagged pixels failed its count clause as a search (\pfReal{} objects with two apparitions on the real sky against \pfCa, \pfCb{} and \pfCc{} in three controls), and a version on the flag masks alone was inconclusive. A join of tracklets of the MPC Isolated Tracklet File across seasons gave no excess over chance. The other tests are listed with their outcomes in a file released with this article (Section~\ref{sec:data_avail}).
\emph{Since version~2.} Five more tests had no positive result. A second version of a search on DECam difference images in a crowded field, meant to reach fainter objects than the first, did not: its linked depth was $r=\rvDfDepth$ where the frozen gate asked for \rvDfGate, and by its rule the line of work stopped there. A search of the single-season candidates of \nscdr{} for a second opposition at the level of the catalogue, in which each candidate's orbit is projected into the other seasons and the rows found there must fit with it, recovered \rvSoTruth{} seasons of designated objects with no false one and no candidate in \rvSoCtl{} control seasons; of the \rvSoUnk{} unidentified candidates that it could test, every one that gained a second opposition turned out to be a designated object, so it adds no new object, and the same search on \dtf{} found only a designated one. Seeds made of pairs of nights in different seasons, meant to reach distant objects that never have three nights close together, failed two of their three criteria: they recovered \rvSxRec{} synthetic objects where half was required, and the whole footprint would have cost, by my estimate, about \rvSxCost{} core-hours against a limit of 600. A search for cometary activity among \rvAcTargets{} objects of the catalogue, all but one with perihelia below 25~au or Centaur orbits, with \rvAcCtl{} bright objects as controls, found no candidate in either group; with these data a coma is seen only if it carries \rvAcComa\% of the light within \rvAcAper, and the best limits on $Af\rho$ are about \rvAcAfLo--\rvAcAfHi~cm. A linker between oppositions on one set of Subaru HSC frames passed its controls and its synthetic test, but none of the objects it could search had a predicted position good enough on the frames of the other season, so it confirmed nothing.
\section{Summary}\label{sec:summary}
\begin{itemize}
\item I searched the single-night detections of the part of \nscdr{} (\nscYears) that I had downloaded and of my own extraction from \ctDtExp{} DECam exposures of 2020--2026. In the cells of Figure~\ref{fig:coverage}, \ctAreaFour~deg$^2$ have at least \dnNmin{} nights; that is more than was read, because for \nscdr{} a cell is filled from a box of four tiles whether or not its own tile has a file: the \nscdr{} tiles with a file and \dnNmin{} or more nights cover \nhFileFour~deg$^2$, against \ctAreaNsc~deg$^2$ in the figure. The search did not cover all of \nscdr: the multi-year linker read \nhSearchedArea{} of the \nhArea~deg$^2$ that have \myNmin{} or more nights inside its block grids. Tiles missing from my copy by accident, most of them after failed downloads, left out \nhAccArea~deg$^2$, and a density cut that was not in the rule file of the production left out \nhDenseArea~deg$^2$.
\item On \raN{} known objects with secure orbits over \nscdr, single-season linking recovers \raDensePct--\raBothPct\%, the multi-year linker \raMyPct\% and all together \raAnyPct\%. Most objects never have the nights that a single-season rule asks for (\lbNscCadPct\% of the objects in \nscdr, \lbDtCadPct\% in \dtf), and block edges put \ctEdgePct\% of those that satisfy the multi-year rule out of reach.
\item The multi-year rule was fixed before its held-out tests. It recovers \ctHoPct\% of DES objects brighter than $r=\hoMagA$ in a strip inside the DES footprint (single-season linking: \ctHoListPct\%) and \exMy{} of \exN{} known objects in a test outside it. Its shifted-sky controls give no candidate with \myNmin{} or more nights on the \pdCtlCollected{} of \pdBlocks{} blocks whose controls are complete. For the \pdProp{} candidates that were not known objects this limits pure chance alignments to under \puPropChancePct\% (\puNewSevenChancePct\% at seven nights); it does not limit links that mix real and chance detections, and with six nights real and control counts cannot be told apart.
\item A check of identity by orbit, at the epochs of the designated objects, found \kcMissed{} identifications that my earlier check by position had missed; repeated on the whole catalogue with a repaired screen, it finds no further identity. One run of the multi-year linker matched \pdKnownDistinct{} designated objects in the catalogue detections, nearly all by a position screen and none examined on images; \ctMkPub{} have nights before their first published observation. For the \ctDesRubin{} Rubin discoveries in the catalogue the archive holds \ctDesRubMin--\ctDesRubMax~yr of earlier arc.
\item The catalogue has \ctReg{} objects with uniform orbits: \ctHead. Of the \ctTierAvTwo{} tier-A candidates of version~2, \ctIndepN{} have been detected on two or more nights that their fits did not use and \ctIndepOne{} more on one such night; for the others the evidence is the orbit, and \ctUnuNone{} of them was tested on \ctUnuNoneNights{} such nights and is not detected at the predicted position; a later search outside the rule of that test finds it and is not counted. The \ctTierAAdded{} tier-A candidates added since version~2 have not yet been through these tests. The image chain of the 2020--2026 channel called \gsPoss{} of the \gsProd{} control links that pass its production rule possible and none real, which puts the chance links expected to pass it, among the \dtFSentA{} links of that rule, at \gsFalseLo--\gsFalseHi, and at fewer than \gsFalseUp{} at \ciLevel\%; that rule is the source of \dtProdSingle{} of the one-opposition links. Tier~A includes \vtTro{} candidate L4 Neptune Trojans whose nominal orbits and clones librate for \ctClsMyr~Myr, one of them reported independently by \citet{jongma2026}, whose orbit rests mostly on data that I did not use and agrees with mine, and matched provisionally by \citet{han2026} to an object linked in TESS images; \ctRetNew{} retrograde candidates; and \ctFar{} candidates with $a>\ctFarLim$~au, which cannot test an alignment of distant orbits without a selection function and with the angles withheld.
\item Searches for slow, distant objects in the same data find no new object and recover three known ones blind, 2014~UZ$_{224}$, Gonggong and (532037) Chiminigagua at \rvChR--\rvUzR~au (Section~\ref{sec:far}).
\item Nothing was followed up, there is no validated selection function, and no population is inferred. Two tier-A objects that DES should have linked according to my orbits and magnitudes are present in the DES data; why DES did not link them is not established. The astrometry of \ctSent{} objects was submitted to the Minor Planet Center; the positions of all undesignated objects are available on request and are committed by one digest per object; my positions of the \vtDesRel{} designated objects of version~2 and of GCT0044 are published with this version.
\end{itemize}
\newcommand{\astrosha}{\texttt{b23e5cfc8d63999b\allowbreak aff2dcfdf37ff934\allowbreak f63b0dda5d56fb13\allowbreak 9d5fa1c2ab6b0bfd}}
\newcommand{\astroshadate}{2026 November 25}
\newcommand{\astroshaobj}{365}
\newcommand{\astroshapos}{6{,}269}
\newcommand{\astroshaRelDes}{352}%
\newcommand{\astroshaRelTroj}{40}%
\newcommand{\astroshaVTwoObj}{251}%
\newcommand{\astroshaVTwoSame}{204}%
\newcommand{\astroshaVTwoChanged}{47}
\section{Data availability and status of positions}\label{sec:data_avail}
\emph{Released with this article.} The three machine-readable tables of Section~\ref{sec:cat_mrt}, the first of which is also printed, without its uncertainties, as Table~\ref{tab:cat271} of Appendix~\ref{app:catalogue}; the list of digests described below; a list of \prFiles{} rule files (Appendix~\ref{app:prereg}), each identified by a label, with its outcome and its SHA-256 digest; the proposal identifiers of the \dtProgrammes{} programmes that took the exposures of 2020--2026, with the number of exposures used from each;  my positions of the \vtDesRel{} designated objects of the catalogue and of GCT0044;  and a README that gives the format of the files and the changes from version 1.

\emph{What the Minor Planet Center has.} I submitted the archival DECam astrometry (station W84) of \ctSent{} objects, \ctSentNew{} that are undesignated today and \ctSentDes{} that have since been identified with designated objects: \vtSentFirst{} objects on \vtSentDates{} (UTC), before the committee on archival observations had answered my first message of \ctSarcContact, and GCT0136 (\vtSentGOneThreeSixRows{} positions) on 2026 November 18. Its review of the observing programme under which these measurements are reported opened on \ctSarcAsk{} and is not decided. The other \ctNotSentNew{} undesignated objects will be submitted after it.

\emph{Positions.}  The positions of the undesignated objects, except those of GCT0044 (below), and their ephemerides are not printed, because they should enter the public record through that procedure. They are available now, on request: I send the file of any object, or of all, to any researcher who asks, on the understanding that the positions are not passed on, published or submitted to the MPC before the MPC has processed them. They will be public when the objects are designated and, as version 1 stated, on \ctReleaseBy{} at the latest. Version 1 offered them to the MPC and to referees only. Not every position in these files is unpublished: \ctRowsOther{} of them are positions of other observers that the MPC has published as isolated tracklets (Section~\ref{sec:cat_tiers}).

\emph{Published astrometry.} My \astroshaRelDes{} positions of the \vtDesRel{} catalogue objects that version~2 identified with designated objects (Table~\ref{tab:designated}) are published again with this article in ADES format, as they are used in the fits of this version, as are my \astroshaRelTroj{} positions of GCT0044, the object also reported by \citet{jongma2026} and matched provisionally by \citet{han2026} to an object linked in TESS images (Section~\ref{sec:rare_trojan}): that identity can now be tested by anyone who holds positions of their object. With the published astrometry of the discoverers, a reader can repeat the joint fits of Table~\ref{tab:designated} and test the positions and the time tags of this pipeline on those objects. My positions of the designated objects identified since version~2 are available on request. Of the \astroshaRelDes{} positions, \relCatDes{} are \nscdr{} catalogue positions that I did not re-measure, as are \relCatTroj{} of the \astroshaRelTroj{} of GCT0044; the README says which. Seven of these positions differ from the file of version~2 by the correction of their time tags (Section~\ref{sec:link_time}): four of 2026~DU$_{71}$ (GCT0235), three of which were at exposure start in version~2, by +60~s; two of GCT0071 by $-14.6$~s; and one of GCT0074 by +6~s. The README lists them, and the positions whose time tags the audit of version~2 could not match to an exposure.

\emph{Commitment.} A file released with this article lists a SHA-256 digest for each of the \astroshaobj{} undesignated objects. Their files hold \astroshapos{} positions in all, \ctRowsMine{} mine and \ctRowsOther{} of other observers. For the \astroshaVTwoObj{} undesignated objects that were already in version~2 the nonces are those of version~2: \astroshaVTwoSame{} digests are the same as in version~2, and the other \astroshaVTwoChanged{} changed because positions were added to those files or corrected. Each digest covers the file of positions of the object, a table in ADES format, preceded by one line that holds a random nonce of 64 hexadecimal digits. The nonce is private: it is sent with the file on request and will be published with the positions, and without it a guessed set of positions cannot be tested against the digest. The files are those of this version, with the time tags corrected (Section~\ref{sec:link_time}): the \ttStartNew{} positions of \ttStartNewObj{} undesignated objects that were at exposure start in version~2 are now at mid-exposure, which is one of the reasons why \astroshaVTwoChanged{} digests changed. The files of version~2 of those \astroshaVTwoChanged{} objects are kept, so the list of version~2 remains checkable; that list also has a line for GCT0103, which is now identified with 2025~ND$_{561}$ and has no line in the present one. The SHA-256 digest of the list itself, computed on \astroshadate, is \astrosha. Whoever receives the file of one object and its nonce can check them against its line of the list.{} The list has a line for GCT0044 as well: the file published with this article is the one that its digest covers, and its nonce is printed in the README, so that this one line can be checked by any reader.{} Positions that reach the public record through the MPC will be in the MPC's format and will not reproduce these digests, so the files themselves will be published with them. A list of the SHA-256 digests of my files of positions as they stood on 2026 November 25 was submitted to the OpenTimestamps calendars (\url{https://opentimestamps.org}) on that day (proof available on request); the SHA-256 of that list is \vtOtsSha. The digests of version 1 were of files in another number format; those files are kept and will be released with them.

\emph{Rule files and code.} The rule files of Appendix~\ref{app:prereg} name sky regions that would help to locate objects not yet designated. They are not released; they are available on request, and the released list gives a label and the digest of each file in it. The linking and vetting code, OrbitSift, is not released with this version either: it is not yet public and is available on request, so the digests of code recorded in the rule files certify nothing today.

\begin{acknowledgments}
I thank Bill Gray for Find\_Orb and Hanno Rein and collaborators for REBOUND.

This research uses services or data provided by the Astro Data Lab, which is part of the Community Science and Data Center (CSDC) Program of NSF NOIRLab. NOIRLab is operated by the Association of Universities for Research in Astronomy (AURA), Inc.\ under a cooperative agreement with the U.S.\ National Science Foundation.

This research draws upon DECam data as distributed by the Astro Data Archive at NSF NOIRLab. NOIRLab is managed by the Association of Universities for Research in Astronomy (AURA) under a cooperative agreement with the U.S.\ National Science Foundation. This project used data obtained with the Dark Energy Camera (DECam), which was constructed by the Dark Energy Survey (DES) collaboration. Funding for the DES Projects has been provided by the US Department of Energy, the US National Science Foundation, the Ministry of Science and Education of Spain, the Science and Technology Facilities Council of the United Kingdom, the Higher Education Funding Council for England, the National Center for Supercomputing Applications at the University of Illinois at Urbana-Champaign, the Kavli Institute for Cosmological Physics at the University of Chicago, Center for Cosmology and Astro-Particle Physics at the Ohio State University, the Mitchell Institute for Fundamental Physics and Astronomy at Texas A\&M University, Financiadora de Estudos e Projetos, Funda\c{c}\~ao Carlos Chagas Filho de Amparo \`a Pesquisa do Estado do Rio de Janeiro, Conselho Nacional de Desenvolvimento Cient\'ifico e Tecnol\'ogico and the Minist\'erio da Ci\^encia, Tecnologia e Inova\c{c}\~ao, the Deutsche Forschungsgemeinschaft and the Collaborating Institutions in the Dark Energy Survey. The Collaborating Institutions are Argonne National Laboratory, the University of California at Santa Cruz, the University of Cambridge, Centro de Investigaciones En\'ergeticas, Medioambientales y Tecnol\'ogicas--Madrid, the University of Chicago, University College London, the DES-Brazil Consortium, the University of Edinburgh, the Eidgen\"ossische Technische Hochschule (ETH) Z\"urich, Fermi National Accelerator Laboratory, the University of Illinois at Urbana-Champaign, the Institut de Ci\`encies de l'Espai (IEEC/CSIC), the Institut de F\'isica d'Altes Energies, Lawrence Berkeley National Laboratory, the Ludwig-Maximilians Universit\"at M\"unchen and the associated Excellence Cluster Universe, the University of Michigan, NSF NOIRLab, the University of Nottingham, the Ohio State University, the OzDES Membership Consortium, the University of Pennsylvania, the University of Portsmouth, SLAC National Accelerator Laboratory, Stanford University, the University of Sussex, and Texas A\&M University.
Based on observations at NSF Cerro Tololo Inter-American Observatory, a program of NOIRLab (NOIRLab Prop.\ IDs 2012B-0001, PI J. Frieman; 2013A-0455, PI S. Sheppard; 2013A-0719, PI A. Saha; 2013A-0741, PI D. Schlegel; 2014A-0283, PI D. Trilling; 2014A-0608, PI F. F\"orster; 2014B-0404, PI D. Schlegel; 2016A-0190, PI A. Dey; 2016B-0288, PI S. Sheppard; 2016B-0301, PI A. Rest; 2017A-0260, PI M. Soares-Santos; 2017B-0279, PI A. Rest; 2017B-0307, PI S. Sheppard; 2018A-0371, PI Malhotra; 2018A-0393, PI S. Sheppard; 2018A-0911, PI F. F\"orster; 2018A-0913, PI A. Zenteno; 2018A-0914, PI M. Makler; 2018B-0122, PI A. Rest; 2019A-0305, PI A. Drlica-Wagner; 2019A-0308, PI I. Dell'Antonio; 2019A-0332, PI S. Sheppard; 2019A-0337, PI D. Trilling; 2019B-0323, PI A. Zenteno; 2020A-0399, PI A. Zenteno; 2020A-0908, PI F. Olivares; 2020A-0909, PI P. Ar\'evalo; 2020B-0053, PI D. Sand; 2020B-0241, PI A. Zenteno; 2020B-0279, PI L. Wang; 2021A-0149, PI A. Zenteno; 2021A-0275, PI A. Rest; 2022A-388025, PI A. Palmese; 2022B-297190, PI A. Palmese; 2022B-563562, PI Pena; 2023A-237157, PI R. Monteiro-Oliveira; 2023A-642922, PI J. Kleyna and R. Wainscoat; 2023A-881453, PI A. Palmese; 2023A-932453, PI Brout; 2023B-184194, PI A. Dey and D. Schlegel; 2023B-735801, PI A. Palmese; 2024B-763968, PI Shen; 2025A-110944, PI I. Dell'Antonio; 2025B-213846, PI Pandey; 2025B-240381, PI Ahumada), which is managed by the Association of Universities for Research in Astronomy (AURA) under a cooperative agreement with the U.S.\ National Science Foundation. The programmes named here are those of the frames used in version 1; \ctPropsInList{} of them are also among the \dtProgrammes{} public DECam programmes of the \ctDtExp{} exposures of 2020--2026. A file released with this article lists the proposal identifiers of all \dtProgrammes, as given by the archive, and the number of exposures used from each. Raw imaging data were processed with the DECam Community Pipeline \citep{valdes2014}.
This research is based on data obtained from the Astro Data Archive at NSF NOIRLab. These data are associated with observing programs 2015A-0801 (PIs Z.\ Xu and X.\ Fan) and 2016A-0453 (PI A.\ Dey). NOIRLab is managed by the Association of Universities for Research in Astronomy (AURA) under a cooperative agreement with the U.S.\ National Science Foundation.

The Legacy Surveys consist of three individual and complementary projects: the Dark Energy Camera Legacy Survey (DECaLS; Proposal ID \#2014B-0404; PIs: David Schlegel and Arjun Dey), the Beijing-Arizona Sky Survey (BASS; NOAO Prop.\ ID \#2015A-0801; PIs: Zhou Xu and Xiaohui Fan), and the Mayall z-band Legacy Survey (MzLS; Prop.\ ID \#2016A-0453; PI: Arjun Dey). DECaLS, BASS and MzLS together include data obtained, respectively, at the Blanco telescope, Cerro Tololo Inter-American Observatory, NSF's NOIRLab; the Bok telescope, Steward Observatory, University of Arizona; and the Mayall telescope, Kitt Peak National Observatory, NOIRLab. Pipeline processing and analyses of the data were supported by NOIRLab and the Lawrence Berkeley National Laboratory (LBNL). The Legacy Surveys project is honored to be permitted to conduct astronomical research on I'oligam Du'ag (Kitt Peak), a mountain with particular significance to the Tohono O'odham Nation. NOIRLab is operated by the Association of Universities for Research in Astronomy (AURA) under a cooperative agreement with the National Science Foundation. LBNL is managed by the Regents of the University of California under contract to the U.S.\ Department of Energy. BASS is a key project of the Telescope Access Program (TAP), which has been funded by the National Astronomical Observatories of China, the Chinese Academy of Sciences (the Strategic Priority Research Program ``The Emergence of Cosmological Structures'' Grant \#XDB09000000), and the Special Fund for Astronomy from the Ministry of Finance. The BASS is also supported by the External Cooperation Program of Chinese Academy of Sciences (Grant \#114A11KYSB20160057), and Chinese National Natural Science Foundation (Grant \#12120101003, \#11433005). The Legacy Survey team makes use of data products from the Near-Earth Object Wide-field Infrared Survey Explorer (NEOWISE), which is a project of the Jet Propulsion Laboratory/California Institute of Technology. NEOWISE is funded by the National Aeronautics and Space Administration. The Legacy Surveys imaging of the DESI footprint is supported by the Director, Office of Science, Office of High Energy Physics of the U.S.\ Department of Energy under Contract No.\ DE-AC02-05CH11231, by the National Energy Research Scientific Computing Center, a DOE Office of Science User Facility under the same contract; and by the U.S.\ National Science Foundation, Division of Astronomical Sciences under Contract No.\ AST-0950945 to NOAO.

This research used the facilities of the Canadian Astronomy Data Centre operated by the National Research Council of Canada with the support of the Canadian Space Agency. Based on observations obtained at the Canada-France-Hawai\okina i Telescope (CFHT) which is operated by the National Research Council of Canada, the Institut National des Sciences de l'Univers of the Centre National de la Recherche Scientifique of France, and the University of Hawai\okina i. CFHT is located on Maunakea on Hawai\okina i Island, a mountain of considerable cultural, natural, and ecological significance. Maunakea is a sacred site to Native Hawaiians, also known as K\={a}naka \okina\={O}iwi. I would like to thank the Canada-France-Hawai\okina i Telescope (CFHT) Operations and Software Groups for their contributions and diligence in maintaining observatory operations; the CFHT Astronomy Group for their observation coordination and data acquisition efforts; and the CFHT Finance \& Administration Group for their contributions to the management and administration of the observatory. Based on observations obtained with MegaPrime/MegaCam, a joint project of CFHT and CEA/DAPNIA. This work is based in part on data products produced at TERAPIX and the Canadian Astronomy Data Centre as part of the Canada-France-Hawai\okina i Telescope Legacy Survey, a collaborative project of NRC and CNRS.

This research is based in part on data collected at the Subaru Telescope, which is operated by the National Astronomical Observatory of Japan. I am honored and grateful for the opportunity of observing the Universe from Maunakea, which has the cultural, historical, and natural significance in Hawaii. Based in part on data collected at Subaru Telescope and obtained from the SMOKA, which is operated by the Astronomy Data Center, National Astronomical Observatory of Japan \citep{baba2002}. HSC is described by \citet{miyazaki2018}.

Funding for the SDSS and SDSS-II has been provided by the Alfred P.\ Sloan Foundation, the Participating Institutions, the National Science Foundation, the U.S.\ Department of Energy, the National Aeronautics and Space Administration, the Japanese Monbukagakusho, the Max Planck Society, and the Higher Education Funding Council for England. The SDSS Web Site is \url{http://www.sdss.org/}. The SDSS is managed by the Astrophysical Research Consortium for the Participating Institutions. The Participating Institutions are the American Museum of Natural History, Astrophysical Institute Potsdam, University of Basel, University of Cambridge, Case Western Reserve University, University of Chicago, Drexel University, Fermilab, the Institute for Advanced Study, the Japan Participation Group, Johns Hopkins University, the Joint Institute for Nuclear Astrophysics, the Kavli Institute for Particle Astrophysics and Cosmology, the Korean Scientist Group, the Chinese Academy of Sciences (LAMOST), Los Alamos National Laboratory, the Max-Planck-Institute for Astronomy (MPIA), the Max-Planck-Institute for Astrophysics (MPA), New Mexico State University, Ohio State University, University of Pittsburgh, University of Portsmouth, Princeton University, the United States Naval Observatory, and the University of Washington.

The Pan-STARRS1 Surveys (PS1) and the PS1 public science archive have been made possible through contributions by the Institute for Astronomy, the University of Hawaii, the Pan-STARRS Project Office, the Max-Planck Society and its participating institutes, the Max Planck Institute for Astronomy, Heidelberg and the Max Planck Institute for Extraterrestrial Physics, Garching, The Johns Hopkins University, Durham University, the University of Edinburgh, the Queen's University Belfast, the Harvard-Smithsonian Center for Astrophysics, the Las Cumbres Observatory Global Telescope Network Incorporated, the National Central University of Taiwan, the Space Telescope Science Institute, the National Aeronautics and Space Administration under Grant No.\ NNX08AR22G issued through the Planetary Science Division of the NASA Science Mission Directorate, the National Science Foundation Grant No.\ AST-1238877, the University of Maryland, Eotvos Lorand University (ELTE), the Los Alamos National Laboratory, and the Gordon and Betty Moore Foundation.

This work has made use of data from the European Space Agency (ESA) mission {\it Gaia} (\url{https://www.cosmos.esa.int/gaia}), processed by the {\it Gaia} Data Processing and Analysis Consortium (DPAC, \url{https://www.cosmos.esa.int/web/gaia/dpac/consortium}). Funding for the DPAC has been provided by national institutions, in particular the institutions participating in the {\it Gaia} Multilateral Agreement.
This research has made use of data and services provided by the International Astronomical Union's Minor Planet Center, and of the JPL Horizons and Small-Body Database services.

\emph{Use of AI.} The code, the analyses and the text were assisted by an AI coding agent, Claude Code (Anthropic), under my direction; I decided the questions and the tests and approved the rules. I read all of the code and all of the text of this version myself.

With the released tables alone a reader can make two checks that do not depend on my pipeline: the rounded elements of GCT0044 can be compared with the orbit of \citet{jongma2026} and with the elements that \citet{han2026} print, and the elements that the catalogue table gives for the \ctDes{} objects identified with designated ones can be compared with the published orbits of those objects. With my positions of \vtDesRel{} of those objects and of GCT0044, which are published with this article (Section~\ref{sec:data_avail}), a reader can refit their orbits with any orbit-fitting program, repeat the joint fits of the identifications with the astrometry published by others, and compare GCT0044 with the objects of those two reports as far as their positions are available. The other checks need the positions of the other undesignated objects, which are available on request: with them a reader can refit those orbits and verify the digest of each file.
The Subaru test of Section~\ref{sec:valid_indep} cannot be checked by a reader: the prediction files and their digests are not published, and only my own files show that they were written first. Everything else (the rules, the held-out tests, the controls, the image measurements) was run on my machines and rests on their logs. I am responsible for every number and statement in this paper.
\end{acknowledgments}

\facilities{Blanco (DECam), Bok (90Prime), Mayall (Mosaic3), CFHT (MegaCam), Subaru (HSC), Sloan, PS1, Astro Data Lab, CADC}
\software{OrbitSift (my linking and vetting code, not yet public; available on request), Find\_Orb \citep{findorb}, REBOUND \citep{rein2012,rein2015,rein2019,reinspiegel2015}, SkyBoT \citep{berthier2006}, \texttt{legacypipe}, Astropy \citep{astropy2013,astropy2018,astropy2022}, photutils \citep{photutils}, NumPy \citep{harris2020}, SciPy \citep{virtanen2020}, jplephem, Matplotlib \citep{hunter2007}, CuPy \citep{okuta2017}}

\appendix
\section{Rules written before the tests}\label{app:prereg}

``Frozen before the held-out read'' means that a file with the rule, the split of the truth objects and the pass criteria was written before the data the test is judged on were read. It is not registration with a third party. The files were written, the runs made and the results read by an AI coding agent working under my instructions, and the times are those printed in the files and their logs. No rule file is under version control and none was deposited anywhere, so no date of a rule file can be checked by a reader. The tests added in version~3 with rules written after 2026 November 17 (the six-night level, the HSC linker with rule Q, the searches for distant objects, the search of the Rubin alerts and the first simulation of the selection function) have rule files that are not in this list; they are available on request. A list of these files was made on \prListDay{} at \prListTime~CET. It holds \prFiles{} files, not only those behind a result: \prPass{} passed, \prFail{} failed, \prNoGo{} ended in a decision not to proceed, \prNotRun{} were not run, \prOpen{} have no verdict, and \prOther{} had no pass criterion or a mixed outcome. It leaves out one draft that this paper does not use, and rule files written later on that day, for work that is not in this paper, are not in it. Table~\ref{tab:prereg} lists the \prUsed{} behind tests that are used or reported in this paper; the list, released with the paper, gives all \prFiles{} with their outcomes (Section~\ref{sec:data_avail}). Two rules have no file of their own: the split of the known-object truth set (the parity of the SHA-1 hash of the designation) and the rule for adopting the second tile grid, written in the log of that study at \sgRuleTime{} on November~\fxDay.

The files name sky regions that would help to locate objects not yet designated, so they are not released with the paper, and neither the table nor the released list gives their names: a file is identified by a label and by its SHA-256 digest. The files are available on request, and a file received later can be checked against its digest.

\begin{deluxetable*}{lllll}
\tabletypesize{\scriptsize}
\tablecaption{Files that state a rule before a test used or reported in this paper, with their outcome\label{tab:prereg}}
\tablehead{\colhead{Label} & \colhead{What is tested} & \colhead{Written or frozen, as printed} & \colhead{Outcome} & \colhead{SHA-256}}
\startdata
R01 & \parbox[t]{0.30\textwidth}{\raggedright Unused-exposure test of version 1\strut} & \parbox[t]{0.25\textwidth}{\raggedright 9 Nov, 17:45; additions to 19:51\strut} & \parbox[t]{0.20\textwidth}{\raggedright rule invalid; post hoc rule used\strut} & \texttt{a68301045887} \\
R02 & \parbox[t]{0.30\textwidth}{\raggedright Over-density mask, pilot tiles\strut} & \parbox[t]{0.25\textwidth}{\raggedright 11 Nov\strut} & \parbox[t]{0.20\textwidth}{\raggedright go\strut} & \texttt{068670e73e35} \\
R03 & \parbox[t]{0.30\textwidth}{\raggedright Second-season join of short links\strut} & \parbox[t]{0.25\textwidth}{\raggedright 15 Nov, 17:03; result appended\strut} & \parbox[t]{0.20\textwidth}{\raggedright fail\strut} & \texttt{df2cea2a2518} \\
R04 & \parbox[t]{0.30\textwidth}{\raggedright Pixel-flag linker, version 1\strut} & \parbox[t]{0.25\textwidth}{\raggedright 14 Nov, 23:00; two amendments\strut} & \parbox[t]{0.20\textwidth}{\raggedright fail\strut} & \texttt{d3c2e1398d29} \\
R05 & \parbox[t]{0.30\textwidth}{\raggedright Pixel-flag linker, version 2\strut} & \parbox[t]{0.25\textwidth}{\raggedright 15 Nov, 07:32\strut} & \parbox[t]{0.20\textwidth}{\raggedright no final verdict\strut} & \texttt{b677052f0101} \\
R06 & \parbox[t]{0.30\textwidth}{\raggedright Pixel-flag linker, masks only\strut} & \parbox[t]{0.25\textwidth}{\raggedright 15 Nov, 13:07\strut} & \parbox[t]{0.20\textwidth}{\raggedright fail; inconclusive\strut} & \texttt{5de120e7abef} \\
R07 & \parbox[t]{0.30\textwidth}{\raggedright Multi-apparition linker on CFHT rows\strut} & \parbox[t]{0.25\textwidth}{\raggedright 15 Nov, 21:10; result appended\strut} & \parbox[t]{0.20\textwidth}{\raggedright fail\strut} & \texttt{f3fb82ac8e3b} \\
R08 & \parbox[t]{0.30\textwidth}{\raggedright Multi-year linker, held-out DES objects\strut} & \parbox[t]{0.25\textwidth}{\raggedright 16 Nov, about 00:42; frozen 03:04\strut} & \parbox[t]{0.20\textwidth}{\raggedright pass (read 04:25)\strut} & \texttt{66df7e4ccefd} \\
R09 & \parbox[t]{0.30\textwidth}{\raggedright Rule A2 outside the DES footprint\strut} & \parbox[t]{0.25\textwidth}{\raggedright 16 Nov, 04:38 (header edited 06:03)\strut} & \parbox[t]{0.20\textwidth}{\raggedright pass (read 05:56)\strut} & \texttt{a208e71e325d} \\
R10 & \parbox[t]{0.30\textwidth}{\raggedright Join of d25 rows to the NSC seeds, one region\strut} & \parbox[t]{0.25\textwidth}{\raggedright 16 Nov, 06:02\strut} & \parbox[t]{0.20\textwidth}{\raggedright fail (read 06:44)\strut} & \texttt{018b1de210b5} \\
R11 & \parbox[t]{0.30\textwidth}{\raggedright Full join of NSC DR2 and d25, one region\strut} & \parbox[t]{0.25\textwidth}{\raggedright 16 Nov, 12:06; amendment 12:27\strut} & \parbox[t]{0.20\textwidth}{\raggedright fail (read 17 Nov, 02:17)\strut} & \texttt{6bded1f985d8} \\
R12 & \parbox[t]{0.30\textwidth}{\raggedright Production on NSC DR2 (stated as the whole catalogue)\strut} & \parbox[t]{0.25\textwidth}{\raggedright 16 Nov, about 09:15 (file time 09:08)\strut} & \parbox[t]{0.20\textwidth}{\raggedright run; no pass criterion\strut} & \texttt{d499d09651f2} \\
R13 & \parbox[t]{0.30\textwidth}{\raggedright Seeded recovery of designated objects\strut} & \parbox[t]{0.25\textwidth}{\raggedright 16 Nov, 19:53\strut} & \parbox[t]{0.20\textwidth}{\raggedright pass\strut} & \texttt{4ebe03cdcc04} \\
\enddata
\tablecomments{All dates are of 2026 November and all times are CET, as printed in the file or in its log; none is certified by a third party. A file is identified by its label, the same as in the released list, and by its digest: the first twelve hexadecimal digits of the SHA-256 of the file as it stood on 2026 November 17 at 04:03 CET, when the list was made. The released list gives the full digests, for these files and for 21 other rule files of the project. The list leaves out one draft that this paper does not use, and rule files written later on that day, for work that is not in this paper, are not in it. The tests added in version~3 with rules written after 2026 November 17 (the six-night level, the HSC linker with rule Q, the searches for distant objects, the search of the Rubin alerts and the first simulation of the selection function) have rule files that are not in this list; they are available on request. The files are available on request.}
\end{deluxetable*}

\emph{Multi-year linker.} All its tests were written, frozen and read on \freezeDate, within hours of each other (Table~\ref{tab:prereg}): the plan at about \myPreregTime, the rule for tiles with many rows at \myAOneTime, the freeze at \myFreezeTime, the held-out read at \myHeldRead. The held-out ephemerides stayed on a separate machine until the freeze. Table~\ref{tab:param} gives the frozen parameters; the check against designated objects is described in Section~\ref{sec:cat_known}.

\begin{deluxetable*}{ll}
\tabletypesize{\footnotesize}
\tablecaption{Frozen parameters of rule A2 and of the stages it uses\label{tab:param}}
\tablehead{\colhead{Parameter} & \colhead{Value}}
\startdata
\parbox[t]{0.25\textwidth}{\raggedright Night\strut} & \parbox[t]{0.68\textwidth}{\raggedright integer part of MJD minus one half\strut} \\
\parbox[t]{0.25\textwidth}{\raggedright Over-density mask\strut} & \parbox[t]{0.68\textwidth}{\raggedright exposure with more than \dnExpCap{} rows in a tile dropped; row dropped if its cell holds more rows of its exposure than the larger of \dnClumpFloor{} and \dnClumpFac{} times the exposure mean, in cells of $\dnCellA''$, then $\dnCellB''$\strut} \\
\parbox[t]{0.25\textwidth}{\raggedright Error of a row, per axis\strut} & \parbox[t]{0.68\textwidth}{\raggedright quadrature sum of \dnSigFloor{} and FWHM $\times$ magnitude error / \dnSigDiv, capped at \dnSigCap\strut} \\
\parbox[t]{0.25\textwidth}{\raggedright Magnitude of a row\strut} & \parbox[t]{0.68\textwidth}{\raggedright brightest over bands of band magnitude plus a fixed colour term\strut} \\
\parbox[t]{0.25\textwidth}{\raggedright Distance range\strut} & \parbox[t]{0.68\textwidth}{\raggedright \myDmin--\myDmax~au\strut} \\
\parbox[t]{0.25\textwidth}{\raggedright Seed\strut} & \parbox[t]{0.68\textwidth}{\raggedright \mySeedN{} nights inside \mySeedWin~days, up to \mySeedK{} seed nights, triplet fit within $\dnKtri\sigma$\strut} \\
\parbox[t]{0.25\textwidth}{\raggedright Seed units\strut} & \parbox[t]{0.68\textwidth}{\raggedright tiles ($\myTileBox^\circ$) and two lattices of $\myBox^\circ$ boxes offset by $\myBoxOff^\circ$\strut} \\
\parbox[t]{0.25\textwidth}{\raggedright Tiles above \myDeepThr{} rows\strut} & \parbox[t]{0.68\textwidth}{\raggedright seeded with \dnNmin{} nights in \dnDtmax~days, \dnK{} seed nights; boxes above it not seeded\strut} \\
\parbox[t]{0.25\textwidth}{\raggedright Seed gate\strut} & \parbox[t]{0.68\textwidth}{\raggedright $\myKgate\sigma$ ellipse, semi-axis at most $\myRcap''$, pad $\myPad''$, within \myDmag~mag of the seed\strut} \\
\parbox[t]{0.25\textwidth}{\raggedright Seeds with more than \myMaxc{} rows in the gate\strut} & \parbox[t]{0.68\textwidth}{\raggedright gate halved, down to $\myRcapMin''$, then the first \myMaxc{} rows in time\strut} \\
\parbox[t]{0.25\textwidth}{\raggedright Four-row fit\strut} & \parbox[t]{0.68\textwidth}{\raggedright normalised rms at most \myNrms\strut} \\
\parbox[t]{0.25\textwidth}{\raggedright Grouping\strut} & \parbox[t]{0.68\textwidth}{\raggedright $(\gamma,\dot\gamma)$ within $\myGroupSig\sigma$; \myMinfar{} nights more than \myFarDt~days from the seed\strut} \\
\parbox[t]{0.25\textwidth}{\raggedright Joint fit\strut} & \parbox[t]{0.68\textwidth}{\raggedright group centre and its \myMaxmem{} nearest members\strut} \\
\parbox[t]{0.25\textwidth}{\raggedright Attachment\strut} & \parbox[t]{0.68\textwidth}{\raggedright $\myKatt\sigma$, floor \myAttFloor, cap $\myRcapAtt''$, \myAttPasses{} passes with refits, one row per exposure\strut} \\
\parbox[t]{0.25\textwidth}{\raggedright Fit during growth\strut} & \parbox[t]{0.68\textwidth}{\raggedright $\chinu\leq\myChiGrow$, orbit bound\strut} \\
\parbox[t]{0.25\textwidth}{\raggedright Final cuts\strut} & \parbox[t]{0.68\textwidth}{\raggedright \myNmin{} nights; arc without any one night longer than \myArc~days; $\chinu\leq\myChi$; $\Pch\leq\myPch$\strut} \\
\parbox[t]{0.25\textwidth}{\raggedright Duplicates\strut} & \parbox[t]{0.68\textwidth}{\raggedright drop a set sharing \myDupRows{} rows with one of more nights, then lower $\chinu$\strut} \\
\parbox[t]{0.25\textwidth}{\raggedright Controls\strut} & \parbox[t]{0.68\textwidth}{\raggedright three; each night moved by $\dnShiftLo$--$\dnShiftHi''$ per axis, uniform size, random sign, one draw per night and block\strut} \\
\hline
\parbox[t]{0.25\textwidth}{\raggedright Seeded mode, differences\strut} & \parbox[t]{0.68\textwidth}{\raggedright \lrDmin--\lrDmax~au; \lrMinfar{} far nights; \lrDmag~mag; no arc condition, no night floor; ellipse uncapped up to \lrMaxRows{} rows; seed-row errors \lrSigOld{} before \lrSigOldYr, \lrSigMid{} to \lrSigNewYr, \lrSigNew{} after\strut} \\
\enddata
\end{deluxetable*}

\emph{Deviations and weaknesses.} The list below is what I know of.
\begin{enumerate}
\item What was tested is the rule frozen at \myFreezeTime, not the plan of \myPreregTime. Four items were added in between: the seeds on $\myBox^\circ$ boxes, which gave most of the gain on the development half; the choice of growth catalogue; two parameters that bound the work per seed; and the definition of $\Pch$. None was changed after the held-out read, and the pass criteria were never changed.
\item The held-out half of the multi-year test contains, by a rule fixed beforehand, \hoOldTest{} objects of the test half of the known-object truth set; Section~\ref{sec:valid_held} gives the result without them. Its criterion (c) uses the position screen as ``known'' and has no power.
\item The header of the file of the test outside the DES footprint was edited after its read: at \exEditTime{} its time was changed from \exHeadOld{} to \exPreregTime. The log gives a file time of \exFileTime; the tile fetch began at \exFetchTime.
\item Both tests of the join with \dtf{} rows and the test of the CFHT linker failed (Section~\ref{sec:null}).
\item The test half of the known-object truth set is not a set read once. It was read for three decisions (the baseline, a looser candidate rule for \dtf{} of which \fxOneVariants{} variants were evaluated together and none adopted, and the two grids). Its first-loss table for \nscdr{} was once printed by mistake before the fixes were measured. The test half was used twice more without tuning (Sections~\ref{sec:valid_edge} and \ref{sec:valid_all}). The summary file of that study still opens with a line saying that the test half was not opened.
\item Three files carry text added after their freeze: the second-season join has two notes and its result, the CFHT test its result, the unused-exposure test of version 1 its post hoc rule. For \prLateFiles{} files of the first group of the table the modification time on disk is later than the printed freeze time; a copy can also change a file time.
\item The production ran on Windows machines for \pdWinBlocks{} of \pdBlocks{} blocks; their results are not bit-identical to the Linux ones, and the comparison rests on one test block and two production blocks with empty lists. The runner removed \pdHeaderOnly{} tile files that held a header and no row.
\item The file written before the production describes itself as a run over the whole \nscdr{} orphan catalogue; the run did not cover all of it (Section~\ref{sec:link_my}). The file does not say that the runner reads only the tiles with a file on my tile server: its \pdTiles{} tiles are boxes of the shifted grid, and the run read \pdFiles{} tile files. It does not contain the density cut of \nhDensCut{} detections per square degree, which came with the download list of version 1. It does not mention the failed downloads, which an audit had found on \nhAuditDay{} and which were not repeated before the production. A second audit on \ctlReadDay, after the run, counted the tiles lost to both.
\item The control files of all production blocks were read on \ctlReadDay{} for Section~\ref{sec:valid_ctl}, including those of the \rxHeldBlocks{} blocks reserved for the test of a relaxed rule, which has thereby lost its blind on the control side.
\item Part~B of Table~\ref{tab:purity}, the count of Section~\ref{sec:valid_edge}, the control links of Section~\ref{sec:vet_dt} and the tests of Section~\ref{sec:valid_indep} had no rule written beforehand, except the Subaru positions and the pass rule of version 1, which failed its validity criterion; they use regions and objects that had been used before.
\item The second version of the pixel-flag linker has a frozen rule and no final verdict, and the lists against which the held-out test compares the multi-year linker are those of the first tile grid only.
\end{enumerate}
\clearpage%
\section{The catalogue in print}\label{app:catalogue}

Table~\ref{tab:cat271} prints, for all \ctReg{} objects of the catalogue, fourteen columns of the machine-readable catalogue of Section~\ref{sec:cat_mrt}, with the values of that file. It is written by the program that writes the file.

\startlongtable
\begin{deluxetable*}{llccrrrcrrrrrl}
\tabletypesize{\scriptsize}
\tablecaption{The \ctReg{} objects of the catalogue\label{tab:cat271}}
\tablehead{\colhead{ID} & \colhead{Status} & \colhead{Tier} & \colhead{Ch.} & \colhead{Opp.} & \colhead{Nights} & \colhead{Arc} & \colhead{rms} & \colhead{$a$} & \colhead{$e$} & \colhead{$i$} & \colhead{$q$} & \colhead{$H$} & \colhead{Class} \\
\colhead{} & \colhead{} & \colhead{} & \colhead{} & \colhead{} & \colhead{} & \colhead{(yr)} & \colhead{(arcsec)} & \colhead{(au)} & \colhead{} & \colhead{(deg)} & \colhead{(au)} & \colhead{(mag)} & \colhead{}}
\startdata
GCT0001 & N & B & TL & 1 & 6 & 0.0 & 0.072 & \nodata & \nodata & 23.833 & \nodata & 6.9 & \nodata \\
GCT0002 & N & A & TL & 7 & 20 & 11.2 & 0.078 & 64.157 & 0.4206 & 4.458 & 37.174 & 7.1 & det \\
GCT0003 & N & B & TL & 1 & 6 & 0.1 & 0.075 & \nodata & \nodata & 26.188 & \nodata & 7.3 & \nodata \\
GCT0004 & N & A & TL & 6 & 15 & 14.9 & 0.084 & 39.267 & 0.2226 & 1.850 & 30.527 & 7.5 & 3:2 \\
GCT0005 & N & A & TL & 5 & 14 & 9.2 & 0.093 & 39.544 & 0.1992 & 3.680 & 31.668 & 7.6 & 3:2 \\
GCT0006 & N & A & TL & 3 & 8 & 6.3 & 0.098 & 78.731 & 0.5375 & 42.066 & 36.411 & 6.6 & det \\
GCT0007 & N & B & TL & 1 & 7 & 0.2 & 0.100 & \nodata & \nodata & 8.713 & \nodata & 6.9 & \nodata \\
GCT0008 & N & A & TL & 7 & 14 & 9.8 & 0.107 & 46.321 & 0.0884 & 4.818 & 42.226 & 6.5 & cc \\
GCT0009 & N & A & TL & 7 & 28 & 8.1 & 0.076 & 113.645 & 0.6395 & 23.756 & 40.965 & 5.8 & det \\
GCT0010 & N & A & TL & 4 & 13 & 7.1 & 0.045 & 55.686 & 0.2698 & 10.689 & 40.661 & 4.4 & det \\
GCT0011 & N & A & TL & 5 & 11 & 9.0 & 0.112 & 45.716 & 0.1831 & 28.310 & 37.344 & 6.0 & ch \\
GCT0013 & N & A & TL & 4 & 14 & 8.4 & 0.096 & 54.674 & 0.3496 & 20.330 & 35.561 & 6.9 & det \\
GCT0014 & N & A & TL & 8 & 22 & 13.2 & 0.098 & 29.988 & 0.0433 & 22.842 & 28.688 & 7.6 & NT \\
GCT0015 & N & A & TL & 11 & 28 & 19.0 & 0.072 & 30.012 & 0.0797 & 25.133 & 27.619 & 7.8 & NT \\
GCT0016 & N & A & TL & 5 & 13 & 9.2 & 0.103 & 53.049 & 0.2398 & 23.744 & 40.326 & 6.5 & det$^\dagger$ \\
GCT0017 & N & A & TL & 3 & 10 & 2.9 & 0.072 & 39.152 & 0.2643 & 21.476 & 28.803 & 7.7 & 3:2 \\
GCT0018 & N & A & TL & 6 & 14 & 9.2 & 0.087 & 38.708 & 0.0033 & 16.711 & 38.582 & 6.5 & ci \\
GCT0019 & N & C & TL & 2 & 6 & 1.2 & 0.073 & 45.550 & 0.0777 & 22.345 & 42.011 & 6.5 & ch$^\dagger$ \\
GCT0020 & N & A & TL & 5 & 12 & 5.9 & 0.078 & 43.842 & 0.0751 & 0.590 & 40.549 & 6.3 & cc \\
GCT0021 & N & C & TL & 6 & 17 & 10.1 & 0.086 & 44.201 & 0.0728 & 4.785 & 40.982 & 6.5 & cc \\
GCT0022 & N & C & TL & 6 & 16 & 9.1 & 0.078 & 50.083 & 0.3023 & 15.608 & 34.944 & 7.2 & det \\
GCT0023 & N & C & TL & 5 & 9 & 6.3 & 0.067 & 105.300 & 0.6767 & 12.780 & 34.046 & 6.3 & sca \\
GCT0024 & N & A & TL & 9 & 29 & 10.2 & 0.104 & 43.911 & 0.0133 & 2.591 & 43.326 & 6.7 & cc \\
GCT0025 & N & A & TL & 7 & 76 & 8.2 & 0.105 & 45.753 & 0.1245 & 22.311 & 40.058 & 5.9 & ch \\
GCT0026 & N & C & TL & 7 & 18 & 10.1 & 0.069 & 47.319 & 0.0866 & 22.304 & 43.220 & 6.5 & ch \\
GCT0027 & N & C & TL & 4 & 15 & 5.4 & 0.120 & 43.966 & 0.1969 & 42.976 & 35.311 & 6.1 & ch \\
GCT0028 & N & C & TL & 6 & 20 & 8.1 & 0.064 & 39.463 & 0.2595 & 22.528 & 29.222 & 7.6 & 3:2 \\
GCT0030 & N & C & TL & 8 & 18 & 7.6 & 0.079 & 119.878 & 0.6615 & 28.762 & 40.582 & 5.2 & det \\
GCT0032 & N & C & TL & 4 & 12 & 9.3 & 0.098 & 42.270 & 0.1449 & 20.377 & 36.144 & 7.1 & 5:3 \\
GCT0040 & N & C & TL & 2 & 13 & 1.2 & 0.084 & 39.224 & 0.2173 & 1.511 & 30.702 & 7.8 & 3:2$^\dagger$ \\
GCT0041 & N & C & TL & 7 & 22 & 10.3 & 0.089 & 36.419 & 0.1318 & 10.914 & 31.620 & 7.8 & 4:3 \\
GCT0042 & N & C & TL & 4 & 14 & 2.9 & 0.070 & 39.415 & 0.2492 & 6.887 & 29.592 & 7.4 & 3:2 \\
GCT0043 & N & A & TL & 6 & 17 & 10.2 & 0.102 & 95.405 & 0.6414 & 2.853 & 34.208 & 6.9 & sca \\
GCT0044 & N & A & XT & 10 & 23 & 16.8 & 0.090 & 30.034 & 0.0646 & 20.346 & 28.093 & 6.8 & NT \\
GCT0045 & N & C & XT & 4 & 17 & 3.3 & 0.073 & 40.380 & 0.1055 & 27.527 & 36.118 & 7.0 & ch \\
GCT0047 & N & A & SS & 5 & 20 & 10.1 & 0.085 & 39.470 & 0.2567 & 11.044 & 29.339 & 8.6 & 3:2 \\
GCT0048 & N & A & XT & 7 & 20 & 9.2 & 0.097 & 48.023 & 0.3189 & 12.329 & 32.710 & 7.2 & 2:1 \\
GCT0049 & N & A & TL & 7 & 14 & 9.3 & 0.063 & 44.540 & 0.1575 & 8.247 & 37.524 & 6.2 & 9:5 \\
GCT0050 & N & A & TL & 6 & 14 & 5.8 & 0.083 & 45.307 & 0.2615 & 16.085 & 33.460 & 7.0 & det \\
GCT0051 & N & A & TL & 7 & 17 & 9.9 & 0.066 & 50.277 & 0.4062 & 21.536 & 29.855 & 7.4 & sca \\
GCT0052 & N & A & XT & 6 & 14 & 10.0 & 0.082 & 60.616 & 0.4080 & 17.586 & 35.887 & 7.6 & det \\
GCT0053 & N & A & XT & 7 & 17 & 8.2 & 0.075 & 69.594 & 0.5307 & 6.446 & 32.663 & 7.3 & 7:2 \\
GCT0054 & N & A & TL & 6 & 12 & 10.1 & 0.070 & 45.967 & 0.1508 & 2.739 & 39.035 & 7.2 & cc \\
GCT0055 & N & A & XT & 4 & 14 & 6.3 & 0.098 & 39.502 & 0.2666 & 15.766 & 28.970 & 7.9 & 3:2 \\
GCT0056 & N & A & XT & 6 & 17 & 9.4 & 0.051 & 53.095 & 0.3069 & 16.369 & 36.798 & 6.9 & det \\
GCT0057 & N & A & SS & 7 & 20 & 10.2 & 0.075 & 39.202 & 0.1532 & 5.260 & 33.194 & 7.9 & 3:2 \\
GCT0058 & N & A & XT & 6 & 16 & 10.0 & 0.089 & 101.913 & 0.6856 & 9.056 & 32.041 & 7.1 & sca \\
GCT0059 & N & A & SS & 4 & 14 & 8.1 & 0.078 & 43.736 & 0.2731 & 5.935 & 31.793 & 7.2 & 7:4 \\
GCT0060 & N & A & SS & 2 & 10 & 1.2 & 0.093 & 32.063 & 0.4076 & 11.620 & 18.994 & 9.4 & sca \\
GCT0061 & N & A & SS & 2 & 7 & 1.3 & 0.029 & 128.691 & 0.8622 & 11.164 & 17.740 & 9.7 & sca \\
GCT0062 & N & A & RV & 6 & 11 & 10.1 & 0.083 & 60.984 & 0.4423 & 14.168 & 34.009 & 7.2 & sca$^\dagger$ \\
GCT0063 & N & A & RV & 6 & 16 & 9.2 & 0.091 & 45.386 & 0.1987 & 23.100 & 36.366 & 6.3 & ch \\
GCT0064 & N & A & RV & 6 & 14 & 8.0 & 0.066 & 39.318 & 0.2246 & 13.038 & 30.486 & 7.9 & 3:2 \\
GCT0065 & N & A & RV & 9 & 22 & 9.1 & 0.086 & 44.048 & 0.1068 & 4.967 & 39.343 & 6.8 & cc \\
GCT0066 & N & A & RV & 4 & 11 & 3.0 & 0.072 & 39.523 & 0.2917 & 9.843 & 27.995 & 8.4 & 3:2 \\
GCT0067 & N & A & TL & 2 & 6 & 5.0 & 0.032 & 42.255 & 0.3151 & 21.284 & 28.940 & 7.7 & 5:3 \\
GCT0068 & N & A & TL & 5 & 11 & 5.2 & 0.094 & 50.300 & 0.2785 & 19.007 & 36.292 & 6.9 & det \\
GCT0069 & N & A & TL & 7 & 14 & 17.3 & 0.081 & 75.920 & 0.3566 & 28.832 & 48.848 & 5.3 & det \\
GCT0070 & N & A & TL & 2 & 6 & 1.0 & 0.052 & 39.115 & 0.1945 & 2.314 & 31.508 & 8.8 & 3:2$^\dagger$ \\
GCT0072 & N & A & RV & 6 & 15 & 9.2 & 0.074 & 43.903 & 0.1758 & 16.237 & 36.185 & 6.3 & ch \\
GCT0073 & N & A & XT & 2 & 7 & 1.0 & 0.099 & 42.926 & 0.0373 & 2.478 & 41.325 & 6.8 & cc \\
GCT0075 & N & A & TL & 2 & 14 & 1.2 & 0.086 & 39.474 & 0.3215 & 10.538 & 26.783 & 8.0 & 3:2 \\
GCT0076 & N & C & SS & 4 & 9 & 4.1 & 0.087 & 39.345 & 0.1811 & 9.172 & 32.220 & 10.7 & sca \\
GCT0077 & N & A & NM & 9 & 22 & 18.2 & 0.078 & 29.942 & 0.0300 & 20.686 & 29.045 & 7.8 & NT \\
GCT0078 & N & A & NM & 5 & 19 & 5.5 & 0.103 & 38.242 & 0.4776 & 11.055 & 19.976 & 9.4 & sca \\
GCT0079 & N & A & RV & 4 & 7 & 9.2 & 0.098 & 39.394 & 0.3230 & 11.626 & 26.671 & 8.9 & 3:2 \\
GCT0080 & N & B & NM & 1 & 7 & 0.2 & 0.086 & \nodata & \nodata & 8.076 & \nodata & 7.7 & \nodata \\
GCT0081 & N & A & XT & 4 & 11 & 8.0 & 0.088 & 44.761 & 0.2512 & 15.410 & 33.517 & 7.2 & det$^\dagger$ \\
GCT0082 & N & A & XT & 5 & 9 & 8.0 & 0.085 & 55.451 & 0.4285 & 10.719 & 31.688 & 7.7 & 5:2 \\
GCT0083 & N & A & NM & 2 & 7 & 3.4 & 0.032 & 29.284 & 0.4841 & 30.236 & 15.109 & 11.3 & Cen \\
GCT0084 & N & A & NM & 7 & 27 & 7.9 & 0.074 & 53.853 & 0.5722 & 138.455 & 23.037 & 8.5 & sca$^\dagger$ \\
GCT0085 & N & A & NM & 5 & 12 & 4.2 & 0.049 & 51.050 & 0.6732 & 44.684 & 16.683 & 9.6 & sca \\
GCT0086 & N & A & NM & 2 & 17 & 1.3 & 0.106 & 20.347 & 0.4844 & 2.625 & 10.491 & 11.9 & Cen \\
GCT0087 & N & A & NM & 5 & 18 & 5.3 & 0.061 & 43.367 & 0.6776 & 71.026 & 13.983 & 10.1 & sca \\
GCT0088 & N & A & NM & 3 & 17 & 10.3 & 0.052 & 24.159 & 0.4884 & 29.194 & 12.360 & 10.7 & Cen \\
GCT0089 & N & B & NM & 1 & 9 & 0.3 & 0.056 & \nodata & \nodata & 19.342 & \nodata & 12.0 & \nodata \\
GCT0090 & N & B & NM & 1 & 7 & 0.1 & 0.079 & \nodata & \nodata & 7.321 & \nodata & 7.3 & \nodata \\
GCT0092 & N & A & DP & 6 & 18 & 16.0 & 0.067 & 39.497 & 0.0464 & 8.435 & 37.665 & 6.7 & 3:2 \\
GCT0094 & N & A & DP & 5 & 15 & 6.5 & 0.093 & 47.649 & 0.2822 & 11.813 & 34.203 & 7.5 & 2:1 \\
GCT0095 & N & A & DP & 2 & 13 & 1.4 & 0.131 & 24.708 & 0.4206 & 16.030 & 14.315 & 9.4 & Cen \\
GCT0096 & N & A & DP & 6 & 27 & 8.5 & 0.094 & 38.401 & 0.0472 & 26.912 & 36.587 & 8.0 & ci \\
GCT0097 & N & A & DP & 6 & 10 & 11.1 & 0.079 & 55.335 & 0.2968 & 17.980 & 38.911 & 7.4 & 5:2 \\
GCT0098 & N & A & DX & 8 & 36 & 10.1 & 0.088 & 273.212 & 0.8801 & 14.802 & 32.762 & 7.3 & sca \\
GCT0099 & N & C & DX & 5 & 12 & 10.1 & 0.102 & 42.500 & 0.1014 & 14.319 & 38.190 & 7.6 & \nodata \\
GCT0100 & N & A & DX & 3 & 16 & 5.0 & 0.103 & 39.364 & 0.1754 & 17.359 & 32.459 & 7.1 & 3:2 \\
GCT0101 & N & B & DX & 1 & 8 & 0.1 & 0.136 & \nodata & \nodata & 6.296 & \nodata & 8.5 & \nodata \\
GCT0102 & N & B & DX & 1 & 8 & 0.1 & 0.069 & \nodata & \nodata & 22.679 & \nodata & 7.6 & \nodata \\
GCT0104 & N & A & DX & 6 & 28 & 10.0 & 0.103 & 280.560 & 0.8683 & 19.592 & 36.949 & 7.3 & sca \\
GCT0105 & N & B & DX & 1 & 12 & 0.2 & 0.100 & \nodata & \nodata & 21.470 & \nodata & 7.1 & \nodata \\
GCT0106 & N & B & DX & 1 & 16 & 0.3 & 0.075 & \nodata & \nodata & 26.943 & \nodata & 6.7 & \nodata \\
GCT0107 & N & A & DX & 4 & 18 & 4.2 & 0.093 & 43.804 & 0.1740 & 27.443 & 36.181 & 7.2 & ch \\
GCT0108 & N & B & DX & 1 & 17 & 0.3 & 0.100 & \nodata & \nodata & 20.359 & \nodata & 7.4 & \nodata \\
GCT0109 & N & A & DX & 6 & 23 & 8.9 & 0.107 & 51.385 & 0.2918 & 15.648 & 36.393 & 5.9 & det \\
GCT0110 & N & B & DX & 1 & 12 & 0.3 & 0.082 & \nodata & \nodata & 26.026 & \nodata & 8.5 & \nodata \\
GCT0111 & N & A & DX & 4 & 11 & 10.3 & 0.091 & 46.760 & 0.2307 & 29.955 & 35.974 & 8.0 & ch \\
GCT0112 & N & A & DX & 4 & 23 & 7.3 & 0.106 & 41.816 & 0.1258 & 22.574 & 36.557 & 7.4 & ch \\
GCT0113 & N & A & DX & 5 & 36 & 9.1 & 0.084 & 43.230 & 0.0702 & 13.807 & 40.197 & 7.2 & ch \\
GCT0114 & N & A & DX & 2 & 24 & 8.3 & 0.102 & 55.328 & 0.3732 & 11.619 & 34.677 & 7.5 & 5:2 \\
GCT0116 & N & A & DX & 3 & 15 & 5.1 & 0.114 & 41.407 & 0.1069 & 17.107 & 36.981 & 7.4 & ch \\
GCT0117 & N & A & DX & 7 & 46 & 11.0 & 0.112 & 43.188 & 0.3525 & 29.312 & 27.966 & 8.5 & sca \\
GCT0118 & N & A & DX & 5 & 28 & 9.1 & 0.096 & 46.645 & 0.1255 & 28.152 & 40.793 & 6.7 & ch \\
GCT0119 & N & A & DX & 4 & 14 & 8.3 & 0.099 & 41.497 & 0.1045 & 15.476 & 37.161 & 7.4 & ch \\
GCT0121 & N & B & DX & 1 & 10 & 0.1 & 0.098 & \nodata & \nodata & 7.096 & \nodata & 7.8 & \nodata \\
GCT0122 & N & A & DX & 2 & 16 & 1.2 & 0.113 & 41.672 & 0.0360 & 19.352 & 40.173 & 6.9 & ch \\
GCT0123 & N & B & DX & 1 & 4 & 0.2 & 0.065 & \nodata & \nodata & 23.799 & \nodata & 8.5 & \nodata \\
GCT0124 & N & A & DN & 8 & 21 & 11.0 & 0.065 & 310.078 & 0.8815 & 20.069 & 36.742 & 6.8 & sca \\
GCT0125 & N & B & DX & 1 & 8 & 0.2 & 0.122 & \nodata & \nodata & 8.909 & \nodata & 8.8 & \nodata \\
GCT0126 & N & A & DX & 2 & 21 & 5.0 & 0.087 & 76.457 & 0.5294 & 16.408 & 35.983 & 7.4 & det \\
GCT0127 & N & A & DX & 2 & 18 & 1.5 & 0.097 & 45.956 & 0.1412 & 17.965 & 39.469 & 7.0 & ch \\
GCT0128 & N & A & DX & 6 & 27 & 10.1 & 0.107 & 41.399 & 0.1129 & 23.613 & 36.724 & 7.0 & ch \\
GCT0129 & N & A & DX & 6 & 28 & 8.1 & 0.098 & 55.944 & 0.3049 & 24.640 & 38.884 & 6.9 & det \\
GCT0130 & N & A & DX & 6 & 20 & 7.3 & 0.068 & 39.201 & 0.2718 & 14.086 & 28.546 & 7.5 & 3:2 \\
GCT0131 & N & A & DX & 3 & 23 & 4.5 & 0.074 & 39.428 & 0.2535 & 15.894 & 29.434 & 7.9 & 3:2 \\
GCT0132 & N & B & DX & 1 & 13 & 0.3 & 0.084 & \nodata & \nodata & 11.024 & \nodata & 6.9 & \nodata \\
GCT0133 & N & A & DX & 2 & 13 & 5.1 & 0.107 & 47.090 & 0.2492 & 18.054 & 35.353 & 7.8 & det \\
GCT0134 & N & A & DX & 4 & 17 & 9.2 & 0.115 & 43.303 & 0.0344 & 13.958 & 41.812 & 6.7 & ch \\
GCT0136 & N & A & DX & 4 & 25 & 8.1 & 0.100 & 207.639 & 0.8902 & 110.280 & 22.794 & 8.3 & sca \\
GCT0137 & N & B & DX & 1 & 11 & 0.2 & 0.067 & \nodata & \nodata & 8.419 & \nodata & 7.0 & \nodata \\
GCT0138 & N & A & DX & 6 & 18 & 11.1 & 0.099 & 41.470 & 0.0478 & 27.632 & 39.485 & 6.6 & ch \\
GCT0139 & N & B & DX & 1 & 10 & 0.2 & 0.121 & \nodata & \nodata & 10.794 & \nodata & 7.5 & \nodata \\
GCT0141 & N & B & DX & 1 & 10 & 0.1 & 0.099 & \nodata & \nodata & 4.264 & \nodata & 7.4 & \nodata \\
GCT0142 & N & C & DX & 1 & 12 & 0.1 & 0.093 & \nodata & \nodata & 21.597 & \nodata & 7.5 & \nodata \\
GCT0144 & N & A & DX & 4 & 13 & 8.2 & 0.101 & 78.230 & 0.5101 & 24.227 & 38.323 & 7.0 & 21:5 \\
GCT0145 & N & A & DX & 4 & 14 & 4.2 & 0.116 & 50.193 & 0.2711 & 13.475 & 36.583 & 6.8 & det \\
GCT0146 & N & B & DX & 1 & 7 & 0.1 & 0.087 & \nodata & \nodata & 20.200 & \nodata & 7.2 & \nodata \\
GCT0147 & N & A & DX & 5 & 17 & 4.0 & 0.081 & 37.552 & 0.3665 & 18.723 & 23.791 & 9.4 & sca \\
GCT0148 & N & A & DX & 3 & 15 & 5.0 & 0.099 & 43.552 & 0.2811 & 16.552 & 31.312 & 7.3 & 7:4 \\
GCT0149 & N & B & DX & 1 & 7 & 0.2 & 0.074 & \nodata & \nodata & 22.887 & \nodata & 6.3 & \nodata \\
GCT0150 & N & B & DX & 1 & 7 & 0.1 & 0.144 & \nodata & \nodata & 27.822 & \nodata & 7.8 & \nodata \\
GCT0151 & N & B & DX & 1 & 7 & 0.1 & 0.067 & \nodata & \nodata & 12.587 & \nodata & 8.0 & \nodata \\
GCT0152 & N & B & DX & 1 & 6 & 0.2 & 0.099 & \nodata & \nodata & 13.206 & \nodata & 8.0 & \nodata \\
GCT0153 & N & B & DX & 1 & 6 & 0.1 & 0.113 & \nodata & \nodata & 29.336 & \nodata & 7.0 & \nodata \\
GCT0154 & N & A & DX & 4 & 14 & 5.1 & 0.104 & 42.975 & 0.1012 & 21.150 & 38.624 & 6.1 & ch \\
GCT0155 & N & B & DX & 1 & 5 & 0.3 & 0.061 & \nodata & \nodata & 12.649 & \nodata & 7.6 & \nodata \\
GCT0156 & N & A & DX & 3 & 24 & 4.4 & 0.098 & 110.981 & 0.6782 & 23.610 & 35.712 & 6.5 & sca \\
GCT0157 & N & A & DX & 5 & 22 & 4.2 & 0.130 & 45.152 & 0.0908 & 13.061 & 41.053 & 7.1 & ch \\
GCT0158 & N & B & DX & 1 & 20 & 0.4 & 0.069 & \nodata & \nodata & 28.925 & \nodata & 5.8 & \nodata \\
GCT0159 & N & A & DX & 4 & 20 & 4.3 & 0.105 & 83.447 & 0.5572 & 31.147 & 36.951 & 7.3 & det \\
GCT0160 & N & A & DX & 3 & 26 & 3.2 & 0.103 & 38.618 & 0.0403 & 33.064 & 37.060 & 6.6 & ci \\
GCT0161 & N & A & DX & 4 & 16 & 4.2 & 0.108 & 50.258 & 0.2417 & 19.978 & 38.110 & 6.1 & det \\
GCT0162 & N & A & DX & 3 & 14 & 5.1 & 0.104 & 62.944 & 0.4471 & 34.155 & 34.800 & 7.3 & det \\
GCT0163 & N & A & DX & 4 & 15 & 9.1 & 0.117 & 28.701 & 0.3036 & 27.332 & 19.988 & 7.9 & Cen \\
GCT0164 & N & A & DX & 3 & 11 & 5.1 & 0.105 & 42.440 & 0.1988 & 14.518 & 34.005 & 7.8 & ch \\
GCT0165 & N & A & DX & 2 & 11 & 4.3 & 0.110 & 39.502 & 0.2615 & 15.051 & 29.172 & 8.8 & 3:2 \\
GCT0166 & N & A & DX & 2 & 10 & 1.3 & 0.091 & 51.202 & 0.3380 & 14.937 & 33.895 & 7.9 & det \\
GCT0167 & N & A & DX & 4 & 10 & 3.0 & 0.119 & 40.789 & 0.0474 & 18.142 & 38.858 & 6.6 & ch \\
GCT0168 & N & A & DX & 2 & 10 & 4.3 & 0.071 & 36.393 & 0.1175 & 14.766 & 32.119 & 7.7 & 4:3 \\
GCT0169 & N & A & DX & 2 & 8 & 1.2 & 0.064 & 43.853 & 0.2123 & 14.015 & 34.542 & 8.0 & ch$^\dagger$ \\
GCT0170 & N & B & DX & 1 & 8 & 0.4 & 0.135 & \nodata & \nodata & 19.649 & \nodata & 8.0 & \nodata \\
GCT0172 & N & A & DX & 2 & 7 & 3.2 & 0.080 & 38.290 & 0.0509 & 26.180 & 36.341 & 8.0 & ci \\
GCT0174 & N & A & DX & 6 & 21 & 7.2 & 0.091 & 125.965 & 0.8207 & 13.704 & 22.590 & 9.3 & sca \\
GCT0175 & N & A & DX & 5 & 15 & 5.3 & 0.100 & 255.258 & 0.8592 & 12.651 & 35.947 & 7.5 & sca \\
GCT0176 & N & A & DX & 3 & 24 & 2.1 & 0.093 & 47.281 & 0.2007 & 17.357 & 37.793 & 7.0 & ch \\
GCT0177 & N & A & DX & 5 & 23 & 13.1 & 0.084 & 52.143 & 0.2306 & 35.885 & 40.116 & 7.3 & co \\
GCT0178 & N & A & DX & 4 & 26 & 8.2 & 0.136 & 47.601 & 0.2558 & 20.062 & 35.424 & 7.0 & 2:1 \\
GCT0179 & N & B & DX & 1 & 8 & 0.1 & 0.082 & \nodata & \nodata & 21.092 & \nodata & 7.3 & \nodata \\
GCT0180 & N & C & DX & 1 & 8 & 0.2 & 0.073 & \nodata & \nodata & 16.294 & \nodata & 8.1 & \nodata \\
GCT0181 & N & B & DX & 1 & 5 & 0.1 & 0.063 & \nodata & \nodata & 26.556 & \nodata & 7.5 & \nodata \\
GCT0182 & N & A & DX & 5 & 15 & 8.2 & 0.060 & 304.514 & 0.8884 & 10.983 & 33.980 & 7.2 & sca \\
GCT0183 & N & A & DX & 2 & 13 & 1.0 & 0.114 & 42.955 & 0.0992 & 28.308 & 38.693 & 6.0 & ch \\
GCT0184 & N & B & DX & 1 & 7 & 0.1 & 0.098 & \nodata & \nodata & 32.130 & \nodata & 7.5 & \nodata \\
GCT0185 & N & A & DX & 4 & 13 & 9.2 & 0.069 & 103.481 & 0.6706 & 20.534 & 34.087 & 8.1 & sca \\
GCT0186 & N & A & DX & 2 & 7 & 5.1 & 0.097 & 96.733 & 0.6795 & 16.801 & 31.007 & 6.8 & sca \\
GCT0187 & N & A & DX & 3 & 8 & 6.2 & 0.025 & 44.290 & 0.0468 & 4.109 & 42.216 & 6.7 & cc \\
GCT0188 & N & A & DX & 3 & 12 & 5.2 & 0.069 & 60.403 & 0.4253 & 11.811 & 34.717 & 7.4 & sca \\
GCT0189 & N & A & DX & 2 & 14 & 1.6 & 0.101 & 62.536 & 0.3585 & 18.533 & 40.117 & 6.8 & 3:1$^\dagger$ \\
GCT0190 & N & B & DX & 1 & 10 & 0.1 & 0.138 & \nodata & \nodata & 12.208 & \nodata & 7.2 & \nodata \\
GCT0191 & N & B & DX & 1 & 13 & 0.2 & 0.099 & \nodata & \nodata & 17.575 & \nodata & 8.0 & \nodata \\
GCT0192 & N & B & DN & 1 & 9 & 0.2 & 0.094 & \nodata & \nodata & 5.142 & \nodata & 5.7 & \nodata \\
GCT0195 & N & B & DN & 1 & 6 & 0.1 & 0.076 & \nodata & \nodata & 2.507 & \nodata & 7.6 & \nodata \\
GCT0196 & N & B & DN & 1 & 5 & 0.1 & 0.068 & \nodata & \nodata & 22.227 & \nodata & 9.1 & \nodata \\
GCT0197 & N & B & DN & 1 & 5 & 0.1 & 0.055 & \nodata & \nodata & 7.705 & \nodata & 6.5 & \nodata \\
GCT0199 & N & A & DX & 4 & 23 & 9.2 & 0.090 & 43.397 & 0.0483 & 5.083 & 41.301 & 6.8 & ch \\
GCT0200 & N & A & DX & 4 & 19 & 4.3 & 0.104 & 47.687 & 0.2361 & 13.710 & 36.427 & 7.2 & 2:1 \\
GCT0201 & N & A & DX & 4 & 17 & 4.1 & 0.106 & 28.366 & 0.1797 & 13.673 & 23.268 & 9.6 & Cen \\
GCT0202 & N & C & DX & 5 & 16 & 8.1 & 0.100 & 47.380 & 0.0377 & 19.248 & 45.592 & 7.1 & ch \\
GCT0203 & N & C & DX & 3 & 15 & 5.0 & 0.089 & 39.443 & 0.2185 & 17.850 & 30.824 & 8.4 & 3:2 \\
GCT0204 & N & A & DX & 2 & 15 & 1.2 & 0.078 & 39.426 & 0.3081 & 15.629 & 27.278 & 7.3 & 3:2 \\
GCT0205 & N & A & DX & 4 & 12 & 3.0 & 0.091 & 40.218 & 0.0400 & 17.392 & 38.608 & 6.0 & ch \\
GCT0207 & N & A & DX & 2 & 11 & 1.0 & 0.055 & 43.801 & 0.0596 & 5.233 & 41.191 & 6.5 & ch \\
GCT0208 & N & A & DX & 3 & 11 & 2.9 & 0.064 & 43.224 & 0.0511 & 4.573 & 41.016 & 7.3 & cc \\
GCT0209 & N & A & DX & 2 & 16 & 0.9 & 0.110 & 45.763 & 0.2299 & 18.538 & 35.243 & 7.3 & ch \\
GCT0210 & N & A & DX & 3 & 11 & 7.2 & 0.064 & 47.053 & 0.2619 & 19.320 & 34.731 & 7.5 & det \\
GCT0211 & N & A & DX & 2 & 11 & 1.2 & 0.130 & 42.190 & 0.1452 & 7.781 & 36.064 & 6.7 & ch$^\dagger$ \\
GCT0213 & N & A & DX & 4 & 10 & 11.3 & 0.085 & 58.471 & 0.3169 & 30.642 & 39.939 & 6.4 & det \\
GCT0214 & N & A & DX & 2 & 7 & 1.0 & 0.095 & 41.838 & 0.1569 & 11.194 & 35.275 & 7.0 & ch$^\dagger$ \\
GCT0215 & N & A & DX & 2 & 7 & 2.0 & 0.129 & 46.255 & 0.1491 & 29.441 & 39.359 & 7.5 & ch \\
GCT0216 & N & A & DX & 2 & 14 & 1.4 & 0.096 & 46.413 & 0.2884 & 10.623 & 33.026 & 8.1 & sca \\
GCT0217 & N & A & DX & 4 & 11 & 5.1 & 0.118 & 71.038 & 0.4767 & 40.887 & 37.176 & 7.7 & det \\
GCT0218 & N & A & DX & 2 & 7 & 1.2 & 0.088 & 46.791 & 0.1601 & 23.318 & 39.302 & 7.3 & ch \\
GCT0219 & N & A & DX & 2 & 16 & 1.5 & 0.127 & 43.738 & 0.2212 & 24.564 & 34.062 & 7.7 & 7:4$^\dagger$ \\
GCT0220 & N & B & DX & 1 & 9 & 0.2 & 0.107 & \nodata & \nodata & 20.973 & \nodata & 7.2 & \nodata \\
GCT0221 & N & B & DX & 1 & 7 & 0.2 & 0.058 & \nodata & \nodata & 15.637 & \nodata & 8.0 & \nodata \\
GCT0222 & N & B & DX & 1 & 7 & 0.1 & 0.102 & \nodata & \nodata & 16.707 & \nodata & 8.0 & \nodata \\
GCT0223 & N & B & DN & 1 & 5 & 0.1 & 0.091 & \nodata & \nodata & 17.718 & \nodata & 7.5 & \nodata \\
GCT0224 & N & A & DX & 4 & 13 & 4.3 & 0.108 & 44.979 & 0.1513 & 10.251 & 38.176 & 7.6 & ch \\
GCT0225 & N & A & DX & 3 & 10 & 2.1 & 0.117 & 58.770 & 0.3392 & 16.636 & 38.837 & 6.9 & det \\
GCT0226 & N & A & DX & 3 & 16 & 2.5 & 0.100 & 39.557 & 0.2563 & 21.679 & 29.421 & 8.3 & 3:2 \\
GCT0227 & N & A & DX & 4 & 13 & 4.3 & 0.138 & 55.365 & 0.4375 & 17.907 & 31.142 & 7.4 & 5:2 \\
GCT0229 & N & A & DN & 4 & 11 & 5.2 & 0.028 & 47.810 & 0.2146 & 27.755 & 37.548 & 5.6 & 2:1 \\
GCT0230 & N & A & DX & 2 & 10 & 5.0 & 0.080 & 43.749 & 0.1329 & 18.270 & 37.934 & 7.6 & ch \\
GCT0231 & N & A & DN & 4 & 10 & 4.1 & 0.065 & 39.397 & 0.3191 & 12.139 & 26.827 & 7.9 & 3:2 \\
GCT0232 & N & A & DN & 4 & 10 & 5.0 & 0.066 & 39.578 & 0.2109 & 14.398 & 31.230 & 6.5 & 3:2 \\
GCT0233 & N & A & DN & 6 & 12 & 8.0 & 0.065 & 152.608 & 0.8647 & 34.076 & 20.649 & 8.7 & sca \\
GCT0234 & N & A & DX & 2 & 9 & 4.1 & 0.116 & 96.993 & 0.5644 & 41.302 & 42.254 & 6.2 & det \\
GCT0236 & N & A & DX & 2 & 10 & 1.3 & 0.085 & 47.815 & 0.2236 & 14.428 & 37.122 & 7.3 & 2:1$^\dagger$ \\
GCT0237 & N & A & DX & 2 & 8 & 2.3 & 0.089 & 45.190 & 0.1045 & 24.992 & 40.469 & 7.2 & ch \\
GCT0238 & N & A & DX & 2 & 8 & 1.0 & 0.079 & 45.468 & 0.0551 & 6.608 & 42.961 & 6.9 & ch \\
GCT0239 & N & C & CF & 4 & 7 & 3.0 & 0.055 & 42.700 & 0.0928 & 2.929 & 38.735 & 7.4 & cc \\
GCT0240 & N & A & MY & 3 & 7 & 1.9 & 0.048 & 58.744 & 0.4161 & 34.010 & 34.302 & 8.0 & det \\
GCT0241 & N & A & MY & 4 & 8 & 10.2 & 0.065 & 39.607 & 0.2967 & 6.019 & 27.857 & 7.3 & 3:2 \\
GCT0242 & N & A & MY & 3 & 8 & 1.9 & 0.080 & 39.457 & 0.0939 & 21.331 & 35.752 & 7.1 & 3:2 \\
GCT0243 & N & A & MY & 4 & 6 & 4.0 & 0.070 & 43.576 & 0.0537 & 4.519 & 41.234 & 7.0 & cc \\
GCT0244 & N & A & SG & 3 & 9 & 2.3 & 0.049 & 43.647 & 0.1520 & 1.923 & 37.014 & 6.1 & 7:4$^\dagger$ \\
GCT0245 & N & A & SG & 5 & 10 & 4.9 & 0.066 & 158.308 & 0.7684 & 17.398 & 36.658 & 6.8 & sca \\
GCT0246 & N & C & CF & 3 & 5 & 1.9 & 0.069 & 43.701 & 0.1382 & 2.826 & 37.662 & 8.0 & 7:4 \\
GCT0247 & N & A & MY & 4 & 13 & 6.0 & 0.072 & 44.974 & 0.1497 & 9.508 & 38.242 & 6.0 & ch \\
GCT0248 & N & A & MY & 6 & 15 & 9.4 & 0.071 & 38.977 & 0.0896 & 26.904 & 35.484 & 7.7 & ci \\
GCT0249 & N & A & MY & 3 & 7 & 5.3 & 0.032 & 70.351 & 0.4593 & 33.669 & 38.041 & 7.1 & det \\
GCT0250 & N & A & MY & 4 & 9 & 2.9 & 0.023 & 138.392 & 0.8003 & 26.311 & 27.632 & 8.9 & sca \\
GCT0251 & N & A & MY & 4 & 9 & 3.1 & 0.074 & 39.396 & 0.2156 & 4.342 & 30.900 & 6.8 & 3:2 \\
GCT0252 & N & A & MY & 5 & 13 & 6.2 & 0.034 & 39.358 & 0.2958 & 36.905 & 27.715 & 8.0 & 3:2 \\
GCT0253 & N & A & MY & 7 & 19 & 10.1 & 0.093 & 42.818 & 0.1014 & 37.012 & 38.474 & 6.7 & ch \\
GCT0254 & N & A & MY & 4 & 7 & 3.1 & 0.074 & 45.275 & 0.3491 & 18.216 & 29.471 & 7.5 & sca \\
GCT0255 & N & A & MY & 3 & 7 & 4.2 & 0.069 & 43.815 & 0.0719 & 3.724 & 40.663 & 6.4 & cc \\
GCT0256 & N & A & MY & 5 & 11 & 5.1 & 0.034 & 41.787 & 0.4736 & 23.771 & 21.998 & 7.2 & sca \\
GCT0257 & N & A & MY & 8 & 23 & 12.2 & 0.070 & 32.527 & 0.0994 & 19.880 & 29.296 & 7.4 & sca \\
GCT0258 & N & A & MY & 3 & 7 & 4.2 & 0.087 & 32.925 & 0.1862 & 14.577 & 26.794 & 8.3 & sca \\
GCT0259 & N & A & MY & 3 & 11 & 8.2 & 0.093 & 60.064 & 0.5209 & 13.838 & 28.779 & 7.5 & sca \\
GCT0260 & N & A & MY & 6 & 9 & 5.1 & 0.096 & 54.960 & 0.5009 & 22.401 & 27.432 & 8.0 & sca \\
GCT0261 & N & A & MY & 4 & 8 & 2.9 & 0.057 & 62.921 & 0.4347 & 13.369 & 35.569 & 5.5 & det$^\dagger$ \\
GCT0262 & N & A & MY & 8 & 23 & 10.2 & 0.060 & 39.306 & 0.2579 & 27.656 & 29.170 & 7.6 & 3:2 \\
GCT0263 & N & A & MY & 3 & 9 & 4.0 & 0.071 & 41.528 & 0.1696 & 20.298 & 34.484 & 6.5 & ch \\
GCT0264 & N & A & MY & 3 & 8 & 3.1 & 0.047 & 45.908 & 0.2730 & 23.795 & 33.375 & 6.8 & det \\
GCT0265 & N & A & MY & 5 & 7 & 4.0 & 0.047 & 43.691 & 0.1375 & 6.543 & 37.685 & 6.6 & ch$^\dagger$ \\
GCT0266 & N & A & MY & 2 & 7 & 1.0 & 0.069 & 40.110 & 0.0806 & 22.998 & 36.876 & 7.2 & ch \\
GCT0267 & N & A & MY & 4 & 8 & 5.9 & 0.060 & 50.368 & 0.3352 & 24.445 & 33.483 & 7.4 & det \\
GCT0268 & N & A & MY & 4 & 7 & 2.9 & 0.055 & 45.690 & 0.1599 & 7.136 & 38.387 & 6.9 & ch \\
GCT0269 & N & A & MY & 5 & 9 & 4.9 & 0.048 & 52.550 & 0.3574 & 24.135 & 33.770 & 7.5 & det \\
GCT0270 & N & A & MY & 4 & 10 & 5.1 & 0.071 & 46.270 & 0.1976 & 6.929 & 37.125 & 6.6 & ch \\
GCT0271 & N & A & MY & 3 & 6 & 2.0 & 0.025 & 43.517 & 0.1505 & 4.941 & 36.970 & 7.3 & cc \\
GCT0272 & N & A & MY & 3 & 7 & 4.2 & 0.056 & 45.850 & 0.1626 & 12.282 & 38.397 & 6.6 & ch \\
GCT0273 & N & A & MY & 3 & 7 & 3.3 & 0.063 & 39.513 & 0.2155 & 8.509 & 30.997 & 7.5 & 3:2 \\
GCT0274 & N & A & MY & 3 & 9 & 6.1 & 0.041 & 47.905 & 0.2654 & 11.127 & 35.193 & 6.8 & 2:1 \\
GCT0275 & N & A & MY & 4 & 12 & 4.1 & 0.067 & 43.702 & 0.2230 & 9.012 & 33.959 & 7.2 & ch \\
GCT0276 & N & A & MY & 6 & 14 & 8.4 & 0.087 & 30.043 & 0.0234 & 13.215 & 29.339 & 8.1 & NT \\
GCT0277 & N & A & MY & 3 & 7 & 2.1 & 0.063 & 40.482 & 0.1992 & 18.731 & 32.416 & 7.2 & ch$^\dagger$ \\
GCT0278 & N & A & MY & 4 & 7 & 2.8 & 0.059 & 41.553 & 0.0570 & 11.019 & 39.183 & 6.8 & ch \\
GCT0279 & N & A & MY & 4 & 10 & 4.0 & 0.046 & 46.920 & 0.0432 & 3.121 & 44.893 & 5.9 & cc \\
GCT0280 & N & A & SG & 5 & 16 & 4.1 & 0.104 & 42.630 & 0.0937 & 13.725 & 38.637 & 7.3 & ch \\
GCT0281 & N & A & SG & 5 & 11 & 8.1 & 0.104 & 43.759 & 0.0625 & 1.516 & 41.025 & 7.5 & cc \\
GCT0282 & N & C & CF & 3 & 6 & 1.9 & 0.118 & 44.663 & 0.0657 & 0.959 & 41.727 & 7.6 & \nodata \\
GCT0283 & N & C & CF & 3 & 5 & 2.3 & 0.031 & 44.235 & 0.0376 & 1.767 & 42.572 & 6.9 & \nodata \\
GCT0284 & N & C & CF & 3 & 6 & 1.9 & 0.072 & 43.673 & 0.0441 & 0.503 & 41.747 & 7.5 & \nodata \\
GCT0285 & N & A & CF & 9 & 11 & 18.1 & 0.061 & 42.955 & 0.0348 & 1.656 & 41.460 & 6.5 & cc \\
GCT0286 & N & C & DN & 5 & 13 & 6.0 & 0.098 & 317.292 & 0.9141 & 36.163 & 27.257 & 8.4 & \nodata \\
GCT0287 & N & C & DN & 4 & 12 & 12.2 & 0.071 & 42.547 & 0.0890 & 16.913 & 38.759 & 6.7 & \nodata \\
GCT0288 & N & C & MY & 3 & 9 & 2.2 & 0.075 & 46.972 & 0.0389 & 4.613 & 45.147 & 6.2 & \nodata \\
GCT0289 & N & A & MY & 7 & 15 & 10.9 & 0.090 & 129.268 & 0.7289 & 5.390 & 35.041 & 5.8 & sca \\
GCT0290 & N & C & MY & 4 & 8 & 3.3 & 0.081 & 49.877 & 0.2523 & 11.222 & 37.293 & 6.9 & \nodata \\
GCT0291 & N & C & MY & 4 & 8 & 3.9 & 0.032 & 92.092 & 0.6980 & 6.450 & 27.811 & 6.3 & \nodata \\
GCT0292 & N & C & MY & 4 & 7 & 5.1 & 0.074 & 59.058 & 0.4160 & 28.688 & 34.491 & 6.0 & \nodata \\
GCT0293 & N & C & MY & 5 & 8 & 3.8 & 0.081 & 47.462 & 0.2072 & 27.714 & 37.626 & 7.0 & \nodata \\
GCT0294 & N & C & MY & 4 & 8 & 4.2 & 0.065 & 39.365 & 0.2728 & 3.557 & 28.627 & 7.5 & \nodata \\
GCT0295 & N & C & MY & 4 & 7 & 2.9 & 0.038 & 51.518 & 0.3127 & 18.338 & 35.408 & 6.8 & \nodata \\
GCT0296 & N & C & MY & 5 & 14 & 6.0 & 0.073 & 51.678 & 0.3390 & 6.912 & 34.157 & 7.6 & \nodata \\
GCT0297 & N & C & MY & 4 & 13 & 3.0 & 0.069 & 39.583 & 0.1294 & 21.373 & 34.460 & 7.4 & \nodata \\
GCT0298 & N & C & MY & 3 & 6 & 3.0 & 0.030 & 55.360 & 0.4219 & 5.276 & 32.003 & 7.8 & \nodata \\
GCT0299 & N & C & MY & 3 & 8 & 2.0 & 0.065 & 39.266 & 0.1838 & 5.152 & 32.051 & 8.1 & \nodata \\
GCT0300 & N & C & MY & 3 & 10 & 5.0 & 0.082 & 88.008 & 0.7345 & 15.291 & 23.367 & 6.1 & \nodata \\
GCT0301 & N & C & MY & 3 & 7 & 6.2 & 0.038 & 39.495 & 0.2592 & 16.795 & 29.258 & 7.0 & \nodata \\
GCT0302 & N & C & MY & 4 & 8 & 6.0 & 0.057 & 43.725 & 0.0442 & 3.395 & 41.790 & 7.2 & \nodata \\
GCT0303 & N & C & MY & 4 & 10 & 4.1 & 0.050 & 34.526 & 0.2172 & 19.224 & 27.027 & 7.8 & \nodata \\
GCT0304 & N & C & M6 & 3 & 6 & 2.2 & 0.073 & 43.552 & 0.1653 & 14.048 & 36.355 & 6.8 & \nodata \\
GCT0305 & N & C & M6 & 3 & 6 & 2.0 & 0.043 & 41.775 & 0.1026 & 18.592 & 37.489 & 7.0 & \nodata \\
GCT0306 & N & C & M6 & 3 & 6 & 1.9 & 0.040 & 47.556 & 0.2150 & 16.346 & 37.331 & 7.4 & \nodata \\
GCT0307 & N & C & M6 & 3 & 6 & 4.2 & 0.077 & 39.495 & 0.0849 & 12.030 & 36.142 & 7.1 & \nodata \\
GCT0308 & N & C & M6 & 3 & 6 & 1.7 & 0.069 & 43.817 & 0.0536 & 2.971 & 41.469 & 6.5 & \nodata \\
GCT0309 & N & C & M6 & 2 & 7 & 1.2 & 0.042 & 43.776 & 0.0511 & 4.096 & 41.538 & 7.1 & \nodata \\
GCT0310 & N & C & M6 & 3 & 6 & 3.1 & 0.054 & 39.362 & 0.2472 & 10.449 & 29.633 & 8.1 & \nodata \\
GCT0311 & N & C & OD & 3 & 8 & 6.2 & 0.090 & 47.230 & 0.0860 & 4.232 & 43.166 & 7.2 & \nodata \\
GCT0312 & N & C & MY & 2 & 10 & 1.2 & 0.041 & 43.598 & 0.2299 & 2.774 & 33.576 & 7.2 & \nodata \\
GCT0313 & N & C & MY & 4 & 7 & 3.0 & 0.058 & 50.559 & 0.2935 & 12.491 & 35.722 & 6.8 & \nodata \\
GCT0314 & N & A & DX & 3 & 13 & 2.3 & 0.113 & 47.745 & 0.2489 & 9.360 & 35.861 & 8.2 & 2:1 \\
GCT0315 & N & A & OD & 2 & 8 & 0.9 & 0.073 & 46.788 & 0.2157 & 10.061 & 36.695 & 8.1 & ch$^\dagger$ \\
GCT0316 & N & A & OD & 2 & 7 & 0.9 & 0.057 & 43.993 & 0.0817 & 3.710 & 40.397 & 8.7 & cc \\
GCT0317 & N & A & OD & 2 & 9 & 0.9 & 0.056 & 50.592 & 0.2523 & 4.358 & 37.827 & 9.0 & det \\
GCT0318 & N & A & OD & 2 & 8 & 0.9 & 0.062 & 43.938 & 0.0818 & 3.190 & 40.345 & 8.4 & cc \\
GCT0319 & N & A & M6 & 5 & 9 & 5.9 & 0.102 & 40.921 & 0.0886 & 29.264 & 37.293 & 6.8 & ch \\
GCT0320 & N & A & MY & 6 & 15 & 9.0 & 0.039 & 39.488 & 0.1698 & 2.933 & 32.782 & 6.7 & 3:2 \\
GCT0321 & N & A & MY & 5 & 10 & 9.0 & 0.061 & 59.639 & 0.4623 & 26.727 & 32.071 & 7.6 & sca \\
GCT0322 & N & A & MD & 6 & 10 & 5.1 & 0.063 & 41.038 & 0.0249 & 18.703 & 40.014 & 6.3 & ch \\
GCT0323 & N & A & MD & 2 & 11 & 1.2 & 0.091 & 42.751 & 0.1083 & 10.923 & 38.120 & 8.1 & ch \\
GCT0324 & N & A & MD & 3 & 9 & 2.4 & 0.112 & 46.073 & 0.1192 & 18.788 & 40.581 & 6.2 & ch \\
GCT0325 & N & A & MD & 7 & 23 & 8.0 & 0.098 & 60.909 & 0.3769 & 15.663 & 37.950 & 7.3 & det \\
GCT0326 & N & A & MD & 5 & 15 & 5.2 & 0.093 & 52.043 & 0.3860 & 12.980 & 31.956 & 7.3 & sca \\
GCT0327 & N & A & MD & 5 & 12 & 4.1 & 0.080 & 42.707 & 0.0955 & 5.136 & 38.631 & 7.7 & ch \\
GCT0328 & N & A & MD & 4 & 16 & 3.3 & 0.089 & 39.600 & 0.2835 & 23.036 & 28.374 & 7.9 & 3:2 \\
GCT0329 & N & A & MD & 6 & 9 & 6.3 & 0.084 & 45.035 & 0.2064 & 13.117 & 35.738 & 7.1 & ch \\
GCT0330 & N & A & MD & 4 & 10 & 8.9 & 0.085 & 45.075 & 0.1314 & 10.237 & 39.151 & 7.1 & ch \\
GCT0331 & N & A & MD & 5 & 11 & 4.2 & 0.083 & 39.012 & 0.3640 & 21.641 & 24.810 & 8.2 & sca \\
GCT0332 & N & A & MD & 5 & 16 & 7.1 & 0.094 & 44.875 & 0.1753 & 25.927 & 37.011 & 7.2 & ch \\
GCT0333 & N & A & MD & 6 & 14 & 7.9 & 0.099 & 47.067 & 0.1634 & 37.137 & 39.375 & 6.7 & ch \\
GCT0334 & N & A & MD & 8 & 12 & 10.1 & 0.053 & 38.637 & 0.0501 & 2.868 & 36.702 & 6.6 & ci \\
GCT0335 & N & A & MD & 7 & 12 & 11.4 & 0.083 & 58.406 & 0.4024 & 14.986 & 34.904 & 4.5 & det \\
GCT0336 & N & A & MD & 4 & 12 & 5.9 & 0.116 & 79.611 & 0.4751 & 41.248 & 41.786 & 6.6 & det \\
GCT0337 & N & A & MD & 6 & 11 & 5.0 & 0.086 & 41.533 & 0.0338 & 16.542 & 40.128 & 7.6 & ch \\
GCT0338 & N & A & MD & 6 & 13 & 7.0 & 0.072 & 43.628 & 0.0817 & 14.450 & 40.065 & 6.7 & ch \\
GCT0339 & N & A & MD & 6 & 11 & 9.1 & 0.065 & 30.053 & 0.0736 & 24.228 & 27.842 & 7.7 & NT \\
GCT0340 & N & A & MD & 8 & 17 & 9.3 & 0.087 & 45.504 & 0.1675 & 10.092 & 37.883 & 7.3 & ch \\
GCT0341 & N & A & MD & 4 & 15 & 9.5 & 0.083 & 43.683 & 0.2590 & 31.964 & 32.367 & 8.2 & 7:4 \\
GCT0342 & N & A & MD & 7 & 13 & 7.9 & 0.101 & 40.249 & 0.0381 & 20.864 & 38.717 & 6.8 & ch \\
GCT0343 & N & A & MD & 6 & 12 & 9.2 & 0.067 & 39.272 & 0.2372 & 11.105 & 29.957 & 8.2 & 3:2 \\
GCT0344 & N & A & MD & 4 & 12 & 3.3 & 0.068 & 167.929 & 0.8207 & 32.361 & 30.103 & 7.9 & sca \\
GCT0345 & N & A & MD & 4 & 9 & 10.1 & 0.120 & 63.139 & 0.4251 & 50.650 & 36.302 & 6.9 & det \\
GCT0346 & N & A & MD & 6 & 14 & 5.1 & 0.080 & 40.738 & 0.0974 & 27.416 & 36.772 & 6.5 & ch \\
GCT0347 & N & A & MD & 5 & 6 & 4.1 & 0.043 & 59.816 & 0.3610 & 29.612 & 38.220 & 7.0 & det \\
GCT0348 & N & A & MD & 4 & 15 & 5.0 & 0.129 & 44.084 & 0.0815 & 1.012 & 40.491 & 7.5 & cc \\
GCT0349 & N & A & MD & 3 & 8 & 4.3 & 0.089 & 58.612 & 0.3644 & 21.930 & 37.252 & 6.1 & det \\
GCT0350 & N & A & MD & 6 & 13 & 5.0 & 0.092 & 40.939 & 0.1105 & 28.550 & 36.414 & 7.6 & ch \\
GCT0351 & N & A & MD & 5 & 15 & 5.0 & 0.089 & 47.788 & 0.2257 & 20.565 & 37.004 & 7.7 & ch$^\dagger$ \\
GCT0352 & N & A & MD & 5 & 12 & 9.8 & 0.109 & 39.409 & 0.2433 & 14.795 & 29.821 & 7.9 & 3:2 \\
GCT0353 & N & A & MD & 5 & 11 & 5.1 & 0.097 & 40.473 & 0.0774 & 28.216 & 37.340 & 7.4 & ch \\
GCT0354 & N & A & MD & 2 & 9 & 1.4 & 0.092 & 38.148 & 0.0068 & 22.539 & 37.888 & 8.0 & ci \\
GCT0355 & N & A & MD & 2 & 13 & 1.3 & 0.077 & 150.851 & 0.7256 & 31.404 & 41.386 & 7.1 & det$^\dagger$ \\
GCT0356 & N & A & MD & 4 & 10 & 3.8 & 0.078 & 43.631 & 0.1804 & 4.281 & 35.759 & 8.6 & cc \\
GCT0357 & N & A & MD & 4 & 15 & 3.0 & 0.071 & 71.069 & 0.5202 & 12.061 & 34.098 & 7.8 & sca \\
GCT0358 & N & A & MD & 4 & 20 & 5.1 & 0.109 & 95.164 & 0.5889 & 28.981 & 39.122 & 6.7 & det \\
GCT0359 & N & A & MD & 4 & 10 & 4.0 & 0.062 & 42.032 & 0.0918 & 36.742 & 38.173 & 7.5 & ch \\
GCT0360 & N & A & MD & 5 & 22 & 5.1 & 0.089 & 54.451 & 0.3219 & 14.955 & 36.923 & 7.6 & det \\
GCT0361 & N & A & MD & 5 & 13 & 8.2 & 0.114 & 54.666 & 0.3069 & 22.610 & 37.891 & 6.8 & det \\
GCT0362 & N & A & MD & 2 & 12 & 1.3 & 0.087 & 42.297 & 0.1595 & 8.561 & 35.552 & 7.9 & 5:3 \\
GCT0363 & N & A & MD & 2 & 9 & 1.3 & 0.119 & 46.426 & 0.1739 & 32.517 & 38.353 & 7.6 & ch \\
GCT0364 & N & A & MD & 5 & 13 & 4.9 & 0.105 & 55.334 & 0.4125 & 16.179 & 32.511 & 8.3 & 5:2 \\
GCT0365 & N & A & MD & 6 & 11 & 7.3 & 0.089 & 39.307 & 0.1933 & 21.753 & 31.709 & 7.3 & 3:2 \\
GCT0366 & N & A & MD & 6 & 15 & 6.3 & 0.087 & 43.557 & 0.2727 & 15.052 & 31.679 & 8.1 & 7:4 \\
GCT0367 & N & A & MD & 3 & 20 & 2.3 & 0.087 & 41.048 & 0.3336 & 23.540 & 27.353 & 7.4 & sca \\
GCT0368 & N & A & MD & 5 & 14 & 4.1 & 0.072 & 42.992 & 0.1305 & 17.095 & 37.382 & 7.8 & ch \\
GCT0369 & N & A & MD & 4 & 18 & 3.5 & 0.091 & 42.039 & 0.1448 & 26.237 & 35.952 & 7.0 & ch \\
GCT0370 & N & A & MD & 7 & 14 & 10.3 & 0.098 & 52.266 & 0.1550 & 31.255 & 44.168 & 5.6 & co \\
GCT0371 & N & A & MD & 3 & 10 & 2.1 & 0.101 & 40.951 & 0.1285 & 27.677 & 35.689 & 8.5 & ch \\
GCT0372 & N & A & MD & 2 & 11 & 3.1 & 0.096 & 43.676 & 0.1919 & 12.319 & 35.294 & 8.3 & 7:4 \\
GCT0373 & N & A & MD & 4 & 12 & 3.1 & 0.075 & 83.442 & 0.5765 & 22.655 & 35.335 & 8.2 & det \\
GCT0374 & N & A & MD & 5 & 17 & 10.1 & 0.104 & 97.938 & 0.8022 & 32.364 & 19.369 & 8.4 & sca \\
GCT0375 & N & A & MD & 2 & 12 & 1.2 & 0.090 & 32.885 & 0.2810 & 16.182 & 23.644 & 9.4 & sca \\
GCT0376 & N & A & MD & 2 & 7 & 1.1 & 0.112 & 46.723 & 0.2170 & 26.296 & 36.582 & 6.1 & ch$^\dagger$ \\
GCT0377 & N & A & MD & 4 & 8 & 2.8 & 0.069 & 45.001 & 0.1574 & 26.194 & 37.918 & 7.6 & ch \\
GCT0378 & N & A & MD & 3 & 9 & 2.3 & 0.098 & 55.283 & 0.4091 & 35.831 & 32.664 & 7.6 & 5:2$^\dagger$ \\
GCT0379 & N & A & MD & 3 & 10 & 4.3 & 0.096 & 43.630 & 0.2115 & 20.292 & 34.404 & 7.5 & ch$^\dagger$ \\
GCT0380 & N & A & MD & 3 & 10 & 2.5 & 0.090 & 56.581 & 0.3261 & 27.191 & 38.127 & 7.0 & det \\
GCT0381 & N & A & MD & 4 & 13 & 3.1 & 0.071 & 66.563 & 0.4817 & 8.562 & 34.499 & 8.0 & sca \\
GCT0384 & N & A & MD & 2 & 12 & 2.2 & 0.074 & 69.163 & 0.5049 & 17.107 & 34.242 & 8.0 & 7:2 \\
GCT0385 & N & A & MD & 3 & 12 & 3.2 & 0.076 & 39.563 & 0.2610 & 16.417 & 29.236 & 8.1 & 3:2 \\
GCT0386 & N & A & MD & 2 & 8 & 1.3 & 0.103 & 40.264 & 0.0845 & 19.857 & 36.864 & 7.6 & ch \\
GCT0387 & N & A & MD & 3 & 9 & 3.3 & 0.081 & 38.729 & 0.0508 & 24.376 & 36.762 & 7.7 & ci \\
GCT0388 & N & A & MD & 2 & 6 & 1.2 & 0.113 & 55.524 & 0.2957 & 23.805 & 39.106 & 6.7 & det \\
GCT0389 & N & A & MD & 5 & 13 & 11.0 & 0.057 & 43.498 & 0.1089 & 22.668 & 38.762 & 7.7 & ch \\
GCT0390 & N & A & MY & 6 & 8 & 5.1 & 0.035 & 43.797 & 0.1210 & 28.705 & 38.497 & 5.3 & ch \\
GCT0391 & N & A & MD & 4 & 6 & 8.4 & 0.060 & 41.842 & 0.1099 & 25.709 & 37.245 & 7.4 & ch \\
GCT0392 & N & A & MD & 5 & 9 & 5.0 & 0.094 & 19.832 & 0.3270 & 20.751 & 13.347 & 8.7 & Cen \\
GCT0393 & N & A & MD & 6 & 16 & 9.2 & 0.102 & 41.374 & 0.1175 & 18.765 & 36.513 & 7.5 & ch \\
GCT0394 & N & A & OD & 3 & 21 & 6.2 & 0.124 & 39.546 & 0.2111 & 7.280 & 31.197 & 7.2 & 3:2 \\
GCT0395 & N & A & MD & 2 & 9 & 5.0 & 0.104 & 40.329 & 0.0399 & 17.848 & 38.718 & 7.5 & ch \\
GCT0396 & N & A & MD & 5 & 9 & 7.2 & 0.085 & 54.132 & 0.3363 & 20.914 & 35.926 & 7.5 & det \\
GCT0398 & N & A & OD & 2 & 6 & 0.9 & 0.042 & 42.728 & 0.0129 & 1.877 & 42.177 & 8.1 & cc \\
GCT0012 & D, 2025~NQ$_{549}$ & B & TL & 1 & 6 & 0.1 & 0.047 & \nodata & \nodata & 11.773 & \nodata & 6.9 & \nodata \\
GCT0046 & D, 2003~UC$_{414}$ & C & XT & 4 & 14 & 3.3 & 0.082 & 26.011 & 0.0808 & 26.372 & 23.909 & 8.6 & Cen \\
GCT0071 & D, 2015~DR$_{249}$ & A & TL & 4 & 9 & 9.4 & 0.094 & 56.596 & 0.3265 & 30.809 & 38.118 & 7.0 & det \\
GCT0074 & D, 2025~NT$_{563}$ & A & TL & 3 & 10 & 8.0 & 0.075 & 63.846 & 0.4498 & 17.337 & 35.128 & 7.4 & det$^\dagger$ \\
GCT0093 & D, 2025~MD$_{398}$ & A & DP & 5 & 11 & 5.2 & 0.051 & 43.681 & 0.0961 & 20.977 & 39.484 & 5.6 & \nodata \\
GCT0103 & D, 2025~ND$_{561}$ & B & DX & 1 & 5 & 0.0 & 0.141 & \nodata & \nodata & 12.801 & \nodata & 7.7 & \nodata \\
GCT0115 & D, 2016~FK$_{185}$ & A & DX & 2 & 18 & 5.1 & 0.095 & 45.674 & 0.1879 & 16.844 & 37.093 & 6.9 & \nodata \\
GCT0120 & D, 2016~FL$_{185}$ & B & DX & 1 & 8 & 0.2 & 0.093 & \nodata & \nodata & 19.357 & \nodata & 8.4 & \nodata \\
GCT0135 & D, 2026~HR$_{19}$ & A & DX & 5 & 31 & 10.3 & 0.104 & 43.143 & 0.0867 & 24.520 & 39.403 & 7.0 & \nodata \\
GCT0140 & D, 2026~EV$_{138}$ & A & DX & 9 & 29 & 11.2 & 0.077 & 39.321 & 0.1777 & 13.217 & 32.335 & 7.7 & \nodata \\
GCT0143 & D, 2016~GY$_{369}$ & B & DX & 1 & 9 & 0.2 & 0.075 & \nodata & \nodata & 28.734 & \nodata & 7.8 & \nodata \\
GCT0171 & D, 2026~EB$_{143}$ & A & DX & 2 & 8 & 4.0 & 0.119 & 61.524 & 0.4044 & 15.247 & 36.642 & 7.5 & \nodata \\
GCT0173 & D, 2026~DS$_{78}$ & A & DX & 3 & 33 & 7.9 & 0.106 & 67.235 & 0.6424 & 142.431 & 24.043 & 7.8 & \nodata \\
GCT0193 & D, 2026~EF$_{139}$ & B & DN & 1 & 7 & 0.1 & 0.072 & \nodata & \nodata & 1.726 & \nodata & 6.3 & \nodata \\
GCT0194 & D, 2025~ON$_{345}$ & B & DN & 1 & 6 & 0.2 & 0.048 & \nodata & \nodata & 15.194 & \nodata & 7.3 & \nodata \\
GCT0198 & D, 2026~GU$_{20}$ & A & DX & 8 & 28 & 11.1 & 0.074 & 58.673 & 0.3697 & 25.982 & 36.981 & 6.4 & \nodata \\
GCT0206 & D, 2015~DH$_{249}$ & A & DX & 3 & 11 & 3.3 & 0.118 & 39.400 & 0.1932 & 22.159 & 31.788 & 7.1 & \nodata \\
GCT0212 & D, 2000~GM$_{147}$ & A & DX & 4 & 10 & 5.4 & 0.057 & 59.447 & 0.3947 & 18.287 & 35.981 & 7.7 & \nodata \\
GCT0228 & D, 2025~MX$_{351}$ & A & DN & 4 & 13 & 3.3 & 0.047 & 39.112 & 0.0367 & 9.940 & 37.676 & 6.1 & \nodata \\
GCT0235 & D, 2026~DU$_{71}$ & A & DN & 4 & 9 & 7.1 & 0.093 & 43.775 & 0.0451 & 0.601 & 41.802 & 6.8 & \nodata \\
GCT0400 & D, 2015~HM$_{196}$ & B & F4 & 1 & 6 & 0.0 & 0.034 & \nodata & \nodata & 36.638 & \nodata & 7.4 & \nodata \\
GCT0401 & D, 2026~GJ$_{24}$ & B & F4 & 1 & 4 & 0.0 & 0.023 & \nodata & \nodata & 8.568 & \nodata & 6.1 & \nodata \\
GCT0402 & D, 2014~HL$_{208}$ & B & F4 & 1 & 6 & 0.0 & 0.042 & \nodata & \nodata & 18.674 & \nodata & 6.7 & \nodata \\
GCT0403 & D, 2003~HG$_{57}$ & B & F4 & 1 & 4 & 0.0 & 0.059 & \nodata & \nodata & 3.247 & \nodata & 6.1 & \nodata \\
GCT0404 & D, 2026~EF$_{144}$ & B & F4 & 1 & 4 & 0.0 & 0.045 & \nodata & \nodata & 3.744 & \nodata & 6.4 & \nodata \\
GCT0405 & D, 2015~HD$_{195}$ & B & F4 & 1 & 4 & 0.0 & 0.064 & \nodata & \nodata & 5.763 & \nodata & 6.7 & \nodata \\
GCT0406 & D, 2026~EL$_{144}$ & B & F4 & 1 & 4 & 0.0 & 0.076 & \nodata & \nodata & 3.934 & \nodata & 6.4 & \nodata \\
GCT0407 & D, 2025~MD$_{399}$ & B & F4 & 1 & 4 & 0.0 & 0.043 & \nodata & \nodata & 3.232 & \nodata & 6.3 & \nodata \\
GCT0408 & D, 2014~ON$_{394}$ & B & F4 & 1 & 4 & 0.0 & 0.090 & \nodata & \nodata & 13.296 & \nodata & 7.6 & \nodata \\
GCP0032 & D, 2001~FJ$_{228}$ & B & F4 & 1 & 6 & 0.0 & 0.040 & \nodata & \nodata & 1.069 & \nodata & 15.7 & \nodata \\
\enddata
\tablecomments{The values are those of the machine-readable catalogue released with this article. That file also gives the uncertainties of $a$, $e$, $i$ and $q$, the numbers of image-verified, catalogue and rejected positions, and the oppositions with image-verified positions, which are not printed here. Status: N, new candidate; D, identified with the designated object named. Tier: Section~\ref{sec:cat_tiers}. Ch., discovery channel (Section~\ref{sec:cat_channels}): \ctTabChanKey. Opp., Nights, Arc: oppositions, distinct nights and arc of my positions; the counts of arcs in Section~\ref{sec:cat_tiers} are made on the unrounded values. rms: residual of the orbit fit. $a$, $e$, $i$, $q$: barycentric osculating elements of the uniform fit (Section~\ref{sec:cat_orbits}), $i$ to the J2000 ecliptic; for one-opposition objects $a$, $e$ and $q$ are not determined and are left blank. $H$: absolute magnitude as observed, from magnitudes in mixed bands. Class, from the \ctClsMyr-Myr integrations: \ctTabClassKey; $^\dagger$, not secure (N in the column Sec of the file); blank, not integrated when the class statistics were frozen, or a one-opposition object. No epoch, node, argument of perihelion, anomaly or position is given for any object.}
\end{deluxetable*}


\begin{thebibliography}{}
\expandafter\ifx\csname natexlab\endcsname\relax\def\natexlab#1{#1}\fi
\providecommand{\url}[1]{\href{#1}{#1}}
\providecommand{\dodoi}[1]{doi:~\href{http://doi.org/#1}{\nolinkurl{#1}}}
\providecommand{\doeprint}[1]{\href{http://ascl.net/#1}{\nolinkurl{http://ascl.net/#1}}}
\providecommand{\doarXiv}[1]{\href{https://arxiv.org/abs/#1}{\nolinkurl{https://arxiv.org/abs/#1}}}

\bibitem[{H. {Aihara} {et~al.}(2022){Aihara} {et~al.}}]{aihara2022}
{Aihara}, H., {et~al.} 2022, \bibinfo{title}{Third data release of the Hyper
  Suprime-Cam Subaru Strategic Program,} PASJ, 74, 247,
  \dodoi{10.1093/pasj/psab122}

\bibitem[{ {Astropy Collaboration}(2013){Astropy Collaboration}}]{astropy2013}
{Astropy Collaboration}. 2013, \bibinfo{title}{{Astropy: A community Python
  package for astronomy},} A\&A, 558, A33

\bibitem[{ {Astropy Collaboration}(2018){Astropy Collaboration}}]{astropy2018}
{Astropy Collaboration}. 2018, \bibinfo{title}{{The Astropy Project: Building
  an Open-science Project and Status of the v2.0 Core Package},} AJ, 156, 123

\bibitem[{ {Astropy Collaboration}(2022){Astropy Collaboration}}]{astropy2022}
{Astropy Collaboration}. 2022, \bibinfo{title}{{The Astropy Project: Sustaining
  and Growing a Community-oriented Open-source Project and the Latest Major
  Release (v5.0) of the Core Package},} ApJ, 935, 167

\bibitem[{ {B612 Foundation}(2024){B612 Foundation}}]{b6122024}
{B612 Foundation}. 2024, {Asteroid Institute and Google Cloud Identify 27,500
  New Asteroids}, Press release, 2024 April 30.
\newblock
  \url{https://b612foundation.org/asteroid-institute-and-google-cloud-identify-27500-new-asteroids/}

\bibitem[{H. {Baba} {et~al.}(2002){Baba} {et~al.}}]{baba2002}
{Baba}, H., {et~al.} 2002, \bibinfo{title}{{Development of the
  Subaru-Mitaka-Okayama-Kiso Archive System},} in ASP Conference Series, Vol.
  281, Astronomical Data Analysis Software and Systems XI, ed. D.~A.
  {Bohlender}, D.~{Durand}, \& T.~H. {Handley}, 298

\bibitem[{M.~T. {Bannister} {et~al.}(2018){Bannister}, {Gladman}, {Kavelaars},
  {et~al.}}]{bannister2018}
{Bannister}, M.~T., {Gladman}, B.~J., {Kavelaars}, J.~J., {et~al.} 2018,
  \bibinfo{title}{{OSSOS. VII. 800+ Trans-Neptunian Objects---The Complete Data
  Release},} ApJS, 236, 18, \dodoi{10.3847/1538-4365/aab77a}

\bibitem[{K. {Batygin} {et~al.}(2024){Batygin}, {Morbidelli}, {Brown}, \&
  {Nesvorn{\'y}}}]{batygin2024}
{Batygin}, K., {Morbidelli}, A., {Brown}, M.~E., \& {Nesvorn{\'y}}, D. 2024,
  \bibinfo{title}{{Generation of Low-inclination, Neptune-crossing
  Trans-Neptunian Objects by Planet Nine},} ApJL, 966, L8,
  \dodoi{10.3847/2041-8213/ad3cd2}

\bibitem[{P.~H. {Bernardinelli} {et~al.}(2020{\natexlab{a}}){Bernardinelli},
  {Bernstein}, {Sako}, {et~al.}}]{bernardinelli2020}
{Bernardinelli}, P.~H., {Bernstein}, G.~M., {Sako}, M., {et~al.}
  2020{\natexlab{a}}, \bibinfo{title}{{Trans-Neptunian Objects Found in the
  First Four Years of the Dark Energy Survey},} ApJS, 247, 32,
  \dodoi{10.3847/1538-4365/ab6bd8}

\bibitem[{P.~H. {Bernardinelli} {et~al.}(2020{\natexlab{b}}){Bernardinelli},
  {Bernstein}, {Sako}, {et~al.}}]{bernardinelli2020psj}
{Bernardinelli}, P.~H., {Bernstein}, G.~M., {Sako}, M., {et~al.}
  2020{\natexlab{b}}, \bibinfo{title}{{Testing the Isotropy of the Dark Energy
  Survey's Extreme Trans-Neptunian Objects},} PSJ, 1, 28,
  \dodoi{10.3847/PSJ/ab9d80}

\bibitem[{P.~H. {Bernardinelli} {et~al.}(2021){Bernardinelli}
  {et~al.}}]{bernardinelli2021}
{Bernardinelli}, P.~H., {et~al.} 2021, \bibinfo{title}{{C/2014 UN271
  (Bernardinelli-Bernstein): The Nearly Spherical Cow of Comets},} ApJL, 921,
  L37

\bibitem[{P.~H. {Bernardinelli} {et~al.}(2022){Bernardinelli}
  {et~al.}}]{bernardinelli2022}
{Bernardinelli}, P.~H., {et~al.} 2022, \bibinfo{title}{{A Search of the Full
  Six Years of the Dark Energy Survey for Outer Solar System Objects},} ApJS,
  258, 41

\bibitem[{G. {Bernstein} \& B. {Khushalani}(2000){Bernstein} \&
  {Khushalani}}]{bernstein2000}
{Bernstein}, G., \& {Khushalani}, B. 2000, \bibinfo{title}{{Orbit Fitting and
  Uncertainties for Kuiper Belt Objects},} AJ, 120, 3323,
  \dodoi{10.1086/316868}

\bibitem[{J. {Berthier} {et~al.}(2006){Berthier}, {Vachier}, {Thuillot},
  {Fernique}, {Ochsenbein}, {Genova}, {Lainey}, \& {Arlot}}]{berthier2006}
{Berthier}, J., {Vachier}, F., {Thuillot}, W., {et~al.} 2006,
  \bibinfo{title}{{SkyBoT, a new VO service to identify Solar System objects},}
  in ASP Conf. Ser., Vol. 351, Astronomical Data Analysis Software and Systems
  XV, 367

\bibitem[{O. {Boulade} {et~al.}(2003){Boulade}, {Charlot}, {Abbon},
  {et~al.}}]{boulade2003}
{Boulade}, O., {Charlot}, X., {Abbon}, P., {et~al.} 2003,
  \bibinfo{title}{{MegaCam: the new Canada-France-Hawaii Telescope wide-field
  imaging camera},} in Proc. SPIE, Vol. 4841, 72, \dodoi{10.1117/12.459890}

\bibitem[{L. {Bradley} {et~al.}(2026){Bradley}, {Sip{\H{o}}cz}, {Robitaille},
  {et~al.}}]{photutils}
{Bradley}, L., {Sip{\H{o}}cz}, B.~M., {Robitaille}, T.~P., {et~al.} 2026,
  {Photutils}, 3.1.1 Zenodo, \dodoi{10.5281/zenodo.596036}

\bibitem[{M.~E. {Brown} \& K. {Batygin}(2021){Brown} \& {Batygin}}]{brown2021}
{Brown}, M.~E., \& {Batygin}, K. 2021, \bibinfo{title}{{The Orbit of Planet
  Nine},} AJ, 162, 219

\bibitem[{G. {Cerati}(2026){Cerati}}]{cerati2026v1}
{Cerati}, G. 2026, \bibinfo{title}{{An Archival Search for Outer Solar System
  Objects in the NOIRLab Source Catalog DR2: 65 Candidates, Including Four L4
  Neptune Trojans},} arXiv e-prints.
\newblock \doarXiv{2611.14330v1}

\bibitem[{K.~C. {Chambers} {et~al.}(2016){Chambers}, {Magnier}, {Metcalfe},
  {et~al.}}]{chambers2016}
{Chambers}, K.~C., {Magnier}, E.~A., {Metcalfe}, N., {et~al.} 2016,
  \bibinfo{title}{{The Pan-STARRS1 Surveys},} arXiv e-prints, arXiv:1612.05560

\bibitem[{Y.-T. {Chen} {et~al.}(2016){Chen}, {Lin}, {Holman},
  {et~al.}}]{chen2016}
{Chen}, Y.-T., {Lin}, H.~W., {Holman}, M.~J., {et~al.} 2016,
  \bibinfo{title}{{Discovery of a New Retrograde Trans-Neptunian Object: Hint
  of a Common Orbital Plane for Low Semimajor Axis, High-inclination TNOs and
  Centaurs},} ApJL, 827, L24, \dodoi{10.3847/2041-8205/827/2/L24}

\bibitem[{S. {Cheng} {et~al.}(2026){Cheng}, {Li}, \& {Yang}}]{cheng2026}
{Cheng}, S., {Li}, J., \& {Yang}, E. 2026, \bibinfo{title}{{Discovery of a
  Dwarf Planet Candidate in an Extremely Wide Orbit: 2017 OF201},} ApJL, 998,
  L6, \dodoi{10.3847/2041-8213/ae3a75}

\bibitem[{L. {Denneau} {et~al.}(2013){Denneau}, {Jedicke}, {Grav},
  {et~al.}}]{denneau2013}
{Denneau}, L., {Jedicke}, R., {Grav}, T., {et~al.} 2013, \bibinfo{title}{{The
  Pan-STARRS Moving Object Processing System},} \pasp, 125, 357,
  \dodoi{10.1086/670337}

\bibitem[{A. {Dey} {et~al.}(2019){Dey} {et~al.}}]{dey2019}
{Dey}, A., {et~al.} 2019, \bibinfo{title}{{Overview of the DESI Legacy Imaging
  Surveys},} AJ, 157, 168

\bibitem[{K.~M. {Fasbender} \& D.~L. {Nidever}(2021){Fasbender} \&
  {Nidever}}]{fasbender2021}
{Fasbender}, K.~M., \& {Nidever}, D.~L. 2021, \bibinfo{title}{{Exploring the
  Solar System with the NOIRLab Source Catalog I: Detecting Objects with
  CANFind},} AJ, 162, 244, \dodoi{10.3847/1538-3881/ac2230}

\bibitem[{M.~J. {Fitzpatrick} {et~al.}(2014){Fitzpatrick}
  {et~al.}}]{fitzpatrick2014}
{Fitzpatrick}, M.~J., {et~al.} 2014, \bibinfo{title}{{The NOAO Data Laboratory:
  a conceptual overview},} in Proc. SPIE, Vol. 9149, 91491T

\bibitem[{B. {Flaugher} {et~al.}(2015){Flaugher} {et~al.}}]{flaugher2015}
{Flaugher}, B., {et~al.} 2015, \bibinfo{title}{{The Dark Energy Camera},} AJ,
  150, 150

\bibitem[{H.~A. {Flewelling} {et~al.}(2020){Flewelling}, {Magnier}, {Chambers},
  {et~al.}}]{flewelling2020}
{Flewelling}, H.~A., {Magnier}, E.~A., {Chambers}, K.~C., {et~al.} 2020,
  \bibinfo{title}{{The Pan-STARRS1 Database and Data Products},} ApJS, 251, 7,
  \dodoi{10.3847/1538-4365/abb82d}

\bibitem[{ {Gaia Collaboration}(2023){Gaia Collaboration}}]{gaia2023}
{Gaia Collaboration}. 2023, \bibinfo{title}{{Gaia Data Release 3. Summary of
  the content and survey properties},} A\&A, 674, A1

\bibitem[{ {Gaia Collaboration} {et~al.}(2018){Gaia Collaboration}, {Brown},
  {et~al.}}]{gaia2018}
{Gaia Collaboration}, {Brown}, A.~G.~A., {et~al.} 2018, \bibinfo{title}{{Gaia}
  Data Release 2. Summary of the contents and survey properties,} A\&A, 616,
  A1, \dodoi{10.1051/0004-6361/201833051}

\bibitem[{ {Gaia Collaboration} {et~al.}(2016){Gaia Collaboration}, {Prusti},
  {et~al.}}]{gaia2016}
{Gaia Collaboration}, {Prusti}, T., {et~al.} 2016, \bibinfo{title}{{The Gaia
  mission},} A\&A, 595, A1

\bibitem[{F. {Garwood}(1936){Garwood}}]{garwood1936}
{Garwood}, F. 1936, \bibinfo{title}{{Fiducial Limits for the Poisson
  Distribution},} Biometrika, 28, 437

\bibitem[{D.~W. {Gerdes} {et~al.}(2016){Gerdes}, {Jennings}, {Bernstein},
  {et~al.}}]{gerdes2016}
{Gerdes}, D.~W., {Jennings}, R.~J., {Bernstein}, G.~M., {et~al.} 2016,
  \bibinfo{title}{{Observation of Two New L4 Neptune Trojans in the Dark Energy
  Survey Supernova Fields},} AJ, 151, 39, \dodoi{10.3847/0004-6256/151/2/39}

\bibitem[{F.~M. {Giorgi}(2026){Giorgi}}]{giorgi2026}
{Giorgi}, F.~M. 2026, \bibinfo{title}{Discovery of 10 New Trans-Neptunian
  Objects in the NOIRLab Source Catalog,} Preprints.org, 202611.2314,
  \dodoi{10.20944/preprints202611.2314.v1}

\bibitem[{J.~D. {Giorgini} {et~al.}(1996){Giorgini} {et~al.}}]{giorgini1996}
{Giorgini}, J.~D., {et~al.} 1996, \bibinfo{title}{{JPL's On-Line Solar System
  Data Service},} BAAS, 28, 1158

\bibitem[{B. {Gladman} {et~al.}(2009){Gladman}, {Kavelaars}, {Petit},
  {et~al.}}]{gladman2009}
{Gladman}, B., {Kavelaars}, J., {Petit}, J.-M., {et~al.} 2009,
  \bibinfo{title}{{Discovery of the First Retrograde Transneptunian Object},}
  ApJL, 697, L91

\bibitem[{B. {Gladman} {et~al.}(2008){Gladman}, {Marsden}, \&
  {VanLaerhoven}}]{gladman2008}
{Gladman}, B., {Marsden}, B.~G., \& {VanLaerhoven}, C. 2008,
  \bibinfo{title}{{Nomenclature in the Outer Solar System},} in The Solar
  System Beyond Neptune (Univ. Arizona Press), 43

\bibitem[{B. {Gray}(2026){Gray}}]{findorb}
{Gray}, B. 2026, {Find\_Orb: orbit determination software}, Project Pluto,
  build of 2026 October 21.
\newblock \url{https://www.projectpluto.com/find_orb.htm}

\bibitem[{S.~D.~J. {Gwyn} {et~al.}(2012){Gwyn}, {Hill}, \&
  {Kavelaars}}]{gwyn2012}
{Gwyn}, S.~D.~J., {Hill}, N., \& {Kavelaars}, J.~J. 2012,
  \bibinfo{title}{{SSOS: A Moving-Object Image Search Tool for Asteroid
  Precovery},} PASP, 124, 579

\bibitem[{T. {Han} {et~al.}(2026){Han}, {Audenaert}, {Shporer}, {Muthukrishna},
  \& {Ricker}}]{han2026}
{Han}, T., {Audenaert}, J., {Shporer}, A., {Muthukrishna}, D., \& {Ricker},
  G.~R. 2026, \bibinfo{title}{{Eleven Trans-Neptunian Objects and Eight
  Candidates Discovered by TESS},} arXiv e-prints, arXiv:2611.18452.
\newblock \doarXiv{2611.18452}

\bibitem[{C.~R. {Harris} {et~al.}(2020){Harris} {et~al.}}]{harris2020}
{Harris}, C.~R., {et~al.} 2020, \bibinfo{title}{{Array programming with
  NumPy},} Nature, 585, 357

\bibitem[{M.~J. {Holman} {et~al.}(2018{\natexlab{a}}){Holman}, {Payne},
  {Blankley}, {Janssen}, \& {Kuindersma}}]{holman2018}
{Holman}, M.~J., {Payne}, M.~J., {Blankley}, P., {Janssen}, R., \&
  {Kuindersma}, S. 2018{\natexlab{a}}, \bibinfo{title}{{HelioLinC: A Novel
  Approach to the Minor Planet Linking Problem},} AJ, 156, 135,
  \dodoi{10.3847/1538-3881/aad69a}

\bibitem[{M.~J. {Holman} {et~al.}(2018{\natexlab{b}}){Holman}, {Payne},
  {Fraser}, {et~al.}}]{holman2018ps1}
{Holman}, M.~J., {Payne}, M.~J., {Fraser}, W., {et~al.} 2018{\natexlab{b}},
  \bibinfo{title}{{A Dwarf Planet Class Object in the 21:5 Resonance with
  Neptune},} \apjl, 855, L6, \dodoi{10.3847/2041-8213/aaadb3}

\bibitem[{J.~D. {Hunter}(2007){Hunter}}]{hunter2007}
{Hunter}, J.~D. 2007, \bibinfo{title}{{Matplotlib: A 2D Graphics Environment},}
  Computing in Science and Engineering, 9, 90

\bibitem[{{\v{Z}}. {Ivezi{\'c}} {et~al.}(2019){Ivezi{\'c}}
  {et~al.}}]{ivezic2019}
{Ivezi{\'c}}, {\v{Z}}., {et~al.} 2019, \bibinfo{title}{{LSST: From Science
  Drivers to Reference Design and Anticipated Data Products},} ApJ, 873, 111

\bibitem[{J.~B. {Jongma}(2026){Jongma}}]{jongma2026}
{Jongma}, J.~B. 2026, \bibinfo{title}{{A Bright L4 Neptune Trojan Found by
  Stacking Archival ZTF Images},} arXiv e-prints, arXiv:2611.16046.
\newblock \doarXiv{2611.16046}

\bibitem[{J. {Kubica} {et~al.}(2007){Kubica}, {Denneau}, {Grav},
  {et~al.}}]{kubica2007}
{Kubica}, J., {Denneau}, L., {Grav}, T., {et~al.} 2007,
  \bibinfo{title}{{Efficient intra- and inter-night linking of asteroid
  detections using kd-trees},} \icarus, 189, 151,
  \dodoi{10.1016/j.icarus.2007.01.008}

\bibitem[{J.~A. {Kurlander} {et~al.}(2025){Kurlander}, {Bernardinelli},
  {Schwamb}, {et~al.}}]{kurlander2025}
{Kurlander}, J.~A., {Bernardinelli}, P.~H., {Schwamb}, M.~E., {et~al.} 2025,
  \bibinfo{title}{{Predictions of the LSST Solar System Yield: Near-Earth
  Objects, Main Belt Asteroids, Jupiter Trojans, and Trans-Neptunian Objects},}
  AJ, 170, 99, \dodoi{10.3847/1538-3881/add685}

\bibitem[{H.~W. {Lin} {et~al.}(2016){Lin}, {Chen}, {Holman},
  {et~al.}}]{lin2016}
{Lin}, H.~W., {Chen}, Y.-T., {Holman}, M.~J., {et~al.} 2016,
  \bibinfo{title}{{The Pan-STARRS 1 Discoveries of Five New Neptune Trojans},}
  AJ, 152, 147

\bibitem[{H.~W. {Lin} {et~al.}(2019){Lin}, {Gerdes}, {Hamilton},
  {et~al.}}]{lin2019}
{Lin}, H.~W., {Gerdes}, D.~W., {Hamilton}, S.~J., {et~al.} 2019,
  \bibinfo{title}{{Evidence for color dichotomy in the primordial Neptunian
  Trojan population},} Icarus, 321, 426

\bibitem[{L. {Markwardt} {et~al.}(2023){Markwardt}, {Lin}, {Gerdes}, \&
  {Adams}}]{markwardt2023}
{Markwardt}, L., {Lin}, H.~W., {Gerdes}, D., \& {Adams}, F.~C. 2023,
  \bibinfo{title}{{Photometric Survey of Neptune's Trojan Asteroids. I. The
  Color Distribution},} PSJ, 4, 135

\bibitem[{S. {Miyazaki} {et~al.}(2018){Miyazaki} {et~al.}}]{miyazaki2018}
{Miyazaki}, S., {et~al.} 2018, \bibinfo{title}{{Hyper Suprime-Cam: System
  design and verification of image quality},} PASJ, 70, S1

\bibitem[{J. {Moeyens} {et~al.}(2021){Moeyens}, {Juri{\'c}}, {Ford},
  {et~al.}}]{moeyens2021}
{Moeyens}, J., {Juri{\'c}}, M., {Ford}, J., {et~al.} 2021,
  \bibinfo{title}{{THOR: An Algorithm for Cadence-independent Asteroid
  Discovery},} AJ, 162, 143, \dodoi{10.3847/1538-3881/ac042b}

\bibitem[{A. {M{\"o}ller} {et~al.}(2021){M{\"o}ller} {et~al.}}]{moller2021}
{M{\"o}ller}, A., {et~al.} 2021, \bibinfo{title}{{FINK}, a new generation of
  broker for the LSST community,} MNRAS, 501, 3272,
  \dodoi{10.1093/mnras/staa3602}

\bibitem[{J. {Murtagh} {et~al.}(2026){Murtagh}, {Schwamb}, {Bernardinelli},
  {et~al.}}]{murtagh2026}
{Murtagh}, J., {Schwamb}, M.~E., {Bernardinelli}, P.~H., {et~al.} 2026,
  \bibinfo{title}{{Predictions of the LSST Solar System Yield: Neptune
  Trojans},} AJ, 171, 90, \dodoi{10.3847/1538-3881/ae27bd}

\bibitem[{K.~J. {Napier} {et~al.}(2021){Napier}, {Gerdes}, {Lin},
  {et~al.}}]{napier2021}
{Napier}, K.~J., {Gerdes}, D.~W., {Lin}, H.~W., {et~al.} 2021,
  \bibinfo{title}{{No Evidence for Orbital Clustering in the Extreme
  Trans-Neptunian Objects},} PSJ, 2, 59, \dodoi{10.3847/PSJ/abe53e}

\bibitem[{R.~G. {Newcombe}(1998){Newcombe}}]{newcombe1998}
{Newcombe}, R.~G. 1998, \bibinfo{title}{{Interval estimation for the difference
  between independent proportions: comparison of eleven methods},} Statistics
  in Medicine, 17, 873

\bibitem[{D.~L. {Nidever} {et~al.}(2021){Nidever}, {Dey}, {Fasbender},
  {et~al.}}]{nidever2021}
{Nidever}, D.~L., {Dey}, A., {Fasbender}, K., {et~al.} 2021,
  \bibinfo{title}{{Second Data Release of the All-sky NOIRLab Source Catalog},}
  AJ, 161, 192

\bibitem[{R. {Nikutta} {et~al.}(2020){Nikutta} {et~al.}}]{nikutta2020}
{Nikutta}, R., {et~al.} 2020, \bibinfo{title}{{Data Lab---A community science
  platform},} Astronomy and Computing, 33, 100411

\bibitem[{ {NSF--DOE Vera C. Rubin Observatory}(2026){NSF--DOE Vera C. Rubin
  Observatory}}]{rubin2026news}
{NSF--DOE Vera C. Rubin Observatory}. 2026, {Early Data from NSF--DOE Vera C.
  Rubin Observatory Reveals Over 11,000 New Asteroids}, News release, 2026
  April 2.
\newblock \url{https://rubinobservatory.org/news/11000-new-asteroids}

\bibitem[{F. {Ochsenbein} {et~al.}(2000){Ochsenbein}, {Bauer}, \&
  {Marcout}}]{ochsenbein2000}
{Ochsenbein}, F., {Bauer}, P., \& {Marcout}, J. 2000, \bibinfo{title}{{The
  VizieR database of astronomical catalogues},} A\&AS, 143, 23,
  \dodoi{10.1051/aas:2000169}

\bibitem[{R. {Okuta} {et~al.}(2017){Okuta}, {Unno}, {Nishino}, {Hido}, \&
  {Loomis}}]{okuta2017}
{Okuta}, R., {Unno}, Y., {Nishino}, D., {Hido}, S., \& {Loomis}, C. 2017,
  \bibinfo{title}{{CuPy: A NumPy-Compatible Library for NVIDIA GPU
  Calculations},} in Proceedings of Workshop on Machine Learning Systems
  (LearningSys) in The Thirty-first Annual Conference on Neural Information
  Processing Systems (NIPS)

\bibitem[{R.~S. {Park} {et~al.}(2021){Park}, {Folkner}, {Williams}, \&
  {Boggs}}]{park2021}
{Park}, R.~S., {Folkner}, W.~M., {Williams}, J.~G., \& {Boggs}, D.~H. 2021,
  \bibinfo{title}{{The JPL Planetary and Lunar Ephemerides DE440 and DE441},}
  AJ, 161, 105

\bibitem[{H. {Rein} {et~al.}(2019){Rein}, {Hernandez}, {Tamayo},
  {et~al.}}]{rein2019}
{Rein}, H., {Hernandez}, D.~M., {Tamayo}, D., {et~al.} 2019,
  \bibinfo{title}{{Hybrid symplectic integrators for planetary dynamics},}
  MNRAS, 485, 5490

\bibitem[{H. {Rein} \& S.-F. {Liu}(2012){Rein} \& {Liu}}]{rein2012}
{Rein}, H., \& {Liu}, S.-F. 2012, \bibinfo{title}{{REBOUND: an open-source
  multi-purpose N-body code for collisional dynamics},} A\&A, 537, A128

\bibitem[{H. {Rein} \& D.~S. {Spiegel}(2015){Rein} \&
  {Spiegel}}]{reinspiegel2015}
{Rein}, H., \& {Spiegel}, D.~S. 2015, \bibinfo{title}{{IAS15: a fast, adaptive,
  high-order integrator for gravitational dynamics, accurate to machine
  precision over a billion orbits},} MNRAS, 446, 1424

\bibitem[{H. {Rein} \& D. {Tamayo}(2015){Rein} \& {Tamayo}}]{rein2015}
{Rein}, H., \& {Tamayo}, D. 2015, \bibinfo{title}{{WHFAST: a fast and unbiased
  implementation of a symplectic Wisdom-Holman integrator for long-term
  gravitational simulations},} MNRAS, 452, 376

\bibitem[{T.~R. {Ruch} {et~al.}(2026){Ruch}, {Napier}, {Gerdes}, \&
  {Lin}}]{ruch2026}
{Ruch}, T.~R., {Napier}, K.~J., {Gerdes}, D.~W., \& {Lin}, H.~W. 2026,
  \bibinfo{title}{Efficient Detection-level Linking of Trans-Neptunian Objects
  via a Spherical-orbit Basis,} PSJ, 7, 228, \dodoi{10.3847/PSJ/aea3f4}

\bibitem[{M.~E. {Schwamb} {et~al.}(2026){Schwamb}, {Bernardinelli}, {Volk},
  {et~al.}}]{schwamb2026}
{Schwamb}, M.~E., {Bernardinelli}, P.~H., {Volk}, K., {et~al.} 2026,
  \bibinfo{title}{{NSF-DOE Vera C. Rubin Observatory's First View of the
  Neptune Trojan L5 (Trailing) Cloud: Two New Long-term Stable Trojans},} ApJL,
  1010, L36, \dodoi{10.3847/2041-8213/aead9a}

\bibitem[{C. {Shankman} {et~al.}(2017){Shankman}, {Kavelaars}, {Bannister},
  {et~al.}}]{shankman2017}
{Shankman}, C., {Kavelaars}, J.~J., {Bannister}, M.~T., {et~al.} 2017,
  \bibinfo{title}{{OSSOS. VI. Striking Biases in the Detection of Large
  Semimajor Axis Trans-Neptunian Objects},} AJ, 154, 50,
  \dodoi{10.3847/1538-3881/aa7aed}

\bibitem[{S.~S. {Sheppard} \& C.~A. {Trujillo}(2006){Sheppard} \&
  {Trujillo}}]{sheppard2006}
{Sheppard}, S.~S., \& {Trujillo}, C.~A. 2006, \bibinfo{title}{{A Thick Cloud of
  Neptune Trojans and Their Colors},} Science, 313, 511

\bibitem[{H. {Smotherman} {et~al.}(2024){Smotherman}, {Bernardinelli},
  {Portillo}, {et~al.}}]{smotherman2024}
{Smotherman}, H., {Bernardinelli}, P.~H., {Portillo}, S.~K.~N., {et~al.} 2024,
  \bibinfo{title}{{The DECam Ecliptic Exploration Project (DEEP). VI. First
  Multiyear Observations of Trans-Neptunian Objects},} AJ, 167, 136,
  \dodoi{10.3847/1538-3881/ad1524}

\bibitem[{S. {Stetzler} {et~al.}(2025){Stetzler}, {Juri{\'c}}, {Bernardinelli},
  {et~al.}}]{stetzler2025}
{Stetzler}, S., {Juri{\'c}}, M., {Bernardinelli}, P.~H., {et~al.} 2025,
  \bibinfo{title}{{An Efficient Shift-and-stack Algorithm Applied to Detection
  Catalogs},} \aj, 170, 352, \dodoi{10.3847/1538-3881/ae0e1a}

\bibitem[{D.~E. {Trilling} {et~al.}(2024){Trilling}, {Gerdes}, {Juri{\'c}},
  {et~al.}}]{trilling2024}
{Trilling}, D.~E., {Gerdes}, D.~W., {Juri{\'c}}, M., {et~al.} 2024,
  \bibinfo{title}{{The DECam Ecliptic Exploration Project (DEEP). I. Survey
  Description, Science Questions, and Technical Demonstration},} AJ, 167, 132,
  \dodoi{10.3847/1538-3881/ad1529}

\bibitem[{C.~A. {Trujillo} {et~al.}(2024){Trujillo}, {Fuentes}, {Gerdes},
  {et~al.}}]{trujillo2024}
{Trujillo}, C.~A., {Fuentes}, C., {Gerdes}, D.~W., {et~al.} 2024,
  \bibinfo{title}{{The DECam Ecliptic Exploration Project (DEEP). II.
  Observational Strategy and Design},} AJ, 167, 133,
  \dodoi{10.3847/1538-3881/ad1523}

\bibitem[{C.~A. {Trujillo} \& S.~S. {Sheppard}(2014){Trujillo} \&
  {Sheppard}}]{trujillo2014}
{Trujillo}, C.~A., \& {Sheppard}, S.~S. 2014, \bibinfo{title}{{A Sedna-like
  body with a perihelion of 80 astronomical units},} Nature, 507, 471,
  \dodoi{10.1038/nature13156}

\bibitem[{F. {Valdes} {et~al.}(2014){Valdes}, {Gruendl}, \& {DES
  Project}}]{valdes2014}
{Valdes}, F., {Gruendl}, R., \& {DES Project}. 2014, \bibinfo{title}{{The DECam
  Community Pipeline},} in ASP Conf. Ser., Vol. 485, Astronomical Data Analysis
  Software and Systems XXIII, 379

\bibitem[{P. {Virtanen} {et~al.}(2020){Virtanen} {et~al.}}]{virtanen2020}
{Virtanen}, P., {et~al.} 2020, \bibinfo{title}{{SciPy 1.0: fundamental
  algorithms for scientific computing in Python},} Nature Methods, 17, 261

\bibitem[{R.~J. {Weryk} {et~al.}(2016){Weryk}, {Lilly}, {Chastel},
  {et~al.}}]{weryk2016}
{Weryk}, R.~J., {Lilly}, E., {Chastel}, S., {et~al.} 2016,
  \bibinfo{title}{{Distant Solar System Objects identified in the Pan-STARRS1
  survey},} arXiv e-prints, arXiv:1607.04895.
\newblock \doarXiv{1607.04895}

\bibitem[{P.~J. {Whidden} {et~al.}(2019){Whidden}, {Kalmbach}, {Connolly},
  {et~al.}}]{whidden2019}
{Whidden}, P.~J., {Kalmbach}, J.~B., {Connolly}, A.~J., {et~al.} 2019,
  \bibinfo{title}{{Fast Algorithms for Slow Moving Asteroids: Constraints on
  the Distribution of Kuiper Belt Objects},} \aj, 157, 119,
  \dodoi{10.3847/1538-3881/aafd2d}

\bibitem[{E.~B. {Wilson}(1927){Wilson}}]{wilson1927}
{Wilson}, E.~B. 1927, \bibinfo{title}{{Probable Inference, the Law of
  Succession, and Statistical Inference},} Journal of the American Statistical
  Association, 22, 209, \dodoi{10.1080/01621459.1927.10502953}

\bibitem[{D.~G. {York} {et~al.}(2000){York}, {Adelman}, {Anderson},
  {et~al.}}]{york2000}
{York}, D.~G., {Adelman}, J., {Anderson}, Jr., J.~E., {et~al.} 2000,
  \bibinfo{title}{{The Sloan Digital Sky Survey: Technical Summary},} AJ, 120,
  1579, \dodoi{10.1086/301513}

\end{thebibliography}
\end{document}